\documentclass[a4size]{aa}
\usepackage{graphicx,natbib,hyperref}
\usepackage[varg]{txfonts}
\usepackage{morefloats}
\usepackage{color}

\newcommand{\av}[1]{\langle{#1}\rangle}
\newcommand{\chisq}{\chi^2/\nu}
\newcommand{\fmi}{\phantom{-}}
\newcommand{\fo}{\phantom{1}}
\newcommand{\ft}{\phantom{11}}

\newcommand{\Hm}{\langle B\rangle}
\newcommand{\Hz}{\langle B_z\rangle}

\newcommand{\Hq}{\av{B_{\rm q}}}
\newcommand{\xover}{\av{X_z}}
\newcommand{\Hav}{\Hm_{\rm av}}
\newcommand{\Hqav}{\Hq_{\rm av}}
\newcommand{\Hzrms}{\Hz_{\rm rms}}
\newcommand{\xoverrms}{\xover_{\rm rms}}

\newcommand{\xHz}{\langle xB_z\rangle}

\newcommand{\vsi}{v\,\sin i}
\newcommand{\Prot}{P_{\rm rot}}
\newcommand{\Porb}{P_{\rm orb}}
\newcommand{\kms}{km\,s$^{-1}$}
\newcommand{\R}[2]{R^{(#1)}_{#2}(\lambda_I)}
\newcommand{\intxy}{\int_{-1}^{+1}dx\int_{-\sqrt{1-x^2}}^{+\sqrt{1-x^2}}dy}
\newcommand{\llo}{\lambda-\lambda_I}
\newcommand{\dlp}{\Delta\lambda_{\rm p}(x,y;\psi)}
\newcommand{\Hvec}{\mbox{\boldmath $B$}}
\newcommand{\Zeeman}{\Delta\lambda_{\rm Z}}
\newcommand{\ew}{W_\lambda}
\newcommand{\C}[2]{\ensuremath{C^{(#1)}_{#2}}}
\newcommand{\OC}{\ensuremath{{\rm O}-{\rm C}}}

\defcitealias{1997A&AS..123..353M}{Paper~I}

\bibpunct[ ]{(}{)}{;}{a}{}{,}

\begin{document}

\title{Ap stars with resolved magnetically split lines: \\
Magnetic field determinations from Stokes $I$ and $V$
spectra\thanks{Based on observations collected at the European
  Southern Observatory, Chile (ESO Programmes
56.E-0688, 56.E-0690, 57.E-0557, 57.E-0637, 58.E-0155, 58.E-0159,
59.E-0372, 59.E-0373, 60.E-0564, 60.E-0565, 61.E-0711, and Period 56
Director Discretionary Time); at
Observatoire de Haute Provence (CNRS), France; at Kitt Peak
National Observatory, National Optical Astronomy Observatory (NOAO
Prop. ID: KP2442; PI: T.~Lanz), which is operated by the Association of Universities
for Research in Astronomy (AURA) under cooperative agreement with the
National Science Foundation; and at the
Canada-France-Hawaii Telescope (CFHT), which is operated from the
summit of Mauna Kea by the National Research Council of Canada, the
Institut National des Sciences de l'Univers of the Centre National de
la Recherche Scientifique of France, and the University of Hawaii. The
observations at the Canada-France-Hawaii Telescope were performed with
care and respect from the summit of Mauna Kea, which is a significant
cultural and historic site.}}

\author{G. Mathys}

\institute{Joint ALMA Observatory \& European Southern Observatory,
  Alonso de Cordova 3107, Santiago, Chile \\ \email{gmathys@eso.org}}

\date{Received $\ldots$ / Accepted $\ldots$}

\titlerunning{Magnetic fields of Ap stars with magnetically resolved lines}
\authorrunning{G. Mathys}

\abstract{}
{We present the results of  a systematic study of the magnetic fields
and other properties of the Ap stars with resolved magnetically split
lines.} 
{This study is based on new measurements of the mean magnetic
field modulus, the mean longitudinal magnetic field, the crossover,
the mean quadratic magnetic field, and the radial velocity of 43 stars,
complemented by magnetic data from the literature for 41 additional
stars.} 
{Stars with resolved magnetically split lines represent a significant
fraction, of the order of several percent, of the whole population of
Ap stars. Most of them are genuine slow rotators, whose consideration
provides new insight into the long-period tail of the distribution of
the periods of the Ap stars. Emerging correlations between rotation
periods and magnetic properties provide important clues for the
understanding of the braking mechanisms that have been at play in the
early stages of stellar evolution. The geometrical structures of the
magnetic fields  of Ap stars with magnetically resolved lines appear
in general to depart slightly, but not extremely, from centred
dipoles. However, there are a few remarkable exceptions, which deserve
further consideration. We suggest that pulsational crossover can be
observed in some stars; if confirmed, this would open the door to the
study of non-radial pulsation modes of degree $\ell$ too high for
photometric or  spectroscopic observations. How the lack of short
orbital periods among binaries containing an Ap component with
magnetically resolved lines is related with their (extremely) slow
rotation remains to be fully understood, but the very existence of a
correlation between the two periods lends support to the merger
scenario for the origin of Ap stars.}
{}

\maketitle

\section{Introduction}
\label{sec:intro}
\citet{1960ApJ...132..521B} was the first to report the observation of
spectral lines resolved into their Zeeman-split 
components by a magnetic field in a
star other than the Sun: the B9p star
HD~215441
(since then known as Babcock's  
star).\footnote{However, the \citet{1960ApJ...132..521B}
  observations of 
  HD~215441, on which his discovery of magnetically resolved lines was
  based, were obtained in October 1959. HD~126515 (also known as
  Preston's star) actually appears to be the first star for which
  spectra showing magnetically resolved lines were recorded, in
  February 1957--- as \citet{1970ApJ...160.1059P} found later, motivated by a
  remark made by \citet{1958ApJS....3..141B} on the unusual appearance of the
  line profiles.} As \citeauthor{1960ApJ...132..521B} realised, the observations of sharp,
resolved line components implies that the star has a low projected
equatorial velocity $\vsi$, and that its magnetic field is fairly
uniform. From consideration of the wavelength separation of the
resolved line components, \citeauthor{1960ApJ...132..521B} inferred
that the average over the 
visible stellar disk of the modulus of its magnetic field (the mean
magnetic field modulus) is of the
order of 34~kG. Quite remarkably, more than half a century later, this
still stands as the highest mean magnetic field modulus measured in a
non-degenerate star. 

\begin{table*}[!ht]
\caption{Ap stars with resolved magnetically split lines: stars for
  which new measurements of the mean magnetic field modulus are
  presented in this paper} 
\label{tab:stars}
\begin{tabular*}{\textwidth}[]{@{}@{\extracolsep{\fill}}rlrllcclc}
%@{\extracolsep{0pt}}@{}}
\hline\hline\\[-4pt]
\multicolumn{1}{c}{HD/HDE}&Other id.&
\multicolumn{1}{c}{$V$}&Sp. type&Period&Ref.&HJD$_0$&Phase
origin&Ref.\\[4pt] 
\hline\\[-4pt]
965&BD $-0$~21&8.624&A8p SrEuCr&$\null>13$~y&1\\
2453&BD $+31$~59&6.893&A1p SrEuCr&521~d&2&2442213.000&$\Hz$ min.&\\
9996&HR 465&6.376&B9p CrEuSi&7936.5~d&3&2433301.360&$\Hz$ min.&3\\
12288&BD $+68$~144&7.750&A2p CrSi&34\fd9&4&2448499.870&$\Hm$ max.&4\\
14437&BD $+42$~502&7.261&B9p CrEuSi&26\fd87&4&2448473.846&$\Hz$ max.&4\\
18078&BD $+55$~726&8.265&A0p SrCr&1358~d&5&2449930.000&$\Hm$ max.&5\\
29578&CPD $-54$~685&8.495&A4p SrEuCr&$\null\gg5$~y\\
47103&BD $+20$~1508&9.148&Ap SrEu&\\
50169&BD $-1$~1414&8.994&A3p SrCrEu&$\null\gg8$~y\\
51684&CoD $-40$~2796&7.950&F0p SrEuCr&371~d&&2449947.000&$\Hm$ max.&\\
55719&HR 2727&5.302&A3p SrCrEu&$\null\gg10$~y\\
59435&BD $-8$~1937&7.972&A4p SrCrEu&1360~d&6&2450580.000&$\Hm$ max.&6\\
61468&CoD $-27$~4341&9.839&A3p EuCr&322~d&&2450058.500&$\Hm$ max.&\\
65339&53 Cam&6.032&A3p SrEuCr&8\fd02681&7&2448498.186&positive crossover&7\\
70331&CoD $-47$~3803&8.898&B8p Si&1\fd9989 or 1\fd9909&&2446987.100%
&arbitrary\\
75445&CoD $-38$~4907&7.142&A3p SrEu&6\fd291?&&2449450.000&arbitrary&\\
81009&HR 3724&7.200&A3p CrSrSi&33\fd984&8&2444483.420%
&max. brightness in $v$&8\\
93507&CPD $-67$~1494&8.448&A0p SiCr&556~d&2&2449800.000&$\Hm$ min.&2\\
94660&HR 4263&6.112&A0p EuCrSi&2800~d&&2447000.000&$\Hm$ min.&\\
110066&HR 4816&6.410&A1p SrCrEu&4900:~d&9\\
116114&BD $-17$~3829&7.026&F0p SrCrEu&27\fd61&&2447539.000&$\Hm$ min.&\\
116458&HR 5049&5.672&A0p EuCr&148\fd39&&2448107.000&$\Hz$ max.&\\
119027&CoD $-28$~10204&\llap{1}0.027&A3p SrEu&\\
126515&BD $+1$~2927&7.094&A2p CrSrEu&129\fd95&2&2437015.000&$\Hm$ max.&10\\
134214&BD $-13$~4081&7.467&F2p SrEuCr&\\
137909&$\beta$ CrB&3.900&A9p SrEuCr&18\fd4868&11&2434204.700&$\Hz$ 
pos. extr.&11\\
137949&33 Lib&6.659&F0p SrEuCr&5195:~d&&2453818.000&$\Hz$ max.\\
142070&BD $-0$~3026&7.966&A0p SrCrEu&3\fd3718&&2449878.200&$\Hz$ max.\\
144897&CoD $-40$~10236&8.600&B8p EuCr&48\fd57&&2449133.700&$\Hm$ min.\\
150562&CoD $-48$~11127&9.848&A5p EuSi?&$\null\ga4.5$~y\\
318107&CoD $-32$~13074&9.355&B8p&9\fd7088&12&2448800.000&$\Hm$ max.\\
165474&HR 6758B&7.449&A7p SrCrEu&$\null\gg9$~y\\
166473&CoD $-37$~12303&7.953&A5p SrEuCr&$\null\ga10$~y&13\\
177765&CoD $-26$~13816&9.155&A5p SrEuCr&$\gg5$~y\\
187474&HR 7552&5.321&A0p EuCrSi&2345~d&14&2446766.000&$\Hz$ pos. extr.&15\\
188041&HR 7575&5.634&A6p SrCrEu&223\fd78&&2432323.000&$\Hz$ min.&16\\
192678&BD $+53$~2368&7.362&A2p Cr&6\fd4193&17&2449113.240&$\Hm$ max.&18\\
335238&BD $+29$~4202&9.242&A1p CrEu&48\fd70&&2447000.000&arbitrary\\
200311&BD $+43$~3786&7.708&B9p SiCrHg&52\fd0084&19%
&2445407.513&$\Hz$ pos. extr.&19\\
201601&$\gamma$ Equ&4.700&A9p SrEu&$\null\ga97$~y&20\\
208217&CPD $-62$~6281&7.196&A0p SrEuCr&8\fd44475&21&2447028.000
&$\Hz$ pos. extr.\\
213637&BD $-20$~6447&9.611&F1p EuSr&$\null>115$~d?\\
216018&BD $-12$~6357&7.623&A7p SrEuCr&$\null\gg6$~y?\\[4pt]
\hline\\[-4pt]
\end{tabular*}
\tablebib{ 
(1)~\citet{2014AstBu..69..427R};
(2)~\citet{1997A&AS..123..353M}; 
(3)~\citet{2014AstBu..69..315M}; 
(4)~\citet{2000A&A...355.1080W}; 
(5)~\citet{2016A&A...586A..85M};
(6)~\citet{1999A&A...347..164W}; 
(7)~\citet{1998MNRAS.297..236H};
(8)~\citet{1997PASP..109....9A};
(9)~\citet{1981A&AS...44..265A};
(10)~\citet{1970ApJ...160.1059P};
(11)~\citet{1989MNRAS.238..261K};
(12)~\citet{2011A&A...535A..25B};
(13)~\citet{2007MNRAS.380..181M}; 
(14)~\citet{1991A&AS...89..121M}; 
(15)~Landstreet (unpublished; cited by \citealt{1991A&AS...89..121M});
(16)~\citet{1969ApJ...158.1231W};
(17)~\citet{2006PASP..118...77A};
(18)~\citet{1996A&A...313..209W};  
(19)~\citet{1997MNRAS.292..748W};
(20)~\citet{2016MNRAS.455.2567B};
(21)~\citet{1997A&A...320..497M}.}
\end{table*}

By the end of 1970, Preston had identified eight more Ap
stars with magnetically resolved lines \citep[see][and references
therein]{1971ApJ...164..309P}. Preston recognised that the mean
magnetic field 
modulus is fairly insensitive to the geometry of the observation,
contrary to the mean longitudinal magnetic field (i.e. the average
over the visible stellar hemisphere of the line-of-sight component of
the magnetic vector). This field moment, which is determined
from the analysis of the circular polarisation of spectral
lines, is the main diagnostic of Ap star magnetic fields, of which
the largest number of measurements have been published. The interest of combining measurements of the 
mean longitudinal field and mean field modulus to derive
constraints on the structure of stellar magnetic fields was also
recognised early on. However, for the next two decades, Ap stars with magnetically
resolved lines, which represented only a very small subset of all Ap
stars, remained hardly more than a curiosity, to which very little
attention was paid. 

\begin{table*}[!ht]
\caption{Ap stars with resolved magnetically split lines: stars with
  magnetic measurements from the literature}
\label{tab:stars_litt}
\begin{tabular*}{\textwidth}[]{@{}@{\extracolsep{\fill}}rlrlclcclc}
%@{\extracolsep{0pt}}@{}}
\hline\hline\\[-4pt]
\multicolumn{1}{c}{HD/HDE}&Other id.&
\multicolumn{1}{c}{$V$}&Sp. type&Ref.&Period&Ref.&HJD$_0$&Phase
origin&Ref.\\[4pt] 
\hline\\[-4pt]
3988&CPD $-83$~10&8.351&A0p CrEuSr&1\\
18610&CPD $-73$~195&8.162&A2p CrEuSr&2\\
33629&CoD $-33$~2151&9.064&A9p SrCr&3\\
42075&CoD $-26$~2736&8.968&A5p EuCrSr&3\\
44226&CoD $-25$~3118&9.492&A5p SrEuCr&3\\
46665&BD $-22$~1450&9.441&A0p SrEu&3\\
47009&BD $-13$~1560&9.070&A0p EuCr&3\\
52847&BD $-22$~1666&8.157&A0p CrEu&3\\
55540&BD $-20$~1779&9.498&A0p EuCr&3\\
57040&CPD $-53$~1304&9.207&A2p EuSr&1&13\fd474&1\\
61513&CoD $-29$~4702&10.161&A0p CrEuSr&1\\
66318&CPD $-60$~1017&9.665&A0p EuCrSr&4\\
69013&BD $-15$~2337&9.456&A2p SrEu&3\\
70702&CoD $-51$~2962&8.572&B9p EuCrSr&1\\
72316&CoD $-33$~5118&8.804&Ap CrEu&3\\
75049&CoD $-50$~3542&9.090&A0p EuCrSi&3&4\fd048267&5&2454509.550&$\Hz$
max. &6\\
76460&CPD $-61$~1106&9.805&A3p Sr&1\\
81588&CoD $-47$~4913&8.445&A5p SrCrEu&1\\
88241&CoD $-39$~6174&7.920&F0p SrEu&1\\
88701&CoD $-36$~6209&9.258&B9p CrEu&3&25\fd765&3\\
92499&CoD $-42$~6407&8.890&A2p SrEuCr&7&$\null>5$~y?&3\\
96237&CoD $-24$~9514&9.434&A4p SrEuCr&3&20\fd91&3\\
97394&CoD $-42$~6806&8.797&A5p EuCrSr&8\\
\tablefootmark{a}&BD $+0$~4535&9.92\fo&Ap SrEu&9\\ 
110274&CoD $-58$~4688&9.328&A0p EuCr&3&265\fd3&3\\
117290&CoD $-48$~8252&9.281&A3p EuCrSr&3&$\null>5.7$~y?&3\\
121661&CPD $-62$~3790&8.556&A0p EuCrSi&3&47\fd0&3\\
135728&CoD $-30$~12099&8.602&A2p SrEuCr&3\\
143487&CoD $-30$~12753&9.420&A3p SrEuCr&3\\
154708&CPD $-57$~8336&8.744&A2p SrEuC&10&5\fd363&11&2454257.740&$\Hz$ max.&12\\
157751&CoD $-33$~12069&7.674&B9p SiCr&7\\
158450&BD $-7$~4448&8.514&A0p SrCrEu&1&8\fd524&1\\
162316&CPD $-75$~1401&9.338&A3p SrEu&1&9\fd304&1\\
168767&CD $-26$~13084&8.674&A0p EuCr&1\\
177268&CD $-34$~13384&9.034&A2p CrEu&1\\
178892&BD $+14$~3811&9.27\fo&B9p
SrCrEu&13&8\fd2478&13&2452708.562&max. brightness in $V$&13\\
179902&BD $-21$~5306&10.35\fo&A1p SrCrEu&1\\
184120&BD $-20$~5601&10.220&A0p CrEu&1\\
185204&CD $-47$~13020&9.534&A2p SrEuCr&1\\
191695&BD $-21$~5644&7.010&A3p SrEuCr&1\\
215441&BD $+54$~2846&8.851&B9p
Si&14&9\fd487574&15&2448733.714&max. brightness in $B$&15\\[4pt]
\hline\\[-4pt]
\end{tabular*}
\tablefoottext{a}{GSC~00510-00339}\\
\tablefoot{HJD$_0$ and the phase origin information appear only when
  available in the literature. In particular,
  \citet{2008MNRAS.389..441F} and \citet{2012MNRAS.420.2727E} used
  photometric databases to obtain reliable period determinations, but
  they give no indication about the phasing of the data.}
\tablebib{
(1)~\citet{2012MNRAS.420.2727E};
(2)~\citet{2003A&A...402..729S};
(3)~\citet{2008MNRAS.389..441F};
(4)~\citet{2003A&A...403..645B};
(5)~\citet{2015AA...574A..79K};
(6)~\citet{2010MNRAS.402.1883E};
(7)~\citet{2007MNRAS.378L..16H}; 
(8)~\citet{2011MNRAS.415.2233E}
(9)~\citet{2010MNRAS.401L..44E};  
(10)~\citet{2005A&A...440L..37H};
(11)~\citet{2014A&A...572A.113L};
(12)~\citet{2009MNRAS.396.1018H};
(13)~\citet{2006A&A...445L..47R};
(14)~\citet{1960ApJ...132..521B};
(15)~\citet{1995A&AS..111...41N}.}
\end{table*}
 
Yet these stars are especially worth studying for various reasons:
\begin{itemize}
\item Thanks to its low sensitivity to the geometry of the
  observation, the mean magnetic field modulus is the observable that
  best characterises the intrinsic strength of the stellar magnetic
  field. By contrast, the mean longitudinal magnetic field depends
  critically on the line of sight, hence on the location of the observer.
\item Furthermore, the determination of the mean field modulus from
  measurements of the wavelength separation of the resolved components
  of Zeeman-split lines is not only very simple and straightforward,
  but also, more importantly, it is model free and almost
  approximation free (see Sect.~\ref{sec:hmdiag}).
\item As already mentioned, the possibility of complementing
  spectropolarimetry-based determinations of magnetic field moments
  (most frequently, of the mean longitudinal field) with mean field
  modulus measurements, and of considering how both vary as the stars
  rotate, allows one to derive significant constraints on the
  structure of the stellar field.
\item Ap stars with resolved magnetically split lines, in their
  majority, have (very) long rotation periods (as discussed in more
  detail in Sect.~\ref{sec:rotation}). They represent extreme
  examples of the slow rotation that is characteristic of Ap stars
  in general (compared to normal main-sequence stars of similar
  temperature). This makes them particularly well suited to
  investigating the mechanisms through which Ap stars have managed
  during their past evolution to shed large amounts of angular
  momentum. 
\end{itemize}
 
These considerations prompted us around 1990 to initiate a
large comprehensive effort to study Ap stars with
resolved magnetically split lines systematically. \citet{1997A&AS..123..353M} (hereafter
\citetalias{1997A&AS..123..353M}) gave an extensive account of the results
obtained 
from the analysis of the observations obtained between May 1988 and
August 1995. By then, 42 Ap stars with magnetically resolved lines
were known, of which 30 had been discovered since 1988; all but one of
these discoveries were made in the framework of our
  project. \citetalias{1997A&AS..123..353M} 
presented 752 measurements of 
the mean magnetic field modulus of 40 of these 42 stars. Their
analysis led to a large number of new results:
\begin{itemize}
\item New or improved rotation periods, or lower limits of such
  periods
\item Variation curves of the mean magnetic field modulus
\item Detection of radial velocity variations indicative of binarity
\item Statistical constraints on the properties of Ap stars with
  magnetically resolved lines, in particular, with respect to their
  magnetic fields
\end{itemize}
We do not present a more detailed summary of those results here, since
they are significantly updated and augmented in the rest of this
paper. 

In 1995, we started a programme of systematic spectropolarimetric
observations of Ap stars with magnetically resolved lines, with a
view towards deriving additional constraints on the properties of their
magnetic fields from analysis of the circular polarisation of their
spectral lines. Here we present the results of this complementary
study, which also includes new magnetic field modulus measurements obtained
from additional high-resolution spectroscopic observations performed
during the same time interval, that is, between October 1995 and
September 1998. Although for a number of the stars of our sample,
more recent high-resolution spectroscopic observations in natural light and/or
spectropolarimetric observations are available in the archives of
several observatories, we do not consider them in this paper for the
sake of the homogeneity of the analysed data. This homogeneity is
particularly important to minimise ambiguities in the study of
small-amplitude, long-term variations (on timescales
of several
years). However, we do fully take new data that have
already been published into account, albeit making sure that these data can always be
distinguished from our own measurements. 

Nowadays, the most detailed and accurate determinations of the
geometrical structure of Ap 
star magnetic fields are achieved via application of the techniques of
Zeeman-Doppler imaging or magnetic Doppler imaging
\citep[e.g.][]{2002A&A...381..736P}. However, the vast majority of Ap
stars with 
resolved magnetically split lines do not lend themselves to the
application of these techniques because their projected equatorial
velocity $\vsi$ is in general too small. In this context, the 
  moment technique \citep{1988A&A...189..179M} represents a powerful
alternative that allows one to characterise the main magnetic
properties of the observed stars with a small set of quantities
(magnetic field moments) and their variations with rotation
phase. This numeric information can then be subjected to statistical
analysis; generic properties of the subset of Ap stars with
magnetically resolved lines, and possible exceptions, can be
inferred. This is the approach that we have followed in this
work. The moments that we have considered characterise the mean
  intensity of the magnetic field over the stellar surface and its
  component along the line of sight; the spread of the distribution of
  the local field values across the disk of the star; and the existing
  correlations between the magnetic field structure and stellar velocity fields that are organised on large scales.

In order to strengthen the statistical significance of our
conclusions, we included, whenever possible while maintaining sufficient
quality and uniformity of the analysed data, results from
other groups that are available in the literature. In particular, we tried
to keep track of all the Ap stars with magnetically resolved lines
that have been discovered since 1998. We reckon that at
present, 84 Ap stars are known to show magnetically resolved
lines. This
represents 2.3\% of the total number of Ap stars in
\citeauthor{2009A&A...498..961R}'s \citeyearpar{2009A&A...498..961R}
catalogue. As these authors stress, this 
catalogue is not homogeneous, nor is the set of known Ap stars
with magnetically resolved lines. The latter, which was assembled from
different sources, is certainly biassed towards the brightest stars,
and it contains a disproportionate number of stars with southern
declination, 68 out of 84 (81\%). Thus 2.3\% should be regarded at
best as a moderately meaningful lower limit to the actual fraction of
Ap stars showing resolved magnetically split lines. The \citet{2012MNRAS.420.2727E}
estimate that this fraction is a little less than 10\%
may be more realistic, given that it is based on a systematic
observational study of a reasonably unbiassed sample of
stars. However, this sample is restricted to (fairly) cool Ap
stars. It is probably safe to assume that the actual fraction of Ap
stars whose spectral lines are resolved in their magnetically split
components is somewhere between 2.3\% and 10\%. Even if this number is
not exactly known, it leaves little doubt
about the fact that, contrary to the feeling that prevailed until two
decades ago, Ap stars with magnetically resolved lines are not isolated odd
specimens, but that instead they constitute a very significant
subset of the Ap star population. 

The main properties of the 43 stars for which new magnetic field
measurements are presented in this paper are summarised in
Table~\ref{tab:stars}. Columns 1 to 4 give, in order, their HD (or
HDE) number, an alternative identification, the visual magnitude $V$, 
and the spectral type according to the {\em Catalogue of Ap, HgMn and
  Am stars\/} \citep{2009A&A...498..961R}. For those stars whose
rotation period is known, or is determined in this paper, its value
appears in Col.~5, with the time origin adopted to phase the magnetic
data in Col.~7, and the property (e.g. specific extremum of a
magnetic field moment) from which this phase was defined in
Col.~8. For the remaining stars, a lower limit of the period is given
in Col.~5 whenever it could be set. Columns 6 and 9 identify the
reference from which the period and phase origin
information is extracted, respectively; when those parameters were determined in the
present work, the corresponding entry is left blank.

The other 41 Ap stars with magnetically resolved lines that are
currently known  are listed in
Table~\ref{tab:stars_litt}. Its structure
is similar to that of Table~\ref{tab:stars}, except for
an additional column (Col.~5) giving the reference of the paper in
which the presence of magnetically resolved lines was first reported.

The new observations whose analysis is reported in this paper are
introduced in Sect.~\ref{sec:obs}. Section~\ref{sec:diag} describes
how they were used for determination of the magnetic field moments of
interest and the stellar radial velocities. The results of these
measurements are presented in Sect.~\ref{sec:results}, which also
explains how the variations of the different derived quantities were
characterised. The implications of the new data obtained in this work
for our knowledge of the physical properties of Ap stars are discussed
in Sect.~\ref{sec:discussion}, and some general conclusions are drawn
in Sect.~\ref{sec:conclusion}. A number of specialised issues that
deserved more detailed consideration outside the main flow of the
paper were moved to the appendices. These include notes on the individual stars for
which new magnetic field 
measurements were obtained (Appendix~\ref{sec:notes}), a revision of
older determinations of 
the mean quadratic magnetic field (Appendix~\ref{sec:hquad_rev}), 
and the
mathematical formalism underlying a proposal for a new physical
mechanism to generate the crossover effect in spectral lines observed
in circular polarisation (Appendix~\ref{sec:pulsxover}).

\section{Observations}
\label{sec:obs}
There are two types of new observations presented in this paper as follows:
\begin{itemize}
\item Additional high-resolution
  ($R=\lambda/\Delta\lambda\sim7\,10^4$--$1.2\,10^5$) spectra recorded
  in natural light, similar to the data of \citetalias{1997A&AS..123..353M}
\item Lower resolution ($R\sim3.9\,10^4$) circularly polarised
  spectra\end{itemize}

 \begin{table}[!t]
\caption{Instrumental configurations used for high-resolution
  spectroscopy observations of this paper and of
  \citetalias{1997A&AS..123..353M}.} 
\label{tab:plot_sym}
\centering
%\begin{tabular}{ccc}
\small{
\begin{tabular}{@{}@{\extracolsep{5pt}}ccc
@{\extracolsep{0pt}}@{}}
\hline\hline\\[-4pt]
Id.&Configuration&Symbol\\[4pt]
\hline\\[-4pt]
1&CAT + CES LC&filled circle (navy blue)\\
2&CAT + CES SC&open circle (steel blue)\\
3&CAT + CES LC + CCD \#38&asterisk (turquoise)\\
4&CAT + CES LC/F200&filled hexagon (sea green)\\
5&CAT + CES SC/fibre&open hexagon (olive)\\
6&3.6 + CES LC/F200&star (orange)\\
7&AURELIE&open square (brown)\\
8&ELODIE&cross (red)\\
9&KPNO coud\'e feed&filled square (salmon)\\
10&CFHT Gecko&filled triangle (violet)\\
11&EMMI&open triangle (dark violet)\\[4pt]
\hline\\[-4pt]
\end{tabular}}\\
\end{table}
 
The unpolarised spectra were obtained between October 1995 and
 September 1998 with a subset of the
 telescope and instrument combinations used in \citetalias{1997A&AS..123..353M}:
\begin{itemize} 
\item The European Southern Observatory's (ESO) Coud\'e
  Echelle Spectrograph (CES) in its Long Camera (LC) configuration,
  fed by the 1.4 m Coud\'e Auxiliary Telescope (CAT)
\item The AURELIE spectrograph of the Observatoire de
  Haute-Provence (OHP), fed by the 1.5 m telescope
\item Kitt Peak National Observatory's (KPNO) coud\'e
  spectrograph at the 0.9 m coud\'e feed
\item The $f/4$ coud\'e spectrograph (Gecko) of the 3.6 m Canada-France-Hawaii
  Telescope (CFHT) at Mauna Kea
\item The ESO Multi-Mode Instrument (EMMI) in its high-resolution
  \'echelle mode, fed by the 3.5 m New Technology Telescope
\end{itemize}

The observing procedures that were followed and the
instrumental configurations that were used have been described in 
more detail in \citetalias{1997A&AS..123..353M}. In April 1998, the
Long Camera (LC) of the CES was decommissioned and replaced by the
Very Long Camera \citep[VLC;][]{1998Msngr..92...18K}. We used the VLC
(with CCD \#38)
for the final three observing runs of 1998. We set the width of the entrance slit of
the CES so as to achieve the same resolving power as with the LC
configuration of our previous runs (with CCD \#34). Thus, in practice,
the VLC+CCD\,\#38 and LC+CCD\,\#34 
configurations of the CES can be regarded as equivalent for our
purpose: we shall not distinguish them further in this paper. 

The full list of the configurations that were used to record the
high-resolution, unpolarised spectra analysed here and in
\citetalias{1997A&AS..123..353M} appears in
Table~\ref{tab:plot_sym}. The third column gives the symbol used to
distinguish them in the figures of
Appendix~\ref{sec:notes}. We used the same 
identification numbers and symbols as in
\citetalias{1997A&AS..123..353M}. As in 
the latter, configurations  
that are practically equivalent were given the same
identification. 

Table~\ref{tab:runs} lists the dates of the individual observing runs
(totalling 50 nights) partly or entirely devoted to the acquisition of
the new high-resolution spectra discussed in this paper. Columns~1 and
2 give the start and end date of the run (in local time of the
observatory, from the beginning of the first night to the end of the
last night). The identification number assigned to the instrumental
configuration used during the run (see Table~\ref{tab:plot_sym})
appears in Col.~3. Each instrumental configuration is summarised in
Cols.~4 to 7, which list, in order, the observatory where the
observations were performed, and the telescope, instrument, and
detector that were used. 

All data were reduced using the ESO image
processing package MIDAS (Munich Image Data Analysis System), 
performing the same steps as described in
\citetalias{1997A&AS..123..353M}, except for the CFHT Gecko spectra,
which were reduced with IRAF (Image Reduction and Analysis Facility).
For a more detailed description of the
reduction of the EMMI \'echelle 
spectra and, in particular, of their wavelength calibration, see
\citet{2006A&A...453..699M}.   

 All the spectropolarimetric data presented here were
obtained with the ESO Cassegrain Echelle Spectrograph (CASPEC) fed by
the ESO 3.6 m telescope. The configuration that was used is the same
as specified by \citet{1997A&AS..124..475M} for the spectra
recorded since 1995. Data reduction was also as described in that
reference. The CASPEC spectropolarimetric observations of this paper were
performed on 28 different nights spread between February 1995 and
January 1998.  

\begin{table*}[t]
\caption{Observing runs and instrumental configurations:  high-resolution spectra in natural light}
\label{tab:runs}
\begin{tabular*}{\textwidth}[]{@{}@{\extracolsep{\fill}}llcllll
@{\extracolsep{0pt}}@{}}
\hline\hline\\[-4pt]
Start date&End date&Id.&Observatory&Telescope&Instrument&Detector\\[4pt]
\hline\\[-4pt]
1995 October 3 &1995 October 4&11&ESO&NTT&EMMI&CCD Tek \#36\\ 
1995 November 30&1995 December 3&7&OHP&152&AURELIE&Barrette Th\\
1995 December 3&1995 December 6&10&CFHT&CFHT&Gecko&CCD Loral \#3\\
1996 January 1&1996 January 2&1&ESO&CAT&CES LC&CCD Loral \#34\\ 
1996 January 17&1996 January 19&7&OHP&152&AURELIE&Barrette Th\\
1996 January 27&1996 January 28&1&ESO&CAT&CES LC&CCD Loral \#34\\
1996 February 1 &1996 February 2&11&ESO&NTT&EMMI&CCD Tek \#36\\
1996 February 18&1996 February 19&1&ESO&CAT&CES LC&CCD Loral \#34\\
1996 March 6&1996 March 7&1&ESO&CAT&CES LC&CCD Loral \#34\\
1996 March 19&1996 March 20&1&ESO&CAT&CES LC&CCD Loral \#34\\
1996 March 28&1996 March 29&1&ESO&CAT&CES LC&CCD Loral \#34\\ 
1996 April 16&1996 April 22&7&OHP&152&AURELIE&Barrette Th\\
1996 May 27&1996 May 28&1&ESO&CAT&CES LC&CCD Loral \#34\\
1996 July 30&1996 August 5&7&OHP&152&AURELIE&Barrette Th\\
1996 September 18&1996 September 24&9&KPNO&Coud\'e feed&Coud\'e spectro&CCD TI \#5\\
1996 September 29&1996 September 30&1&ESO&CAT&CES LC&CCD Loral \#34\\
1996 December 9&1996 December 10&1&ESO&CAT&CES LC&CCD Loral \#34\\
1997 March 14&1997 March 15&1&ESO&CAT&CES LC&CCD Loral \#34\\
1997 May 28&1997 May 29&1&ESO&CAT&CES LC&CCD Loral \#34\\
1997 September 10&1997 September 11&1&ESO&CAT&CES LC&CCD Loral \#34\\
1997 December 12&1997 December 16&7&OHP&152&AURELIE&Barrette Th\\
1998 January 3&1998 January 4&1&ESO&CAT&CES LC&CCD Loral \#34\\
1998 March 27&1998 March 28&1&ESO&CAT&CES LC&CCD Loral \#34\\
1998 June 6&1998 June 7&1&ESO&CAT&CES VLC&CCD Loral \#38\\
1998 August 16&1998 August 17&1&ESO&CAT&CES VLC&CCD Loral \#38\\
1998 September 27&1998 September 29&1&ESO&CAT&CES VLC&CCD Loral \#38\\[4pt]
\hline\\[-4pt]
\end{tabular*}
\end{table*}
 
\section{Magnetic field and radial velocity determinations}
\label{sec:diag}

\subsection{Mean magnetic field modulus}
\label{sec:hmdiag}
The mean magnetic field modulus $\Hm$ is the average over the
visible stellar disk of the modulus of the magnetic field vector,
weighted by the local emergent line intensity. In the present study,
as in \citetalias{1997A&AS..123..353M},
the mean magnetic field modulus was determined from measurement of the wavelength separation of the
magnetically split components of the Fe~{\sc ii} $\lambda\,6149.2$
line, a Zeeman doublet, by application of the following formula:
\begin{equation}
\lambda_{\rm r}-\lambda_{\rm b}=g\,\Zeeman\,\Hm\,.
\label{eq:Hm}
\end{equation}
In this equation, $\lambda_{\rm r}$ and $\lambda_{\rm b}$ are the wavelengths of the red and blue split line
components,
respectively; $g$ is the Land\'e factor of the split level of the
transition ($g=2.70$; \citealt{1985aeli.book.....S}); 
$\Zeeman=k\,\lambda_0^2$, with
$k=4.67\,10^{-13}$\,\AA$^{-1}$\,G$^{-1}$; and $\lambda_0=6149.258$\,\AA\
is the nominal wavelength of the considered transition.

The motivation for the usage of Fe~{\sc ii} $\lambda\,6149.2$ as the
diagnostic line, the involved limitations and approximations, and the
methods used to measure the wavelengths of its split components have
been exhaustively described in \citetalias{1997A&AS..123..353M}.
Like in \citetalias{1997A&AS..123..353M}, the actual measurements of
$\lambda_{\rm r}$ and 
$\lambda_{\rm b}$ are carried out either by direct integration of the
split line component profiles, or by fitting them (and possible
blending lines) by multiple Gaussians, as described by
\citet{1992A&A...256..169M}.  

The difficulty in estimating the uncertainties affecting our mean
magnetic field modulus measurements was explained in
\citetalias{1997A&AS..123..353M} (see in 
particular Sect. 6, which also includes a detailed discussion of
  the systematic errors). For the stars that were studied in that paper, we
adopt here the same values of these uncertainties. For HD~47103,
HD~51684, and HD~213637, which did not feature in \citetalias{1997A&AS..123..353M}, the way in
which the measurement undertainties were estimated is described in the
respective sections of Appendix~\ref{sec:notes}. 

\subsection{Mean longitudinal magnetic field, crossover, and mean
  quadratic magnetic field} 
\label{sec:moments}
The moment technique
(\citealt{1988A&A...189..179M,1989FCPh...13..143M}; see also
\citealt{2000Mathys}) was used to extract, from the CASPEC Stokes $I$
and $V$ 
spectra, three parameters characterising the magnetic fields of the 
studied stars: the mean longitudinal magnetic field, the crossover,
and the quadratic field. Hereafter, we give an operational definition
of these parameters, based on a description of their determination. We
then briefly discuss their physical interpretation. This approach is
slightly different from the one we used in previous papers. Adopting this approach is
justified by the fact that, as becomes apparent later (see in
particular Sect.~\ref{sec:xdisc}), some of the magnetic parameters
that we consider may correspond to different physical processes in
different stars. 

Let $\lambda_I$ be the
wavelength of the centre of gravity of a 
spectral line observed in the Stokes parameter $I$. In what follows,
we refer to $\lambda_I$ as the line centre. The moment of order
$n$ of the profile of this line about its 
centre, in the Stokes parameter $X$ ($X=I,Q,U,V$), is defined as
\begin{equation}
\R{n}{X}={1\over\ew}\,\int r_{{\cal F}_X}(\llo)\,(\llo)^n\,d\lambda\,,
\label{eq:moment}
\end{equation}
where $\ew$ is the equivalent width of the line, and $r_{{\cal F}_X}$ is
its normalised profile in the Stokes parameter $X$,
\begin{equation}
r_{{\cal F}_X}=({\cal F}_{X_{\rm c}}-{\cal F}_X)/{\cal F}_{I_{\rm
    c}}\,.
\end{equation}
The notations ${\cal F}_X$ and ${\cal F}_{X_{\rm c}}$ represent the
integral over the visible stellar disk of the emergent intensity (in
the considered Stokes parameter)
 in the line and the neighbouring continuum, respectively, as follows:
\begin{eqnarray}
{\cal F}_X&=&\intxy\,X(x,y)\,dy\,,\\
{\cal F}_{X_{\rm c}}&=&\intxy\,X_{\rm c}(x,y)\,dy\,.
\end{eqnarray}
where $(x,y)$ are the coordinates of a point on the visible stellar
disk, in a reference system where the $z$-axis is parallel to the line
of sight, the $y$-axis lies in the plane defined by the line of sight
and the stellar rotation axis, and the unit length is the stellar
radius. The integral over the wavelength in Eq.~(\ref{eq:moment})
extends over the whole observed line (see \citealt{1988A&A...189..179M} for
details).

The mean longitudinal field $\Hz$ is derived from measurements of the
first-order moments $\R{1}{V}$ of Stokes $V$ line profiles about the
respective line centre, by application of the formula \citep{1995A&A...293..733M}, 
\begin{equation}
\R{1}{V}=\bar g\,\Zeeman\,\Hz\,,
\label{eq:Hz}
\end{equation}
where $\bar g$ is the effective Land\'e factor of the transition. One
can note that the 
first-order moment of the Stokes $V$ line profile is a measurement of
the wavelength shift of a spectral line between observations in 
opposite circular polarisations,
\begin{equation}
\R{1}{V}=(\lambda_{\rm R}-\lambda_{\rm L})/2\,,
\end{equation}
where $\lambda_{\rm R}$ (resp. $\lambda_{\rm L}$) is the wavelength of
the centre of gravity of the line in right (resp. left) circular
polarisation. Under some assumptions that in most cases represent a
good first approximation
\citep{1989FCPh...13..143M,1991A&AS...89..121M,1995A&A...293..733M},
$\Hz$ can be 
interepreted as being the average over the visible stellar disk of the
component of the magnetic vector along the line of sight, weighted by
the local emergent line intensity. 

In practice, $\R{1}{V}$ is measured
for a sample of lines. The required integration, for this and the
other line moments discussed below, is performed as described in
Sect.~2 of \citet{1994A&AS..108..547M}. Then, in application of Eq.~(\ref{eq:Hz}),
$\Hz$ is determined from a linear least-squares fit as a function of
$\bar g\,\Zeeman$, forced through the origin. Following
\citet{1994A&AS..108..547M}, this fit is weighted by 
the inverse of the mean-square error of the $\R{1}{V}$ measurements
for the individual lines, $1/\sigma^2[\R{1}{V}]$. The standard error
$\sigma_z$ of the longitudinal field that is derived from this
least-squares analysis is used as an estimate of the uncertainty
affecting the obtained value of $\Hz$. 

The crossover $\xover$ is derived from measurements of the second-order
moments $\R{2}{V}$ of the Stokes $V$ line profiles about the
respective line centre, by application of the formula
\begin{equation}
\R{2}{V}=2\,\bar g\,\Zeeman\,(\lambda_0/c)\,\xover\,.
\label{eq:xover}
\end{equation}
The second-order moment of the Stokes $V$ line profile is
a measurement of the difference of the width of a spectral line
between observations in opposite circular polarisations. Under the
same assumptions as for the longitudinal field
\citep{1995A&A...293..733M}, and as 
long as the only macroscopic velocity field contributing to the line
profile is the stellar rotation (a condition that may not always be
fulfilled; see Sect.~\ref{sec:xdisc}), $\xover$ can
be interpreted as being the product of the projected equatorial
velocity of the star, $\vsi$, and of the mean asymmetry of the
  longitudinal magnetic field, $\xHz$. The latter is defined as the
first-order moment about the plane defined by the stellar rotation
axis and the line of sight of the component of the magnetic vector
parallel to the line of sight, weighted by the local emergent line
intensity.  

In practice, $\R{2}{V}$ is measured
for a sample of lines. Then, in application of Eq.~(\ref{eq:xover}),
$\xover$ is determined from a linear least-squares fit as a function of
$2\bar g\,\Zeeman\,(\lambda_0/c)$, forced through the origin. This fit
is weighted by 
the inverse of the mean-square error of the $\R{2}{V}$ measurements
for the individual lines, $1/\sigma^2[\R{2}{V}]$. The standard error
$\sigma_x$ of the crossover that is derived from this
least-squares analysis is used as an estimate of the uncertainty
affecting the obtained value of $\xover$. 

The mean quadratic field $\Hq$ is derived from measurements of the
second-order moments $\R{2}{I}$ of the Stokes $I$ line profiles about
the respective line centre by application of the following
formula \citep{2006A&A...453..699M}:  
\begin{equation}
\R{2}{I}=a_1\,{1\over5}\,{\lambda_0^2\over
  c^2}+a_2\,{3S_2+D_2\over4}\,\Zeeman^2+a_3\,\ew^2\,{\lambda_0^4\over c^4}\,,
\label{eq:Hq}
\end{equation}
where $a_2=\Hq^2$. In this equation,
$S_2$ and $D_2$ are atomic parameters characterising the Zeeman
pattern of the considered transition. They are defined in terms of the
coefficients $\C 12$ and $\C 02$ introduced by \citet{1987A&A...171..368M}, 
\begin{eqnarray}
S_2&=&\C 12+\C 02\,,\label{eq:s2}\\
D_2&=&\C 12-\C 02\,.\label{eq:d2}
\end{eqnarray}
The second-order moment of the Stokes $I$ line profile is
a measurement the spectral line width in natural light. It results
from the combination of contributions of various effects: 
natural line width and instrumental profile (both of which cancel out
in the Stokes $V$ profiles), Doppler effect of various origins
(thermal motion, stellar rotation, possibly microturbulence, or
pulsation, $\ldots$), and magnetic broadening. For the interpretation
of the physical meaning of $\Hq$, the basic assumptions are similar to 
those used for $\Hz$ and $\xover$, but their relevance is somewhat
different \citep{1995A&A...293..746M}. \citet{2006A&A...453..699M}
discussed in detail the physical exploitation of the information
contents of $\R{2}{I}$ via a semi-empirical approach, on which
Eq.~(\ref{eq:Hq}) is based. Within this context, $\Hq$ is interpreted
as the square root of the sum of the following two field moments:
\begin{itemize}
\item The average over the visible stellar disk of the square of the
  modulus of the magnetic vector, $\av{B^2}$
\item  The average over the visible stellar disk of the square of the
  component of the magnetic vector along the line of sight,
  $\av{B_z^2}$
\end{itemize}
Both averages are weighted by the local emergent line intensity. The
form adopted here for Eq.~(\ref{eq:Hq}) is based on the assumption
that $\av{B_z^2}=\av{B^2}/3$. The reason why this assumption is
needed and its justification have been presented in previous works
on the quadratic field
\citep{1995A&A...293..746M,2006A&A...453..699M}. The adequacy of this
approximation is discussed further in Sect.~\ref{sec:Hqdisc}.

Practical application of Eq.~(\ref{eq:Hq}) is somewhat more
complicated than for Eqs.~(\ref{eq:Hz}) and (\ref{eq:xover}). The same
principle as for the latter is applied: $\R{2}{I}$ is measured for a 
sample of lines and a linear least-squares fit to these data of the
form given by Eq.~(\ref{eq:Hq}) is computed. However this is now a 
multi-parameter fit, with three independent variables, so that three
fit parameters ($a_1$, $a_2$, and $a_3$) must be determined. In most
studied stars, the number 
of diagnostic lines is of the order of 10, sometimes less, and it
exceeds 20 in only one case. As a result, the derived parameters may
be poorly constrained; in particular, there may be significant
ambiguities between their respective contributions. 

In order to
alleviate the problem, we used an approach similar to that of
\citet{1995A&A...293..746M}. This approach is based on the
consideration that the 
$a_2$ term is the only term on the right-hand side of
Eq.~(\ref{eq:Hq}) that is expected to show significant variability
with stellar rotation phase, so that the fit parameters $a_1$ and
$a_3$ should have the same values at all epochs of observation. 
Accordingly, for each star that has been
observed at least at three different epochs, we compute for each
diagnostic line the average $[\R{2}{I}]_{\rm av}$ of the measurements
of $\R{2}{I}$ for this line at the various epochs. We derive the
values of $a_1$ and $a_3$ through a least-squares fit of
$[\R{2}{I}]_{\rm av}$ by a function of the form given in the
right-hand side of Eq.~(\ref{eq:Hq}). This fit is weighted by
$1/\sigma^2\lbrace[\R{2}{I}]_{\rm av}\rbrace$, where
$\sigma\lbrace[\R{2}{I}]_{\rm av}\rbrace$ is calculated by application of error
propagation to the root-mean-square errors of the individual
measurements of $\R{2}{I}$ for the considered line at the different
epochs of 
observations. Then we use these 
values of $a_1$ and $a_3$ to compute  the magnetic part of $\R{2}{I}$ for each line
and at each epoch as follows:
\begin{equation}
[\R{2}{I}]_{\rm mag}=\R{2}{I}-a_1\,{1\over5}\,{\lambda_0^2\over
  c^2}-a_3\,\ew^2\,{\lambda_0^4\over c^4}\,.
\label{eq:R2I_mag}
\end{equation}
Finally, we compute a linear least-squares fit, forced through the
origin, of $[\R{2}{I}]_{\rm mag}$ as a function of
$(3S_2+D_2)\,\Zeeman^2/4$ to derive $\Hq$, in
application of the relation,
\begin{equation}
[\R{2}{I}]_{\rm mag}=\Hq^2\,{3S_2+D_2\over4}\,\Zeeman^2\,.
\label{eq:Hq_av}
\end{equation}
This
least-squares fit is weighted by the inverse of the mean-square error
of the $\R {2}{I}$ measurements
for the individual lines at the considered epoch,
$1/\sigma^2[\R{2}{I}]$. The standard error 
$\sigma_{\rm q}$ of the mean quadratic field that is derived from this
least-squares analysis is used as an estimate of the uncertainty
affecting the obtained value of $\Hq$. 

Previous mean quadratic field determinations
\citep{1995A&A...293..746M,1997A&AS..124..475M}
were not based on Eq.~(\ref{eq:Hq}), but
on the simpler form,
\begin{equation}
\R{2}{I}=a_0+a_2\,{3S_2+D_2\over4}\,\Zeeman^2\,.
\label{eq:Hqold}
\end{equation}
In other words, the dependences of $\R{2}{I}$ on the line wavelength
and equivalent width were ignored, except for the wavelength
dependence appearing in the magnetic term. The obvious shortcoming of
the application of Eq.~(\ref{eq:Hqold}) instead of Eq.~(\ref{eq:Hq})
for determination of the quadratic field is that part of the wavelength 
dependence of $\R{2}{I}$, corresponding to the $a_1$ and $a_3$ terms
of Eq.~(\ref{eq:Hq}), is likely to end up absorbed into the $a_2$
term of Eq.~(\ref{eq:Hqold}), rather than into its $a_0$ term. In other
words, neglecting the wavelength (and equivalent width) dependence of
the non-magnetic contributions to the Stokes $I$ line width must in
general introduce systematic errors (typically, overestimates) in the
determination of the mean 
quadratic magnetic field. 

Accordingly, to be able to combine, in the most consistent way, the
quadratic field data based on our previous observations with the
measurements obtained
from the new observations of this paper, whenever possible we used
revised $\Hq$ values determined by application of Eq.~(\ref{eq:Hq_av}) to
the spectra of \citet{1995A&A...293..746M}. These values are given in
Appendix~\ref{sec:hquad_rev}. 

This approach cannot be
used for the observations of \citet{1997A&AS..124..475M}, which for
each star have
been obtained at different epochs with different
instrumental configurations, 
for which the fit parameters $a_1$ and $a_3$ in general
have different values. On the other hand, the number of diagnostic lines in
those spectra is too small for direct application of
Eq.~(\ref{eq:Hq}), so that in the rest of this paper, the values of
the quadratic field derived by \citet{1997A&AS..124..475M} through
application of Eq.~(\ref{eq:Hqold}) are used unchanged. When
assessing their implications, one should keep in mind that they may in
general be less accurate than the most recent determinations of the
present study or the revised values derived from the older observations of
\citet{1995A&A...293..746M}.

\subsection{Radial velocities}
\label{sec:rv}
Our observations, both in natural light and in circular polarisation,
were not originally meant to be used 
  for the determination of stellar radial velocities, whether they were obtained in the framework of the present project (for this paper
and \citetalias{1997A&AS..123..353M}) or our earlier works
\citep{1991A&AS...89..121M,1994A&AS..108..547M,1995A&A...293..733M,1995A&A...293..746M,1997A&AS..124..475M}. In particular, no
  observations of radial velocity standards were obtained in any of
  our observing  runs. Nevertheless, most of our spectra proved 
  extremely well suited to the determination of accurate radial
  velocities, as became apparent in the analysis of our data. This
  allowed us to report in \citetalias{1997A&AS..123..353M} the discovery of radial velocity
  variations in eight stars that were not previously known to be
  spectroscopic binaries, some of which had observed amplitudes that do not
  exceed a few km\,s$^{-1}$. Consideration of these stars, together with
  the other members of our sample that were already known as
  binaries, suggested the possibility of a systematic difference of the
  lengths of orbital periods between binaries involving Ap stars with
  magnetically resolved lines and other Ap binaries. In view of these
  results, it appeared justified to exploit the potential of our data in a more systematic
  manner, to publish the radial velocities
  obtained from their analysis, and to carry out determinations of the
  orbital elements of the observed spectroscopic binaries whenever
  the number and distribution of the available
  measurements made it possible. 

\begin{table*}[!ht]
\caption{Mean magnetic field modulus}
\label{tab:hm}\small{
\begin{tabular*}{\textwidth}[]{@{}@{\extracolsep{\fill}}cccc
@{\extracolsep{0pt}}@{}}
\parbox[t]{4.2cm}{
\begin{tabular}[t]
{@{}@{\extracolsep{-2.5pt}}lrr@{\extracolsep{0pt}}@{}}
\hline
\hline\\[-4pt]
HJD&$\Hm$&Id.\\
{\tiny$\null-2400000.$}&(G)\\[2pt]
\hline
\hline\\[-6pt]
\multicolumn{3}{c}{HD 965}\\[2pt]
\hline\\[-6pt]
50231.914& 4199& 1\\
50296.576& 4199& 7\\
50971.914& 4249& 1\\
51042.807& 4249& 1\\
51084.731& 4251& 1\\[2pt]
\hline\\[-6pt]
\multicolumn{3}{c}{HD 2453}\\[2pt]
\hline\\[-6pt]
50056.945& 3704&10\\
50100.269& 3752& 7\\
50297.610& 3658& 7\\
50345.868& 3712& 9\\
50797.259& 3647& 7\\[2pt]
\hline\\[-6pt]
\multicolumn{3}{c}{HD 9996}\\[2pt]
\hline\\[-6pt]
50055.809& 4572&10\\
50297.587& 4107& 7\\
50345.758& 4220& 9\\[2pt]
\hline\\[-6pt]
\multicolumn{3}{c}{HD 12288}\\[2pt]
\hline\\[-6pt]
50052.563& 7559& 7\\
50100.348& 8440& 7\\
50101.388& 8458& 7\\
50347.880& 8196& 9\\
50348.956& 8496& 9\\
50349.850& 8370& 9\\
50350.889& 8203& 9\\
50796.373& 8068& 7\\[2pt]
\hline\\[-6pt]
\multicolumn{3}{c}{HD 14437}\\[2pt]
\hline\\[-6pt]
50052.439& 7544& 7\\
50052.523& 7253& 7\\
50054.974& 7479&10\\
50055.825& 7634&10\\
50056.967& 7741&10\\
50100.422& 7376& 7\\
50101.275& 7458& 7\\
50345.804& 7093& 9\\
50346.815& 7424& 9\\
50347.782& 7498& 9\\
50348.844& 7573& 9\\
50349.799& 7701& 9\\
50350.833& 7632& 9\\
50796.286& 6535& 7\\
50797.372& 7322& 7\\[2pt]
\hline\\[-6pt]
\multicolumn{3}{c}{HD 18078}\\[2pt]
\hline\\[-6pt]
50055.010& 4203&10\\
50295.588& 3278& 7\\
50346.897& 2884& 9\\
50349.936& 3146& 9\\
50797.471& 2682& 7\\[2pt]
\hline
\end{tabular}}
&\parbox[t]{4.2cm}{
\begin{tabular}[t]
{@{}@{\extracolsep{-2.5pt}}lrr@{\extracolsep{0pt}}@{}}
\hline
\hline\\[-4pt]
HJD&$\Hm$&Id.\\
{\tiny$\null-2400000.$}&(G)\\[2pt]
\hline
\hline\\[-6pt]
\multicolumn{3}{c}{HD 29578}\\[2pt]
\hline\\[-6pt]
50084.578& 2890& 1\\
50110.663& 2827& 1\\
50132.583& 2932& 1\\
50149.564& 2880& 1\\
50427.581& 3127& 1\\
50817.564& 3683& 1\\
51025.903& 3964& 1\\
51042.895& 4079& 1\\
51084.768& 4107& 1\\[2pt]
\hline\\[-6pt]
\multicolumn{3}{c}{HD 47103}\\[2pt]
\hline\\[-6pt]
50055.056&17452&10\\
50099.546&17898& 7\\
50110.565&17131& 1\\
50115.615&17766&11\\
50427.689&17013& 1\\
50817.626&16938& 1\\
51084.871&17051& 1\\[2pt]
\hline\\[-6pt]
\multicolumn{3}{c}{HD 50169}\\[2pt]
\hline\\[-6pt]
50057.065& 5266&10\\
50084.633& 5220& 1\\
50110.622& 5220& 1\\
50132.541& 5285& 1\\
50149.522& 5354& 1\\
50427.744& 5470& 1\\
50817.678& 5724& 1\\
51084.833& 5812& 1\\[2pt]
\hline\\[-6pt]
\multicolumn{3}{c}{HD 51684}\\[2pt]
\hline\\[-6pt]
50162.661& 5852& 1\\
50171.500& 5807& 1\\
50231.468& 6078& 1\\
50356.789& 6275& 1\\
50427.623& 5977& 1\\
50817.840& 5900& 1\\
50900.506& 5780& 1\\
50971.463& 6089& 1\\
51084.797& 6260& 1\\
51085.781& 6254& 1\\[2pt]
\hline\\[-6pt]
\multicolumn{3}{c}{HD 55719}\\[2pt]
\hline\\[-6pt]
50054.952& 6852&10\\
50084.709& 6489& 1\\
50110.525& 6214& 1\\
50132.513& 6405& 1\\
50149.495& 6607& 1\\
50162.490& 6443& 1\\
50171.481& 6267& 1\\
50231.445& 6330& 1\\
50427.647& 6437& 1\\
50522.568& 6334& 1\\
50900.483& 6462& 1\\
50971.443& 6086& 1\\
51042.924& 6162& 1\\
51085.797& 6403& 1\\[2pt]
\hline
\end{tabular}}
&\parbox[t]{4.2cm}{
\begin{tabular}[t]
{@{}@{\extracolsep{-2.5pt}}lrr@{\extracolsep{0pt}}@{}}
\hline
\hline\\[-4pt]
HJD&$\Hm$&Id.\\
{\tiny$\null-2400000.$}&(G)\\[2pt]
\hline
\hline\\[-6pt]
\multicolumn{3}{c}{HD 59435}\\[2pt]
\hline\\[-6pt]
50057.102& 2327& 1\\
50084.685& 2307& 1\\
50110.708& 2333& 1\\
50132.633& 2454& 1\\
50149.613& 2585& 1\\
50171.539& 2522& 1\\
50522.625& 3989& 1\\
50817.800& 3417& 1\\
51085.870& 2320& 1\\[2pt]
\hline\\[-6pt]
\multicolumn{3}{c}{HD 61468}\\[2pt]
\hline\\[-6pt]
50055.108& 7850&10\\
50084.757& 7831& 1\\
50110.780& 7267& 1\\
50132.700& 6982& 1\\
50149.680& 6439& 1\\
50162.539& 6185& 1\\
50171.605& 6189& 1\\
50427.809& 7374& 1\\
50817.737& 6070& 1\\
50900.570& 5925& 1\\
51085.828& 7248& 1\\[2pt]
\hline\\[-6pt]
\multicolumn{3}{c}{HD 65339}\\[2pt]
\hline\\[-6pt]
50052.642&16424& 7\\
50053.617&16225& 7\\
50055.138&12901&10\\
50057.123& 8700&10\\
50100.633&16198& 7\\
50101.461&15849& 7\\
50190.484&16160& 7\\
50192.450&12169& 7\\
50193.455& 8494& 7\\
50194.418& 8958& 7\\
50195.389& 9467& 7\\
50796.589& 8708& 7\\
50797.551&12146& 7\\
50798.552&15262& 7\\[2pt]
\hline\\[-6pt]
\multicolumn{3}{c}{HD 70331}\\[2pt]
\hline\\[-6pt]
50084.825&13100& 1\\
50110.846&13398& 1\\
50132.769&12978& 1\\
50162.612&13029& 1\\
50171.693&12030& 1\\
50522.727&13135& 1\\
50971.506&12494& 1\\[2pt]
\hline\\[-6pt]
\multicolumn{3}{c}{HD 75445}\\[2pt]
\hline\\[-6pt]
50057.135& 3035&10\\
50084.860& 2991& 1\\
50162.730& 3079& 1\\
50171.659& 2848& 1\\
50522.772& 3016& 1\\
50971.538& 2938& 1\\
51085.888& 3020& 1\\[2pt]
\hline
\end{tabular}}
&\parbox[t]{4.2cm}{
\begin{tabular}[t]
{@{}@{\extracolsep{-2.5pt}}lrr@{\extracolsep{0pt}}@{}}
\hline
\hline\\[-4pt]
HJD&$\Hm$&Id.\\
{\tiny$\null-2400000.$}&(G)\\[2pt]
\hline
\hline\\[-6pt]
\multicolumn{3}{c}{HD 81009}\\[2pt]
\hline\\[-6pt]
50115.681& 8117&11\\
50900.627& 7787& 1\\
51084.890& 8511& 1\\[2pt]
\hline\\[-6pt]
\multicolumn{3}{c}{HD 93507}\\[2pt]
\hline\\[-6pt]
50900.672& 6844& 1\\
50971.583& 6932& 1\\[2pt]
\hline\\[-6pt]
\multicolumn{3}{c}{HD 94660}\\[2pt]
\hline\\[-6pt]
50055.144& 6055&10\\
50115.721& 6158&11\\
50132.803& 6147& 1\\
50171.729& 6143& 1\\
50231.499& 6137& 1\\
50427.855& 6279& 1\\
50522.792& 6288& 1\\
50900.644& 6380& 1\\[2pt]
\hline\\[-6pt]
\multicolumn{3}{c}{HD 110066}\\[2pt]
\hline\\[-6pt]
50053.710& 4088& 7\\
50191.514& 4107& 7\\
50796.687& 4039& 7\\[2pt]
\hline\\[-6pt]
\multicolumn{3}{c}{HD 116114}\\[2pt]
\hline\\[-6pt]
50110.879& 5982& 1\\
50115.774& 5954&11\\
50162.861& 6015& 1\\
50231.517& 5954& 1\\
50817.864& 5982& 1\\
51042.495& 5996& 1\\[2pt]
\hline\\[-6pt]
\multicolumn{3}{c}{HD 116458}\\[2pt]
\hline\\[-6pt]
50115.790& 4677&11\\
50162.903& 4667& 1\\
50597.669& 4738& 1\\
50702.487& 4631& 1\\
51042.477& 4698& 1\\
51085.480& 4671& 1\\[2pt]
\hline\\[-6pt]
\multicolumn{3}{c}{HD 119027}\\[2pt]
\hline\\[-6pt]
50231.571& 3203& 1\\[2pt]
\hline\\[-6pt]
\multicolumn{3}{c}{HD 126515}\\[2pt]
\hline\\[-6pt]
50132.822&16260& 1\\
50971.558&10480& 1\\[2pt]
\hline\\[-6pt]
\multicolumn{3}{c}{HD 134214}\\[2pt]
\hline\\[-6pt]
50132.844& 3073& 1\\
50162.757& 3161& 1\\
50171.748& 3083& 1\\
50971.613& 3064& 1\\[2pt]
\hline
\end{tabular}}\\
\end{tabular*}}
\end{table*}

\addtocounter{table}{-1}

\begin{table*}[!ht]
\caption{continued.}
\small{
\begin{tabular*}{\textwidth}[]{@{}@{\extracolsep{\fill}}cccc
@{\extracolsep{0pt}}@{}}
\parbox[t]{4.2cm}{
\begin{tabular}[t]
{@{}@{\extracolsep{-2.5pt}}lrr@{\extracolsep{0pt}}@{}}
\hline
\hline\\[-4pt]
HJD&$\Hm$&Id.\\
{\tiny$\null-2400000.$}&(G)\\[2pt]
\hline
\hline\\[-6pt]
\multicolumn{3}{c}{HD 137909}\\[2pt]
\hline\\[-6pt]
50100.682& 5210& 7\\
50190.540& 5331& 7\\
50191.549& 5243& 7\\
50191.635& 5285& 7\\
50192.537& 5237& 7\\
50193.498& 5300& 7\\
50194.477& 5233& 7\\
50195.442& 5398& 7\\
50345.598& 5476& 9\\
50346.591& 5598& 9\\
50347.588& 5652& 9\\
50348.586& 5692& 9\\
50349.583& 5661& 9\\
50350.584& 5600& 9\\[2pt]
\hline\\[-6pt]
\multicolumn{3}{c}{HD 137949}\\[2pt]
\hline\\[-6pt]
50115.806& 4675&11\\
50132.865& 4633& 1\\
50231.624& 4675& 1\\
50522.810& 4685& 1\\
50900.705& 4658& 1\\
51042.524& 4637& 1\\
51084.482& 4690& 1\\[2pt]
\hline\\[-6pt]
\multicolumn{3}{c}{HD 142070}\\[2pt]
\hline\\[-6pt]
50900.732& 4669& 1\\
50971.634& 4589& 1\\[2pt]
\hline\\[-6pt]
\multicolumn{3}{c}{HD 144897}\\[2pt]
\hline\\[-6pt]
50971.668& 8765& 1\\
51085.524& 8660& 1\\[2pt]
\hline\\[-6pt]
\multicolumn{3}{c}{HD 150562}\\[2pt]
\hline\\[-6pt]
50171.802& 4755& 1\\
50231.675& 5000& 1\\
50900.795& 4830& 1\\
50971.740& 4799& 1\\
51084.538& 4725& 1\\[2pt]
\hline
\end{tabular}}
&\parbox[t]{4.2cm}{
\begin{tabular}[t]
{@{}@{\extracolsep{-2.5pt}}lrr@{\extracolsep{0pt}}@{}}
\hline
\hline\\[-4pt]
HJD&$\Hm$&Id.\\
{\tiny$\null-2400000.$}&(G)\\[2pt]
\hline
\hline\\[-6pt]
\multicolumn{3}{c}{HDE 318107}\\[2pt]
\hline\\[-6pt]
48841.606&13846& 1\\
48842.593&13266& 1\\
49078.880&13970& 1\\
49079.861&14696& 1\\
49100.853&16468& 1\\
49101.750&15306& 1\\
49130.807&16338& 1\\
49145.692&13437& 6\\
49161.624&13731& 1\\
49210.743&13438& 2\\
49214.680&13630& 1\\
49215.608&14006& 1\\
49397.862&13584& 1\\
49398.858&14247& 1\\
49418.860&14132& 1\\
49419.827&13855& 1\\
49436.867&13429& 2\\

49437.740&13396& 2\\
49456.754&13440& 1\\
49457.760&14090& 1\\
49494.769&13259& 1\\
49534.757&13513& 1\\
49535.799&14113& 1\\
49828.882&16988& 1\\
49829.835&15822& 1\\
49853.829&13721& 1\\
49881.798&13851& 1\\
49882.691&13568& 1\\
49908.716&14136& 1\\
49909.733&13805& 1\\
49947.705&14010& 3\\
49949.577&13725& 3\\
50149.807&17160& 1\\
50162.807&13754& 1\\
50971.788&13513& 1\\
51085.565&13656& 1\\[2pt]
\hline
\end{tabular}}
&\parbox[t]{4.2cm}{
\begin{tabular}[t]
{@{}@{\extracolsep{-2.5pt}}lrr@{\extracolsep{0pt}}@{}}
\hline
\hline\\[-4pt]
HJD&$\Hm$&Id.\\
{\tiny$\null-2400000.$}&(G)\\[2pt]
\hline
\hline\\[-6pt]
\multicolumn{3}{c}{HD 165474}\\[2pt]
\hline\\[-6pt]
50162.880& 6726& 1\\
50171.863& 6879& 1\\
50192.564& 6479& 7\\
50193.587& 6783& 7\\
50194.584& 6804& 7\\
50345.634& 6739& 9\\
50346.800& 6709& 9\\
50347.619& 6709& 9\\
50348.616& 6957& 9\\
50349.613& 6760& 9\\
50350.627& 6797& 9\\
50522.871& 6902& 1\\
50900.886& 6967& 1\\
50971.700& 6900& 1\\[2pt]
\hline\\[-6pt]
\multicolumn{3}{c}{HD 166473}\\[2pt]
\hline\\[-6pt]
50132.885& 5969& 1\\
50149.872& 5883& 1\\
50171.896& 5789& 1\\
50231.800& 5820& 1\\
50522.835& 5661& 1\\
50702.538& 5770& 1\\
50900.853& 6068& 1\\
50971.819& 6095& 1\\
51042.606& 6038& 1\\
51084.582& 6118& 1\\[2pt]
\hline\\[-6pt]
\multicolumn{3}{c}{HD 177765}\\[2pt]
\hline\\[-6pt]
50231.751& 3414& 1\\
51084.614& 3427& 1\\[2pt]
\hline\\[-6pt]
\multicolumn{3}{c}{HD 187474}\\[2pt]
\hline\\[-6pt]
49994.519& 6093&11\\
50149.891& 6200& 1\\
50162.911& 6216& 1\\
50171.915& 6223& 1\\
50231.820& 6258& 1\\
50356.519& 6185& 1\\
50522.904& 5868& 1\\
50702.569& 5340& 1\\
50900.908& 4981& 1\\
50971.944& 4872& 1\\
51042.554& 4874& 1\\
51084.636& 4939& 1\\[2pt]
\hline\\[-6pt]
\multicolumn{3}{c}{HD 188041}\\[2pt]
\hline\\[-6pt]
50149.902& 3536& 1\\
50297.499& 3622& 7\\
50347.651& 3658& 9\\[2pt]
\hline
\end{tabular}}
&\parbox[t]{4.2cm}{
\begin{tabular}[t]
{@{}@{\extracolsep{-2.5pt}}lrr@{\extracolsep{0pt}}@{}}
\hline
\hline\\[-4pt]
HJD&$\Hm$&Id.\\
{\tiny$\null-2400000.$}&(G)\\[2pt]
\hline
\hline\\[-6pt]
\multicolumn{3}{c}{HD 192678}\\[2pt]
\hline\\[-6pt]
50055.701& 4767&10\\[2pt]
\hline\\[-6pt]
\multicolumn{3}{c}{HDE 335238}\\[2pt]
\hline\\[-6pt]
50295.425& 7997& 7\\
50348.709& 8907& 9\\[2pt]
\hline\\[-6pt]
\multicolumn{3}{c}{HD 200311}\\[2pt]
\hline\\[-6pt]
50052.295& 8037& 7\\
50053.285& 7800& 7\\
50054.284& 8148& 7\\
50055.724& 7609&10\\
50294.509& 8997& 7\\
50345.689& 8972& 9\\
50349.747& 9440& 9\\[2pt]
\hline\\[-6pt]
\multicolumn{3}{c}{HD 201601}\\[2pt]
\hline\\[-6pt]
50055.741& 3918&10\\
50231.826& 3977& 1\\
50300.531& 4060& 7\\
50345.728& 3987& 9\\
50702.511& 3966& 1\\
50971.831& 4042& 1\\
51085.507& 3974& 1\\[2pt]
\hline\\[-6pt]
\multicolumn{3}{c}{HD 208217}\\[2pt]
\hline\\[-6pt]
50084.533& 9029& 1\\
50231.844& 6909& 1\\
50702.663& 6883& 1\\
51042.669& 6892& 1\\
51084.501& 7196& 1\\
51085.494& 7641& 1\\
51085.762& 7892& 1\\[2pt]
\hline\\[-6pt]
\multicolumn{3}{c}{HD 213637}\\[2pt]
\hline\\[-6pt]
50971.857& 5321& 1\\
51042.727& 5252& 1\\
51084.687& 5132& 1\\
51085.613& 5052& 1\\[2pt]
\hline\\[-6pt]
\multicolumn{3}{c}{HD 216018}\\[2pt]
\hline\\[-6pt]
49829.908& 5568& 1\\
49882.892& 5604& 1\\
49908.934& 5604& 1\\
50231.872& 5591& 1\\
50297.544& 5495& 7\\
50971.888& 5600& 1\\
51042.858& 5562& 1\\
51084.653& 5606& 1\\[2pt]
\hline
\end{tabular}}\\
\end{tabular*}}
\end{table*}

Radial velocities were determined from the new spectropolarimetric
observations of this paper as well as from the similar CASPEC spectra
of \citet{1991A&AS...89..121M} and \citet{1997A&AS..124..475M} by analysing
the same sets of spectral lines as for the measurement of the mean
longitudinal magnetic field. We computed a least-squares fit, forced
through the origin, of the differences between the wavelength
$\lambda_I$ of their centre of gravity in Stokes $I$ and the nominal
wavelength of the corresponding transitions, $\lambda_0$. This fit is
weighted by the inverse of the mean-square error of the $\lambda_I$
measurements of the individual lines. The radial velocity is derived
from its slope in the standard manner. The standard
deviation of the velocity that is obtained from this analysis
correctly characterises the precision of the measurements, but
not their accuracy. This is discussed further in
Sect.~\ref{sec:binarity}. 

\begin{table*}[!ht]
\caption{Mean longitudinal magnetic field, crossover, and 
mean quadratic magnetic field}
\label{tab:hzxq}\small{
\begin{tabular*}{\textwidth}[]{@{}@{\extracolsep{\fill}}cc
@{\extracolsep{0pt}}@{}}
\parbox[t]{8.8cm}{
\centering 
\begin{tabular}[t]
{@{}@{\extracolsep{-2.5pt}}lrrrrrrr@{\extracolsep{0pt}}@{}}
\hline
\hline\\[-4pt]
HJD%
&$\Hz$&$\sigma_z$%
&$\xover$&$\sigma_x$%
&$\Hq$&$\sigma_{\rm q}$%
&$n$\\
{\tiny$\null-2400000.$}%
&\multicolumn{2}{c}{(G)}%
&\multicolumn{2}{c}{(km$\,$s$^{-1}$\,G)}%
&\multicolumn{2}{c}{(G)}\\[4pt]
\hline
\hline\\[-6pt]
\multicolumn{8}{c}{HD 965}\\[2pt]
\hline\\[-6pt]
49916.907&  $-$574& 113&  $-$424& 509&  7570& 200& 6\\
49974.744&  $-$691& 141&  $-$911&1014&  7668& 191& 6\\
50039.588&  $-$863&  44&  $-$829& 531&  7664& 309& 6\\
50294.905& $-$1063& 125& $-$1458&1105&  7439& 232& 6\\
50615.928& $-$1057&  52& $-$3065&1503&  7909& 390& 4\\
50629.884& $-$1159& 142& $-$2798& 997&  7817& 319& 6\\
50784.596& $-$1270& 139& $-$2708&1175&  7737& 428& 5\\[2pt]
\hline\\[-6pt]
\multicolumn{8}{c}{HD 2453}\\[2pt]
\hline\\[-6pt]
49974.766& $-$1011&  67& $-$1033& 393&  4628& 479&14\\
50039.541& $-$1155&  43&  $-$235& 459&  4191& 536&12\\
50294.858&  $-$553&  41&  $-$851& 361&  4369& 321&13\\
50629.947&  $-$831&  53&  $-$788& 352&  4150& 328&15\\
50784.520&  $-$592&  51&  $-$659& 676&  3991& 668&10\\[2pt]
\hline\\[-6pt]
\multicolumn{8}{c}{HD 29578}\\[2pt]
\hline\\[-6pt]
49916.930&  $-$889& 106&  $-$545& 912&  3734& 325&13\\
49974.816& $-$1068&  61&   344& 409&  3650& 363&17\\
50039.634&  $-$982&  46&   133& 314&  3309& 263&17\\
50107.539&  $-$935&  53&   219& 388&  3438& 227&17\\
50294.930&  $-$780&  77&  $-$371& 426&  3734& 282&16\\
50505.563&  $-$668&  57&  $-$202& 327&  3508& 352&13\\
50629.916&  $-$682&  65&   330& 513&  3468& 413&14\\
50784.622&  $-$490&  61&  $-$187& 290&  4435& 266&16\\
50833.723&  $-$135&  75&   132& 426&  4809& 268&17\\[2pt]
\hline\\[-6pt]
\multicolumn{8}{c}{HD 47103}\\[2pt]
\hline\\[-6pt]
50107.596& $-$2241& 327& $-$1344&4200& 16582&2097& 6\\
50183.500& $-$2507& 331&  4524&2973& 15114&2064& 6\\
50497.547& $-$2226& 490& $-$4232&2936& 15793&2280& 6\\
50505.529& $-$3378& 450&  1778&5236& 20595&1874& 6\\
50784.775& $-$3253& 427&  2123&3750& 15971&2431& 6\\
50831.731& $-$2835& 478&  2566&5809& 16481&2369& 5\\[2pt]
\hline\\[-6pt]
\multicolumn{8}{c}{HD 50169}\\[2pt]
\hline\\[-6pt]
49830.504&  1035&  70&   889& 596&  5894& 240&14\\
49974.857&  1002&  85&  $-$423&1068&  6085& 371&11\\
50039.684&   763&  39&   229& 469&  5995& 267&14\\
50111.554&   911&  64&  $-$354& 615&  5984& 280&13\\
50183.543&   891&  52&   356& 715&  6323& 258&11\\
50497.591&   471&  75&  $-$451& 684&  6002& 340&10\\
50784.745&   188&  60&  $-$323& 619&  6453& 164&14\\
50832.727&   115&  94&  1132& 894&  7085& 359&13\\[2pt]
\hline\\[-6pt]
\multicolumn{8}{c}{HD 51684}\\[2pt]
\hline\\[-6pt]
50183.577& $-$1713&  90&   144& 534&  6718& 453&10\\
50476.639& $-$1304&  89&  1233& 726&  6516& 150&10\\
50497.627& $-$1324& 105& $-$1400&1100&  6053& 614& 8\\
50505.599& $-$1472& 127&    83& 752&  6634& 591& 8\\
50784.657& $-$1206& 150&   268& 688&  7072& 498& 9\\
50831.755& $-$1266& 131&  $-$903& 896&  6836& 363&10\\
50833.751& $-$1201&  90&  $-$261& 705&  6407& 538&10\\[2pt]
\hline
\end{tabular}}
&\parbox[t]{8.8cm}{
\centering 
\begin{tabular}[t]
{@{}@{\extracolsep{-2.5pt}}lrrrrrrr@{\extracolsep{0pt}}@{}}
\hline
\hline\\[-4pt]
HJD%
&$\Hz$&$\sigma_z$%
&$\xover$&$\sigma_x$%
&$\Hq$&$\sigma_{\rm q}$%
&$n$\\
{\tiny$\null-2400000.$}%
&\multicolumn{2}{c}{(G)}%
&\multicolumn{2}{c}{(km$\,$s$^{-1}$\,G)}%
&\multicolumn{2}{c}{(G)}\\[4pt]
\hline
\hline\\[-6pt]
\multicolumn{8}{c}{HD 55719}\\[2pt]
\hline\\[-6pt]
49830.473&   943&  44&  3039& 797&  7274& 295&11\\
49974.831&   533&  87&  1713& 876&  7074& 436&11\\
50039.653&   648&  99&   773& 724&  7020& 354&12\\
50107.619&   742&  87&  $-$846&1028&  6804& 299&10\\
50183.470&   693&  83& $-$1161& 955&  7429& 185&11\\
50476.616&   783&  96& $-$1033& 908&  7161& 361&11\\
50504.555&   881& 117&  $-$801&1148&  7316& 316&10\\
50784.640&   591&  82&  $-$600& 917&  7641& 322&11\\
50832.745&   608&  57&  $-$164& 502&  7123& 482&10\\[2pt]
\hline\\[-6pt]
\multicolumn{8}{c}{HD 61468}\\[2pt]
\hline\\[-6pt]
50039.787& $-$2168& 150&  $-$453& 988&  7528& 354& 9\\
50111.608& $-$2376& 151&     1&1399&  7429& 440& 8\\
50183.623& $-$1298& 131&  $-$975&1264&  5156& 463&12\\
50784.699& $-$2110& 138& $-$1291& 645&  6197& 590& 7\\
50833.796& $-$1122& 110& $-$1439& 710&  5864& 382& 9\\[2pt]
\hline\\[-6pt]
\multicolumn{8}{c}{HD 70331}\\[2pt]
\hline\\[-6pt]
49830.578& $-$2504& 157&  2267&2765& 16291& 877&12\\
49974.895& $-$2440& 165& $-$2338&1947& 16154&1425& 9\\
50039.732& $-$2893& 214& $-$2415&2506& 16105&1935&10\\
50107.701& $-$2889& 101& $-$2725&1216& 17528&1184&12\\
50111.654& $-$2309& 183&  2519&2750& 16619&1046&11\\
50183.672& $-$2586& 167& $-$4865&2054& 14751&1665&11\\
50468.817& $-$2610& 181&  $-$913&2590& 17194&1141& 8\\
50469.780& $-$2770& 183&   536&2667& 17396& 823&12\\
50476.681& $-$1996& 452&  1581&3429& 18101&2124& 9\\
50497.689& $-$2533& 128&  3633&1958& 15986&1578&10\\
50505.760& $-$2405& 163&  4929&1864& 15076&1719& 8\\
50784.802& $-$2486& 146&  4691&3335& 16491&1249&10\\
50831.784& $-$2558& 210&  7251&3177& 16732&2164&10\\
50832.766& $-$2737& 154&  3469&4314& 15722&1963&12\\
50833.838& $-$2433& 289&  4089&3111& 16059&1446& 8\\[2pt]
\hline\\[-6pt]
\multicolumn{8}{c}{HD 75445}\\[2pt]
\hline\\[-6pt]
50107.728&   147&  52&   277& 521&  4022& 243&18\\
50505.715&  $-$143&  76&   661& 509&  4376& 340&15\\
50784.828&  $-$112&  37&   $-$39& 247&  4428& 214&22\\[2pt]
\hline\\[-6pt]
\multicolumn{8}{c}{HD 81009}\\[2pt]
\hline\\[-6pt]
49830.550&  2040& 233& $-$2494& 938& 10086& 238&10\\
50039.823&  2120& 226& $-$2637&1695& 10578& 365&10\\
50107.747&  2294& 155& $-$1146&1041& 10392& 291&10\\
50111.683&  2377& 156&  $-$222&1128& 10234& 296&10\\
50183.702&  1984& 256&  2321&1075&  9669& 264& 8\\
50468.790&  1387& 179& $-$2593&1149&  9371& 201&10\\
50476.715&  1967& 139&  $-$612&1830& 10008& 328&10\\
50497.718&  1372& 153&  $-$282&1352&  9158& 256& 9\\
50504.688&  1563& 201&   $-$99&1209&  9378& 217&10\\
50629.452&  1735& 175&  1174&1209&  9742& 228&10\\
50784.842&  1849& 192& $-$2660&1175& 10335& 288&10\\
50831.810&  1390& 216&   668&1250& 10672& 485& 8\\[2pt]
\hline
\end{tabular}}\\
\end{tabular*}}
\end{table*}

\addtocounter{table}{-1}

\begin{table*}[!ht]
\caption{continued.}
\small{
\begin{tabular*}{\textwidth}[]{@{}@{\extracolsep{\fill}}cc
@{\extracolsep{0pt}}@{}}
\parbox[t]{8.8cm}{
\centering 
\begin{tabular}[t]
{@{}@{\extracolsep{-2.5pt}}lrrrrrrr@{\extracolsep{0pt}}@{}}
\hline
\hline\\[-4pt]
HJD%
&$\Hz$&$\sigma_z$%
&$\xover$&$\sigma_x$%
&$\Hq$&$\sigma_{\rm q}$%
&$n$\\
{\tiny$\null-2400000.$}%
&\multicolumn{2}{c}{(G)}%
&\multicolumn{2}{c}{(km$\,$s$^{-1}$\,G)}%
&\multicolumn{2}{c}{(G)}\\[4pt]
\hline
\hline\\[-6pt]
\multicolumn{8}{c}{HD 93507}\\[2pt]
\hline\\[-6pt]
49830.687&  2662& 167&  2899&1158&  6288& 375& 9\\
50039.841&  1274& 292& $-$3834&2081&  9136&1180& 7\\
50107.765&  1422& 263&  $-$448&2099&  7184&1024& 6\\
50183.723&  2342& 201&  1452& 533&  8216& 505& 6\\
50294.475&  2895&  77&  2136&1498&  7174& 356& 9\\
50476.732&  2613& 202&  1108&1952&  6739& 336& 8\\
50497.654&  2259& 142&   587&1095&  7031& 555& 9\\
50629.480&   949& 331&  $-$925&1776&  9244& 879& 7\\
50783.844&  2365& 173&  2157& 764&  6547& 597& 6\\
50831.822&  2570& 100&  2414&1660&  7083& 338& 8\\[2pt]
\hline\\[-6pt]
\multicolumn{8}{c}{HD 94660}\\[2pt]
\hline\\[-6pt]
49830.605& $-$1934&  66&  $-$377& 507&  5944& 211&17\\
50039.857& $-$1866&  91&  $-$538& 591&  5987& 240&17\\
50107.785& $-$1931&  83& $-$1248& 640&  5946& 246&16\\
50294.454& $-$1837&  79&  $-$791& 688&  6039& 195&16\\
50476.753& $-$1846& 102& $-$1133& 557&  6319& 210&17\\
50504.724& $-$1812&  72& $-$1046& 409&  6152& 184&17\\
50616.466& $-$1836&  59&    25& 620&  6678& 254&17\\
50784.850& $-$1848& 103&  $-$979& 606&  6622& 148&15\\[2pt]
\hline\\[-6pt]
\multicolumn{8}{c}{HD 110066}\\[2pt]
\hline\\[-6pt]
49830.634&  $-$129&  47&  $-$478& 212&  1773& 895&13\\
50497.852&   $-$85&  60&  $-$288& 447&  2689& 365&13\\
50832.873&  $-$131&  76&    69& 396&  2558& 507&12\\[2pt]
\hline\\[-6pt]
\multicolumn{8}{c}{HD 116114}\\[2pt]
\hline\\[-6pt]
49830.621& $-$1832&  48& $-$1207& 385&  7497& 170&17\\
50111.719& $-$1822&  72& $-$1970& 426&  7985& 165&15\\
50294.512& $-$1794&  68& $-$1047& 495&  7582& 251&16\\
50497.734& $-$1769&  77&  $-$993& 547&  7699& 146&15\\
50629.503& $-$1844&  74&   272& 408&  7966& 164&18\\
50832.803& $-$1714&  64&   $-$67& 489&  7577& 171&17\\[2pt]
\hline\\[-6pt]
\multicolumn{8}{c}{HD 116458}\\[2pt]
\hline\\[-6pt]
49763.898& $-$1530&  44&  $-$350& 576&  5435& 257&14\\
49912.606& $-$1371&  56&  $-$328& 408&  5362& 224&15\\
49972.487& $-$1850&  50& $-$1231& 756&  5145& 248&11\\
50107.811& $-$1741&  59&  $-$431& 589&  5187& 199&13\\
50294.526& $-$1586&  88&  $-$206& 651&  4924& 198&14\\
50476.761& $-$1365&  65&  $-$911& 419&  5019& 189&16\\
50497.745& $-$1459&  76&    54& 658&  5295& 243&16\\
50504.612& $-$1502&  90&   $-$11& 542&  5299& 128&13\\
50629.517& $-$1266&  61&  $-$894& 394&  5455& 176&17\\
50784.857& $-$1391&  50&  $-$648& 623&  5381& 221&14\\
50832.815& $-$1875&  65&   $-$31& 834&  5270& 243&14\\[2pt]
\hline\\[-6pt]
\multicolumn{8}{c}{HD 119027}\\[2pt]
\hline\\[-6pt]
50111.760&   510&  55&    38& 547&  3655& 756& 9\\[2pt]
\hline
\end{tabular}}
&\parbox[t]{8.8cm}{
\centering 
\begin{tabular}[t]
{@{}@{\extracolsep{-2.5pt}}lrrrrrrr@{\extracolsep{0pt}}@{}}
\hline
\hline\\[-4pt]
HJD%
&$\Hz$&$\sigma_z$%
&$\xover$&$\sigma_x$%
&$\Hq$&$\sigma_{\rm q}$%
&$n$\\
{\tiny$\null-2400000.$}%
&\multicolumn{2}{c}{(G)}%
&\multicolumn{2}{c}{(km$\,$s$^{-1}$\,G)}%
&\multicolumn{2}{c}{(G)}\\[4pt]
\hline
\hline\\[-6pt]
\multicolumn{8}{c}{HD 126515}\\[2pt]
\hline\\[-6pt]
50111.802&  $-$792& 105&  $-$204&1232& 15667& 861& 7\\
50183.749&  1459& 158&  2632&1984& 12822& 227& 8\\
50294.538&  $-$741&  79&  $-$946&2575& 16613& 541& 7\\
50468.852&   823& 103& $-$1270&3674& 10448& 762& 7\\
50469.852&   392& 151& $-$1812& 698& 10703& 543& 8\\
50476.802&   357&  95&  $-$788&1097& 11124& 433& 9\\
50497.787&  $-$692&  93&  3939&1828& 15469& 615& 8\\
50504.744& $-$1078& 141&  $-$158&3034& 17080& 687& 7\\
50629.542&  $-$590& 137&  $-$110&1232& 15976& 593& 8\\
50832.832&  1467& 135&  2911& 766& 12455& 472& 8\\[2pt]
\hline\\[-6pt]
\multicolumn{8}{c}{HD 134214}\\[2pt]
\hline\\[-6pt]
50111.818&  $-$356&  44&  $-$441& 321&  4085& 208&19\\
50294.557&  $-$441&  46&   $-$61& 531&  3816& 225&16\\
50476.819&  $-$402&  56&   295& 345&  3689& 140&19\\
50504.763&  $-$355&  54&  $-$634& 332&  3983& 110&16\\
50629.560&  $-$282&  39&  $-$333& 256&  4103& 204&19\\
50832.849&  $-$336&  37&  $-$338& 237&  3951& 194&17\\[2pt]
\hline\\[-6pt]
\multicolumn{8}{c}{HD 137909}\\[2pt]
\hline\\[-6pt]
49830.771&  $-$189&  75&  1422& 418&  7714& 398& 7\\
49916.486&   537&  91& $-$1762& 626&  6850& 222& 7\\
50183.796&  $-$370&  81&   780&1278&  7174& 203& 7\\
50294.498&  $-$382&  95&   879&1016&  7221& 159& 7\\
50615.620&  $-$223&  99& $-$2677& 649&  7282& 308& 7\\
50616.590&   180&  74& $-$2346& 418&  7278& 389& 7\\[2pt]
\hline\\[-6pt]
\multicolumn{8}{c}{HD 137949}\\[2pt]
\hline\\[-6pt]
49830.662&  1520& 131&  $-$655& 561&  4535& 356&12\\
49916.541&  1425&  92&   134& 676&  4620& 396&13\\
49974.488&  1545& 105&    63& 681&  4994& 331&14\\
50107.822&  1675&  85&   254& 498&  4271& 334&14\\
50294.596&  1671&  98&  1806& 676&  4586& 454&13\\
50497.801&  1631&  85&  $-$185& 540&  4419& 301&14\\
50616.675&  1468& 100&   151& 887&  4923& 255&14\\[2pt]
\hline\\[-6pt]
\multicolumn{8}{c}{HD 142070}\\[2pt]
\hline\\[-6pt]
49830.787&   778&  64& $-$1937& 324&  5006& 362&14\\
49974.509&  $-$144&  76& $-$3066& 618&  4805& 333&14\\
50107.873&   596&  74&  2972& 526&  4972& 227&16\\
50111.870&   179&  63&  4781& 470&  4152& 361&12\\
50183.780&   $-$34&  66& $-$3918& 437&  4391& 248&15\\
50294.616&  $-$196&  64&   312& 586&  4615& 209&15\\
50468.874&   225& 123&  3133& 796&  5267& 217&14\\
50469.872&  $-$160&  68&  $-$178& 627&  4486& 220&12\\
50476.872&  $-$251&  76& $-$1823& 496&  4685& 231&17\\
50497.867&   202&  47& $-$6203& 590&  5061& 213&17\\
50504.786&   483&  47& $-$4459& 505&  4636& 355&16\\
50505.830&   529&  71&  2854& 605&  4684& 351&13\\
50629.583&   306&  84& $-$4566&1159&  4771& 268&14\\[2pt]
\hline
\end{tabular}}\\
\end{tabular*}}
\end{table*}

\addtocounter{table}{-1}

\begin{table*}[!ht]
\caption{continued.}
\small{
\begin{tabular*}{\textwidth}[]{@{}@{\extracolsep{\fill}}cc
@{\extracolsep{0pt}}@{}}
\parbox[t]{8.8cm}{
\centering 
\begin{tabular}[t]
{@{}@{\extracolsep{-2.5pt}}lrrrrrrr@{\extracolsep{0pt}}@{}}
\hline
\hline\\[-4pt]
HJD%
&$\Hz$&$\sigma_z$%
&$\xover$&$\sigma_x$%
&$\Hq$&$\sigma_{\rm q}$%
&$n$\\
{\tiny$\null-2400000.$}%
&\multicolumn{2}{c}{(G)}%
&\multicolumn{2}{c}{(km$\,$s$^{-1}$\,G)}%
&\multicolumn{2}{c}{(G)}\\[4pt]
\hline
\hline\\[-6pt]
\multicolumn{8}{c}{HD 144897}\\[2pt]
\hline\\[-6pt]
49830.815&  1635&  88&  1958& 980& 10893& 313&12\\
49972.556&  1631& 130&  1255&1634& 10815& 429& 9\\
50107.843&  2248& 110&    17&1076&  9594& 226&12\\
50111.840&  2045& 132&   978&1461&  9812& 243&12\\
50183.815&  1638&  88&  $-$435&1762& 10984& 363&11\\
50294.647&  2056& 101&  1646&1467& 10027& 351&12\\
50469.827&  1310& 122&  1005&1845& 10958& 301&11\\
50476.842&  1740&  98&   746& 913& 10188& 258&11\\
50497.822&  2192& 115&  $-$181& 902&  9658& 154&14\\
50505.801&  2143&  71&  2166& 860&  9890& 220&13\\
50629.613&  1977& 110&  3362& 767&  9735& 212&13\\
50831.866&  1681& 159&  2813&1052&  9609& 369&10\\[2pt]
\hline\\[-6pt]
\multicolumn{8}{c}{HD 150562}\\[2pt]
\hline\\[-6pt]
50629.689&  1206&  91&   347& 447&  5710& 837&13\\[2pt]
\hline\\[-6pt]
\multicolumn{8}{c}{HDE 318107}\\[2pt]
\hline\\[-6pt]
49830.738&  3082& 165&  $-$576&3650& 21029& 612& 7\\
49974.626&  1067&  54& $-$4934&2919& 21087& 695& 7\\
50294.741&   910& 257& $-$3748&4404& 24519& 520& 8\\
50505.860&  3054& 132&  3116&3751& 20173& 981& 8\\
50629.747&  3785& 119&  2924&3833& 19419& 492& 8\\[2pt]
\hline\\[-6pt]
\multicolumn{8}{c}{HD 165474}\\[2pt]
\hline\\[-6pt]
49830.854&   102& 115&  $-$178& 774&  8507& 267&11\\
49974.536&   177& 107&   724& 806&  8672& 453&11\\
50183.848&   334& 135&   447&1542&  8486& 331&11\\
50294.674&   405& 147&  $-$491& 848&  8505& 589&11\\
50629.643&   190& 112&  $-$620&1105&  8499& 427&11\\[2pt]
\hline\\[-6pt]
\multicolumn{8}{c}{HD 166473}\\[2pt]
\hline\\[-6pt]
49830.876&   245& 129&  $-$430& 873&  7434& 282& 7\\
49916.826&   711& 119&  $-$372&1589&  7556& 155& 9\\
49972.637&   841& 213&  $-$510&1396&  6482& 316& 7\\
50183.873&  1458& 185&  1316& 981&  6840& 211& 9\\
50294.775&  1679& 133&    13&1108&  6630& 348& 9\\
50497.889&  1746& 155&   964&1041&  6188& 160& 9\\
50616.882&  1822& 116&   494& 494&  6183& 162& 9\\[2pt]
\hline\\[-6pt]
\multicolumn{8}{c}{HD 187474}\\[2pt]
\hline\\[-6pt]
49830.911& $-$1404&  69&  $-$675& 482&  6904& 186&11\\
49916.859& $-$1587&  72&  $-$173& 873&  6948& 224&11\\
49974.664& $-$1872&  75&   512& 673&  7019& 288&12\\
50039.522& $-$1885&  94&  $-$796& 706&  7541& 240&12\\
50183.888& $-$1773&  81&  $-$794& 862&  7042& 246&12\\
50294.832& $-$1854&  65&  $-$114& 837&  6821& 269&10\\
50505.904& $-$1670& 124& $-$1227&1031&  7271& 254&12\\
50615.875& $-$1305&  99& $-$1497&1079&  6780& 440&11\\
50783.509&  $-$316&  60&  $-$507& 423&  5905& 230&11\\[2pt]
\hline
\end{tabular}}
&\parbox[t]{8.8cm}{
\centering 
\begin{tabular}[t]
{@{}@{\extracolsep{-2.5pt}}lrrrrrrr@{\extracolsep{0pt}}@{}}
\hline
\hline\\[-4pt]
HJD%
&$\Hz$&$\sigma_z$%
&$\xover$&$\sigma_x$%
&$\Hq$&$\sigma_{\rm q}$%
&$n$\\
{\tiny$\null-2400000.$}%
&\multicolumn{2}{c}{(G)}%
&\multicolumn{2}{c}{(km$\,$s$^{-1}$\,G)}%
&\multicolumn{2}{c}{(G)}\\[4pt]
\hline
\hline\\[-6pt]
\multicolumn{8}{c}{HD 188041}\\[2pt]
\hline\\[-6pt]
50294.815&   977&  54&   360& 380&      &    &14\\
50629.822&   928&  63&    64& 276&      &    &15\\[2pt]
\hline\\[-6pt]
\multicolumn{8}{c}{HDE 335238}\\[2pt]
\hline\\[-6pt]
49974.589& $-$1828& 272&  2708&3031& 14032& 910&10\\
50294.703&   859& 144&  $-$921&1300& 11103& 830&16\\[2pt]
\hline\\[-6pt]
\multicolumn{8}{c}{HD 201601}\\[2pt]
\hline\\[-6pt]
49916.849& $-$1007&  55&  $-$333& 260&  4535& 134&26\\
50039.516&  $-$965&  58&   177& 348&  4626& 137&26\\
50183.922& $-$1023&  46&   293& 316&  4291& 127&26\\
50294.823&  $-$925&  34&  $-$105& 315&  4467& 174&26\\
50615.858&  $-$980&  39&  $-$388& 244&  4822& 167&27\\
50784.530&  $-$962&  42&    81& 220&  4647& 155&27\\[2pt]
\hline\\[-6pt]
\multicolumn{8}{c}{HD 208217}\\[2pt]
\hline\\[-6pt]
49830.891&   965& 131&  1316&2101& 10236&1371& 6\\
49916.869&   586& 120&  7907&2203& 10053&1329& 8\\
49974.784&   668& 187&  1088&2431& 11264&1132& 9\\
50039.605& $-$1178& 259& $-$5377&1386&  9481&1284& 6\\
50183.901&  $-$817& 157& $-$6032&1171& 10489& 872& 9\\
50294.844&   466& 177& $-$3318&1307& 10477&1099& 9\\
50629.835& $-$1843& 214& $-$4202&2055&  9136& 887& 7\\
50784.570&   $-$66& 114& $-$4326&1289& 11048&1006& 9\\[2pt]
\hline\\[-6pt]
\multicolumn{8}{c}{HD 213637}\\[2pt]
\hline\\[-6pt]
50782.554&   230&  63&   431& 490&  5375& 886&14\\[2pt]
\hline\\[-6pt]
\multicolumn{8}{c}{HD 216018}\\[2pt]
\hline\\[-6pt]
49830.918&  1407& 105&  1613& 833&  6414&1008& 9\\
49916.887&  1386& 112&  1708& 687&  8030&1045& 7\\
50039.559&  1380& 175&  1800& 918&  6335&1508& 9\\
50294.882&  1546& 137&   $-$89&1080&  5497&1538& 9\\
50629.860&  1374& 146&  $-$142& 977&  6653&1223& 9\\
50784.550&  1377& 138&   446& 858&  6810&1081& 9\\[2pt]
\hline
\end{tabular}}\\
\end{tabular*}}
\end{table*}

The narrow wavelength range covered by most of our high-resolution,
natural light spectra does not allow us to use a similar approach to determine
radial velocities from their analyses. In these spectra, the
unpolarised wavelength $\lambda_I$ of the Fe~{\sc ii}
$\lambda\,6149.2$ line was computed by averaging the
wavelengths of its blue and red split components as measured for the
determination of the mean magnetic field modulus as follows:
\begin{equation}
\lambda_I=(W_{\lambda,{\rm b}}\,\lambda_{\rm b}+W_{\lambda,{\rm
    r}}\,\lambda_{\rm r})/(W_{\lambda,{\rm b}}+W_{\lambda,{\rm r}})\,.
\label{eq:rv6149}
\end{equation}
Then, the wavelength difference $\lambda_I-\lambda_0$, where
$\lambda_0=6149.258$\,\AA\ is the nominal wavelength of the transition,
was converted to a radial velocity value in the standard
manner.
The notations $W_{\lambda,{\rm b}}$ and $W_{\lambda,{\rm r}}$ refer
to the equivalent widths of the measured parts of each line
component. Indeed, the values of those equivalent widths depend on the
limits between which the 
direct integration of the line was performed, or between which the
multiple Gaussians were fitted to the observed profile. These limits
in turn depend on line blends possibly affecting the Fe~{\sc ii}
$\lambda\,6149.2$ line. As a result, the measured equivalent widths of
the split components may be significantly different from (in general,
smaller than) their actual values (which would be measured in an ideal
case). The approach described in this paragraph was applied to the
high-resolution unpolarised spectra analysed in the present paper and to the observations presented in
\citetalias{1997A&AS..123..353M}. A few of the latter 
were omitted, since they had been obtained as part of observing runs
lacking absolute wavelength calibrations suitable for reliable radial
velocity determinations. 

Like for the mean field modulus, as the radial velocity is determined
from consideration of a single line, it is not straightforward to
estimate the measurement uncertainties. The discussion of this issue
is postponed to Sect.~\ref{sec:binarity}.

\section{Results}
\label{sec:results}
\subsection{Individual measurements}
\label{sec:res_ind}
The results of our measurements of the mean magnetic field modulus are
presented Table~\ref{tab:hm}. Column~1 gives the heliocentric Julian
date of mid-observation. The value of the field modulus appears in
Col.~2, and the code of the instrumental configuration used for the
observation (as per Table~\ref{tab:runs}) is listed in Col.~3.

The values of the three magnetic field moments determined from the
analysis of our CASPEC spectropolarimetric observations are given in
Table~\ref{tab:hzxq}: the mean longitudinal field in Col.~2, the
crossover in Col.~4, and the mean quadratic field in Col.~6. Their
standard errors appear in Cols.~3, 5, and 7, respectively. Column~8
gives the number of spectral lines that were used in their determination.

\stepcounter{table}
 
Table~\arabic{table}, available at the CDS, presents all the radial
velocity measurements from the 
spectra analysed in this paper as well as from the observations of
\citet{1991A&AS...89..121M}, \citet{1997A&AS..124..475M} and
\citetalias{1997A&AS..123..353M}. 
Columns~1 to 3 contain the heliocentric Julian date of
the observation, the heliocentric radial velocity value, and the
adopted value of its uncertainty, respectively (see Sect.~\ref{sec:binarity}). 

\begin{table*}[!t]
\caption{Variation of the mean magnetic field modulus: least-squares
  fit parameters (measurements from this paper and from \citetalias{1997A&AS..123..353M})}
\label{tab:mfit}
\begin{tabular*}{\textwidth}[]{@{}@{\extracolsep{\fill}}rcccccrcc
@{\extracolsep{0pt}}@{}}
\hline\hline \\[-4pt]
\multicolumn{1}{c}{HD/HDE}&$M_0\pm\sigma$&$M_1\pm\sigma$
&$\phi_{M_1}\pm\sigma$&$M_2\pm\sigma$
&$\phi_{M_2}\pm\sigma$&\multicolumn{1}{c}{$\nu$}&$\chi^2/\nu$&$R$\\
&(G)&(G)&&(G)\\[4pt]
\hline \\[-4pt]
  2453&$\fo3742\pm\fo10$&$\ft73\pm\fo12$&$0.012\pm0.029$&&&10&2.0&0.90\\
 12288&$\fo8004\pm\fo36$&$\fo399\pm\fo37$&$0.979\pm0.022$&$\fo{\it
   77}\pm\fo{\it 45}$&${\it 0.688}\pm{\it 0.088}$&23&1.2&0.94\\
 14437&$\fo7575\pm\fo45$&$\fo395\pm\fo62$&$0.078\pm0.025$&&&29&1.5&0.78\\
 18078&$\fo3450\pm\fo60$&$\fo993\pm105$&$0.000\pm0.012$&&&6&2.1&0.97\\
 51684&$\fo6044\pm\ft9$&$\fo260\pm\fo14$&$0.999\pm0.008$&&& 7&1.3&0.99\\
 59435&$\fo3036\pm\fo22$&$\fo985\pm\fo33$&$0.997\pm0.005$&$221\pm\fo43$&$0.007\pm0.021$&23&1.0&0.99\\
 61468&$\fo6772\pm\fo34$&$1103\pm\fo37$&$0.001\pm0.006$&&&12&7.8&0.99\\
 65339&$12301\pm224$&$4006\pm275$&$0.730\pm0.014$&${\it 713}\pm{\it
   306}$&${\it 0.353}\pm{\it 0.067}$&25&1.5&0.95\\
 70331&$12385\pm\fo69$&$\fo538\pm\fo93$&$0.664\pm0.027$&${\it
   185}\pm\fo{\it 90}$&${\it 0.445}\pm{\it 0.081}$&33&1.5&0.74\\
 70331&$12392\pm\fo62$&$\fo567\pm\fo92$&$0.062\pm0.025$&&&35&1.6&0.70\\
 75445&$\fo2974\pm\ft7$&$\ft70\pm\fo11$&$0.273\pm0.021$&$\fo69\pm\fo10$&$0.885\pm0.023$&11&0.8&0.92\\
 81009&$\fo8420\pm\fo20$&$\fo858\pm\fo29$&$0.478\pm0.005$&&&39&1.3&0.98\\
 93507&$\fo7154\pm\fo20$&$\fo304\pm\fo28$&$0.500\pm0.015$&&&27&1.0&0.90\\
 94660&$\fo6232\pm\ft9$&$\fo163\pm\fo14$&$0.464\pm0.011$&$\fo{\it
   35}\pm\fo{\it 12}$&${\it 0.572}\pm{\it 0.053}$&20&1.6&0.95\\
116114&$\fo5961\pm\ft6$&$\ft33\pm\ft9$&$0.999\pm0.038$&&&21&1.3&0.64\\
126515&$12726\pm\fo54$&$3258\pm\fo69$&$0.015\pm0.004$&${\it
  233}\pm\fo{\it 80}$&${\it 0.827}\pm{\it 0.050}$&17&3.9&1.00\\
137909&$\fo5457\pm\ft8$&$\fo216\pm\fo10$&$0.298\pm0.008$&$\fo50\pm\fo10$&$0.508\pm0.034$&41&1.1&0.97\\
142070&$\fo4919\pm\fo16$&$\fo131\pm\fo23$&$0.898\pm0.028$&$143\pm\fo21$&$0.994\pm0.026$&19&1.4&0.95\\
144897&$\fo8986\pm\fo34$&$\fo561\pm\fo51$&$0.500\pm0.014$&&&25&1.1&0.91\\
318107&$14468\pm\fo68$&$1343\pm100$&$0.000\pm0.011$&$642\pm\fo92$&$0.001\pm0.025$&31&1.6&0.94\\
187474&$\fo5417\pm\fo10$&$\fo627\pm\fo14$&$0.458\pm0.003$&$232\pm\fo14$&$0.951\pm0.009$&35&5.3&0.99\\
188041&$\fo3660\pm\fo10$&$\ft{\it 37}\pm\fo{\it 14}$&${\it
  0.052}\pm{\it 0.059}$&&&15&1.7&0.58\\
192678&$\fo4664\pm\fo14$&$\fo108\pm\fo19$&$0.926\pm0.029$&$\fo{\it
  32}\pm\fo{\it 19}$&${\it 0.944}\pm{\it 0.095}$&30&1.0&0.74\\
335238&$\fo9328\pm\fo91$&$1818\pm105$&$0.968\pm0.013$&$997\pm112$&$0.938\pm0.018$&13&0.6&0.98\\
200311&$\fo8404\pm152$&$\fo603\pm102$&$0.986\pm0.077$&${\it
  398}\pm{\it 173}$&${\it 0.084}\pm{\it 0.053}$&30&1.4&0.86\\
208217&$\fo7690\pm\fo66$&$\fo640\pm\fo96$&$0.851\pm0.022$&$427\pm\fo92$&$0.913\pm0.034$&33&1.2&0.83\\[4pt]
\hline
\end{tabular*}
\end{table*}

\begin{table*}
\caption{Variation of the mean longitudinal magnetic field: least-squares
  fit parameters for stars with $\Hz$ data in this paper}
\label{tab:zfit}
\begin{tabular*}{\textwidth}[]{@{}@{\extracolsep{\fill}}rcccccrcc
@{\extracolsep{0pt}}@{}}
\hline\hline \\[-4pt]
\multicolumn{1}{c}{HD/HDE}&$Z_0\pm\sigma$&$Z_1\pm\sigma$
&$\phi_{Z_1}\pm\sigma$&$Z_2\pm\sigma$
&$\phi_{Z_2}\pm\sigma$&\multicolumn{1}{c}{$\nu$}&$\chi^2/\nu$&$R$\\
&(G)&(G)&&(G)\\[4pt]
\hline \\[-4pt]
  2453&$\fo{-}745\pm\fo19$&$\fo293\pm\fo25$&$0.500\pm0.016$&&&30&1.6&0.91\\
 51684&$-1548\pm\fo60$&$\fo360\pm\fo40$&$0.312\pm0.030$&&& 4&0.2&0.98\\
 61468&$-1564\pm\fo48$&$\fo981\pm\fo55$&$0.589\pm0.011$&&& 2&0.2&1.00\\
 81009&$\fmi1781\pm\fo46$&$\fo451\pm\fo63$&$0.538\pm0.023$&&&10&0.8&0.91\\
 93507&$\fmi2188\pm\fo71$&$\fo799\pm118$&$0.987\pm0.018$&${\it
   188}\pm{\it 95}$&${\it 0.462}\pm{\it 0.089}$& 7&1.4&0.95\\
 94660&$-1890\pm\fo36$&$\ft{\it 80}\pm\fo{\it 46}$&${\it 0.415}\pm{\it
   0.084}$&&& 9&1.0&0.54\\
116114&$-1763\pm\fo24$&$\ft{\it 84}\pm\fo{\it 33}$&${\it 0.466}\pm{\it
  0.043}$&&& 4&0.3&0.85\\
116458&$-1627\pm\fo29$&$\fo280\pm\fo33$&$0.001\pm0.027$&&&18&2.7&0.89\\
126515&$\fo{-}434\pm\fo51$&$1922\pm\fo79$&$0.457\pm0.005$&$568\pm75$&$0.648\pm0.020$&14&1.6&0.99\\
137909&$\fo{-}102\pm\fo30$&$\fo705\pm\fo44$&$0.991\pm0.009$&&&18&1.7&0.97\\
137949&$\fmi1618\pm\fo31$&$\fo256\pm\fo41$&$0.000\pm0.028$&&&31&4.4&0.76\\
142070&$\fmi\fo246\pm\fo27$&$\fo445\pm\fo36$&$0.015\pm0.013$&&& 8&1.8&0.98\\
144897&$\fmi1840\pm\fo44$&$\fo365\pm\fo68$&$0.023\pm0.025$&${\it
  189}\pm{\it 69}$&${\it 0.441}\pm{\it 0.050}$& 8&2.0&0.90\\
318107&$\fmi2624\pm218$&$1496\pm252$&$0.456\pm0.041$&&& 3&\llap{1}5.5&0.96\\
187474&$\fo{-}149\pm\fo50$&$1973\pm\fo67$&$0.971\pm0.006$&$196\pm65$&$0.018\pm0.056$&15&4.4&0.99\\
188041&$\fmi\fo710\pm\fo24$&$\fo284\pm\fo37$&$0.570\pm0.018$&&&74&1.4&0.65\\
208217&$\fo{-}491\pm159$&$1333\pm175$&$0.001\pm0.022$&&& 5&3.1&0.96\\[4pt]
\hline
\end{tabular*}
\end{table*}

\begin{table*}[!t]
\caption{Variation of the crossover: least-squares fit parameters} 
\label{tab:xfit}
\begin{tabular*}{\textwidth}[]{@{}@{\extracolsep{\fill}}rcccccrcc
@{\extracolsep{0pt}}@{}}
\hline\hline \\[-4pt]
\multicolumn{1}{c}{HD/HDE}&$X_0\pm\sigma$&$X_1\pm\sigma$
&$\phi_{X_1}\pm\sigma$&$X_2\pm\sigma$
&$\phi_{X_2}\pm\sigma$&\multicolumn{1}{c}{$\nu$}&$\chi^2/\nu$&$R$\\
&(km\,s$^{-1}$\,G)&(km\,s$^{-1}$\,G)&&(km\,s$^{-1}$\,G)\\[4pt]
\hline \\[-4pt]
 81009&$\fo{-}601\pm\fo317$&$1848\pm\fo436$&$0.807\pm0.039$&&& 9&0.8&0.83\\
 93507&${\fo}\fmi948\pm\fo336$&$1598\pm\fo473$&$0.920\pm0.043$&&& 9&0.7&0.73\\
 94660&$\fo{-}169\pm\fo191$&$\fo770\pm\fo221$&$0.713\pm0.047$&&& 9&0.4&0.75\\
137909&$\fo{-}684\pm\fo227$&$1911\pm\fo229$&$0.283\pm0.028$&$\fo{\it
  884}\pm\fo{\it 327}$&${\it 0.706}\pm{\it 0.046}$&16&0.9&0.92\\
142070&$\fo{-}383\pm\fo156$&$4975\pm\fo240$&$0.249\pm0.006$&&& 8&1.0&0.99\\
144897&$\fmi1551\pm\fo225$&&&$1364\pm\fo309$&$0.852\pm0.020$& 9&0.5&0.85\\
318107&$\fmi\fo606\pm1335$&$4376\pm1457$&$0.554\pm0.059$&&& 3&0.4&0.87\\
208217&$\fmi\fo996\pm\fo758$&$7594\pm\fo934$&$0.188\pm0.014$&&& 5&0.5&0.97\\[4pt]
\hline
\end{tabular*}
\end{table*}

\begin{table*}
\caption{Variation of the mean quadratic magnetic field: least-squares
  fit parameters}
\label{tab:qfit}
\begin{tabular*}{\textwidth}[]{@{}@{\extracolsep{\fill}}rcccccrcc
@{\extracolsep{0pt}}@{}}
\hline\hline \\[-4pt]
\multicolumn{1}{c}{HD/HDE}&$Q_0\pm\sigma$&$Q_1\pm\sigma$
&$\phi_{Q_1}\pm\sigma$&$Q_2\pm\sigma$
&$\phi_{Q_2}\pm\sigma$&\multicolumn{1}{c}{$\nu$}&$\chi^2/\nu$&$R$\\
&(G)&(G)&&(G)\\[4pt]
\hline \\[-4pt]
  2453&$\fo4349\pm\fo69$&$\fo{\it 331}\pm\fo{\it 131}$&${\it
    0.771}\pm{\it 0.038}$&&& 2&0.1&0.87\\
 61468&$\fo6474\pm301$&$1226\pm\fo346$&$0.016\pm0.066$&&& 2&1.2&0.95\\
 81009&$\fo9829\pm\fo67$&$\fo581\pm\ft94$&$0.537\pm0.026$&&& 9&0.7&0.90\\
 93507&$\fo7484\pm180$&$1168\pm\fo287$&$0.524\pm0.030$&$\fo{\it
   407}\pm{\it 233}$&${\it 0.868}\pm{\it 0.090}$& 7&0.8&0.85\\
 94660&$\fo7212\pm185$&$1456\pm\fo226$&$0.668\pm0.029$&$\fo{\it
   385}\pm{\it 131}$&${\it 0.454}\pm{\it 0.074}$& 7&0.6&0.95\\
126515&$16071\pm656$&$5308\pm1043$&$0.993\pm0.022$&${\it 2237}\pm{\it
  940}$&${\it 0.868}\pm{\it 0.067}$&14&1.3&0.82\\
144897&$10215\pm\fo76$&$\fo715\pm\fo109$&$0.484\pm0.021$&$\fo{\it
  248}\pm{\it 110}$&${\it 0.930}\pm{\it 0.065}$& 8&0.8&0.92\\
%187474&$\fo7081\pm176$&$\fo811\pm\fo247$&$0.210\pm0.049$&$\fo477\pm239$&$0.186\pm0.074$&15&3.7&0.64\\
208217&$\fo9824\pm210$&$1053\pm\fo246$&$0.885\pm0.040$&&& 5&0.1&0.91\\[4pt]
\hline
\end{tabular*}
\end{table*}

\subsection{Variation curves of the moments of the magnetic field}
\label{sec:hvar}
For those stars with a known rotation period, or for which a probable
value of this period could be derived with reasonable confidence, we
fitted our magnetic measurements against rotation phase $\phi$ with a
cosine curve, 
\begin{equation}
A_B(\phi)=A_0+A_1\,\cos[2\pi\,(\phi-\phi_1)]\,, 
\label{eq:fit1}
\end{equation}
or with the superposition of a cosine and of its first harmonic,
\begin{eqnarray}
A_B(\phi)=A_0&+&A_1\,\cos[2\pi\,(\phi-\phi_1)]\nonumber\\
&+&A_2\,\cos[2\pi\,(2\phi-\phi_2)]\,.
\label{eq:fit2}
\end{eqnarray}
In these formulae, $A_B$ represents any of the magnetic moments of
interest ($\Hm$, $\Hz$, $\xover$, $\Hq$). The rotation phase is
calculated using the values of the phase origin, HJD$_0$, and of the
rotation period, $\Prot$, that appear in
Table~\ref{tab:stars}. 
The mean value $A_0$, the amplitude(s) $A_1$ (and $A_2$),
and the phase(s) $\phi_1$ (and $\phi_2$) of the variations are
determined through a least-squares fit of the field moment
measurements by a function of the form given in either
Eq.~(\ref{eq:fit1}) or (\ref{eq:fit2}). This fit is weighted by the
inverse of the square of the uncertainties of the individual
measurements. 

\begin{table*}
\caption{New or revised orbital elements of spectroscopic
  binaries. The second row for each entry gives its standard deviation.
}
\label{tab:orbits}
\begin{tabular*}{\textwidth}[]{@{}@{\extracolsep{\fill}}rrrrrrrrrrr
@{\extracolsep{0pt}}@{}}
\hline\hline \\[-4pt]
\multicolumn{1}{c}{HD/HDE}&\multicolumn{1}{c}{$P_{\rm orb}$}&\multicolumn{1}{c}{$T_0$}&\multicolumn{1}{c}{$e$}&\multicolumn{1}{c}{$V_0$}&\multicolumn{1}{c}{$\omega$}&\multicolumn{1}{c}{$K$}&\multicolumn{1}{c}{$f(M)$}&\multicolumn{1}{c}{$a\,\sin
  i$}&\multicolumn{1}{c}{$N$}&\multicolumn{1}{c}{${\rm O}-{\rm C}$}\\
&\multicolumn{1}{c}{(d)}&\multicolumn{1}{c}{(HJD)}&&\multicolumn{1}{c}{(km\,s$^{-1}$)}&\multicolumn{1}{c}{($^\circ$)}&\multicolumn{1}{c}{(km\,s$^{-1}$)}&$(M_{\sun})$&\multicolumn{1}{c}{($10^6$\,km)}&&\multicolumn{1}{c}{(km\,s$^{-1}$)}\\[4pt]
\hline \\[-4pt]
9996&272.81&2444492.8&0.512&0.36&19.5&11.51&0.0273&37.1&57&1.37\\
&0.15&2.0&0.021&0.19&3.1&0.28&0.0024&1.1&&\\[4pt]
12288&1546.5&2444488.3&0.342&$-$53.29&122.5&9.16&0.102&183.1&59&0.65\\
&4.6&9.3&0.013&0.11&3.4&0.15&0.051&3.2&&\\[4pt]
29578&926.7&2449148.3&0.377&15.66&225.3&9.74&0.070&115.0&27&0.60\\
&2.8&8.1&0.018&0.14&8.1&0.28&0.021&3.4&&\\[4pt]
50169&1764&2448529&0.47&13.35&44.3&1.90&0.0009&40.7&42&0.51\\
&49&36&0.10&0.12&9.1&0.32&0.0223&7.4&&\\[4pt]
55719&46.31803&2441672.98&0.1459&$-$6.46&195.0&48.27&0.522&30.42&76&2.15\\
&0.00078&0.38&0.0071&0.25&3.0&0.37&0.012&0.24&&\\[4pt]
61468&27.2728&2445994.20&0.1723&32.04&181.7&19.35&0.1309&13.47&20&0.61\\
&0.0014&0.29&0.0057&0.17&2.2&0.14&0.0029&0.29&&\\[4pt]
94660&848.96&2445628.6&0.4476&18.53&264.5&17.94&0.3631&187.3&47&0.36\\
&0.13&1.7&0.0049&0.06&0.8&0.11&0.0075&1.2&&\\[4pt]
116458&126.233&2441422.3&0.143&2.68&70.7&14.13&0.0357&24.27&65&1.08\\
&0.012&2.1&0.014&0.15&6.4&0.19&0.0015&0.33&&\\[4pt]
187474&689.68&2435139.0&0.4856&$-$1.771&7.1&11.06&0.0646&91.7&108&0.75\\
&0.10&2.3&0.0092&0.074&1.3&0.14&0.0027&1.2&&\\[4pt]
\hline
\end{tabular*}
\end{table*}

The forms adopted for the fit functions are the same as in our
previous studies of magnetic fields of Ap stars
\citep[e.g.][]{1997A&AS..124..475M}. This is not an arbitrary choice
in the sense 
that these functions generally represent well the observed
behaviour of the various field moments. Nevertheless they should be
considered only as convenient tools to characterise the field
variations in a simple way that lends itself well to inferring
statistical properties of the magnetic fields of the studied
stars, but they should not be overinterpreted in terms of the actual
physical properties of the field of any single object. 

The results of the fits are presented in Table~\ref{tab:mfit}, for
$\Hm$, the amplitudes are denoted as $M_i$, $i=0$, 1, 2, and the phases are listed as
$\phi_{M_i}$, $i=1$, 2; Table~\ref{tab:zfit}, for $\Hz$, the amplitudes are denoted as
$Z_i$ and the phases are listed as $\phi_{Z_i}$; in Table~\ref{tab:xfit}, for $\xover$,
the amplitudes are denoted as $X_i$ and the phases are listed as $\phi_{X_i}$; and in Table~\ref{tab:qfit},
for $\Hq$, the amplitudes are denoted as $Q_i$ and the phases are listed as $\phi_{Q_i}$. Columns~7 to 9
of these tables give the number of degrees of freedom about the fit
$\nu$, the reduced $\chi^2$ of the fit $\chi^2/\nu$, and the multiple
correlation coefficient $R$. Fits were only computed when at least five
or seven measurements of the field moment (for fits of the form given by
Eqs.~(\ref{eq:fit1}) or (\ref{eq:fit2}), respectively), sufficiently spread over
the rotation period, were obtained for the considered star. The
following criteria were applied to decide whether or not the first
harmonic should be included in the fits whose results
appear in Tables~\ref{tab:mfit} to \ref{tab:qfit}:
\begin{enumerate}
\item The harmonic was included if the value of the coefficient $A_2$
  in a fit of the form given in Eq.~(\ref{eq:fit2}) was formally significant at
  the $3\sigma$ level.
\item In a number of instances, the first harmonic was included
    even though the significance level of $A_2$ is below $3\sigma$, for
    reasons specified on a case-by-case basis in the subsection of
    Appendix~\ref{sec:notes} devoted to the star of interest.
\end{enumerate}
Fitted amplitudes that are not formally significant and their associated
phases appear in italics in Tables~\ref{tab:mfit} to
\ref{tab:qfit}. 

In a few cases (the $\Hz$ variations of HD~94660 and HD~116114, the
$\Hq$ curve of HD~2453 and the $\Hm$ curve of HD~188041), the
parameters corresponding to a fit by a 
single cosine are given in the respective tables, even though its
amplitude is not formally significant. These exceptions are discussed
individually in the respective subsections of
Appendix~\ref{sec:notes}. Finally, for the variation of the crossover
of HD~144897, the only significant term of the fit is the first
harmonic, so that the fundamental was not included. 

Plots of the fitted curves for each star are shown in the figures of
the respective subsection of Appendix~\ref{sec:notes}. When all the
parameters of the fit, as given in Tables~\ref{tab:mfit} to
\ref{tab:qfit}, are formally significant, the fitted curve is
represented by a red, solid line. A blue, long-dashed line is used
when the fit includes a first harmonic that is not formally
significant. Non-significant fits by the fundamental only appear as
green, short-dashed lines.

\subsection{Orbital solutions}
\label{sec:orbits}
For those stars showing radial velocity variations for which enough
data suitably distributed across the orbital phases were available, we
computed orbital solutions using the Li\`ege Orbital Solution
Package\footnote{\texttt{http://www.stsci.edu/\textasciitilde{}hsana/losp.html}}
(LOSP). The derived elements are given in Table~\ref{tab:orbits}. 

Curves representing the orbital solutions of this table are plotted in
the figures of the respective subsections of
Appendix~\ref{sec:notes}. The lower panel of these figures shows the
deviations of the individual measurements from the computed radial
velocity curve.

\section{Discussion}
\label{sec:discussion}
We presented here 271 new measurements of the mean magnetic field
modulus of 43 Ap stars with resolved magnetically split lines, which
complement the 752 measurements obtained in
\citetalias{1997A&AS..123..353M} for 40 of these 
stars. For 34 stars, we recorded at least one spectrum
in right and left circular polarisation. From the analysis of these
data, we carried out 231 determinations of the mean longitudinal
magnetic field and of the crossover, and 229 determinations of the
mean quadratic magnetic field, or of an upper limit of this
  magnetic moment. (In one
star, HD~188041, the quadratic field is too small to allow any
meaningful quantitative information to be derived about it.) For many of the
studied stars, the longitudinal field had never been systematically
measured before. 

Measurements of the field modulus covering a whole rotation period
have now been obtained for more than half of our sample. For more than
a third of our sample,
the variations of the other field moments considered here have also
been characterised.

Thus we accumulated an unprecedented sample of homogeneous data
that lends itself to the derivation of information of statistical
nature on the magnetic fields of the Ap stars, and on their relation
with their other physical properties. The rest of this section is 
devoted to the discussion of these issues.

\begin{table*}
\caption{Magnetic parameters of Ap stars with resolved magnetically
  split lines: stars with $\Hm$ data in this paper or in
  \citetalias{1997A&AS..123..353M}} 
\label{tab:summary}
\begin{tabular*}{\textwidth}[]{@{}@{\extracolsep{\fill}}rrrrcrrrrrr
@{\extracolsep{0pt}}@{}}
\hline\hline\\[-4pt]
\multicolumn{1}{c}{HD/HDE}&\multicolumn{1}{c}{$N_{\rm
    m}$}&\multicolumn{1}{c}{$\Hav$}&\multicolumn{1}{c}{$\sigma_B$}&$q$&\multicolumn{1}{c}{$N_z$}&\multicolumn{1}{c}{$\Hzrms$}&\multicolumn{1}{c}{$r$}&\multicolumn{1}{c}{$N_{\rm
    q}$}&\multicolumn{1}{c}{$\Hqav$}&\multicolumn{1}{c}{$p$}\\
&&\multicolumn{1}{c}{(G)}&\multicolumn{1}{c}{(G)}&&&\multicolumn{1}{c}{(G)}&&&\multicolumn{1}{c}{(G)}\\[4pt]
\hline\\[-4pt]
   965&11&{\it 4315}&30&{\it 1.08}&7&982&&7&7686\\
  2453&14& 3722&20&1.03&6&808&0.44&6&3962&1.16\\
  9996&14&{\it 4717}&30&{\it 1.34}&{\it 63}&{\it 713}&{\it $-$0.35}\\
 12288&28& 7978&120&1.11&{\it 20}&{\it 1643}&{\it 0.15}\\
 14437&32& 7549&200&1.11&{\it 36}&{\it 1930}&{\it 0.35}\\
 18078& 9&3508&120&1.81&{\it 25}&{\it 692}&{\it $-$0.92}\\
 29578&18&{\it 3085}&40&{\it 1.53}&9&785&&9&3787\\
 47103& 7&{\it 17321}&50&{\it 1.06}&6&2778&&6&16756\\
 50169&21&{\it 5030}&30&{\it 1.31}&9&832&&9&6431\\
 51684&10& 6027&50&1.09&7&1366&0.62&7&6605\\
 55719&43&{\it 6465}&60&{\it 1.13}&12&719&&12&7813\\
 59435&19& 3059&115&1.87\\
 61468&15& 6976&30&1.39&5&1884&0.23&5&6435&1.47\\
 65339&30&12770&900&1.95&{\it 19}&{\it 3005}&{\it $-$0.72}\\
 70331&38&12416&300&1.10&16&2570&0.90&16&16290\\
 75445&16& 2987&30&1.08&3&135&&3&4275\\
 81009&42& 8382&115&1.23&13&1873&0.60&13&10356&1.13\\
 93507&30& 7146&110&1.09&12&2217&0.43&12&7679&1.43\\
 94660&25& 6183&30&1.06&12&1911&0.92&12&6880&1.53\\
110066& 7&{\it 4088}&30&{\it 1.03}&3&117\\
116114&24& 5960&25&1.01&7&1815&0.91&7&7646\\
116458&21& 4677&30&1.00&21&1684&0.71&21&5549\\
119027&13&{\it 3166}&100&{\it 1.22}&1&510&&1&3655\\
126515&22&12417&120&1.69&19&1660& $-$0.64&19&16339&2.17\\
134214&30&{\it 3092}&40&{\it 1.07}&8&391&&7&3977\\
137909&46& 5476&45&1.09&21&604&$-$0.64&21&6536\\
137949&20&{\it 4672}&25&{\it 1.02}&9&1629&&9&4680\\
142070&24& 4898&55&1.10&13&376&$-$0.23&13&4733\\
144897&28& 8992&170&1.13&13&1892&0.59&13&10163&1.16\\
150562&12&{\it 4860}&35&{\it 1.06}&1&1206&&1&5710\\
318107&36&14192&300&1.22&6&2551&0.27&6&21209\\
165474&37&{\it 6625}&25&{\it 1.14}&8&233&&8&8349\\
166473&33& 7126&80&1.45&10&1626&&10&7923\\
177765& 8&{\it 3415}&20&{\it 1.02}\\
187474&40& 5423&25&1.27&20&1476&$-$0.98&20&7128\\
188041&18& 3653&25&1.02&9&1196&0.43\\
192678&35& 4764&80&1.05&{\it 14}&{\it 1451}&{\it 0.81}\\
335238&18& 8678&300&1.54&3&1399&&3&11895\\
200311&35& 8540&300&1.22&{\it 24}&{\it 1410}&{\it $-$0.99}\\
201601&28&{\it 3881}&35&{\it 1.10}&18&1000&&16&5241\\
208217&38& 7872&350&1.27&8&961&$-$0.46&8&10273&1.24\\
213637& 4&{\it 5189}&50&{\it 1.05}&1&230&&1&5375\\
216018&23&{\it 5602}&40&{\it 1.04}&9&1381&&9&6769\\[4pt]
\hline
\end{tabular*}
\tablefoot{Each star is identified by its HD
  or HDE number. $N_{\rm m}$ is the number of measurements of the mean
magnetic field modulus from which its average $\Hav$ is computed. For
stars for which a full rotation cycle has not yet been observed, the
latter appears in italics, as does the ratio $q$ between 
the highest and lowest value of $\Hm$. The $\sigma_B$ is the
estimated uncertainty of the $\Hm$ measurements (see
Sect.~\ref{sec:hmdiag} for details). $N_z$ is the number of
measurements from which the rms longitudinal field $\Hzrms$ is
computed, and $r$ is the ratio of the extrema of $\Hz$ (defined in
Sect.~\ref{sec:Hzdisc}); the value of $r$ is given only for those
stars for which $\Hz$ measurements have been obtained throughout a
full rotation cycle. Values of $\Hzrms$ and $r$ computed from mean
longitudinal field measurements from the literature (rather than from
this paper and from \citetalias{1997A&AS..123..353M}) appear in
italics. $N_{\rm q}$ is the number of measurements of the mean
quadratic field from which its average $\Hqav$ is computed; $p$ is the
ratio between the extrema of $\Hq$, which is given only for stars for
which $\Hq$ determinations have been obtained throughout a full
rotation cycle.} 
\end{table*}

\begin{table*}
\caption{Magnetic parameters of Ap stars with resolved magnetically
  split lines: stars with $\Hm$ data from the literature.} 
\label{tab:summary2}
\begin{tabular*}{\textwidth}[]{@{}@{\extracolsep{\fill}}rrrlcrrcr
@{\extracolsep{0pt}}@{}}
\hline\hline\\[-4pt]
\multicolumn{1}{c}{HD/BD}&\multicolumn{1}{c}{$N_{\rm
    m}$}&\multicolumn{1}{c}{$\Hav$}&Ref.&$q$&\multicolumn{1}{c}{$N_z$}&\multicolumn{1}{c}{$\Hzrms$}&Ref.&\multicolumn{1}{c}{$r$}\\
&&\multicolumn{1}{c}{(G)}&&&&\multicolumn{1}{c}{(G)}\\[4pt]
\hline\\[-4pt]
  3988&4&{\it 2650}&1&{\it 1.08}\\
 18610&1&{\it 5700}&2\\
 33629&1&{\it 4760}&3\\
 42075&3&{\it 8540}&3&{\it 1.00}\\
 44226&1&{\it 4990}&3\\
 46665&2&{\it 4630}&3&{\it 1.00}\\
 47009&1&{\it 7360}&3\\
 52847&2&{\it 4440}&3&{\it 1.00}\\
 55540&4&{\it 12730}&3&{\it 1.05}\\
 57040&1&{\it 7500}&1\\
 61513&1&{\it 9200}&1\\
 66318&3&{\it 14500}&4&{\it 1.00}&2&4339&5\\
 69013&1&{\it 4800}&3\\
 70702&1&{\it 15000}&1\\
 72316&1&{\it 5180}&3\\
 75049&14&27149&6&1.20&13&6659&6&0.03\\
 76460&1&{\it 3600}&1\\
 81588&1&{\it 2400}&1\\
 88241&1&{\it 3600}&1\\
 88701&1&{\it 4380}&3\\
 92499&5&{\it 8350}&3 7&{\it 1.04}&3&1144&8\\
 96237&2&{\it 2900}\rlap{:}&3&&1&720&3\\
 97394&1&{\it 3100}&9\\
+0~4535&2&{\it 20900}&10&{\it 1.00}\\ 
110274&3&{\it 4020}&3&{\it 1.17}\\
117290&3&{\it 6380}&3&{\it 1.00}\\
121661&2&{\it 6160}&3&{\it 1.28}\\
135728&2&{\it 3630}&3&{\it 1.00}\\
143487&5&{\it 4480}&3 11 12&{\it 1.12}\\
154708&3&{\it 24500}&13&{\it 1.00}&16&8023&14&0.79\\
157751&1&{\it 6600}&7&&2&4016&8\\
158450&2&{\it 11550}&1&{\it 1.06}&4&{\it 1570}&15\\
162316&1&{\it 6000}&1\\
168767&1&{\it 16500}&1\\
177268&3&{\it 3867}&1&{\it 1.08}\\
178892&4&{\it 17450}&16&{\it 1.05}&18&5513&16&0.28\\
179902&2&{\it 3800}&1&{\it 1.05}\\
184120&2&{\it 5750}&1&{\it 1.02}\\
185204&3&{\it 5600}&1&{\it 1.06}\\
191695&2&{\it 3200}&1&{\it 1.13}\\
215441&11&33592&17&1.08&14&16876&18&0.54\\[4pt]
\hline
\\[-4pt]
\end{tabular*}
\tablefoot{Each star is identified by its HD
  or BD number. $N_{\rm m}$ is the number of measurements of the mean
magnetic field modulus from which its average $\Hav$ is computed. For
stars for which the available $\Hm$ measurements do not adequately
sample the rotation cycle, $\Hav$ and the ratio $q$ between 
the highest and the lowest value of $\Hm$ appear in italics. $N_z$ is the number of
measurements from which the rms longitudinal field $\Hzrms$ is
computed, and $r$ is the ratio of the extrema of $\Hz$ (defined in
Sect.~\ref{sec:Hzdisc}); the value of $r$ is given only for those
stars for which $\Hz$ measurements well distributed across the
rotation phases are available. The
references for the 
$\Hm$ (resp., $\Hz$) data are listed in Col.~4 (resp., Col.~8).}
\tablebib{
(1)~\citet{2012MNRAS.420.2727E};
(2)~\citet{2003A&A...402..729S};
(3)~\citet{2008MNRAS.389..441F};
(4)~\citet{2003A&A...403..645B};
(5)~\citet{2006A&A...450..777B};
(6)~\citet{2010MNRAS.402.1883E};
(7)~\citet{2007MNRAS.378L..16H};
(8)~\citet{2006AN....327..289H};
(9)~\citet{2011MNRAS.415.2233E};
(10)~\citet{2010MNRAS.401L..44E}; 
(11)~\citet{2010MNRAS.404L.104E};
(12)~\citet{2013MNRAS.431.2808K};
(13)~\citet{2009MNRAS.396.1018H};
(14)~\citet{2015A&A...583A.115B};
(15)~\citet{2006MNRAS.372.1804K}; 
(16)~\citet{2006A&A...445L..47R};
(17)~\citet{1969ApJ...156..967P};
(18)~\citet{1978ApJ...222..226B}.}
\end{table*}

\begin{table*}[t]
\caption{Variation of the mean magnetic field modulus: least-squares
  fit parameters for data from the literature.}
\label{tab:mfit-litt}
\begin{tabular*}{\textwidth}[]{@{}@{\extracolsep{\fill}}rcccccrcc
@{\extracolsep{0pt}}@{}}
\hline\hline \\[-4pt]
\multicolumn{1}{c}{HD/HDE}&$M_0\pm\sigma$&$M_1\pm\sigma$
&$\phi_{M_1}\pm\sigma$&$M_2\pm\sigma$
&$\phi_{M_2}\pm\sigma$&\multicolumn{1}{c}{$\nu$}&$\chi^2/\nu$&$R$\\
&(G)&(G)&&(G)\\[4pt]
\hline \\[-4pt]
 75049&$27639\pm112$&$2368\pm153$&$0.488\pm0.011$&$613\pm159$&$0.579\pm0.041$&8&3.0&0.99\\
215441&$33349\pm203$&$1257\pm272$&$0.986\pm0.037$&&&8&4.3&0.86\\[4pt] 
\hline
\end{tabular*}
\end{table*}

\subsection{The mean magnetic field modulus}
\label{sec:Hmdisc}
We use the average of all our measurements of the
mean magnetic field modulus, $\Hav$, to characterise the intensity of
the stellar magnetic field with a single number. This is justified by
the fact that $\Hm$ depends little on the geometry of the observation
(the angle between the magnetic axis and the line of sight), and that
in most cases, its amplitude of variation is fairly small compared to
its average value (see below). For stars for which the $\Hm$ data cover a full
rotation cycle, we could in principle use the independent term $M_0$
of the fitted variation curve instead of $\Hav$. We did not do so, so
that we deal in the same manner with all stars. In any event, the
difference between $M_0$ and $\Hav$ is small: less than 100\,G in most
cases. Even the highest value of this difference, 650\,G, obtained for
HDE~335238 owing to the very unfortunate phase coverage of our data, and
accordingly exceptional, amounts only to less than 5\% of $\Hav$ and
is mostly irrelevant for the following 
discussion. The values of $\Hav$ for all the Ap stars with resolved
magnetically split lines   for which we present magnetic field
measurements in this paper are given in Col.~3 of 
Table~\ref{tab:summary}. They are based exclusively on the measurements
of the mean magnetic field modulus of this paper and of
\citetalias{1997A&AS..123..353M}; the 
number $N_{\rm m}$ of these measurements for each star appears in Col.~2. 
For those stars for which a full rotation
period has not yet been observed, the  value of $\Hav$ is italicised. For
these stars with incomplete phase coverage, one should keep in mind
that $\Hav$ is only a preliminary order of magnitude of the actual
field strength.

The values of $\Hav$ for the 41 stars with resolved magnetically split
lines that were not studied in detail by us (i.e. the stars from
Table~\ref{tab:stars_litt}) are summarised in
Table~\ref{tab:summary2}. These values, which appear in Col.~3, were
computed from $\Hm$ measurements published in the references listed in
Col.~4; the number $N_{\rm m}$ of these measurements is given in
Col.~2. As in Table~\ref{tab:summary}, italics are used to
distinguish those values of $\Hav$ that are based on data with
incomplete phase coverage. Field modulus measurements with good
sampling of the rotation period are available only for 2 of  the 41
stars of Table~\ref{tab:summary2}: HD~75049 and HD~215441. One should
also bear in mind that the $\Hm$ determinations for the stars of this
table were based on much less homogeneous observational material than
our measurements, among them many spectra of considerably lower
spectral resolution than those that we analysed. Furthermore, 
different methodologies and different diagnostic lines were
used to determine $\Hm$.

In some works
  \citep[e.g.][]{2008MNRAS.389..441F}, several values of 
  $\Hm$ are given for a specific observation, corresponding to
  different analyses of different sets of lines. In such cases, for
  the compilation of Table~\ref{tab:summary2}, we used preferentially
  the value of $\Hm$ obtained by using Eq.~(\ref{eq:Hm}) to interpret the
  measured wavelength shift between the blue and red components of the
  Fe~{\sc ii} $\lambda\,6149$ doublet, whenever it was
  available. Indeed this value should be almost approximation free and
  model independent (see \citetalias{1997A&AS..123..353M});
  furthermore, its consideration improves 
  consistency with our field modulus measurements. However this
  diagnostic line cannot be used in HD~75049, HD~154708, and
  BD~+0~4535 because their very strong magnetic fields put the
    transition too 
  much into the partial Paschen-Back regime to allow the Zeeman regime
  approximation of Eq.~(\ref{eq:Hm}) to be reliably applicable
  \citep[e.g.][]{1990A&A...232..151M}. For HD~75049, we use the
  values of $\Hm$ determined 
  from analysis of the Fe~{\sc ii} $\lambda\,5018$ line. For
  HD~154708, the Nd~{\sc iii} $\lambda\,6145.1$ line and a small set
  of iron-peak lines were used \citep{2005A&A...440L..37H}. For
  BD~+0~4535, we adopt the average $\Hm$ values of Table~1 of
  \citet{2010MNRAS.401L..44E}, based on a set of rare earth
  lines. Also, Fe~{\sc ii} $\lambda\,6149$ was not resolved
  in the \citet{2012MNRAS.420.2727E}
  spectrum of HD~3988, so that $\Hm$ 
  in this star was determined from the Fe~{\sc i} $\lambda\,6336.8$
  line instead. 

Despite these limitations, the
consideration of the magnetic data of Table~\ref{tab:summary2}  to
complement those of our  main set proves useful to strengthen the 
conclusions that we draw from the latter. For the two stars of this
table for which field modulus measurements that provide a good
sampling of the
rotation period are available in the literature, HD~75049
\citep{2010MNRAS.402.1883E} and HD~215441 \citep{1969ApJ...156..967P},
the parameters of the 
fit of the $\Hm$ variation curve by functions of the form of
Eqs.~(\ref{eq:fit1}) and (\ref{eq:fit2}) are given in
Table~\ref{tab:mfit-litt}. The format of this table is the same as
that of Table~\ref{tab:mfit}. We used the value of the rotation
period and the phase origin listed in Cols.~6 and 8 of
Table~\ref{tab:stars_litt}. 

\begin{figure}[t]
\resizebox{\hsize}{!}{\includegraphics{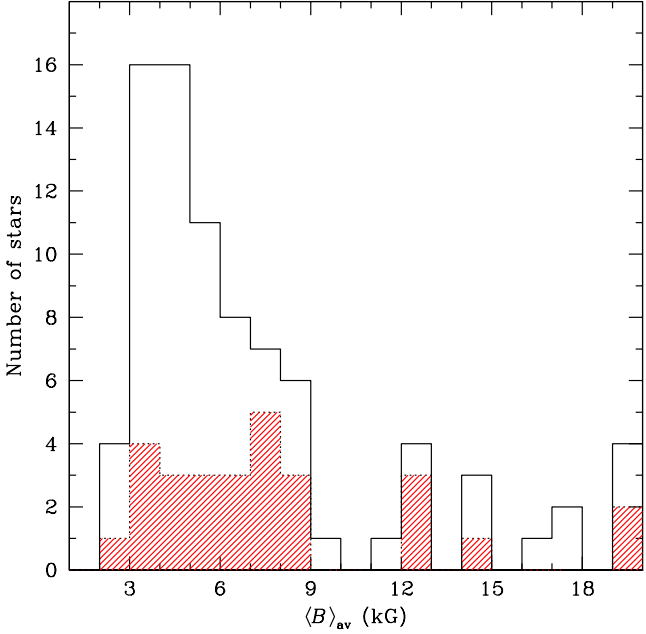}}
\caption{Histogram showing the distribution of the phase-averaged mean
  magnetic field moduli of the 84 Ap stars with resolved magnetically
  split lines presently known. The shaded part of the histogram
  corresponds to those stars whose field modulus has been measured
  throughout their rotation cycle (see text).}
\label{fig:hmhist}
\end{figure}

Figure~\ref{fig:hmhist} shows the distribution of $\Hav$ over the
sample of all known Ap stars with magnetically resolved lines. The
shaded part of the histogram corresponds to those stars for which
measurements provide a good sampling of the whole rotation period. 
The figure is an updated version of Fig.~47 of \citetalias{1997A&AS..123..353M}. The
most intriguing result derived from consideration of that
\citetalias{1997A&AS..123..353M} figure, the 
evidence for the 
existence of a discontinuity at the low end of the distribution, 
($\Hav\sim2.8$\,kG), is strengthened by the addition of the new
data. As discussed in \citetalias{1997A&AS..123..353M}, this evidence arises from the fact that
while the majority of Ap stars with resolved magnetically split lines
have a value of $\Hav$ comprised between 3 and 9\,kG, and the
distribution of the field strengths within this interval is skewed
towards its lower end; there is a sharp cut-off at the latter
(approximately at $\Hav=2.8$\,kG). This cut-off cannot be fully
accounted for by an observational bias, since we expect to be able to
observe resolved splitting of the line \ion{Fe}{ii}~$\lambda\,6149.2$
down to a field modulus of approximately 1.7\,kG (see \citetalias{1997A&AS..123..353M} for the
detailed argument). Individual measurements of the field modulus lower
than 2.8\,kG have actually been obtained over a fraction of the rotation
cycle of some stars, but the average of the values measured over a
whole rotation cycle is always greater than or equal to 2.8\,kG.

One of the questions that remained open at the time of
\citetalias{1997A&AS..123..353M} was 
whether Ap stars with very sharp spectral lines, in which the line
\ion{Fe}{ii}~$\lambda\,6149.2$ shows no hint of magnetic broadening or
splitting, are actually non-magnetic. Since then, the presence of weak
magnetic fields has been reported for a number of such stars: 
\begin{itemize}
\item HD~133792: 1.1\,kG
  \citep{2004IAUS..224..580R,2006A&A...453..699M,2006A&A...460..831K} 
\item HD~138633: 0.7\,kG \citep{2013AstL...39..347T}
\item HD~176232: 1.0\,kG \citep{2000A&A...357..981R}; 1.5\,kG
  \citep{2002MNRAS.337L...1K}; 1.4\,kG \citep{2003A&A...409.1055L}
\item HD~185256: 1.4\,kG \citep{2013MNRAS.431.2808K}
\item HD~204411: 750\,G \citep{2005A&A...438..973R}
\end{itemize}
The quoted field values were obtained by various methods and most of
these values do not a priori represent measurements of the field moments
discussed in this paper, but their order of magnitude should be
similar to the mean magnetic field modulus or to the mean quadratic
magnetic field of the considered stars. Several of these values are actually
upper limits, but for stars in which magnetic fields are definitely
present. The uncertainties affecting the determinations of the
magnetic field strengths are not quantified in several of the
above-mentioned references, and their interpretation would not be
straightforward since there is often significant ambiguity left as to
the exact physical meaning of the field value that is reported. But in
all cases, convincing evidence is presented that a magnetic field of
kG order is indeed detected.

The field strengths obtained in the above-mentioned references are
consistent with our estimate of a lower limit of 1.7\,kG for
resolution of the magnetic splitting of the line
\ion{Fe}{ii}~$\lambda\,6149.2$. Admittedly, each of these field strengths refers to an
observation at a single epoch, and it is very plausible that, for any
of the considered stars, the mean field modulus may reach a somewhat
higher value at other phases of its rotation cycle. However, as
discussed below, we did not find any evidence that the ratio between
the extrema of $\Hm$ may considerably exceed 2.0 in any star with
resolved magnetically split line. Actually, in most of those stars, this
ratio is smaller than 1.3. Moreover, on physical grounds, very exotic
field structures, of a type that has not been found in any Ap star
until now,  would be required to cause variations of the mean field
modulus with stellar rotation phase by a factor much greater than
2. Thus it appears reasonable to assume that the maximum value
of the field modulus of each of the five stars listed above may not
significantly exceed twice the lowest published value of their field
strength. This in turn allows us to derive an upper limit of $\Hav$
for each of those five stars, of about 1.5 times its lowest published field
strength. This is a conservative estimate, since it is very unlikely
that the magnetic fields of all five stars were determined close
to the phase of minimum of their mean field modulus. This estimate yields values
of $\Hav$ that range from 1.05 to 2.1\,kG, which are all well below the 2.7\,kG
lower limit of the $\Hav$ distribution for stars with magnetically
resolved lines. 

Those results point to the existence of a subgroup of Ap stars, most
of which are likely (very) slow rotators, and whose magnetic
fields, averaged over a rotation period, are weaker than
$\sim2$\,kG. HD~8441 is another member of this group, where
  $\Prot\simeq69$\,d and no magnetic enhancement of the spectral lines
  is detected in a spectrum recorded in natural light,
  but in which \citep{2012AstL...38..721T} could measure a
    definite, weak longitudinal field. 

This subgroup appears to constitute a 
distinct, separate 
population from that of the Ap stars with resolved magnetically split
lines, most of which are also (very) slow rotators, but for which
$\Hav$ is always greater than $\sim2.7$\,kG. In other words, the sharp
drop at the lower end of the distribution of the phase-averaged mean
field modulus values in
stars with magnetically resolved lines, shown  in
Fig.~\ref{fig:hmhist}, reflects the existence of a gap in the
distribution of the field strengths of slowly rotating Ap stars, which
accordingly appears bimodal. This bimodality may possibly be related
to a similar feature in the distribution of the rare earth abundances, in which the weakly magnetic stars are rare earth poor and the strongly
magnetic stars are rare earth rich
\citep{2012AstL...38..721T,2013AstL...39..347T}. But more data are
needed to establish this connection on a firm basis. 

\begin{figure}[t]
\resizebox{\hsize}{!}{\includegraphics{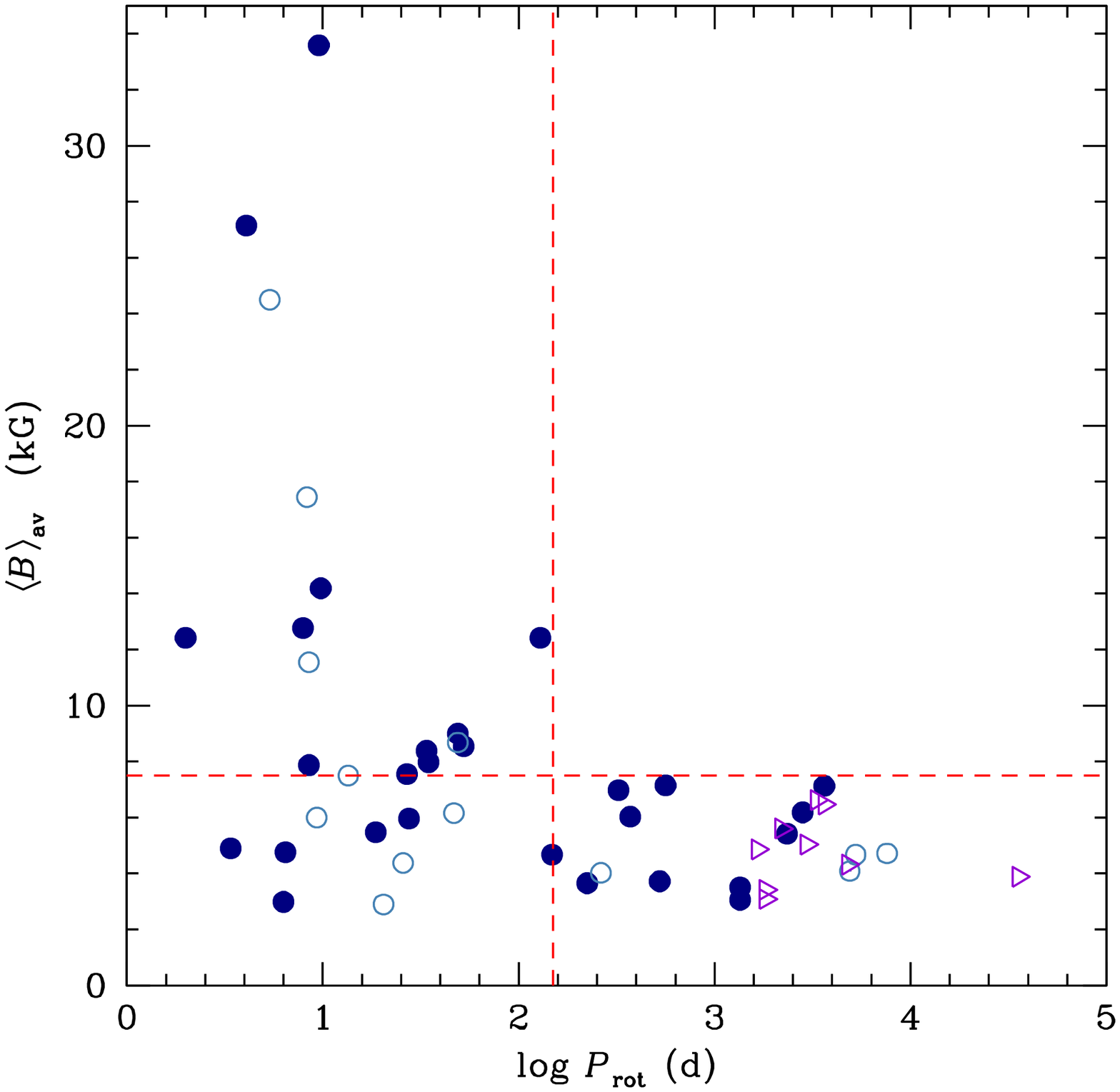}}
\caption{Observed average of the mean magnetic field modulus against
  rotation period. Dots: stars with known rotation periods;
  triangles: stars for which only the lower limit of the
  period is known. Open symbols are used to distinguish those
  stars for which existing measurements do not cover the whole rotation
  cycle. The horizontal and vertical dashed lines, corresponding
   to $\Hav=7.5$\,kG and  $\Prot=150$\,d, respectively, emphasise the absence
of very strong magnetic fields in the stars with the slowest rotation
(see text for details).}
\label{fig:hm_vs_P}
\end{figure}

More observations should also be obtained to refine the
characterisation of the 
low end of the $\Hav$ distribution, so as to provide a firm basis for
its theoretical understanding. In particular, we do not know at
present if the bimodal character of the field strength distribution is
restricted to slowly rotating stars, or if it is general feature of
the magnetism of Ap stars, independent of their rotation. On
  the other hand, the cut-off 
that \citet{2007A&A...475.1053A} inferred from a systematic search for
weak longitudinal fields in Ap stars could represent the ultimate
lower limit of the field strength distribution. However, at the low dipole
strength  ($\la300$\,G) at which the $\Hz$ cut-off is found,
measuring the mean 
magnetic field modulus represents a
major challenge.

On the strong field side, while compared to \citetalias{1997A&AS..123..353M}, 
a few additional stars
have populated the high end of the distribution of $\Hav$, their
meaning for the characterisation of this distribution is limited,
since several were observed at high resolution as the
result of the 
detection of their exceptionally strong longitudinal fields as
part of spectropolarimetric surveys. In other words, they are
not a priori representative of the distribution of magnetic field
strengths that one would derive from the study of an unbiased sample of Ap
stars with low $\vsi$.

In Fig.~\ref{fig:hm_vs_P}, we plotted $\Hav$ against the rotation
period $\Prot$ for those stars from Tables~\ref{tab:stars} and
\ref{tab:stars_litt} for which the latter, or at least 
a lower limit of it, could be determined. This figure, which is an
updated version of Fig.~50 of \citetalias{1997A&AS..123..353M}, fully
confirms the result 
inferred from the latter, that very strong magnetic fields
($\Hav\ga7.5$\,kG) are found only in stars with rotation periods
shorter than $\sim150$ days. This result is visually emphasised in the
figure by dashed lines: the horizontal line corresponds to
$\Hav=7.5$\,kG, and the vertical line to $\Prot=150$\,d. The
representative points of 50 stars appear in Fig.~\ref{fig:hm_vs_P}. Twenty-seven of them
correspond to stars with $\Prot<150$\,d, of which 17 have
$\Hav\geq7.5$\,kG. By contrast, none of the 23 stars with $\Prot>150$\,d
have $\Hav\geq7.5$\,kG. The difference between the two groups is highly
significant: a Kolmogorov-Smirnov test indicates that the
distributions of $\Hav$ between the stars with a rotation period shorter
than 150 days and those with a longer rotation period are different at
the 100.0\% confidence level.

This result receives further support from consideration of the stars
with magnetically resolved lines whose 
period is unknown and for which the average
of the mean magnetic field modulus over this period  may exceed
$\sim7.5$\,kG. Among the stars of Tables~\ref{tab:summary} and
\ref{tab:summary2}, this is 
almost certainly the case for HD~47103, HD~55540, HD~66318, HD~70702, 
BD~+0~4535, and HD~168767. The spectral lines of HD~70702 and
HD~168767 show considerable rotational broadening, so that their
periods must be of the order of a few days \citep{2012MNRAS.420.2727E}, while
\citet{2010MNRAS.404L.104E} infer from their estimate of $\vsi$ for
BD~+0~4535 that its period must be shorter than $\sim60$ days. The mean
field modulus of HD~55540 shows a variation of $\sim600$~G
between two observations taken one month apart
\citep{2008MNRAS.389..441F}; its period should
probably be of the order of months. The situation is less clear for
HD~47103 (Appendix~\ref{sec:hd47103}) and HD~66318
\citep{2003A&A...403..645B}. Both of these stars have a low $\vsi$ and neither shows
definite variations of the magnetic field; either one of the angles
$i$ or $\beta$ is small for these two stars or their periods are
longer than one year. Obtaining more and better observations to
establish if HD~47103 and HD~66318 are actually variable, and if so,
to determine their periods, 
would represent a further test of the existence and nature of a
difference in the distribution of the field strengths below and above
$\Prot=150$\,d. 

There are several other stars in Table~\ref{tab:summary2} for which
$\Hav$, as computed from the few available measurements, is of the
order of or slightly greater than 7.5\,kG. It would be interesting to
determine their periods (if they are not already known) and to obtain
more $\Hav$ measurements  distributed well
throughout their rotation 
cycle, but it appears highly improbable that any of these stars could
represent a significant exception to the conclusion that very strong
fields do not occur in very long-period stars. 

For two of the stars that appear to rotate extremely slowly, HD~55719
and HD~165474, the scatter of the individual $\Hm$ values around a
smooth, long-term variation trend is much greater than we would expect
from the appearance of the \ion{Fe}{ii}~$\lambda\,6149.2$ diagnostic
line in their spectrum. This is unique: for
all the other stars studied  
here, the scatter of the individual measurements of the mean field
modulus about its variation curve or trend is consistent with the
observed resolution of the \ion{Fe}{ii}~$\lambda\,6149.2$ line, the
definition of its components, and the amount of blending to which they
are subject. By
contrast, the radial velocity measurements for HD~55719 
and HD~165474, do not show any anomalous scatter about the
corresponding variation curves. On the contrary, as stressed in
Sect.~\ref{sec:hd165474}, the exquisite precision of the radial
velocity measurements in HD~165474 enabled us to detect minute
long-term variations. Since those radial velocity values are computed
from the very measurements of the split components of the
\ion{Fe}{ii}~$\lambda\,6149.2$ line from which the $\Hm$ values are
obtained (see \ref{sec:rv} and Eq.~(\ref{eq:rv6149})), their high quality
suggests that the unexpectedly high scatter of the mean field modulus
values does not result from measurement errors nor does it have an instrumental
origin. The remaining interpretation, that the observed scatter has a
stellar origin, is intriguing, as it calls for additional variability
on timescales considerably shorter than the rotation periods of the Ap
stars of interest. This might possibly represent a new, previously
unobserved type of variability for an Ap star, or those variations
could originate from an unresolved companion. The data available at
present are inconclusive, but this is an issue that will definitely
deserve further follow up in the future.

To characterise the variability of the mean magnetic field modulus, we
use the ratio $q$ between its extrema. This ratio is given in the
fifth column of Tables~\ref{tab:summary} and \ref{tab:summary2}. For
those stars for which we 
fitted a curve to the measurements (that is, the stars appearing in
Tables~\ref{tab:mfit} and \ref{tab:mfit-litt}), the adopted value of
$q$ is the ratio between 
the maximum and minimum of the fitted curve. For the other stars,
the adopted value of $q$ is the ratio of the 
highest to the lowest value of $\Hm$ that we measured until now (for
those stars that we studied) or that we found in the literature. 

\begin{figure}[t]
\resizebox{\hsize}{!}{\includegraphics{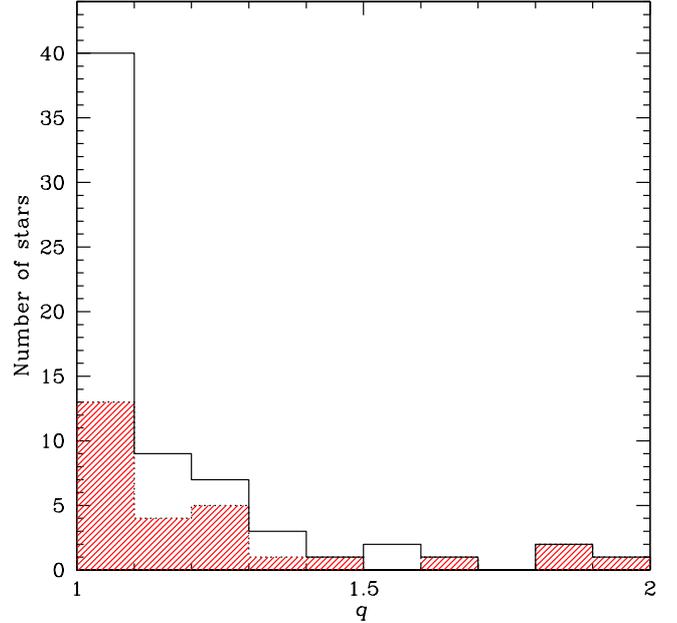}}
\caption{Distribution of the ratio $q$ between
  the observed extrema of mean
  magnetic field moduli of the 66 Ap stars with resolved magnetically
  split lines presently known for which more than one measurement of
  $\Hm$ is available. Although HD~96237 was observed twice by
  \citet{2008MNRAS.389..441F}, its spectral lines were not
  resolved in the first epoch spectrum. The shaded part of the histogram
  corresponds to those stars that have been observed throughout their
  rotation cycle (see text).}
\label{fig:qhist}
\end{figure}

\begin{figure}[t]
\resizebox{\hsize}{!}{\includegraphics{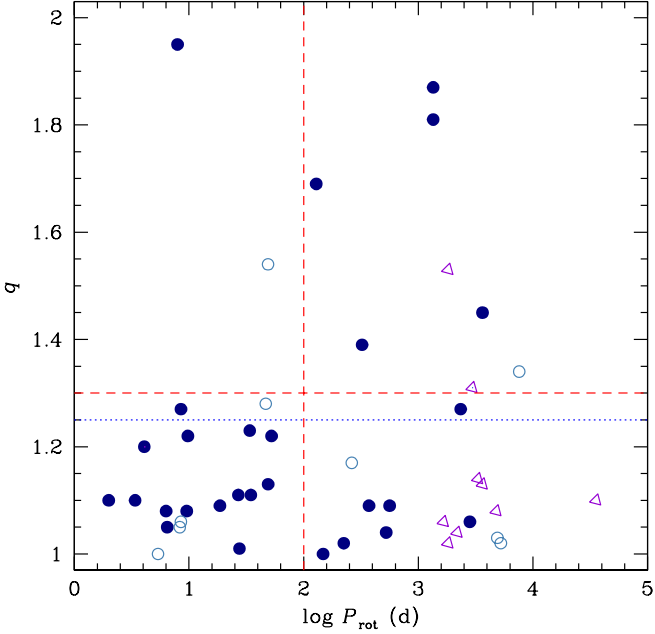}}
\caption{Ratio $q$ of the observed extrema of the mean magnetic field
  modulus against 
  rotation period. Dots: stars with known rotation periods;
  triangles: stars for which only the lower limit of the
  period is known. Open symbols are used to distinguish those
  stars for which existing measurements do not cover the whole rotation
  cycle. The horizontal and vertical dashed (red) lines, corresponding
to $q=1.3$ and $\Prot=100$\,d, respectively, emphasise the tendency
  for higher values of $q$ to have a higher rate of occurrence in the
  stars with the slowest rotation. The horizontal dotted (blue) line
  corresponds to the upper limit of $q$ for a centred dipole (see text).}
\label{fig:q_vs_P}
\end{figure}

 The distribution of the values of $q$ is shown in
Fig.~\ref{fig:qhist} for the 66 stars for which more than one $\Hm$
measurement has been obtained. In 41 of these stars, $q$ does not exceed
1.10. This shows that the amplitude of variation of the mean magnetic
field modulus is in most cases small compared to its value and
supports the view that the field modulus is very useful to
characterise the magnetic fields of Ap stars in statistical studies. 
The fact that for about two-thirds of the stars with $q\le1.10$, the
existing data do
not sample (yet) the whole rotation cycle does not seriously question
this conclusion, if one considers that also in $\sim50$\% of  the stars for which
good phase coverage is achieved (shaded part of the histogram), $q$
does not exceed 1.10. However, for about one-fifth of the
known Ap stars with magnetically resolved lines (and one quarter of
those that have been observed throughout their rotation period), $q$
is greater than 1.25, which is the upper limit for a centred dipole
\citep{1969ApJ...158.1081P}. This 
points to the occurrence of significant departures from this simple
field geometry. The highest value of $q$ obtained so far in our study
is 1.95, in the star HD~65339. More recent observations by
\citet{2004A&A...423..705R} suggest a similar, or even higher value
of this ratio in HD~29578. 

On the other hand, one can find in
Fig.~\ref{fig:q_vs_P} some hint that high relative amplitudes of
variation of $\Hm$ are found more frequently in stars with very long
periods. In order to help the eye visualise this, two dashed
lines were drawn in Fig.~\ref{fig:q_vs_P}: the first is horizontal, at
$q=1.3$ and the second is vertical at $\log\Prot=2.0$. Among the 21 stars
with $\Prot<100$\,d, only 2 have $q>1.3$. By contrast, 8 of the 25
stars with $\Prot>100$\,d have $q>1.3$. The evidence is less
compelling than for the above-mentioned absence of very strong fields
in very slowly rotating stars, but the high rate of occurrence of
large relative amplitudes of $\Hm$ variations among the latter 
nevertheless seems significant, especially if one considers that for
the stars that have not been observed yet throughout a whole rotation
cycle, the values of $q$ adopted here are lower limits.  In
particular, among the stars with $q<1.3$, phase coverage is
incomplete for 10 of 17 with $\Prot>100$\,d, but only for 4 of 19 with 
$\Prot<100$\,d: thus there is significantly more possibility for
future studies to increase the fraction of stars with high values of
the ratio $q$ among the long-period members of this sample than among
their short-period counterparts. The proximity of the $q=1.3$ dividing
line  to the upper limit of $q$ for a centred dipole may further
suggest that there exists a real difference in the structure of the
magnetic fields below and above $\Prot=100$\,d. The $q=1.25$
value of this upper limit is only approximate; in particular, it may vary
slightly according to the applicable limb-darkening law
\citep{1969ApJ...158.1081P}.

\begin{figure}
\resizebox{\hsize}{!}{\includegraphics{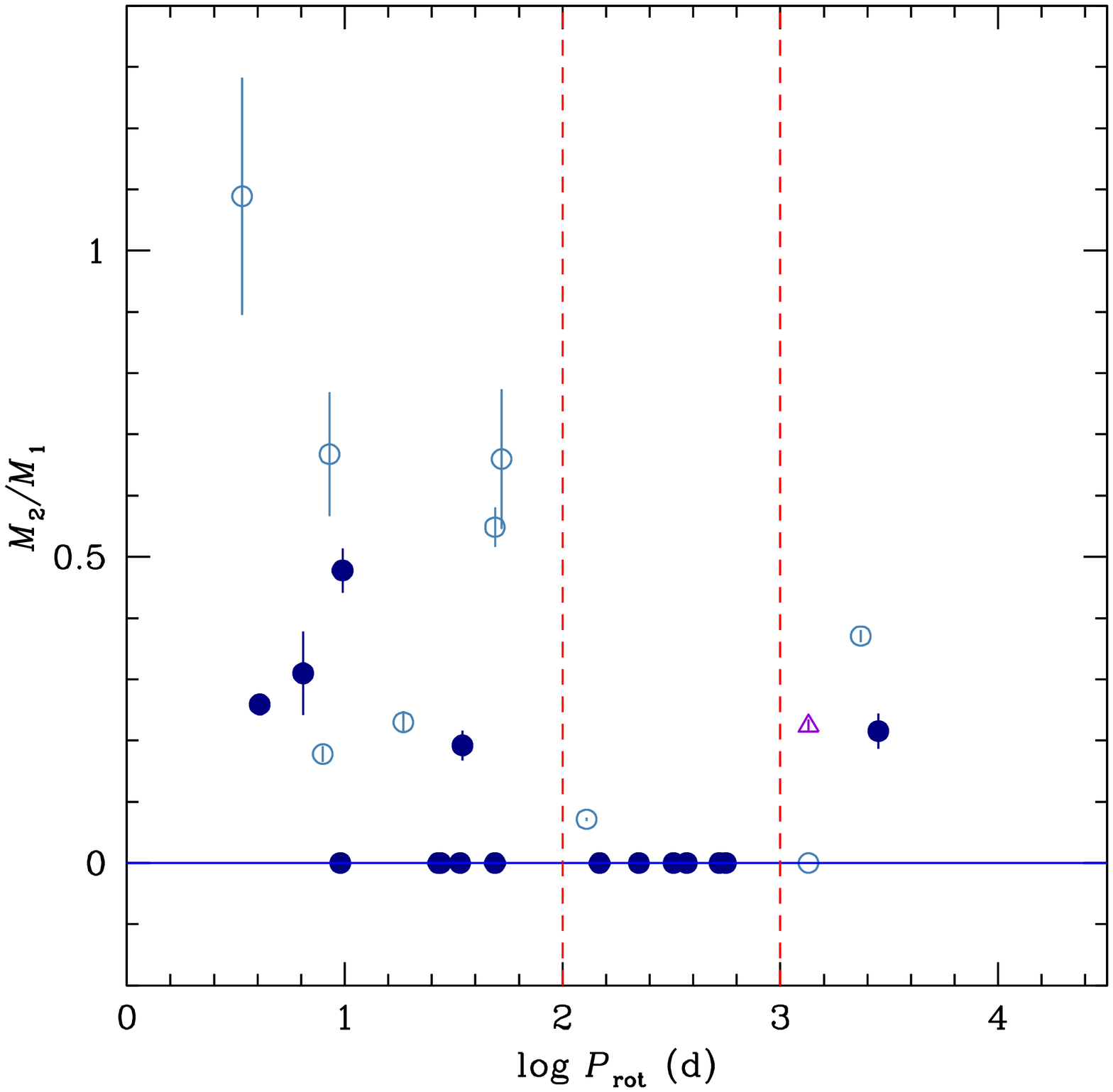}}
\caption{Ratio $M_2/M_1$ of the fit coefficients given in
  Tables~\ref{tab:mfit}  and \ref{tab:mfit-litt} against rotation
  period. Open circles: 
  stars showing $\Hz$ reversal; dots: stars in which $\Hz$ has
  a constant sign; open triangle: star for which no $\Hz$
  measurements exist. For a fraction of the stars, the error bar is
  shorter than the size of the representative symbol. The horizontal
  (solid) line 
  corresponds to the 
  stars for which the curve of variation of $\Hz$ does not
  significantly depart from a sinusoid ($M_2=0$). The two vertical
  (dashed) lines
  emphasise the tendency for anharmonicity of the mean field modulus
  variations to be restricted to rotation periods shorter than $\Prot=100$\,d or
  greater than $\Prot=1000$\,d.} 
\label{fig:m2overm1_vs_P}
\end{figure}

In order to explore the dependency on rotation period not only of the
amplitude of the curve of variation of the field modulus, but also of
its shape, in Fig.~\ref{fig:m2overm1_vs_P} we plotted against the
rotation period, the ratio $M_2/M_1$ of the amplitude of the
harmonic to that of the fundamental in a fit of the $\Hm$ data by a
function of the form given in Eq.~(\ref{eq:fit2}). Zero values of this ratio
correspond to the cases where the measurements are adequately
represented by a single sinusoid (Eq.~(\ref{eq:fit1})). The pattern
that appears in 
Fig.~\ref{fig:m2overm1_vs_P} is intriguing. The curves of variation of
$\Hm$ show a significant degree of anharmonicity for the majority of
the stars with $\Prot<100$\,d (but not for all of them). For rotation
periods between 100 and 1000 days, the variation curves are mostly
sinusoidal. For three of the four stars with $\Prot>1000$\,d for which we accumulated enough data to fully define the shape of the $\Hm$
variation curve, a significant first harmonic is present in the
fits. Furthermore, there are at least three more stars whose rotation
periods exceed 1000 days and for which the data acquired so far are
insufficient to characterise the variation curve of the
mean field modulus  completely, for which this curve is definitely anharmonic:
\begin{itemize}
\item The variation curve of the field modulus of HD~965, which has a
  rotation period considerably longer than 10 years, shows a very
  broad, almost flat minimum \citep{2005MNRAS.358.1100E}.
\item The
  $\Hm$ variations in HD~9996, which has a rotation period of $\sim22$
  years, must have a high degree of anharmonicity, according to the arguments given in Appendix~\ref{sec:hd9996}.
\item The $\Hm$ variations in HD~166473, whose rotation period may be
  of the order of 10 years, also show definite anharmonicity
  \citep{2007MNRAS.380..181M}.  
\end{itemize}

The interpretation of the behaviour pattern that is observed in
Fig.~\ref{fig:m2overm1_vs_P} is
not straightforward at this stage. We postpone its discussion to
Sect.~\ref{sec:Hzdisc}. 

On the other hand, among the
fitted $\Hm$ curves, 7 show two (generally different) minima and two
(generally different) maxima per rotation period. We introduce the
notations $\Hm_{\rm max,1}$ and $\Hm_{\rm max,2}$ to refer to, respectively,
the primary (greater) and secondary maxima, and $\Hm_{\rm min,1}$ and
$\Hm_{\rm min,2}$ for the primary (smaller) and secondary
minima. Interestingly, in all but one of the stars for which these
quantities are defined, the difference $\Hm_{\rm max,1}-\Hm_{\rm max,2}$
between the primary and secondary maxima of the field modulus is
significantly larger than the difference between its secondary and
primary minima, $\Hm_{\rm min,2}-\Hm_{\rm min,1}$. This must be related to the
observation in a number of stars of broad, more or less flat minima of
the $\Hm$ variation curve, mentioned on several occasions in
Appendix~\ref{sec:notes}. In other words, it is rather usual for the mean
magnetic field modulus to be close to its minimum value over a large
portion of the rotation period, and to show a rather steep variation
from near minimum to an absolute maximum and back within a fairly
narrow phase range. The common character of this behaviour is further
supported by consideration of other stars, whose $\Hm$ curves have only
one minimum and one maximum per cycle, but show shallower
variations about their phase of minimum than around their maximum:
HD~59435, probably HD~166473 (if its rotation period is close to 10
years), and HD~192678. HD~965 is another star that almost
certainly shows a similar behaviour. By contrast, we did not observe
any star showing a broad, shallow maximum of the field modulus and a
narrower, steeper minimum. Also, for the only star found so far in
which $(\Hm_{\rm max,1}-\Hm_{\rm max,2})<(\Hm_{\rm min,2}-\Hm_{\rm
  min,1})$, HD~200311, the
difference between the two sides of the inequality is more moderate
than in most stars in which
$(\Hm_{\rm max,1}-\Hm_{\rm max,2})>(\Hm_{\rm min,2}-\Hm_{\rm
  min,1})$. Actually, owing to the incomplete phase coverage of the
field modulus determinations for this star, it is not implausible that
the value of $(\Hm_{\rm min,2}-\Hm_{\rm  min,1})$ derived from our
observations overestimates the actual field strength
difference between the two minima (see Fig.~\ref{fig:hd200311}).
%, and
%more observations of this star in the phase range 0.6--0.9 might
%possibly shift its representative point below the diagonal dashed line
%in Fig.~\ref{fig:twoextrema}. 

Regardless, we have not observed any star in which the difference
between the minima of the field modulus is much greater than
the difference between its maxima.  
This probably indicates that it is frequent for Ap stars to have a
fairly uniform magnetic field (such as a global dipole) covering the
largest part of their 
surface, with one rather limited region (a kind of large
spot) where the
field is much stronger. In this spot the small-scale structure
  of the field must be considerably more complex
than in the rest of the star, since its
signature is not readily seen in the longitudinal field variations (see
Sect.~\ref{sec:Hzdisc}). The opposite topology, a star with a large
spot that 
has a lower field than the surroundings, does not seem to occur
commonly, if at all. If this interpretation is indeed correct, it suggests that
models based on low-order multipole expansions are unlikely to provide
a very good representation of the actual field geometries, even though
they may provide useful first order approximations. 

The frequent occurrence in the $\Hm$ variation curves of nearly flat
minima extending over a broad phase range implies that the probability
of observing a star when its field modulus is close to its minimum value
is, in general, significantly higher than close to its maximum
value. This, in 
turn, means that for those stars for which we have observed so far a
constant, or nearly constant value of $\Hm$, this value is likely to
correspond roughly to the minimum of this field moment. It can be
expected that, if the star varies at all, at some point it will start
to show a comparatively steep increase towards higher values of the
field modulus; by contrast, it seems improbable that $\Hm$ starts to 
decrease after it remained nearly constant for an extended period
of time.

\begin{figure}[t]
\resizebox{\hsize}{!}{\includegraphics{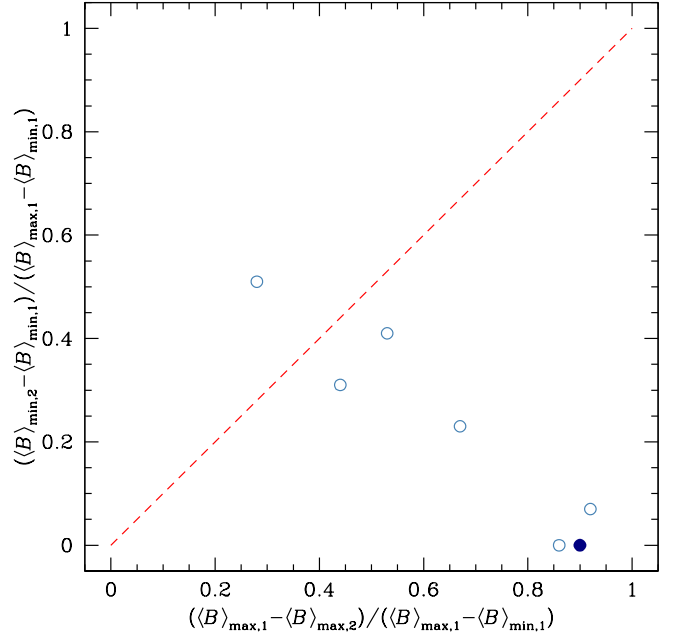}}
\caption{Difference between the primary and secondary maxima
  of the mean magnetic field modulus against the difference
  between its two minima; both differences are normalised via division
  by the difference between the primary maximum and primary
  minimum. Open circles represent stars whose longitudinal
  field reverses its sign over a rotation period; the dot
  corresponds to a star with non-reversing longitudinal field. The
  dashed line is drawn to help the eye visualise the fact that the
  difference between the minima is generally smaller, and never much
  greater than the difference between the maxima. The only star above
this line is HD~200311, for which no $\Hm$ data could be obtained
between phases 0.6 and 0.9. More observations of the star in this
phase range might possibly shift its representative point below the
dashed line.}
\label{fig:twoextrema}
\end{figure}

Another intriguing result is shown in Fig.~\ref{fig:twoextrema}, where
the difference between the primary and secondary minima of the
mean magnetic field modulus is plotted against the difference between
its two maxima, for those stars where $\Hm$ has two maxima and two
minima per period. Both differences are normalised via division by the
difference between the primary maximum and  primary minimum, so as
to allow for comparison between stars with different field strengths. 
The dashed diagonal line, which corresponds to
$\Hm_{\rm min,2}-\Hm_{\rm min,1}=\Hm_{\rm max,1}-\Hm_{\rm max,2}$, is
shown for easier visualisation of the above-mentioned result, that the 
difference between the $\Hm$ minima is in general smaller, and never
much greater, than the difference between its maxima. Very
unexpectedly, a rather tight anti-correlation appears, namely,
\begin{displaymath}
{\Hm_{\rm min,2}-\Hm_{\rm min,1}\over\Hm_{\rm max,1}-\Hm_{\rm min,1}}
\propto
{\Hm_{\rm max,1}-\Hm_{\rm min,1}\over\Hm_{\rm max,1}-\Hm_{\rm max,2}}\,.
\end{displaymath}
In other words, for a given peak-to-peak amplitude of variation of the
field modulus, the greater the difference between the two minima is,
the smaller the difference between the two maxima. We know of no
obvious reason why this should be so; as a matter of fact, if the
field were a centred dipole (which is not an implausible geometry),
both differences would be zero. Thus the observed behaviour provides a
constraint on the systematics of the field structure of magnetic Ap
stars. Admittedly, this result is deduced from consideration of only seven
stars, so that its statistical significance is still moderate, but the
observed anti-correlation is very obvious.

\subsection{The mean longitudinal magnetic field}
\label{sec:Hzdisc}
We have presented 231 new measurements of the mean longitudinal
magnetic field of 34 Ap stars with resolved magnetically split lines. 
For 10 of these stars, these are the first determinations of this
field moment ever published. 

The $\Hz$ measurements reported here are the most precise ever
obtained with the CASPEC spectrograph. This is best seen by comparing
the median of the standard errors $\sigma_z$ of all the measurements
appearing in Table~\ref{tab:hzxq}, 98\,G, with their median for the 44
measurements of $\Hz$ in Ap stars with magnetically resolved lines
previously published by \citet{1994A&AS..108..547M} and
\citet{1997A&AS..124..475M}, 159\,G. Such a comparison is more
meaningful than a 
comparison with all the previous $\Hz$ determinations based on CASPEC
observations, since errors increase quickly with $\vsi$. The better
precision of the present data results from the combination of several
factors:
(1) The spectra were obtained at a resolving power $R\sim3.9\,10^4$,
which is more than twice as high as that used by \citet{1994A&AS..108..547M} and for a
  fraction of the observations of \citet{1997A&AS..124..475M},
  $R\sim1.6\,10^4$.
(2) Their spectral coverage, continuous over a range of the order of
  1250\,\AA, is at least comparable to that of the lower resolution
  spectra of the above-mentioned papers, and much larger than that 
  ($\sim600$\,\AA, once the inter-order gaps have been subtracted) of
  the higher resolution ($R\sim3.5\,10^4$) spectra of
  \citet{1997A&AS..124..475M}.  
(3) The combination of a new Zeeman analyser (used by Mathys \&
  Hubrig for only a few of their observations), a CCD with higher
  sensitivity, and improvements in the optical alignment of CASPEC and
  in its stability, enabled us to achieve a considerably higher
  signal-to-noise ratio in the reduced spectra.

The precision of the present $\Hz$ measurements compares well with that
of the contemporaneous determinations of \citet{2000MNRAS.313..851W},
which were achieved through application of the least-squares
deconvolution (LSD) technique to spectra recorded with the
spectrograph MuSiCoS. The median standard error of the MuSiCoS
  measurements is
48\,G. This is a factor of 2 smaller than what we achieve here, which
can be accounted for by the fact that 
the spectral coverage of MuSiCoS is
about four times larger than that of CASPEC in the configuration that we
used. Admittedly, the comparison is valid only for the low values
of $\vsi$ typical of our sample; for stars with higher projected
equatorial velocities, the LSD technique is more robust, in particular
against line blending. 

Because the mean longitudinal magnetic field is very sensitive to the
geometry of the observation, so that, in particular, it may reverse
its sign as the star rotates, the average of all the obtained
measurements of $\Hz$, or the independent term $Z_0$ of the fit of
their variation curves, is inadequate to characterise this field
moment in a given star by a single number, as we did for the mean
magnetic field modulus. To this effect, following
\citet{1993A&A...269..355B}, we use instead the root-mean-square (rms) 
longitudinal field, defined as
\begin{equation}
\Hzrms=\left({1\over N_z}\,\sum_{i=1}^{N_z}\Hz_i^2\right)^{1/2}\,,
\end{equation}
where $N_z$ is the number of individual measurements of $\Hz$ that we
obtained, and $\Hz_i$ is the $i$-th such measurement. The values
of $N_z$ and $\Hzrms$ are given in Cols.~6 and 7 of
Tables~\ref{tab:summary} and \ref{tab:summary2}. For most stars of
Table~\ref{tab:summary}, as in the rest of this paper, 
these values were computed by combining the new $\Hz$ determinations
presented here with the published CASPEC-based values of
\citet{1994A&AS..108..547M} and \citet{1997A&AS..124..475M}. For a few
northern 
stars, which could not be observed with CASPEC and 
for which recent, reliable measurements of
the longitudinal field distributed well throughout the rotation cycle
are available in the literature, the values of $N_z$ and $\Hzrms$
appearing in Table~\ref{tab:summary} (in italics) were 
computed using these 
published data: HD~9996 \citep{2012AcA....62..297B}, HD~12288 and
HD~14437 \citep{2000A&A...355.1080W}, HD~18078 \citep{2016A&A...586A..85M},
HD~65339 \citep{2000MNRAS.313..851W,2001A&A...369..889B},
HD~192678 \citep{1996A&A...313..209W}, and HD~200311
\citep{1997MNRAS.292..748W,2000A&A...355..315L}. Finally, the $N_z$
and 
$\Hzrms$ data of Table~\ref{tab:summary2} are based on the $\Hz$
measurements from the references listed in Col.~8 of that table.

\begin{figure}[t]
\resizebox{\hsize}{!}{\includegraphics{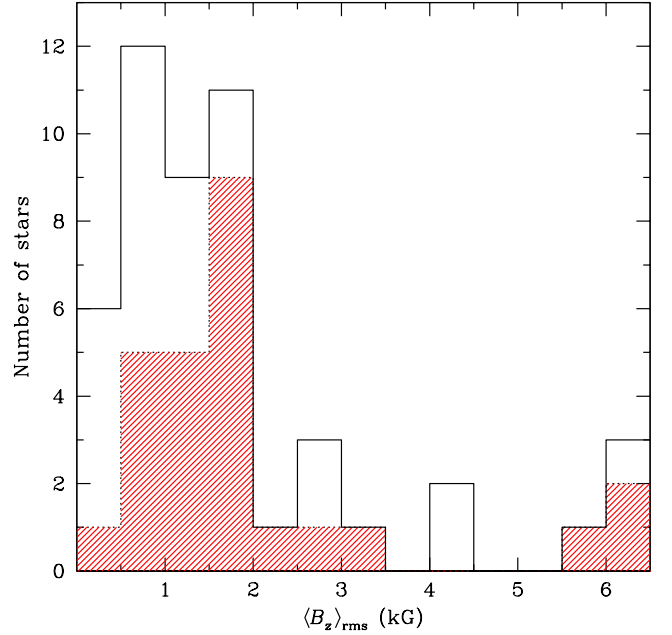}}
\caption{Histogram showing the distribution of the root-mean-square
  longitudinal magnetic field of the 49 Ap stars with resolved magnetically
  split lines for which this field moment was measured at least
  once. The shaded part of the histogram 
  corresponds to those stars whose longitudinal field has been measured
  throughout their rotation cycle (see text).}
\label{fig:hzhist}
\end{figure}

\begin{figure}[t]
\resizebox{\hsize}{!}{\includegraphics{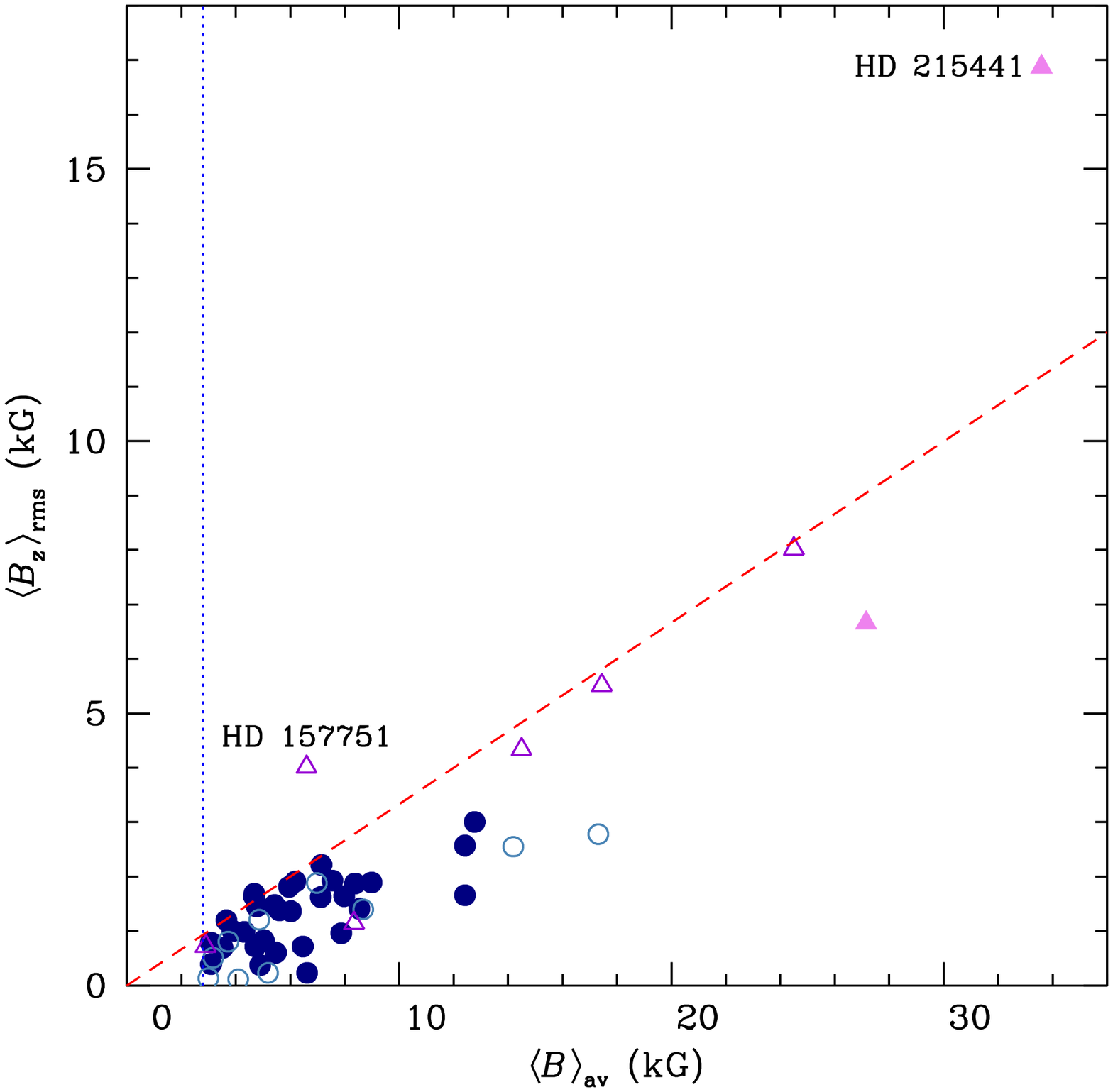}}
\caption{Observed root-mean-square longitudinal field against the
  observed average of the mean magnetic field modulus. Open
    symbols are used to distinguish those stars for which either or
  both of $\Hzrms$ and $\Hav$ were computed using less than 7
  individual measurements of, resp., $\Hz$ and $\Hm$; 
    triangles are used for field modulus values from the literature. The
  dashed line 
  corresponds to $\Hzrms=\Hav/3$. For only two stars (identified in
  the figure), the representative points are clearly above this
  line. However this is significant only for HD~215441, since the
  number of measurements of HD~157751 is insufficient ({\it see
    text\/}). The dotted vertical line corresponds
  to $\Hav=2.8$\,kG; it emphasises the discontinuity at the low end of
  the distribution of $\Hav$, discussed in Sect.~\ref{sec:Hmdisc}.}
\label{fig:hz_vs_hm}
\end{figure}

The distribution of the values of $\Hzrms$ for Ap stars with resolved
magnetically split lines, shown in Fig.~\ref{fig:hzhist},  differs
markedly from their distribution for all Ap stars \citep[see Fig.~1
of][]{2003A&A...407..631B}. The deficiency of stars with
$\Hzrms<0.5$\,kG among the stars of the present sample starkly
contrasts with the fact that 50\% of all Ap stars have $\Hzrms<525$\,G
\citep[see Table~1 of][]{2003A&A...407..631B}. This difference is
further emphasised if one focuses attention on the stars for which
longitudinal field measurements were obtained throughout the full
rotation cycle (i.e. the shaded part of the histogram in
Fig.~\ref{fig:hzhist}). One can reasonably expect that, as more data
are obtained for the stars for which phase coverage is so far
incomplete, the shape of the unshaded part of the histogram progressively becomes more similar to the shaded part, at least up to
2\,kG, so that the lack
of low $\Hzrms$ values stands out even more conspicuously. This
scarcity of weak longitudinal fields in our sample is not unexpected,
since magnetic fields that are strong enough to produce line splitting
sufficiently large
to be resolved observationally should, in their majority, also have a
sizeable mean longitudinal component. However, as we show below,
there is no one-to-one relation between $\Hzrms$ and $\Hav$. 

In Fig.~\ref{fig:hz_vs_hm}, we plotted $\Hzrms$ against
$\Hav$. One can see that only two points are found significantly above a
(dashed) line corresponding to $\Hzrms=\Hav/3$. Although the slope
that we adopted for this line is purely empirical, based only on the
observational points, it is interesting to note the following
similarity. As shown by \citet{1970ApJ...160.1059P}, for a centred dipole,
the maximum of the ratio between the absolute value of $\Hz$ and
$\Hm$, $(|\Hz|/\Hm)_{\rm max}$, can be at most 
of the order of 0.4; this value is reached only when the dipole
axis becomes parallel to the line of sight at some rotation phase. 
This phase also corresponds to an extremum of $\Hz$. Since the
amplitude of the variations of $\Hm$ is typically small, in first
approximation $(|\Hz|/\Hm)_{\rm max}\sim|\Hz|_{\rm max}/\Hav$, where
$|\Hz|_{\rm max}$ is the maximum of the absolute value of $\Hz$. On
the other hand, $\Hzrms$ is always smaller than (or equal to)
$|\Hz|_{\rm max}$. In particular, for a star
in which $\Hz$ shows sinusoidal variations with an amplitude 
much greater than the absolute value of its mean ($Z_1\gg|Z_0|$),
$\Hzrms$ is of the order of $|\Hz|_{\rm max}/\sqrt{2}$. Thus, for a
centred dipole, one would expect the ratio $\Hzrms/\Hav$ not to be
significantly greater than $0.4/\sqrt{2}=0.28$. The similarity between
this number and the value 1/3 represented by the dashed line in
Fig.~\ref{fig:hz_vs_hm} suggests that, even though actual field structures
often show significant departures from a centred dipole (as indicated
in particular by other arguments presented elsewhere in this paper), the latter
represents a good first approximation in most cases. 

\begin{table*}[t]
\caption{Variation of the mean longitudinal magnetic field: least-squares
  fit parameters for data from the literature.}
\label{tab:zfit-litt}
\begin{tabular*}{\textwidth}[]{@{}@{\extracolsep{\fill}}rcccccrcc
@{\extracolsep{0pt}}@{}}
\hline\hline \\[-4pt]
\multicolumn{1}{c}{HD/HDE}&$Z_0\pm\sigma$&$Z_1\pm\sigma$
&$\phi_{Z_1}\pm\sigma$&$Z_2\pm\sigma$
&$\phi_{Z_2}\pm\sigma$&\multicolumn{1}{c}{$\nu$}&$\chi^2/\nu$&$R$\\
&(G)&(G)&&(G)\\[4pt]
\hline \\[-4pt]
  9996&$\fo{-}414\pm\fo30$&$1308\pm\fo45$&$0.559\pm0.005$&$285\pm41$&$0.549\pm0.021$&58&4.4&0.97\\
 12288&$-1628\pm\fo74$&$1202\pm103$&$0.377\pm0.014$&&&17&5.8&0.95\\
 14437&$-1850\pm\fo62$&$\fo881\pm\fo69$&$0.998\pm0.019$&&&33&2.7&0.91\\
 18078&$\fmi\fo158\pm\fo14$&$1070\pm\fo20$&$0.330\pm0.003$&$217\pm20$&$0.196\pm0.013$&20&1.5&1.00\\
 65339&$\fo{-}406\pm\fo61$&$4426\pm\fo91$&$0.256\pm0.003$&$324\pm87$&$0.021\pm0.039$&14&5.4&1.00\\
 75049&$-4834\pm\fo29$&$4274\pm\fo44$&$0.995\pm0.001$&$169\pm37$&$0.458\pm0.062$&9&6.5&1.00\\
154708&$\fmi7984\pm\fo29$&$\fo923\pm\fo30$&$0.995\pm0.008$&&&13&1.2&0.99\\
178892&$\fmi4687\pm158$&$2620\pm206$&$0.003\pm0.015$&&&15&2.9&0.96\\
200311&$\fmi\fo\ft4\pm154$&$1175\pm167$&$0.986\pm0.033$&&&21&4.4&0.84\\
215441&$\fmi\llap{1}5743\pm498$&$4757\pm576$&$0.021\pm0.025$&&&11&1.6&0.93\\[4pt]
\hline
\end{tabular*}
\end{table*}

Babcock's star, HD~215441, represents a remarkable exception to this
general 
conclusion. Its location in Fig.~\ref{fig:hz_vs_hm}, well above the
$\Hzrms=\Hav/3$ line, distinguishes it very clearly from the
bulk of the considered sample, and emphasises its exceptional
character. Not only  has
it stood for more than 50 years as the star with the strongest known
mean magnetic field modulus since its discovery by \citet{1960ApJ...132..521B}, but also the structure of its magnetic
field now appears to be very different from that of almost all other
magnetic Ap stars. HD~215441 definitely cannot be considered a
representative member of 
the class. Whether the outstanding structure of its magnetic field is
related to the unusual field strength, or whether the conjunction
of these two features is the result of a mere coincidence cannot be
definitely established on the basis of the existing observational
evidence. One may note, however, that in HD~75049, which among the Ap stars
with magnetically resolved lines has the second highest value of
$\Hav$, $\Hzrms/\Hav=0.25$. Moreover, it also appears highly unlikely that
this ratio may significantly exceed 1/3 in HD~154708 (which has the
third highest value of $\Hav$), even though the variations of its
mean field modulus are not fully characterised. These considerations
support the view that Babacock's star has an extremely unusual magnetic
field structure.

For the other star whose representative point in
Fig.~\ref{fig:hz_vs_hm} lies significantly above the $\Hzrms=\Hav/3$
line, HD~157751, this departure from the general trend most likely is
purely circumstantial since the values of $\Hzrms$ and $\Hav$ are
based on only 2 and 1 measurements, respectively. As such, these values cannot
be expected (especially the former) to represent accurate estimates of
the actual values of the considered quantities, which are meant to be
averages over the stellar rotation period. 

On the other hand, the discontinuity at 2.8\,kG at the low end of the
distribution of the values of $\Hav$ is clearly seen in
Fig.~\ref{fig:hz_vs_hm}, where it is emphasised by a dotted line parallel
to the ordinate axis. No similar discontinuity is apparent in the
distribution of $\Hzrms$ \citep[but see][]{2007A&A...475.1053A}. 

For 17 stars, the phase sampling of our spectropolarimetric observations
lends itself to carrying out a quantitative
characterisation of the shape of the variation curve of the
  longitudinal field by fitting
the $\Hz$ measurements by a function of the form given in
Eqs.~(\ref{eq:fit1}) or 
(\ref{eq:fit2}). The fit 
parameters are given in Table~\ref{tab:zfit}. In the case of HD~2453,
HD~137949, and HD~188041, they were derived by combining our $\Hz$ data
with measurements from the literature: see
Appendices~\ref{sec:hd2453}, \ref{sec:hd137949}, and \ref{sec:hd188041}
for details. Furthermore, the curves
of variation of the longitudinal field of the 6 northern stars HD~9996,
HD~12288, HD~14437, HD~18078, HD~65339, and HD~200311 are well
constrained by 
the data of the references cited above. The seventh northern star of our 
study that could not be observed with CASPEC but for which a good
recent set of $\Hz$ measurements exists in the literature, HD~192678,
does not show variations of this field moment above the noise level,
so that no significant fit can be computed. This star should be
regarded as having a constant longitudinal field at the achieved
precision. On the other hand, an additional 4 stars from
Table~\ref{tab:summary2}, which were not part of our
observing sample, also have well defined $\Hz$ variation curves; these are
HD~75049 \citep{2015AA...574A..79K}, HD~154708
\citep{2015A&A...583A.115B}, HD 178892 \citep{2006A&A...445L..47R},
and HD~215441 \citep{1978ApJ...222..226B}. The
fit parameters computed from the published measurements of the
longitudinal field of the 10 stars identified above are presented in
Table~\ref{tab:zfit-litt}. Its format is identical to that of
Table~\ref{tab:zfit}. The phase origins and the period values used for
these fits are listed in Tables~\ref{tab:stars} and \ref{tab:stars_litt}.

Thus, in total, there are now 28 stars with magnetically
resolved lines for which the variations of $\Hz$ with rotation phase are characterised,
against 10 at the time of publication of \citetalias{1997A&AS..123..353M}.

Anharmonicity in the $\Hz$ variation curves occurs less frequently and
is in general less pronounced than in the $\Hm$ variation curves. The
fraction of Ap stars with magnetically resolved lines in which a fit
of the observations by a cosine curve and its first harmonic is found
to be better than a fit by a single cosine is not significantly
different from this fraction in the samples of Ap stars considered in
previous CASPEC studies
\citep{1994A&AS..108..547M,1997A&AS..124..475M}. In other words, on
the basis of 
this (admittedly limited) comparison, there is no evidence of a
relation between the anharmonicity of the $\Hz$ variations of Ap stars
and the length of their rotation period. The comparison
  with longitudinal field determinations 
obtained by other authors is not meaningful since systematic 
differences are frequent between $\Hz$ measurements obtained with
different instruments and/or methods.

The relative amplitude of variation of the longitudinal field is
customarily characterised by the parameter $r=\Hz_{\rm s}/\Hz_{\rm
  g}$, where $\Hz_{\rm s}$ is the smaller and $\Hz_{\rm g}$ the
greater (in absolute value) of the observed extrema $\Hz_{\rm max}$
and $\Hz_{\rm min}$ of $\Hz$ (see \citetalias{1997A&AS..123..353M} for details about the
physical meaning of $r$). 

\begin{figure}[t]
\resizebox{\hsize}{!}{\includegraphics{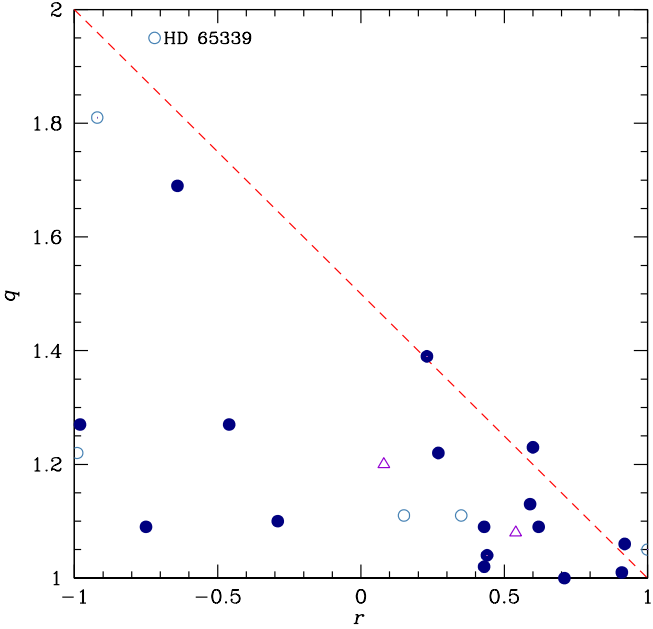}}
\caption{Relative amplitude $q$ of the variation of the mean magnetic field modulus
  against the ratio $r$ of the extrema of the mean longitudinal field
  ({\it see text\/}). Dots identify values of $q$ computed
  from the $\Hm$ data of this paper and of \citetalias{1997A&AS..123..353M}; triangles
  are used for field moduli from the literature. Filled
    symbols correspond to stars for 
  which the $\Hz$ measurements are based on CASPEC observations
  \citep[this paper;][]{1994A&AS..108..547M,1997A&AS..124..475M}, 
    open 
    symbols to other good $\Hz$ measurements from the literature
  ({\it references given in the text\/}). The dashed diagonal line,
  $q=(3-r)/2$, emphasises the concentration of the majority of the stars
  of the sample in the lower left part of the figure: namely, high
  amplitudes of variation of $\Hm$ are observed when both the positive
and negative magnetic poles of the star come into sight during its
rotation cycle.}
\label{fig:q_vs_r}
\end{figure}

The ratio $q=\Hm_{\rm max}/\Hm_{\rm min}$ of
the extrema of $\Hm$ is 
plotted against $r$ in Fig.~\ref{fig:q_vs_r}. The values of $r$ and
$q$ correspond to the extrema of the fitted curves of the variation
of the $\Hz$ and $\Hm$ measurements, respectively, using the fit
parameters of Tables~\ref{tab:mfit}, \ref{tab:zfit},
\ref{tab:mfit-litt}, and \ref{tab:zfit-litt}. There are 22 stars for
which both fits appear in these tables. In addition, we consider the
field modulus as constant in HD~116458, for which a fit to the $\Hz$
variation curve is available; and conversely in HD~192678, the $\Hm$
variation curve is well defined while $\Hz$ is not significantly
variable. 

Figure~\ref{fig:q_vs_r} is an
updated version of Fig.~53 of \citetalias{1997A&AS..123..353M}. With
respect to this \citetalias{1997A&AS..123..353M} figure, the 
number of stars for which the relevant information is available has
doubled. Consideration of this sample of increased statistical
significance strengthens the result found in \citetalias{1997A&AS..123..353M} that the
representative points of all the stars are concentrated towards the
lower left half of the figure. This is emphasised by the dashed line
crossing the figure in diagonal. The star most significantly above
this line, HD~65339 
(identified in the figure), is also the
star with the highest value of $q$ confirmed so far. However, its
location in Fig.~\ref{fig:q_vs_r} remains rather close to the area
where the majority of the studied stars are concentrated.
The conclusion that a
significant difference generally exists between the two poles of the studied
stars, which was drawn in \citetalias{1997A&AS..123..353M} from
consideration of the distribution of the representative points in the
$q$ versus $r$ diagram, is strengthened.  

\begin{figure}[t]
\resizebox{\hsize}{!}{\includegraphics{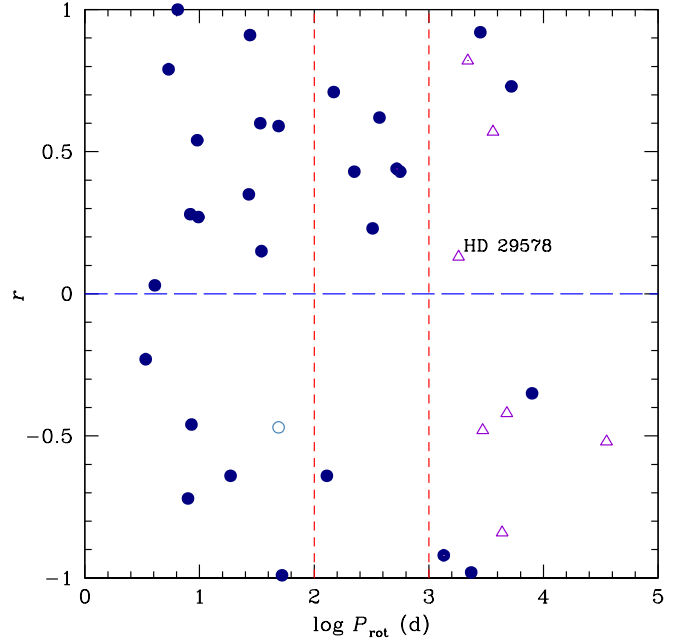}}
\caption{Ratio $r$ of the extrema of the mean longitudinal magnetic field
  ({\it see text\/}) against 
  rotation period. Dots: stars with known rotation periods;
  triangles: stars for which only the lower limit of the
  period is known. Open symbols are used to distinguish those
  stars for which existing measurements do not cover the whole rotation
  cycle. Stars with representative points below the long-dashed
  horizontal line have reversing mean longitudinal fields. The two
  vertical (dashed) lines 
  emphasise the tendency for longitudinal fields showing sign reversals
  to be restricted to rotation periods shorter than $\Prot=100$\,d or
  longer than $\Prot=1000$\,d. }
\label{fig:r_vs_P}
\end{figure}

Not surprisingly, very long rotation periods are under-represented in
Fig.~\ref{fig:q_vs_r}, compared to the whole group of known stars with
resolved magnetically split lines. As a matter of fact, only three stars
with rotation periods longer than 1000\,d have representative points; these are
HD~18078, HD~94660, and HD~187474. If
we assume that the rotation period of HD~166473 is close to the
time span over which its mean magnetic field modulus has been studied so
far, we can use the same
tentative value of the period as 
\citet{2000A&A...359..213L} ($\Prot=4400$\,d) to obtain good fits of
the variation 
curves of $\Hz$ and $\Hm$  \citep[including all measurements of
Table~1 of][]{2007MNRAS.380..181M} with functions of the form given in
Eq.~(\ref{eq:fit2}). From these fits, we derive $r=-0.84$ and $q=1.50$. These
values should represent good approximations of the actual values of
the $r$ and $q$ parameters for this star, as long as its rotation
period is not much longer than 12 years. The corresponding location of
the representative point of the star in Fig.~\ref{fig:q_vs_r} would be
below the $q=(3-r)/2$ line, consistent with the rest of the studied
sample. For HD~9996 ($\Prot\sim22$\,yr), an exact value of $r=-0.35$
is derived from the fully constrained curve of variation of $\Hz$, but
only a lower limit can be set for $q$ ($q>1.34$) since measurements of $\Hm$ have
been obtained only over a fraction of the rotation period. This is
compatible with the location of the representative point of the
star below the $q=(3-r)/2$ line, but it is not impossible that the
actual ratio of the extrema of the field modulus may be greater than
the maximum predicted from this line, $q=1.675$. This would only
require $\Hm$ to become of the order of 3\,kG or smaller over part of
the rotation cycle, which is not implausible at all since the
\ion{Fe}{ii}~$\lambda\,6149.2$ line is definitely not resolved at some
phases. Thus the representative point of
HD~9996 in Fig.~\ref{fig:q_vs_r} may well be above the line drawn
through it, which would represent another exception to the tendency of
the majority of the stars to concentrate in the lower left part of the
$q$ versus $r$ diagram. This is not entirely surprising, since it was
already apparent from the shapes of the spectral lines that the
magnetic field of HD~9996 must have a rather unusual structure (see
Appendix~\ref{sec:hd9996}). 

Figure~\ref{fig:r_vs_P} illustrates the behaviour of $r$ as a function
of the stellar rotation period. Besides the 27 stars listed in
Tables~\ref{tab:zfit} and \ref{tab:zfit-litt} for which the
longitudinal field variations were quantified by fits of the forms
given in Eqs.~(\ref{eq:fit1}) or (\ref{eq:fit2}), this figure includes
representative points of 
\begin{enumerate}
\item HD~192678, for which $\Hz$ does not show
any significant variation. 
\item HD~166473, for which we assumed that
$\Prot=4400$\,d and $r=-0.84$ are suitable, although inexact,
representative values (see above). 
\item HD~335238, whose period is reliably determined but only sampled
  in a very incomplete manner by the existing $\Hz$ measurements.
\item Seven stars for which only a lower limit of the period has been
  established so far. For all of these stars, this lower limit exceeds 1000
  days; it is used as the abscissa of their representative point in
  Fig.~\ref{fig:r_vs_P}.  
\end{enumerate}
For the stars identified under items 3 and 4 above, we computed $r$
adopting for $\Hz_{\rm min}$ and $\Hz_{\rm max}$ the observed
extreme values of the longitudinal field. Given the nature of the
variations of this field moment, these observed extreme values may
greatly differ from the actual extrema of variation of $\Hz$; they may
even have a different sign. Thus the preliminary values of $r$ derived
here for those stars for which the available $\Hz$ data cover only a
fraction of the variation curve can also be strongly different from
their actual values. However, if the
preliminary value of $r$ obtained here is negative, its actual value,
based on the full variation curve of the longitudinal field of the
star, must definitely be negative as well. By contrast, a positive
preliminary value of $r$ does not constrain its final sign, since the
possibility of a sign reversal of $\Hz$ at a phase not yet observed
remains open. In particular, HD~29578, whose
representative point is labelled, appears in the figure with $r>0$
only because of the incomplete phase coverage of the available
longitudinal field measurements. But its incomplete $\Hz$ variation
curve (Fig.~\ref{fig:hd29578})
indicates unambiguously that the mean longitudinal field must 
reverse its sign in that star, so that it actually has $r<0$. On
the other hand, the mean longitudinal field of HD~335238, which is
represented in Fig.~\ref{fig:r_vs_P} by an open circle, also
definitely reverses its sign, but not enough measurements have been
obtained to characterise its variations fully  over the
well-established rotation period. 

Two short-dashed, vertical lines in
Fig.~\ref{fig:r_vs_P} at $\Prot=100$\,d and $\Prot=1000$\,d emphasise the following apparent correlation. All but one of the 7 stars
with a period in the range $[100\ldots1000]$\,d have a non-reversing
mean longitudinal field ($r>0$). By contrast, $r<0$ for 6 of the 17 stars
with $\Prot<100$\,d and for 8 of the 12 stars with $\Prot>1000$\,d
(taking into account the remark above about HD~29578). For rotation
periods exceeding 100 days, the stellar 
equatorial velocity $v$ is low enough so that the observation
of resolved magnetically split lines does not set any constraint on the
inclination angle $i$ of the rotation angle to the line of sight. Thus
one should expect the distribution of the values of $i$ to be random:
any difference in the distribution of the values of $r$ between the
period ranges $100\leq\Prot\leq1000$\,d and $\Prot>1000$\,d must
reflect a difference in the distribution of the values of the angle
$\beta$ between the magnetic and rotation axes. The behaviour of $r$
illustrated in Fig.~\ref{fig:r_vs_P} suggests that the rate of
occurrence of large values of $\beta$ is higher for rotation periods
longer than 1000 days than for periods between 100 and 1000 days. On
the other hand, for the stars with the shortest rotation periods of
the sample (say, $\Prot<30$\,d), $i$ must be small enough so that
Doppler broadening does not smear out the magnetic splitting of
the spectral lines. Thus again the difference in the rate of
occurrence of negative values of $r$ suggests that $\beta$ must in
general be greater for stars with rotation periods shorter than 30
days than for stars with rotation periods between 100 and 1000 days.

Admittedly, the moderate size of the sample limits the statistical
significance of the conclusions that can be drawn from inspection of
Fig.~\ref{fig:r_vs_P}. Those conclusions cannot be
  meaningfully submitted to quantitative statistical tests. The high rate of
occurrence of large values of 
$\beta$ in stars with either the shortest periods ($\Prot<100$\,d) or very long
periods ($\Prot>1000$\,d) seems definite, even though the
  best locations of the dividing lines between the three period ranges
  of interest may be debated. But with only 7 stars with
rotation periods in the 
$[100\ldots1000]$\,d range, one may legitimately wonder if the paucity
of reversing mean longitudinal magnetic fields in this interval is not
a mere statistical fluke. This is a possibility that cannot be easily
ruled out, since the majority of the known Ap stars with
$100\leq\Prot\leq1000$\,d appear in Fig.~\ref{fig:r_vs_P}. HD~221568 ($\Prot=159$\,d) is the only
other star with a well-established period in that range that is listed
in the \citet{1998A&AS..127..421C} catalogue and its supplement
\citep{2001A&A...378..113R}. Among the
Ap stars with 
resolved magnetically split lines, HD~110274 has $\Prot=265\fd3$ (see
Table~\ref{tab:stars_litt}). There are not enough
$\Hz$  data available for either of these two stars to decide whether $r$ is positive or negative. 

The pattern apparent in Fig.~\ref{fig:r_vs_P} can 
be related to the conclusion reached by \citet{2000A&A...359..213L}
that magnetic Ap stars with long rotation periods 
(greater than a month) have their magnetic and rotation axes nearly
aligned, unlike shorter period Ap stars in which the angle between
these two axes is usually large. This conclusion was reached from
computation of models of the fields of the studied stars consisting of
the superposition of collinear, centred, dipole, quadrupole, and
octupole components. It was supported by the results of a separate
study by \citet{2002A&A...394.1023B}, who assumed that the magnetic
topology of the stars of interest is described by the superposition of
a dipole and a quadrupole field at the centre of the star with
arbitrary orientations; in this case, the angle under consideration is
that between the dipole axis and the rotation axis. With respect to
these earlier results, with more information now available about the
longest-period stars, a new trend that had been previously unnoticed
appears in Fig.~\ref{fig:r_vs_P}. Namely, the tendency for the 
magnetic and rotation axes to be aligned seems to be restricted to
an intermediate period range, while the angle between them appears to
be predominantly large both in stars with relatively short periods and
in stars with extremely long periods. If anything, the additional
evidence presented here rather questions the robustness of the conclusion
reached in earlier studies about the existence of a systematic difference in
the magnetic geometries between stars with rotation periods shorter than
30--100 days and stars with periods above this value. 

In light of the discussion presented in this section, the
interpretation of the dependence on the rotation period of the
anharmonicity of the variations of the mean magnetic field modulus,
illustrated in Fig.~\ref{fig:m2overm1_vs_P}, becomes clear. This can be
understood by using different
symbols in Fig.~\ref{fig:m2overm1_vs_P} to distinguish those stars
where a sign reversal of $\Hz$ is observed ($r<0$; open circles) from
those where $\Hz$ is found to have always the same sign ($r>0$; 
  dots). For one star, HD~59435 (open triangle), no
longitudinal field determinations were obtained. 

Then, a clear-cut result
appears: all but one of the stars in which the field modulus variations show no
significant deviation from a sinusoid have a non-reversing
longitudinal field. The exception is HD~18078, whose magnetic field
appears to have a very unusual structure (see below). This implies
that anharmonicity in the 
$\Hm$ curve generally occurs when both poles come into sight as the star
rotates. 

This is not surprising. For the simplest magnetic geometry, a
centred dipole, if both poles come alternatively into view as the star
rotates, the $\Hm$ variation curve is a perfect double wave,
that is, a sinusoid with twice the rotation frequency of the star
with no contribution of the fundamental. By contrast, if the angle
between the rotation axis and line of sight is small, or if the
angle between the magnetic and rotation axes of the star is small so
that the same pole remains visible at all times, the shape of the
variation of $\Hm$ is a sinusoid with the rotation period of the
star. In summary, Fig.~\ref{fig:m2overm1_vs_P}, as 
Fig.~\ref{fig:r_vs_P}, suggests that the angle $\beta$ between the
magnetic and rotation axes is statistically large in stars with
rotation periods shorter than 100 days or longer than 1000 days, while
the two axes tend to be aligned in the period range
$[100\ldots1000]$\,d. 

In addition,  
it is remarkable and highly significant that none of the
mean field modulus variation curves for the stars of the
studied sample are perfect
double sinusoids. This strongly indicates that the two magnetic poles
of (slowly rotating) Ap stars are different: the  
magnetic field structures significantly depart from centred dipoles. This
conclusion had already been reached in
\citetalias{1997A&AS..123..353M}; it is strengthened 
here because of the increase in the number of stars with  fully characterised mean field modulus variations. 

\begin{figure}
\resizebox{\hsize}{!}{\includegraphics{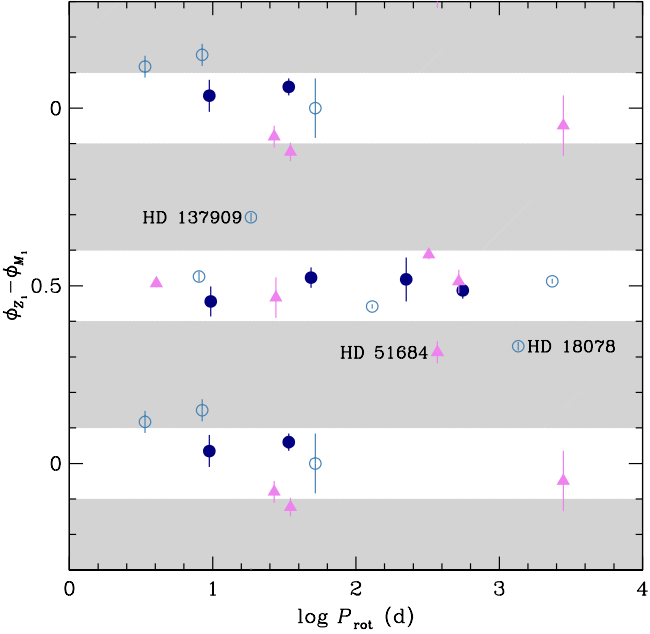}}
\caption{Phase difference $\phi_{Z_1}-\phi_{M_1}$ (modulo 1) between
  the fundamentals of the fits of the measurements of $\Hz$ and $\Hm$
  by functions of the forms given in Eqs.~(\ref{eq:fit1}) or
  (\ref{eq:fit2}), against stellar rotation period. Open
    symbols: stars showing $\Hz$ reversal; filled symbols:
  stars in which $\Hz$ has a constant sign. Among the latter, stars
  with positive values of $\Hz$ are represented by dots, while
  triangles correspond to negative values of $\Hz$. For a
  fraction of the stars, the error bar is 
  shorter than the size of the representative symbol. 
    Shaded zones correspond to forbidden 
  regions where representative points of stars with dipole-like fields
  are not expected to be found. We identified three stars that depart from this
  expected behaviour.}
\label{fig:phi_zm}
\end{figure}

For the first time, we built a statistically significant sample
of stars for which both the curves of variation of the mean magnetic
field modulus and of the mean longitudinal magnetic field are fully characterised. This gives us an unprecedented opportunity to study the
relation between the variations of these two field moments. In
Fig.~\ref{fig:phi_zm}, which is based on Tables~\ref{tab:mfit},
\ref{tab:zfit}, \ref{tab:mfit-litt}, and \ref{tab:zfit-litt}, the phase
lag $\phi_{Z_1}-\phi_{M_1}$ (modulo 1) between the fundamental 
components of the fits by a function of the form given by
Eqs.~(\ref{eq:fit1})  or (\ref{eq:fit2}) of the $\Hz$ measurements, on the
one hand, and of the $\Hm$ data, 
on the other hand, is plotted against stellar rotation
period. Admittedly, a significant number
of the studied stars show 
some degree of anharmonicity in the variations of either or both of
the considered field moments. But in the vast majority of these stars, the
amplitude of the harmonic is considerably smaller than that of the
fundamental, so that in first approximation, the actual phases of
extrema do not differ much from those of the fundamental. Moreover, if
(as is
widely believed) the field geometry is dominated by a 
nearly dipolar component encompassing the whole star, on which
higher order, smaller scale 
structures are superimposed, consideration of the fundamental in the
study of phase relations between various field moments makes
sense on physical grounds, since the large-scale structure to which it
corresponds seems more likely to lend itself to the study of
star-to-star systematics in the field properties. 

For a centred dipole and for some of the most commonly used
axisymmetric field models, such as a dipole offset along its axis
that passes through the centre of the star
\citep[e.g.][]{1970ApJ...160.1059P} or the superposition of collinear
centred dipole, 
quadrupole, and octupole \citep[e.g.][]{2000A&A...359..213L}, one
expects the fundamental components of the variations of $\Hz$ and
$\Hm$ to be in phase or in anti-phase. For fields showing modest
departures from these simple models, $\phi_{Z_1}-\phi_{M_1}$ should 
accordingly remain close to 0 or to 0.5. To reflect this expectation, in Fig.~\ref{fig:phi_zm} 
we shaded the bands corresponding to
$0.1\leq\phi_{Z_1}-\phi_{M_1}\leq0.4$ and
$0.6\leq\phi_{Z_1}-\phi_{M_1}\leq0.9$, which should be forbidden
regions for the representative points of stars with nearly dipolar
fields. One can see that, consistent with the view that Ap stars
have magnetic structures dominated by a dipole-like component, the
vast majority of the stars of the present sample are confined to the
allowed (unshaded) regions of the figure, or at least are not
significantly outside them. HD~18078 and HD~137909 are clear
exceptions. HD~51684 is a marginal exception, $2.6\sigma$
  below the strict limit $\phi_{Z_1}-\phi_{M_1}=0.4$, which however is
fairly arbitrary. These three stars are labelled in the
figure. 

The poor phase
distribution of the existing measurements of the magnetic field of
HD~51684, especially of the 
longitudinal field, implies that the shapes of their variations are
poorly constrained, possibly to an extent worse than indicated by the
formal standard errors of the fits. Accordingly, it would be tempting
to regard the apparent departure of the star from
normality in Fig.~\ref{fig:phi_zm} as spurious. However,
consideration of Fig.~\ref{fig:hd51684} leaves little doubt about the
reality of the existence of a very significant phase shift between the
extrema of the longitudinal field and the field modulus. But the
available observations are insufficient to lend themselves to a more
detailed discussion of this unusual and interesting case. 

By contrast, the curves of variation of $\Hz$ and
$\Hm$ in HD~137909 are defined beyond any ambiguity (not only in the
present study, but also in other published works; see
Appendix~\ref{sec:hd137909} for some recent references), and while the
anharmonicity of the field modulus variations is significant, it can
by no means account for the high value of
$\phi_{Z_1}-\phi_{M_1}$. Hence the location of the
representative point of HD~137909 in one of the ``forbidden'' regions
of Fig.~\ref{fig:phi_zm} is highly significant; it earmarks the star
as extremely anomalous. This is not altogether surprising. HD~137909
has long since been one of the prototypes of the group of the Ap
stars, and among its members is one that shows some of the
most striking spectral and 
chemical peculiarities. The discrepancies between the field modulus
determinations in this star obtained from AURELIE spectra and those
based on other spectrographs are unique; this is the only star of the
present sample in which the two groups of measurements could not be
reconciled in a simple manner (e.g. by applying a constant offset). The
uniqueness of the star in this respect most likely arises from the
fact that the structure of its magnetic field is very different from
that of most other stars. The failure by \citet{2000A&A...358..929B}
to find a 
model that adequately represents the behaviour of all the magnetic 
moments of the star, despite using a fairly sophisticated
representation of the field structure, is another sign that points to
a very unusual geometry \citep[see also][]{2001A&A...369..889B}. 

HD~18078 has been much less studied. But the phase
shift between the variation curves of its longitudinal field and its field modulus, which is as significant as for HD~137909, is not
the only observable manifestation of the unusual structure of
its magnetic field. This unusual structure is also revealed by the significant
broadening of its resolved line components, which restricts our
ability to determine its field modulus close to its phase of minimum
(see Appendix~\ref{sec:hd18078}), and by the fact that this star is one of only
a handful of stars to shown anharmonic $\Hz$ variations while its
$\Hm$ variation curves does not significantly depart from a sinusoid. 

Leaving those three exceptions aside, another remarkable feature appears in
Fig.~\ref{fig:phi_zm}, when one distinguishes stars with reversing
longitudinal fields (represented by open symbols) from stars with
longitudinal fields keeping the same sign throughout the rotation
period (filled symbols). Focusing attention on the latter, we
furthermore distinguish those whose longitudinal field is positive
throughout the rotation period (identified by dots) from those for
which this field moment is negative at all phases (triangles). One
can see in Fig.~\ref{fig:phi_zm} that both among the stars in
which the $\Hm$ and $\Hz$ variations occur (nearly) in phase and among 
those where they occur (nearly) in anti-phase, the subset of stars
whose longitudinal field shows no reversal is fairly evenly split
between stars  with $\Hz>0$ and stars with $\Hz<0$. On the other hand,
for a centred dipole geometry such that $\Hz$ keeps the same sign
throughout the stellar rotation period, the longitudinal field
variations of stars with $\Hz>0$ should always occur in phase with
their field modulus variations, while for stars with $\Hz<0$, the
$\Hz$ and $\Hm$ variation curves should be in anti-phase. Thus, about
half of the studied stars whose longitudinal field is not reversing
show a behaviour that is the opposite the predicted behaviour for a
centred dipole. This behaviour can be achieved with axisymmetric
configurations where the dipole is offset along its axis, or where
moderate collinear, centred higher order multipole components are
superimposed to a centred dipole. In Fig.~\ref{fig:phi_zm}, we
identify seven stars whose magnetic field structure shows definite
departure from a centred dipole in the above-described manner (filled
triangles in the $\phi_Z-\phi_M\simeq0$ permitted band, filled
circles in the $\phi_Z-\phi_M\simeq0.5$ permitted band); these seven stars are HD~12288,
HD~14437, HD~93507, HD~94660, HD~144897, HD~188041, and HD~318107.  

In summary, all but three of the stars for which enough data exist 
show longitudinal field and field modulus variations nearly in
phase or nearly in anti-phase. This is consistent with
a magnetic field 
structure dominated by a single dipole. A significant number of the
studied stars show definite departures from a centred dipole geometry,
but in general their field structures appear to bear a fair
degree of resemblance with two of the simple, 
axisymmetric models that have long been used to represent them, namely
a dipole offset along its axis, or the superposition of collinear,
centred low-order multipoles. Furthermore, the fact that the behaviour
of the phase difference illustrated in Fig.~\ref{fig:phi_zm} shows no
dependence on the rotation period suggests that this conclusion about
the field geometry can probably be
generalised to all Ap stars (and not only to the strongly magnetic,
slowly rotating subsample considered here). While the phase lag
between the longitudinal field and field modulus variation curves of
HD~137909 has been known for a long time \citep{1970ApJ...160.1049W},
what emerges here is how 
exceptional this situation is. Accordingly, HD~137909 cannot be
regarded as typical of the behaviour of the majority of the
stars of its class, and one should be cautioned against generalising
conclusions drawn from its study.

\subsection{The crossover}
\label{sec:xdisc}
Resolved magnetically split lines can be observed in stars where the
rotational Doppler broadening of the spectral lines is sufficiently
small compared to the separation of their Zeeman components. Up to now,
the observed crossover effect has been interpreted as resulting from 
the existence of a
correlation between the rotational Doppler shift of the contributions
to the observed, disk-integrated line coming from different parts of
the stellar disk, on the one hand, and the different Zeeman shifts of
their right and left circularly polarised components, corresponding to
the local magnetic field strength and orientation, on the other
hand \citep{1995A&A...293..733M}. This correlation results from the
large-scale structure of the 
field and its inclination with respect to the stellar rotation axis:
it is intrinsic to the oblique rotator model. In what follows, we
refer to this process as rotational crossover to distinguish
it from other effects that may also create difference of line widths
between right and left circular polarisation observations. Rotational
crossover is, to first order, proportional to $\vsi$, so that in stars
with magnetically 
resolved lines, it is typically small, often vanishingly so. Thus, one
expects 
crossover measurements to have a limited, but definite, value to
characterise the magnetic fields of these stars.

In all stars where its variation throughout the rotation period has
been determined prior to this study, the crossover was found to reverse
its sign as the star rotates
\citep{1995A&A...293..733M,1997A&AS..124..475M}. Therefore, to
characterise this crossover with a single  
number for each star, such as for the longitudinal field, we use its rms
value,
\begin{equation}
\xoverrms=\left({1\over N_{\rm x}}\,\sum_{i=1}^{N_{\rm x}}\xover_i^2\right)^{1/2}\,,
\end{equation}
where $N_{\rm x}$ is the number of individual measurements of $\xover$ that
we obtained, and $\xover_i$ is the $i$-th such
measurement. Furthermore, since measurements of the crossover in the
stars studied here are often close to the limit of significance, we
also consider their reduced $\chi^2$,
\begin{equation}
\chisq={1\over N_{\rm x}}\,\sum_{i=1}^{N_{\rm x}}
\left(\xover_i\over\sigma_{x,i}\right)^2\,,
\end{equation}
where $\sigma_{x,i}$ is the error of the $i$-th measurement of the
crossover. The reduced $\chi^2$ is useful in marginal cases to
decide if crossover has actually been detected. 

As for the longitudinal field, the crossover measurements presented
here are considerably more precise than the older determinations of
\citet{1995A&A...293..733M} and \citet{1997A&AS..124..475M}. Indeed, the
median of the standard errors $\sigma_x$ of all the measurements
appearing in Table~\ref{tab:hzxq} is 848~\kms\,G, while the median of
the standard errors of the 44 measurements of the crossover in Ap
stars with magnetically resolved lines of the above-mentioned previous
studies is 1557~\kms\,G. Close to the detectability limit, this
difference of precision may be very significant. Accordingly, in
Table~\ref{tab:xoverrms}, we give the values of $N_{\rm x}$, $\xoverrms$,  and
$\chisq$ both for the new measurements of this paper (in Cols.~2 to 4)
and for all the existing measurements of the crossover (in Cols.~5 to
7). For the sake of clarity, data in Cols.~5 to 7 appear only for
those stars for which previous measurements of the crossover do exist.

\begin{table}[t]
\caption{Summary of crossover measurements in Ap stars with resolved
  magnetically split lines. Italics identify those stars where
  crossover is detected at the 99\% confidence level (see text).}
\label{tab:xoverrms}
\centering
\small{
\begin{tabular}{rrrrrrr}
\hline\hline\\[-4pt]
&\multicolumn{3}{c}{This paper}&\multicolumn{3}{c}{All measurements}\\
\multicolumn{1}{c}{HD/HDE}&\multicolumn{1}{c}{$N_{\rm x}$}&\multicolumn{1}{c}{$\xoverrms$}&\multicolumn{1}{c}{$\chisq$}&\multicolumn{1}{c}{$N_{\rm x}$}&\multicolumn{1}{c}{$\xoverrms$}&\multicolumn{1}{c}{$\chisq$}\\
&&\multicolumn{1}{c}{(km\,s$^{-1}$\,G)}&&&\multicolumn{1}{c}{(km\,s$^{-1}$\,G)}\\[4pt]
\hline\\[-4pt]
   965&{\it  7}&{\it 2013}&{\it  3.3}\\
  2453&{\it  5}&{\it  762}&{\it  3.7}\\
 29578& 9& 302& 0.4\\
 47103& 6&3012& 0.9\\
 50169& 8& 598& 0.7& 9& 631&0.7\\
 51684& 8& 754& 0.7\\
 55719&{\it  9}&{\it 1371}&{\it  2.7}&{\it 12}&{\it 2093}&{\it 2.9}\\
 61468& 4&1106& 2.2\\
 70331&{\it 15}&{\it 3636}&{\it  2.3}&{\it 16}&{\it 3561}&{\it 2.3}\\
 75445& 3& 414& 0.7\\
 81009&{\it 12}&{\it 1732}&{\it  2.3}&{\it 13}&{\it 3598}&{\it 3.2}\\
 93507&{\it 10}&{\it 2071}&{\it  3.0}&{\it 12}&{\it 1930}&{\it 2.6}\\
 94660& 8& 863& 2.5&12& 821&1.7\\
110066& 3& 325& 1.8\\
116114&{\it  6}&{\it 1118}&{\it  6.6}&{\it  7}&{\it 1118}&{\it 5.8}\\
116458&11& 605& 1.4&21&1226&1.4\\
119027& 1&  38&    \\
126515&{\it 10}&{\it 1934}&{\it  2.8}&{\it 19}&{\it 5264}&{\it 2.2}\\
134214& 6& 390& 1.7& 8&1494&2.1\\
137909&{\it  6}&{\it 1788}&{\it 11.5}&{\it 21}&{\it 2212}&{\it 6.0}\\
137949& 7& 740& 1.3& 9&1827&2.4\\
142070&{\it 13}&{\it 3517}&{\it 40.9}\\
144897&{\it 12}&{\it 1702}&{\it  3.3}&{\it 13}&{\it 1642}&{\it 3.1}\\
150562& 1& 347&    \\
318107& 5&3376& 1.0& 6&3517&1.0\\
165474& 5& 526& 0.3& 8&1064&0.4\\
166473& 7& 706& 0.6&10&1020&1.2\\
187474& 9& 818& 1.1&20&1436&1.2\\
188041& 2& 259& 0.5& 9&1602&1.4\\
335238& 2&2023& 0.7& 3&3495&4.3\\
201601& 6& 257& 0.9&18& 810&0.8\\
208217&{\it  8}&{\it 4718}&{\it  9.6}\\
213637& 1& 431& 0.8\\
216018& 6&1224& 2.3& 9&1040&1.7\\[4pt]
\hline
\end{tabular}}
\end{table}

When the $\chi^2$ statistic indicates that crossover is detected
at the 99\% confidence level, the corresponding entry appears in
italics in the table. Such a detection is achieved in 12 of the 34
stars for which CASPEC Stokes $V$ spectra were recorded. Seven of
these 12 stars have rotation periods that are shorter than 50 days, similar
to all the stars in which crossover had been observed in previous studies
\citep{1995A&A...293..733M,1997A&AS..124..475M}. We were able to
compute, for five of these (relatively)
short period stars---namely, 
HD~81009, HD~137909, HD~142070, HD~144897, and HD~208217---a formally
significant fit of the observed values of $\xover$ against 
rotation phase, of one of the forms given in Eqs.~(\ref{eq:fit1}) or
(\ref{eq:fit2}) (see Table~\ref{tab:xfit}). The
independent term of the fit, $X_0$, is not significantly different
from 0 in HD~81009, HD~142070, and HD~208217, and only marginally
non-zero (at the $3.0\sigma$ level) in HD~137909. Also, in these
four stars, the difference $\phi_{X_1}-\phi_{Z_1}$ between the phase
origins of the fundamental components of the fits of $\xover$ and $\Hz$, respectively,
is equal to 0.25 within their uncertainties. In
other words, the variations of the crossover are lagging behind those
of the longitudinal field by a quarter of a cycle. Thus, these stars 
display the typical behaviour of the rotational crossover, as observed
so far in 
all Ap stars in which the variation of this field moment throughout the 
rotation period was studied, but HD~147010
\citep{1995A&A...293..733M,1997A&AS..124..475M}. 

The behaviour  
observed in HD~144897 is very different. The $\xover$ measurements
are best fitted by a sinusoid with twice the rotation frequency of the
star without a fundamental component. Although not visually compelling
(see Fig.~\ref{fig:hd144897}), this fit appears to be very significant
with a value of $X_2$ determined at the $4.4\sigma$ level. The $\Hz$
variation curve also seems to show some degree of anharmonicity: while
the fit coefficient $Z_2$ is not formally significant (at the
$2.7\,\sigma$ level), a fit including it is visually better than a fit
by the fundamental alone (see Fig.~\ref{fig:hd144897}). Within
  the errors, the derived
phase difference $\phi_{X_2}-\phi_{Z_2}$ is 
consistent with a phase lag of 0.125 rotation cycle of the crossover
variations with respect to the longitudinal field, as previously
observed for the first harmonic in other cases \citep{1995A&A...293..733M}. 
What is
unusual in HD~144897, though, is that the independent term of the fit,
$X_0$, is positive, at a high level of significance,
$6.9\sigma$. As a matter of fact, the crossover in HD~144897 is
never signficantly negative. This is a behaviour that had never been 
seen in any other star. It is definitely incompatible with
the process of rotational crossover in a star in which the 
magnetic field is symmetric about a plane containing the rotation
axis, which is a condition fulfilled by most analytic models successfully used
until now to
represent Ap star magnetic fields. Actually, it is not clear that the
behaviour observed in HD~144897 can
be explained by the occurrence of rotational crossover for any plausible
magnetic configuration.  

In the other two short period stars where the $\chi^2$ test
indicates that crossover is detected above the 99\% confidence level,
HD~70331 and HD~116114, its variability is ill-defined, as is that of
the longitudinal field. In HD~70331, the individual measurements
are distributed between negative and positive values. The apparent
predominance of negative $\xover$ values in HD~116114 may just reflect
their marginal significance and their uneven distribution over the
stellar rotation period; see below, however. 

There is one more short period star, for which a fit of the $\xover$ 
variations at the threshold of formal significance
($X_1/\sigma_{X_1}=3.0$) can be obtained, although crossover is
detected only at a confidence level of 70\%, according to the $\chi^2$
statistic; this star is HD~318107. The $\xover$
fit is not inconsistent with the standard crossover behaviour,
$X_0=0$ and $\phi_{X_1}-\phi_{Z_1}=0.25$. However one should in keep
in mind that we obtained only six determinations of the crossover
in this star, so that the fit has only 3 degrees of freedom.

No crossover was detected in the last star with period shorter than
50~d that was observed with CASPEC as part of this project, HD~335238,
for which, however, only three determinations were performed.

In  three stars with intermediate periods in
Table~\ref{tab:xoverrms}, crossover is detected at the 99\%
confidence level: these are HD~126515 ($P=130$~d), HD~2453
($P=521$~d), and 
HD~93507 ($P=556$~d). Crossover had not been previously detected in stars
with such long periods, but the following order of magnitude estimates
suggest that the observed effect may still be consistent with the
rotational crossover interpretation. It is easy to show from the
definitions of the rotational crossover and of the mean magnetic field
modulus, that for a given star, at any phase, the
following inequality is verified:
\begin{equation}
|\xover|\leq\vsi\,\Hm\,.
\end{equation}
Thus, in order for the observed crossover to be consistent with the
rotational crossover interpretation, a necessary condition is that 
\begin{equation}
\vsi\geq{|\xover|_{\rm max}\over\Hm_{\rm max}}\,,
\label{eq:xover_cond}
\end{equation}
where the subscripts refer to the maximum  
of the considered moments over the rotation period. For the
crossover, the absolute value is used. For practical application of
this condition, we adopt the largest of the individual measurements of
the crossover (in absolute value) for $|\xover|_{\rm max}$, and of the mean field modulus
for $\Hm_{\rm max}$. For the crossover, we restrict ourselves
to the consideration of the measurements of this paper, since as
discussed above, earlier measurements may be considerably less
precise. In this way, the minimum values that we derive for $\vsi$
are 0.24~\kms for HD~126515, 0.27~\kms for HD~2453, and 0.50~\kms for
HD~93507. The necessary condition (\ref{eq:xover_cond}) can be expressed in terms of the
stellar radius $R$ (in units of solar radii), by application of the
well-known relation between the latter, $\vsi$ (in \kms) and the
stellar rotation period $P$ (in d), i.e. \begin{equation}
R\,\sin i=P\,\vsi/50.6.
\end{equation}
We find that for HD~126515, $R\,\sin i\geq 0.6~R_\odot$; for HD~2453, 
$R\,\sin i\geq 2.8~R_\odot$; and for HD~93507, $R\,\sin i\geq
5.5~R_\odot$. These lower limits for HD~126515 and HD~2453 are
consistent with the typical values of the radii of stars of their
spectral types. They also compare well with the radius estimates
obtained from Hipparcos-based luminosity determinations by
\citet{2000ApJ...539..352H}: $R=(2.97\pm0.46)~R_\odot$ for HD~2453,
and 
$R=(2.35\pm0.42)~R_\odot$ for HD126515. For HD~93507, the derived
lower limit of the radius is 
somewhat high for a star of spectral type A0. However taking the involved uncertainties into account, this does not decisively rule out
rotational crossover as the source of the observed behaviour. Indeed, 
if one uses for $|\xover|_{\rm max}$ and $\Hm_{\rm max}$ the values
computed from the fits of 
the variations of the crossover (Table~\ref{tab:xfit}) and of the
field modulus (Table~\ref{tab:mfit}) instead of the largest
measurement of each of these moments, the resulting lower limit on
$R\,\sin i$ decreases to $(3.8\pm1.2)~R_\odot$, which is marginally
compatible with the spectral type of HD~93507. 

Thus HD~126515, HD~2453, and HD~93507 fulfil at least one necessary
condition for their observed crossover to be consistent with the
classical interpretation of the effect in terms of rotational 
crossover.
In the case of HD~126515, this interpretation receives further
support from the observation by \citet{2002A&A...394..151C} of
radial velocity variations due to approaching and receding
inhomogeneities on the stellar surface. By contrast, for HD~93507, the
fact that the curves fitted to our longitudinal field data, on the one
hand, and to our crossover measurements, on the other hand, vary in
phase rather than in quadrature, raises some difficulties. We shall come
back to this point below. 

The remaining two stars of Table~\ref{tab:xoverrms} in which
crossover is detected at the 99\% confidence level, HD~965 and
HD~55719, have very long periods. For those stars, identification of the origin
of the effect is challenging. The same radius lower
limit test as above applied to these two extremely slow rotators
(with the values of their rotation periods set to a conservative 
common lower limit of 10~y) definitely rules out rotational crossover
as the 
source of the observed effect: $R\,\sin i\geq49~R_\odot$ for HD~965
and $R\,\sin i\geq32~R_\odot$ for HD~55719.

HD~94660 may be another example of an extremely slow rotator showing
crossover. The $\chi^2$ statistic supports the reality of the
detection of the effect at a confidence level between 90\%, if all
the available crossover measurements are considered, and 98\%, if 
the $\chi^2$ test is restricted to the better determinations presented
in this paper. In addition, the amplitude of a fit of the $\xover$
data by a function of the form given in Eq.~(\ref{eq:fit1}) is
significant 
at the $3.5\sigma$ level. The correlation coefficient $R$ also
supports the existence of a correlation between $\xover$ and the
rotation phase at a confidence level of 99\%. Using the value of the
stellar radius published by \citet{2000ApJ...539..352H},
$R=(3.15\pm0.40)~R_\odot$, we can invert the argument of the previous
paragraphs to derive an upper limit of the crossover that is compatible with
the field modulus of the star, 
$\xover\leq(50.6\,R/P)\,\Hm_{\rm max}=(364\pm46)$~\kms. Strikingly,
this limit is within a factor of $\sim2$ of the actually observed
values. Taking the uncertainties and approximations
involved in this comparison into account, this actually suggests that it is not
implausible, although rather unlikely, that rotational crossover is 
detected in HD~94660. This is further supported by the
fact that, based on the fit parameters shown in Tables~\ref{tab:zfit}
and \ref{tab:xfit}, the phase lag between the crossover variation and
that of the longitudinal field amounts, within errors, to a quarter of 
a cycle. 

As pointed out in Appendix~\ref{sec:notes}, for stars for which a
sufficient number of measurements of the crossover have been obtained
as part of the present study,\footnote{The older data from  \citet{1995A&A...293..733M} and \citet{1997A&AS..124..475M} are not
  considered 
  in this part of the discussion because of their lower
  precision.}
the fact that all those measurements, but possibly one or two, yield
values of $\xover$ with the same sign also represents an indication
that the effect is actually detected, even though none of these values
taken individually reach the threshold of significance. The stars in
which this applies are HD~965, 2453, 116114, 116458, and 187474, in
which (almost) all $\xover$ determinations are negative, and HD~93507
and 144897, where the crossover is found to be (almost) always
positive. For five of these seven stars, the $\chi^2$ test also
indicates that crossover was detected at the 99\% confidence
level. The exceptions are HD~187474, for which the $\chi^2$ statistic
indicates a non-zero crossover at the 70\% confidence level (based on
all available measurements; this drops to 50\% if only the new
determinations of this paper are considered), and HD~116458, where the
$\chi^2$ value corresponds to a detection at the 70\% confidence level
(for either the whole set of existing $\xover$ measurements, or a
subset restricted to the new data of this paper). 

The usage of the $\chi^2$ statistic to decide in marginal cases if
crossover has indeed been detected rests on the implicit assumption
that the only significant errors affecting the measurements are of
statistical nature, to the exclusion of systematic errors. Moreover,
systematic errors could plausibly account for the fact that all the
crossover measurements in a star in which this effect is very small
yield values of the same sign. The following arguments support the
view that the crossover determinations of this paper are not
significantly affected by systematic errors. 

No crossover is detected
in more than half of the studied stars. For a number of these stars, the rms
crossover is considerably smaller than in any of the stars in which
detection is achieved. For instance, in the stars HD~29578, 134214, and
201601, which have all been measured six times or more, $\xoverrms$ is 302, 390, and 257~\kms\,G,
respectively. By contrast, all the stars
in which the $\chi^2$ test indicates the presence of a crossover at
the 99\% confidence level, have $\xoverrms$ twice as large or (much)
more. 

Also, for those stars in which (almost) all the crossover data
have the same sign, we consider the average $\xover_{\rm av}$ of
all the crossover measurements in a given star, weighted according to
their relative uncertainties $\sigma_x$ \citep[see Eq.~(16)
of][]{1994A&AS..108..547M}. For the most marginal cases of HD~116458
and HD~187474, 
this average is $(-514\pm118)$~\kms\,G and
$(-505\pm165)$~\kms\,G, respectively. These values can again be compared with those
obtained for HD~29578, which is $\xover_{\rm av}=(0\pm87)$~\kms\,G,
HD~134214, which is
$\xover_{\rm av}=(-296\pm118)$~\kms\,G, and HD~201601, which is $\xover_{\rm
  av}=(-80\pm112)$~\kms\,G. This comparison represents a strong
indication that the systematic measurement of a negative crossover in
HD~116458 and HD~187474, and the formally significant non-zero average
$\xover_{\rm av}$ in each of them, are not the result of systematic
errors in the measurements, and that they instead reflect the
occurrence of some physical process. This conclusion also applies of course to
the other stars in which (almost) all the measurements of the
crossover have the same sign and in which, furthermore, the $\chi^2$
test supports the reality of the detection of a crossover. 

The fact
that the constant sign of the crossover is positive in some of these
stars, and negative in others, represents an additional argument
against the suspicion that the constant sign in any star may be due to
systematic measurement errors. 

Thus we established that the crossover measurements presented
here are unlikely to be significantly affected by systematic
errors large enough to lead to spurious detections. We built an
inventory of the stars in which our CASPEC 
observations yield definite or probable crossover detections. We discussed the reliability of these detections. We showed that in
most stars, the observations can be interpreted in terms of rotational
crossover. Remarkably, some of the stars to
which this interpretation is applicable rotate very slowly. In
previous studies, the star with the longest rotation period in which
crossover had been detected at a high level of confidence was HD~81009
\citep{1997A&AS..124..475M} with a period of 34~d. With the better
data analysed in this study, we showed that rotational
crossover is almost certainly detected in a star with period longer
than 100 days (HD~126515, $P=130$~d); that it may be detectable in
stars with periods of the order of 500~d (such as HD~2453); and that
it may even plausibly have been observed in HD~94660, which has a
period of 7.7~y. Our data are insufficient to draw definitive
conclusions for either HD~2453 or HD~94660, but the discussion of these
cases in the previous paragraph clearly indicates that it is not
unrealistic with current observational facilities to observe
rotational crossover in stars with rotation periods of several
years. 

Yet, some of the fairly definite crossover detections reported here
are inconsistent with the rotational crossover mechanism. The most
challenging examples are HD~965 and HD~55719, which rotate too slowly
to generate Doppler shifts of the order of magnitude required to
account for the measured values of $\xover$. As long as the local
emergent Stokes $V$ line profiles at any point of the stellar surface are
anti-symmetric, the apparence of crossover in disk-integrated
observations implies the presence of some macroscopic velocity field
that is variable across the stellar surface, in a way that some
large-scale correlation pattern exists between the Doppler shift that
it generates and the Zeeman effect across the stellar surface. The
effects that can generate departures from Stokes $V$ anti-symmetry at a
given point of the stellar surface have been discussed by
\citet{1995A&A...293..733M}. Two of these effects, non-LTE populations of the
magnetic 
substates of the levels involved in the observed transitions and
the partial Paschen-Back effect in these transitions, can almost certainly 
be ruled out for the studied spectral lines in the stars of
interest. The third effect, which is the existence of velocity gradients along the
line of sight in the line-forming region \citep{1983SoPh...87..221L},
represents another form of 
Doppler-Zeeman combination. Thus the need for a velocity field of some
kind appears inescapable.

One well-known type of velocity field that is
definitely present in (cool) Ap stars is non-radial pulsation, which
furthermore has a large-scale structure that makes it a priori suitable
for the generation of correlations across the stellar surface between
the Doppler shifts resulting from it and the local Zeeman
effect. At first sight, there are several arguments against the
hypothesis that this mechanism may be responsible for the observed
crossover. The effective temperature of HD~55719, an A3 star, is
close to, or 
above the upper limit of the temperature range to which all rapidly
oscillating Ap (roAp) stars known so far are confined. By contrast,
HD~965 has all the characteristics of typical roAp stars, but all
efforts so far to detect pulsation in this star have been unsuccessful
\citep{2005MNRAS.358.1100E}. Also, the integration times that were used
to record the CASPEC spectra of HD~965 analysed here, ranging from 25
to 40~min, are longer than the typical pulsation periods of roAp
stars, which for the stars known so far range from 5 to
24~min. Thus one would expect the disk-integrated radial velocity
variations to average out to a large extent. This does not apply to
HD~55719, which is much brighter than HD~965, and for which the
exposure times of our CASPEC observations exceeded 5~min only on a
single cloudy night. 

However, the pulsation amplitude in roAp stars varies across the depth of
the photosphere (e.g. the introduction of \citealt{2006MNRAS.370.1274K}
for an overview). The most obvious
observational manifestation of this depth effect is the existence of
systematic amplitude differences in the radial velocity variations of
lines of different ions or of lines of different intensities of a
given ion. In particular, it is not unusual for a subset of spectral
lines not to show any radial velocity variations due to pulsation;
this indicates the presence of at least one pulsation node in the 
photospheres of roAp stars. In first
approximation, the photospheric depth dependence of pulsation is
inferred from the observed radial velocity amplitudes by assuming that
each line forms at a single depth. This is, of course, an
oversimplification: line formation actually takes place over a range
of optical depths. 

For the purpose of the present discussion, the
important consequence is that in roAp stars, there are indeed radial
velocity gradients along the line of sight in the line-forming region,
which may generate local emergent Stokes $V$ line profiles that are
not anti-symmetric. This departure from anti-symmetry is a radiative
transfer effect whose calculation is non-trivial
\citep{1996SoPh..164..191L}. The details of  
this calculation are unimportant here. The relevant point is that one
can show very generally that such a departure from anti-symmetry of
the local emergent Stokes $V$ line profiles can produce a non-zero
second-order moment of the observable, disk-integrated Stokes $V$
profiles even in a non-rotating star, which does not cancel out in
exposures integrated over a pulsation period (see 
Appendix~\ref{sec:pulsxover}
for details). Hence this is a mechanism that can generate crossover in
stars that have negligible rotation. In what follows, we refer
to this as pulsational crossover. 

The
defining feature for the appearance and the sign of this crossover is
the location in the photospheric layer of the region of formation of the
diagnostic lines with respect to that of the pulsation nodes, which in
general should not vary much across the stellar surface. Hence, one can
expect that, in most cases, pulsational crossover keeps the same
sign throughout the 
rotation period. Some rotational modulation could plausibly occur; in
stars where the magnetic field is approximately symmetric about an
axis passing through the stellar centre, this modulation should be
in phase (or in antiphase) with the longitudinal field
variation. Finally, since the 
departure from anti-symmetry of the local Stokes $V$ line profiles does
not cancel out when averaged over the stellar disk, pulsational
crossover observations may actually be sensitive to pulsations
corresponding to spherical harmonics $Y_{\ell m}$ of higher degree
$\ell$ than 
those that can be detected through photometric or radial velocity
observations.   

Thus, with the pulsational crossover, we identified
a mechanism,
which at least qualitatively can account for all the observed
features of the crossover data presented here that cannot be explained
by the classical rotational crossover:
\begin{enumerate}
\item The detection, at a high level of confidence, of non-zero
  crossover in extremely slow rotators (HD~965, HD~55719)
\item The occurrence of non-reversing crossover (HD~2453, HD~93507,
  HD~116114, HD~144897; possibly also HD~965, HD~116458, and HD~187474)
\item The absence of a phase lag between the longitudinal field and
  crossover variations (HD~93507)\end{enumerate}
However, this interpretation is sustainable only for stars where
non-radial pulsations similar to those of roAp stars occur. Among the
examples given above, only HD~116114 is an established roAp star
\citep{2005MNRAS.358..665E}. 
HD~187474 showed no evidence of pulsation in the the Cape Photometric
Survey \citep{1994MNRAS.271..129M}, and neither did HD~2453 in the
Nainital-Cape Survey \citep{2009A&A...507.1763J}. We did not find in the
literature any report of attempts to detect pulsation in 
HD~93507 or HD~144897. The null results for HD~965, HD~2453, and
HD~187474 do 
not necessarily rule out the interpretation of their crossover as
being due to pulsation, as pulsational crossover is sensitive to
harmonics of higher degree $\ell$ than photometry or radial
velocities. Also, the fact that HD~2453, HD~55719, HD~93507,
HD~116458, HD~144897, and HD~187474 are hotter than all roAp stars
known to date does not represent a definitive argument against the
occurrence of non-radial pulsations in these stars. As a matter of
fact, some of these stars have effective temperatures and luminosities
\citep[e.g.][]{2000ApJ...539..352H} that place them within the
boundaries of the \citet{2002MNRAS.333...47C} theoretical roAp instability 
strip. 

As mentioned previously, numerical modelling of the radiative transfer
effects that are invoked in the pulsational crossover interpretation
is nontrivial. Given 
the uncertainties involved on the structure of the magnetic field and
of the pulsation, such modelling may not allow one to
establish on a quantitative basis if the proposed pulsational
crossover mechanism 
is adequate to explain the type of effect found in the data presented
here. Further observations should provide more critical tests of this 
mechanism. For instance, in slowly rotating roAp stars, determinations
of the crossover from sets of
diagnostic lines of different ions, formed in optical depth ranges
located differently with respect to photospheric pulsation nodes,
should yield different results, possibly even of opposite sign in some
cases. Or, in time series of Stokes $V$ spectra of roAp stars with
a time resolution that is short compared to their pulsation period, the
crossover should be found to vary with the pulsation frequency.

Of course, rotational crossover and pulsational crossover are not
mutually exclusive, and in stars with moderate, but non-negligible
rotation, the observed effect may actually be a combination of both.  
Of the stars studied here, HD~137909 may possibly show the result of
such a combination, with a dominant rotational crossover component
varying in phase quadrature with the longitudinal field, and a weaker
pulsational crossover component responsible for the formally
significant non-zero value of $X_0$. We mention this possibility here
for illustrative purposes, rather than as a firm statement that both
rotational and pulsational crossovers are definitely detected in the
star. Such a statement would be an overinterpretation of a marginally
significant observational result. 
 
As hinted above, even though observation of pulsational crossover is
mostly restricted to very slowly rotating stars, this effect could
potentially be a valuable diagnostic to detect in magnetic Ap stars 
non-radial pulsations that
are not detectable via more traditional approaches such as photometric
or radial velocity variations, for example because they correspond to
harmonics with a higher degree $\ell$, for which local photometric or
radial velocity variations average out in disk-integrated
observations. Such higher degree pulsations in Ap stars represent
unexplored territory.

% More generally, this indicates
% that in the presence of magnetic fields of the order of 5~kG, it is
% definitely possible with the kind of observations presented 
% here, which have a resolution of the order of 0.75~\kms, to detect in
% Stokes $V$ line profiles effects due to rotation in 
% stars with periods of the order of 5 years, or equatorial velocities
% of the order of 0.1~\kms. 

\begin{figure}
  \resizebox{\hsize}{!}{\includegraphics{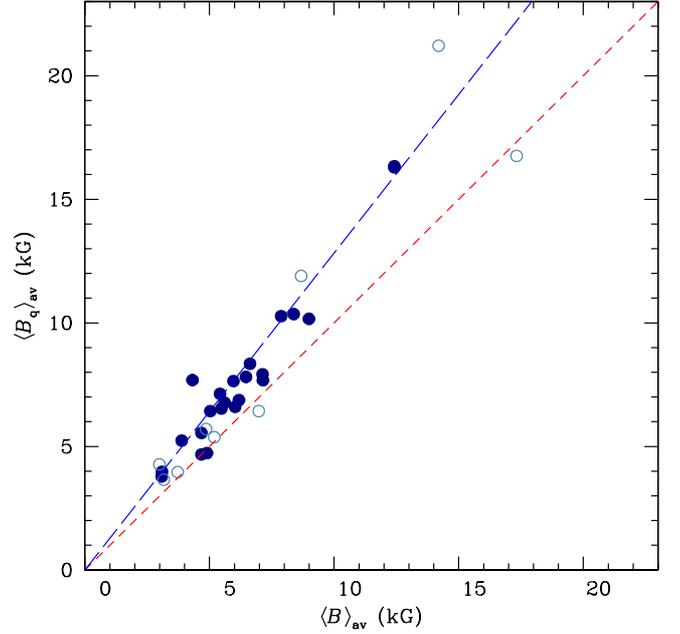}}
  \caption{Observed average of the mean quadratic magnetic field
    against the observed average of the mean magnetic field
    modulus. Open symbols are used to distinguish those stars
    for which either or both of $\Hqav$ and $\Hav$ were computed using
    less than 7 individual measurements of $\Hq$ and
    $\Hm$, respectively. The short-dashed line corresponds to $\Hqav=\Hav$. The
    long-dashed line represents the least-squares fit of the observed
    relation between $\Hqav$ and $\Hav$; it is based only on the stars
    for which at least 7 measurements of both $\Hq$ and $\Hm$ are
    available.}
  \label{fig:hq_vs_hm}
\end{figure}

\subsection{The mean quadratic magnetic field}
\label{sec:Hqdisc}
As for the longitudinal field and the crossover, the precision of
the new mean quadratic magnetic field measurements of  this study is
significantly better than the precision of older determinations of this field
moment from CASPEC spectra. The median of the standard errors
$\sigma_{\rm q}$ of the  229 $\Hq$ measurements reported in
Table~\ref{tab:hzxq} is 336~G. For the 40 measurements of this field
moment in Ap stars with resolved magnetically split lines that were
published by \citet{1995A&A...293..733M} and \citet{1997A&AS..124..475M},
this median was 581~G. 

In Col.~10 of Table~\ref{tab:summary}, we give for each star the
average value 
$\Hqav$ of the mean quadratic field measurements of this study
and of the previous works of \citet{1995A&A...293..746M} and
\citet{1997A&AS..124..475M}. The total number of these measurements,
$N_{\rm q}$, 
appears in Col.~9. For HD~134214 and HD~201601,
\citet{1997A&AS..124..475M} could not determine a physically meaningful
value of $\Hq$ at some epochs, so that $N_{\rm q}<N_{\rm
  z}$. Moreover, as already mentioned, the quadratic field of
HD~188041 is too small to allow any meaningful quantitative information
to be derived about it at any epoch. 

In Fig.~\ref{fig:hq_vs_hm}, we plot $\Hqav$ against
the average of the mean magnetic field modulus values $\Hav$. The
short-dashed line corresponds to $\Hqav=\Hav$. At any given phase, the
mean quadratic field must be greater than or equal to the mean field
modulus: $\Hq>\Hm$, by definition of the quadratic field
\citep[see][]{2006A&A...453..699M}. This does not strictly imply that
a similar 
relation should be 
satisfied between the averages of our measurements of these two field
moments for each star individually because the field modulus and
quadratic field data considered here sample different sets of rotation
phases of the studied stars.  But on a statistical basis, given that the
epochs of the considered observations can be regarded as random with
respect to the rotation phases of all the stars of our sample, the
relation $\Hqav>\Hav$ should be fulfilled. The fact that, in
Fig.~\ref{fig:hq_vs_hm}, the representative points of the majority of
the stars fall above the $\Hqav=\Hav$ line (short-dashed line) is
fully consistent with this expectation. 

Actually, considering only the
stars for which at least seven determinations of both $\Hq$ and $\Hm$
are available (which
are represented by filled symbols), one can note in the 
figure the existence of a nearly linear relation between $\Hav$ and
$\Hqav$, with a steeper slope than the $\Hqav=\Hav$ line. We estimated
this slope to be 1.28 from a least-squares fit of the data; in
Fig.~\ref{fig:hq_vs_hm}, this fit is represented by the long-dashed
line. This value is reasonably consistent with the assumption
$\av{B_z^2}=\av{B^2}/3$ that underlies the quadratic field
determinations. Indeed the latter implies that
$\Hq^2=1.33\,\av{B^2}\geq1.33\,\Hm^2$, hence $\Hq\geq1.15\,\Hm$. This
consistency, and the fact that the deviations of individual points
from the best fit line are moderate at most, give us confidence that
the quadratic field values that we obtain are realistic. In
particular, these results obtained from the analysis of a large number
of $\Hq$ determinations indicate that the tendency of the applied
method to underestimate the quadratic field, identified by
\citet{2006A&A...453..699M}, is moderate at most. Conversely, there is
no 
indication in the present results that the quadratic field may have
been significantly overestimated in any star. These conclusions, which
are based on the largest set of $\Hq$ determinations ever
performed, confirm that the quadractic field is a good diagnostic of the
intrinsic strength of Ap star magnetic fields, at least on a
statistical basis. This diagnostic should be particularly useful in
stars with moderate rotation in which spectral lines are not
magnetically resolved because 
their Doppler broadening exceeds their Zeeman splitting. 

For 8 stars, we obtained $\Hq$ measurements distributed sufficiently well
across the rotation period so as to allow the variation of
this field moment to be characterised. We fitted the variation curves
of the quadratic field of these stars by functions of the forms given
in Eqs.~(\ref{eq:fit1}) or (\ref{eq:fit2}). The fit parameters are given in
Table~\ref{tab:qfit}. The formal significance of the
amplitude $Q_2$ of the first harmonic of the fit is below the
$3\sigma$ level for the four stars of this table for which it was
included. This is the case even though for all of these stars, the
inclusion of the first harmonic significantly
improves both the reduced $\chi^2$ and the multiple correlation
coefficient $R$ with respect to a fit by a single cosine with the
stellar rotation frequency. Furthermore, for HD~2453, even the fit by
the fundamental is not formally significant, since
  $Q_1/\sigma(Q_1)=2.5$. This fit was included in Table~\ref{tab:qfit}
only on account of the fact that visually it appears rather
convincing (see Fig.~\ref{fig:hd2453}).

For these 8 stars, we computed from the fits the ratio
$p=\Hq_{\rm max}/\Hq_{\rm min}$ of the maximum to the minimum of the
quadratic field. The resulting values appear in the last column of
Table~\ref{tab:summary}. For those stars for which it was not possible to
compute a significant fit of the variations of this field moment, contrary
to the field modulus and to the longitudinal field, we
do not regard the ratio of the highest to the lowest measured value of
$\Hq$ as a reliable indicator of its amplitude of variation. Indeed,
a formally significant fit of the variations of the quadratic field could only be
computed for  7 of the 16 stars for which this could be performed for
the variation of the longitudinal field  because the relative
uncertainties of the $\Hq$ measurements with respect to the amplitude
of their variations are much greater than for the $\Hz$
data. Comparison of the fit parameters $Z_1$ and $Q_1$ in
Tables~\ref{tab:zfit} and \ref{tab:qfit} shows that in their majority,
their orders of magnitude are fairly similar. By contrast, the median
of the standard errors of the field moment determinations is $\sim7$
times as large for the  quadratic field as for the
longitudinal field. Accordingly, in many cases, the difference between
the highest and lowest individual values of $\Hq$ in a given star is
more representative of the measurement errors than of the actual
variation of this field moment. 

\begin{figure}[t]
  \resizebox{\hsize}{!}{\includegraphics{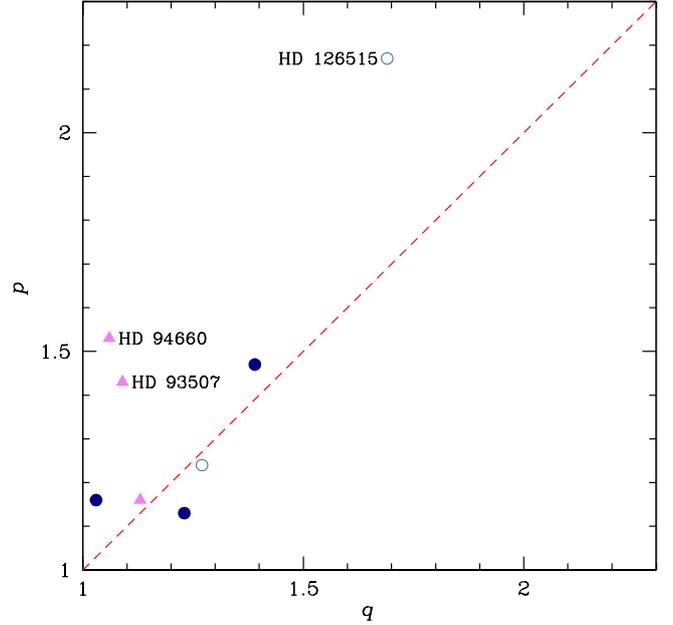}}
  \caption{Relative amplitude $p$ of the mean quadratic magnetic field
    variations against that of the mean magnetic field modulus
    variations ($q$). Open symbols: stars showing $\Hz$
    reversal; filled symbols: stars in which $\Hz$ has a
    constant sign. Among the latter, triangles are used to
    distinguish those stars in which the phase of maximum $\Hm$
    coincides approximately with the phase of lowest absolute value
    of $\Hz$, while dots are used when $\Hm$ maximum occurs
    close to the phase of highest absolute value of $\Hz$.}
  \label{fig:p_vs_q}
\end{figure}

 Given the correlation between quadratic field and field modulus
illustrated in Fig.~\ref{fig:hq_vs_hm} \citep[see
also][]{2006A&A...453..699M}, we would expect the amplitudes 
of variation of these two quantities to be related to each other. To
check this, we have plotted $p=\Hq_{\rm max}/\Hq_{\rm min}$ against
$q=\Hm_{\rm max}/\Hm_{\rm min}$ in Fig.~\ref{fig:p_vs_q}. The dashed
diagonal line represents the locus $p=q$. While most points cluster
around it, three appear considerably above, corresponding to the
stars HD~93507, HD~94660, and HD~126515. Both HD~93507 and HD~94660
have non-reversing longitudinal fields with the phase of the lowest
absolute value of $\Hz$ roughly coinciding with the maximum of the
field modulus. As discussed in Sect.~\ref{sec:Hzdisc}, this indicates
that the magnetic fields of these stars have structures significantly
departing from centred dipoles. Similarly, HD~126515 is one of very
few stars known whose longitudinal field variation curve shows pronounced
anharmonicity, reflecting a considerable departure from a centred
dipole. As a result of such a departure, the variation of the magnetic
field strength across the stellar surface may be larger than for a
centred dipole. 

The quadratic field is defined as
\begin{eqnarray}
\Hq&=&(\av{B^2}+\av{B_z^2})^{1/2}\nonumber\\
&=&\left[\Hm^2+\sigma^2(B)+\Hz^2+\sigma^2(B_z)\right]^{1/2}\,,
\label{eq:Hqdef}
\end{eqnarray}
where $\sigma^2(B)$ and $\sigma^2(B_z)$ are the
standard deviations across the visible stellar disk of the local
values of the modulus and the line of sight components
of the magnetic vector, respectively. In general, these standard deviations become
greater when the field structure shows significant departures from a centred
dipole. The resulting increase of their contribution to the quadratic
field, which is phase dependent, may possibly account for the large
amplitude of variation of $\Hq$ compared to $\Hm$ in HD~93507,
HD~94660, and HD~126515. However, in the latter
two stars, the maximum of the quadratic field is only constrained by
old measurements of \citet{1995A&A...293..746M} and
\citet{1997A&AS..124..475M}, which are much  
less precise and reliable than those of the present 
study. Consideration of Figs.~\ref{fig:hd94660} and \ref{fig:hd126515}
suggests indeed that the value of $p$ for these two stars may be
affected by significant uncertainties. The phase sampling of our new
$\Hq$ measurements of HD~93507 is better, but as noted in
Appendix~\ref{sec:hd93507}, the distortion of the spectral lines around
the phase of maximum of $\Hm$ increases the measurement errors of all
field moments, and hence, the uncertainties affecting both $p$ and
$q$ as well. On the other hand, the line distortion is also suggestive
of the presence of 
a significant small-scale structure in the local magnetic field in the
corresponding region of the stellar surface, which in turn could
account for a large amplitude of variation of the quadratic
field, compared to the mean field modulus. 

\begin{figure}[t]
  \resizebox{\hsize}{!}{\includegraphics{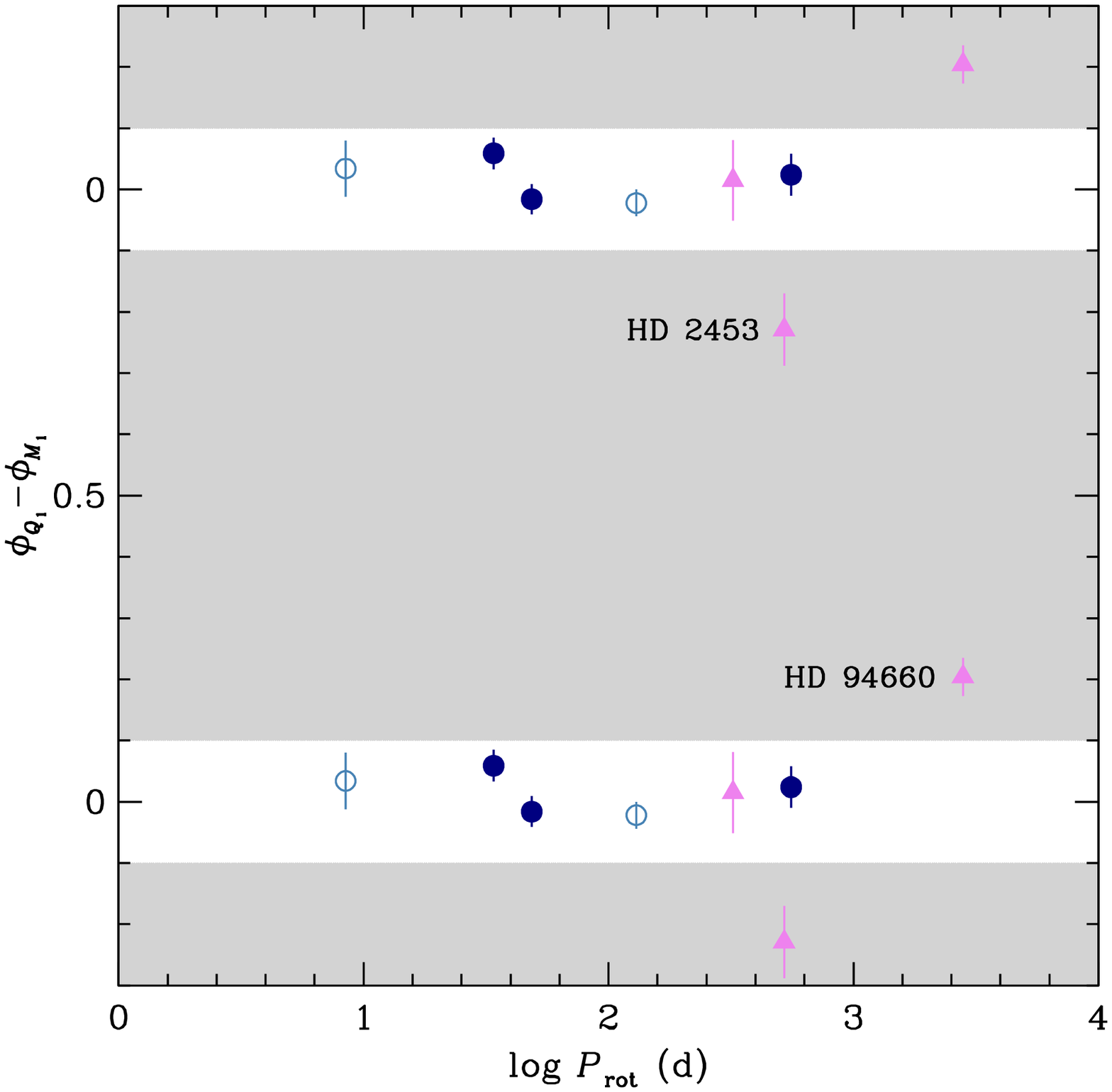}}
  \caption{Phase difference $\phi_{Q_1}-\phi_{M_1}$ between the
    fundamenals of the fits of the measurements of $\Hq$ and $\Hm$ by
    functions of the forms given in Eqs.~(\ref{eq:fit1}) and
    (\ref{eq:fit2}), against stellar rotation period. The meaning of the
    symbols is the same as in Fig.~\ref{fig:p_vs_q}. Shaded
      zones correspond to forbidden regions where points are
    not expected to be found ({\it see text for details}). Two stars
    that depart from this expected behaviour (one of them marginally)
    are identified.}
  \label{fig:phi_qm_vs_P}
\end{figure}
 
The phase difference $\phi_{Q_1}-\phi_{M_1}$ between the fundamental
components of the fits to the variation curves of $\Hq$ and $\Hm$ (see
Table~\ref{tab:qfit}) is plotted against the stellar rotation period
in Fig.~\ref{fig:phi_qm_vs_P}. For magnetic field structures symmetric
about an axis going through the centre of the star, the quadratic
field should vary in phase with the mean field modulus. To visualise
this in Fig.~\ref{fig:phi_qm_vs_P}, the forbidden region corresponding to
$0.1\leq\phi_{Q_1}-\phi_{M_1}\leq0.9$ is shaded. Of the two
representative points lying in this shaded region, only the one
corresponding to HD~94660 identifies a 
phase lag in the $0.1\leq\phi_{Q_1}-\phi_{M_1}\leq0.9$ band at a
formally significant level ($3.4\,\sigma$). As mentioned above, this star
has a magnetic field whose
structure significantly departs from a centred dipole. Actually, the
variations of its field modulus indicate without any doubt
that its field is not symmetric about an axis passing through the
centre of the star. Thus despite the poor definition of the phase of
maximum of the quadratic field of this star resulting from the lack of new,
precise $\Hq$ measurements around it, it is quite plausible that the
non-zero $\phi_{Q_1}-\phi_{M_1}$ difference is real and significant.
On the other hand, 
while the non-zero value of this difference for HD~2453 may also appear
significant from consideration of its error bar, one must keep in mind
that the amplitude of the $\Hq$ fit is not formally
significant. 

Thus the statistical conclusions that can be drawn from the quadratic
field measurements are limited by the incomplete phase coverage of
the new determinations of this paper in many stars (especially, those with very
long rotation periods) and by the moderate precision of the older data
of \citet{1995A&A...293..746M} and \citet{1997A&AS..124..475M}. 

On the
other hand, while in principle one can also determine the projected
equatorial velocity from the
analysis of the second-order moments of the Stokes $I$ line profiles
\citep[see][]{2006A&A...453..699M},  the contribution of the
rotational Doppler effect to the line profiles of the majority of the
stars with magnetically resolved lines is too small to allow
significant constraints on $\vsi$ to be derived from observations at
the moderately high resolution of CASPEC. 

\subsection{Rotation}
\label{sec:rotation}
Based on the new magnetic field determinations presented in this
paper, we were able to derive new constraints on the rotation periods
of 21 stars. For four of these stars, the rotation period was determined
for the first time, and for a fifth star, a
first lower limit of its 
value was tentatively obtained. The recent determination of the period of
HD~18078 by \citet{2016A&A...586A..85M} also relies in part on our new
$\Hm$ data. 
For six stars, the new $\Prot$ value reported here
only corresponds to an improvement of the accuracy of a previously
published value, without significant impact on, for example, the statistical
distribution of the rotation periods. By contrast, the lengthening of
the time span covered by our 
measurements, resulting from the combination of the data of \citetalias{1997A&AS..123..353M}
with those discussed here, allowed a considerably increased lower limit
of the rotation period to be set for seven stars. For two more
  stars, HD~965 and HD~166473, $\Prot$ lower limits substantially
  higher than in \citetalias{1997A&AS..123..353M}, partly based on our new magnetic field
  measurements, have already been published in separate papers
  \citep{2005MNRAS.358.1100E,2007MNRAS.380..181M}.

\begin{figure}
\resizebox{\hsize}{!}{\includegraphics{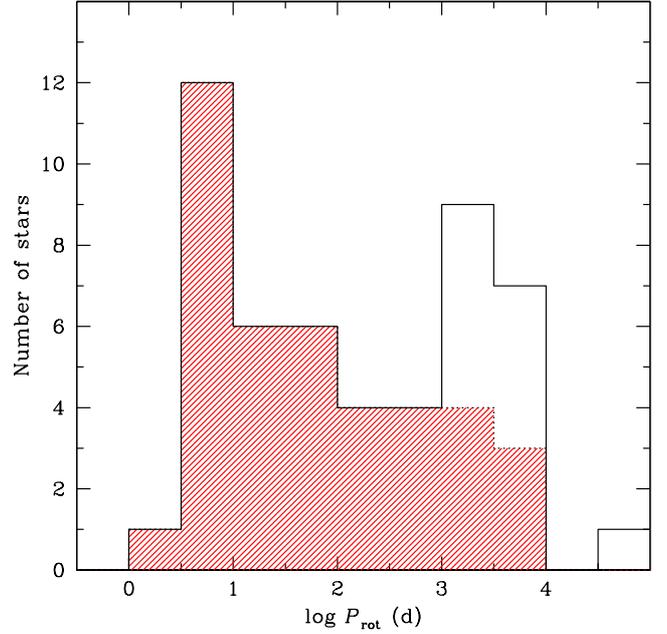}}
\caption{Histogram showing the distribution of the rotation periods of
  the 50 Ap stars with resolved magnetically
  split lines for which their value has been determined (shaded part of the
  histogram) or for which a meaningfully lower limit has been obtained
  (see text).} 
\label{fig:Phist}
\end{figure}

Figure~\ref{fig:Phist} shows the distribution of the rotation periods
of the Ap stars with resolved magnetically split lines (see
Tables~\ref{tab:stars} and \ref{tab:stars_litt}). The shaded part of
the histogram corresponds to the 40 stars for which the exact value of
the rotation period is known; for the remaining ten
stars, only a lower
limit could be determined until now. The tentative lower limits quoted in
    Table~\ref{tab:stars} for HD~213637 and in
    Table~\ref{tab:stars_litt} for HD~92499 and HD~117290 are regarded
    as too uncertain for inclusion in this accounting. 

Unsurprisingly, while up to $\sim70$\% of
all Ap stars have rotation periods between 1 and 10 days
\citep{2004IAUS..215..270M}, for {$\sim75$\%} of the present sample, the
rotation period 
is longer than 10 days, and all stars but one (for which the value of
the period still involves some ambiguity) have rotation periods
longer than 3 days. This is obviously a selection effect, as the magnetic
splitting of the spectral lines can be observed only in stars with
small enough $\vsi$. 

Conversely, magnetically split lines should be resolved in the vast
majority of the Ap stars with $\Prot>10$\,d, provided that they have a
strong enough field ($\Hm\ga2.5$\,kG). This suggests that, for
rotation periods above $\sim10$ days, the distribution seen in
Fig.~\ref{fig:Phist} should be reasonably representative of the
long-period tail of the actual distribution of the rotation periods of
(strongly magnetic) Ap stars. For more than half of the stars of the
rightmost four bins of this distribution (corresponding to
$\Prot>1000$\,d), the available magnetic data sample a time interval
shorter than the rotation period; this time interval represents a
lower limit of the actual period. Accordingly, the secondary peak of
the period distribution, between 1000 and 10000 days, may not be real,
but it may just reflect the limited time coverage of the observations of most
long-period stars. As time passes by and more measurements are
acquired, stars that are currently contributing to that peak may shift
to higher bins. Ultimately, the long-period tail of the $\log\Prot$
distribution may actually prove flat, corresponding to an exponential
decrease with the period. Such a decrease would be consistent with a
statistical distribution. The actual existence of a secondary peak in
the distribution (in the 1000--10000 day range or higher) cannot definitely be
ruled out, but understanding its origin would require
some additional, unknown mechanism to be invoked. 

For a flat distribution in $\log\Prot$ scale to be achieved above
$\Prot=100$\,d, about 2/3 
of the stars currently appearing in the $[3\ldots3.5[$ and
$[3.5\ldots4[$ $\log\Prot$ bins of the Fig.~\ref{fig:Phist} histogram
on the basis of the current lower limits of their periods must
actually have rotation periods longer than $10^4$ days. This in turn
requires among the stars for which only a lower limit of
the rotation period is known so far, about half populate the
$[4.5\ldots5[$ bin of the histogram. Thus following this line of
reasoning strongly suggests that a small but
non-negligible number of Ap stars must have rotation periods of the
order of 1--3 centuries. The argument developed here represents the
first concrete piece of evidence of the probable occurrence of such long
rotation periods, which have never been directly
measured. As a matter of fact, \citet{1913AN....195..159B} achieved
  the first determination of the period of an Ap star, $\alpha^2$~CVn
  ($\Prot=5.5$\,d), just one century ago, so that for no such star
  have the variations been studied over a time span significantly exceeding 100
  years.

There is no reason a priori why the $\log\Prot$
distribution should become flat at $\Prot=100$\,d. It could plausibly
continue to decrease monotonically up to a higher value of the period,
then become flat, or even just tend asymptotically towards zero. In
either of these alternatives, some stars should eventually populate
bin $[5\ldots5.5[$, or even higher bins, of the histogram. This leads
to the conjecture that some Ap stars may have rotation periods ranging
from 300 to 1000 years, or even longer. Obviously, obtaining
meaningful constraints on such periods, if they do indeed occur, will
represent a major challenge.

An additional constraint on the rate of occurrence of (very) long
rotation periods in Ap stars can be derived from consideration of the
whole sample of 84 Ap stars with resolved magnetically split lines
from Tables~\ref{tab:stars} and \ref{tab:stars_litt}. Out of the 50 of
these stars for which a reliable value, or lower limit, of the
rotation period was obtained, 31 have $\Prot>30$\,d. The
remaining 34 stars with magnetically resolved lines were not studied
enough to constrain their rotation periods. But there is nothing in the
way in which they were discovered that would suggest that the
distribution of their periods should be significantly different. Thus,
among all the known Ap stars with resolved magnetically split lines, approximately 54 should have rotation periods longer than 30 days. 

Conversely, the vast majority of the stars known to have
$\Prot>30$\,d show magnetically resolved lines. Of the Ap stars listed
in \citet{1998A&AS..127..421C} and \citet{2001A&A...378..113R} for
which the period definitely exceeds 30 days, only two, HD~8441 and
HD~221568, do not show resolved magnetically split lines. Period
values greater than 30 days were derived more recently for two more
stars with unresolved lines, HD~123335 \citep{2007MNRAS.379..349H} and
HD~184471 \citep{2012AN....333...41K}. But there are a considerable
number of Ap stars whose spectral lines show very little, if any, hint
of rotational or magnetic broadening, even when observed at a
resolving power of the order of $10^5$, and whose period is
unconstrained. It is very plausible that a sizeable portion of these stars may have
(very) long rotation periods, hence further contributing to increasing
the population of the long-period tail of the distribution of the
rotation periods of the Ap stars.  

The list of critically evaluated values of rotation periods
extracted by \citet{2004IAUS..215..270M} from the \citet{1998A&AS..127..421C}
catalogue and its 2001 supplement
\citep{2001A&A...378..113R} comprised 23 entries in the range 10--300
days, and eight
periods longer than 300 days. Updating it with the reliable new period
values or lower limits of Tables~\ref{tab:stars} and
\ref{tab:stars_litt} raises these counts to 31 stars with a rotation
period between 10 and 300 days (including the above-mentioned
HD~123335 and HD~184471), and 21 stars whose period exceeds 300
days. This corresponds to an increase of $\sim35$\% in the number of
known intermediate length (i.e. 10--300~d) periods and to doubling
the number of constrained very long ($>300$~d)
periods. Despite the small number statistics,  the much greater
relative increase in the population of the group of stars with very
long periods represents a significant difference, especially keeping
in mind that 2 of the stars of Tables~\ref{tab:stars} and
\ref{tab:stars_litt} for which the lower limit of the period is too
uncertain for inclusion in the above accounting (HD~92499
and HD~117290) likely have $\Prot>300$~d, while HD~213637 is
  the only star in these tables for which the value of the period
is still unknown and could be between 10 and 300 days. Also, while
for nine of the stars of Table~\ref{tab:stars}, the period is definitely of the
order of 10 years or longer, only three of these stars were reported to have
such slow variations in the \citet{1998A&AS..127..421C}
catalogue. 

In summary, consideration of the new constraints on the rotation
periods of Ap stars that were obtained as part of the systematic study
of those stars with resolved magnetically split lines, whose results
are reported in \citetalias{1997A&AS..123..353M} and in this paper, indicates that the
long-period tail of their distribution is considerably more populated
and likely extends to significantly higher values than was apparent only
15 years ago.

\subsection{Binarity}
\label{sec:binarity}
As explained in Sect.~\ref{sec:rv}, realistic estimates of the
measurement uncertainties cannot be inferred by the methods used to
determine the radial velocity. The external
errors (due e.g. to flexures) cannot be evaluated from a comparison of
different diagnostic lines for CASPEC observations, while  the fact that the radial velocity is obtained from a
single line does not even allow the internal errors to be constrained
for high-resolution
spectra in
natural light; this limitation is similar to the one encountered for the mean
magnetic field modulus, which was discussed in detail in Sect.~6 of
\citetalias{1997A&AS..123..353M}. 

As a workaround, we used the individual differences
between the observed values of the radial velocity and those computed
from the fitted orbits for the nine stars for which we computed orbital
solutions (see Table~\ref{tab:orbits}), as follows. We did this separately for the
CASPEC data, on the one hand, and for the measurements obtained from
high-resolution, unpolarised observations, on the other hand. For both
subsets, we considered all nine stars together (only six of these stars were
actually observed with CASPEC), and we calculated the rms of the
individual differences between measurements and fit in each of
them: 0.95\,\kms\ in the CASPEC case (from 62 differences) and
0.97\,\kms\ in the high-resolution case (from 216
differences). In the
latter case, the derived rms likely correctly reflects the actual
uncertainties affecting our 
radial velocity measurements. Accordingly, we adopt 1\,\kms\ as the
value of the measurement errors for all 
the instrumental configurations of Table~\ref{tab:plot_sym}. Indeed we did 
not identify any obvious difference between them with respect to the
precision of the radial velocity measurements. However, the
vast majority of those measurements were obtained with either
the CAT + CES LC 
configuration at ESO, the AURELIE spectrograph at OHP, or the KPNO
coud\'e feed. The number of data resulting from observations recorded
with other instrumental configurations is too small to allow
differences in their precision to be assessed beyond ruling out major
discrepancies. 

By contrast, we believe that the actual uncertainty 
of the radial velocity values determined from CASPEC observations is
underestimated by their rms difference with the orbital fits; this is probably
out of a fortunate coincidence (in particular, such measurements were
available for only six of the nine binaries of
Table~\ref{tab:orbits}). Indeed, visual comparison of the CASPEC-based radial
velocity values with those obtained from high-resolution spectra in
some stars suggests the existence of systematic
differences greater than 1\,\kms\ (e.g. HD~29578 in
Fig.~\ref{fig:hd29578_rv}) as well as a larger scatter than for
the high-resolution points (e.g. HD~144897 in
Fig.~\ref{fig:hd144897_rv}). Based on these considerations, we adopt 
a value of 2\,\kms\ for the errors of the radial velocity measurements
obtained from CASPEC spectra. 

Both in the CASPEC case and
for the high-resolution unpolarised spectra, we assume that the
measurement errors are the same for all stars. This approach partly
reflects the limitations of our study, such as the fact that
exploitation of the radial velocity information was an
afterthought. However, it is borne out by the fact that consideration
of the various radial velocity curves of Appendix~\ref{sec:notes} 
shows that the scatter of the individual measurement points about the
fitted orbits, or about reasonably smooth variation trends, for those
stars for which orbital elements could not be computed, is fairly
similar in all stars. 

\begin{table}[t]
\caption{Ap stars with magnetically resolved lines with a known orbital
  period (or a meaningful lower limit thereof).}
\label{tab:rot_orb}
\centering
\small{
\begin{tabular}{@{}@{\extracolsep{5pt}}rrrrrc
@{\extracolsep{0pt}}@{}}
\hline\hline\\[-4pt]
\multicolumn{1}{c}{HD/HDE}&$P_{\rm rot}$&$P_{\rm
  orb}$&$e$&$f(M)$&Ref.\\ 
&(d)&(d)&&$(M_{\sun})$&\\[4pt]  
\hline\\[-4pt]
2453&521&$\null\gg2850$\\
9996&7962&227.81&0.512&0.0273&\\
12288&34.9&1546.5&0.342&0.102&\\
18078&1358&978&&&1\\
29578&$\null\gg1800$&926.7&0.377&0.070\\
50169&$\null\gg2900$&1764&0.47&0.0009\\
55719&$\null\gg3650$&46.31803&0.1459&0.522\\
59435&1360&1386.1&0.285&\tablefootmark{a}&2\\
61468&322&27.2728&0.1723&0.1309\\
65339&8.02681&2422.04&0.718&0.149&1\\
81009&33.984&10700&0.718&\tablefootmark{b}&3\\
94660&2800&848.96&0.4476&0.331\\
116114&27.61&$\sim4000$&&&1\\
116458&148.39&126.233&0.143&0.0357\\
137909&18.4868&3858.13&0.534&0.1955&4\\
142070&3.3718&$\null\ga2500$&&&1\\
144897&48.57&$\null\ga2200$\\
165474&$\null\gg3300$&$\null\gg2200$&\\
187474&2345&689.68&0.4856&0.0646\\
200311&52.0084&$\null\gg1800$\\
201601&$\null\ga35400$&100200&0.56&\tablefootmark{c}&5\\[4pt]
\hline\\[-4pt]
\end{tabular}}
\tablefoot{\tablefoottext{a} {SB2: $M_1\,\sin^3i=2.809~M_{\sun}$, $M_2\,\sin^3i=2.614~M_{\sun}$.} 
\tablefoottext{b}{Visual binary: $M_1=2.6~M_{\sun}$, $M_2=1.6~M_{\sun}$.}
\tablefoottext{c}{Visual binary.}}
\tablebib{
(1)~\citet{2002A&A...394..151C};
(2)~\citet{1999A&A...347..164W};
(3)~\citet{2000A&A...361..991W};
(4)~\citet{1998A&AS..130..223N};
(5)~\citet{2011A&A...529A..29S}.}
\end{table}

We identified three additional spectroscopic binaries (SB) among
the 43 Ap stars with resolved magnetically split lines for which new
observations are presented in this paper; these SBs are HD~144897, HD~165474, and
HD~200311. HD~165474 was already flagged as a visual binary in
\citetalias{1997A&AS..123..353M}. Furthermore, our radial velocity measurements confirm and
strengthen the inference by \citet{2002A&A...394..151C}
that HD~2453 and 
HD~18078 are also spectroscopic binaries, which were not recognised as
such 
in \citetalias{1997A&AS..123..353M}. By contrast, we point out that the apparent variability of
the radial velocity of HD~216018 that we reported in that paper
was actually due to spurious determinations. In addition, HD~201601 is
a visual binary for which a first, preliminary orbital solution was
recently computed \citep{2011A&A...529A..29S}. Thus 22 stars out of our
sample of 43 are binaries, compared to 18 out of 41 in
\citetalias{1997A&AS..123..353M}. The rate of occurrence of binarity in our sample, 51\%, is
slightly higher than the values of 43\%\ reported by
\citet{2002A&A...394..151C} for Ap stars in general, and of 46\%\
derived by 
\citet{1985A&A...146..341G} for the cool Ap stars. This difference
is marginal, but it may to some extent reflect the fact that 
our study is particularly
suited to the identification of systems with long orbital periods
because of its character (high spectral resolution observations
repeated at semi-regular 
intervals over a time span of several years)  and of the
nature of our sample (stars with low $\vsi$, for which radial
velocities can be determined very precisely). Such
systems may escape detection when the primary is rotating faster or in
projects carried out on a shorter time span.

\begin{table*}[t]
\caption{Ap stars with published reliable orbital and rotation periods,
  not known to show magnetically resolved lines.}
\label{tab:rot_orb_litt}
\begin{tabular*}{\textwidth}[]{@{}@{\extracolsep{\fill}}rllrcrrrc
@{\extracolsep{0pt}}@{}}
\hline\hline\\[-4pt]
\multicolumn{1}{c}{HD/HDE}&Other id.&Sp. type&$P_{\rm
  rot}$&Ref.&$P_{\rm orb}$&$e$&$f(M)$&Ref.\\
&&&(d)&&(d)&&$(M_{\sun})$&\\[4pt]  
\hline\\[-4pt]
5550&HR~273&A0p Sr&6.84&1&6.82054&0.006&\tablefootmark{a}&1 2\\
8441&BD~$+42$~293&A2p Sr&69.2&3&106.357&0.122&0.209&4\\
15089&$\iota$~Cas&A4p Sr&1.74050&5&17200&0.626&\tablefootmark{b}&6\\
{\it 25267}&$\tau^9$~Eri&A0p Si&{\it 5.954}&7&5.9538&0.13&0.00414&8\\
25823&HR~1268&B9p SrSi&7.227&9&7.227424&0.18&0.0033&10\\
98088&HR~4369&A8p SrCrEu&5.9051&11&5.9051&0.1840&\tablefootmark{c}&11\\
{\it 123335}&HR~5292&{\it Bp He wk SrTi}&55.215&12&35.44733&\tablefootmark{d}&\tablefootmark{d}&12\\
125248&CS~Vir&A1p EuCr&9.295&13&1607&0.21&0.065&14\\
184471&BD~$+32~$3471&A9p SrCrEu&50.8&15&429.17&0.2017&0.86&2\\
{\it 191654}&BD~$+15$~4071&A2p SrCr&1.857&16&{\it 2121}&{\it 0.48}&0.00140&2\\
216533&BD~$+58$~2497&A1p SrCrSi&17.2&17&1413.1&0.437&0.0137&2\\
219749&HR~8861&B9p Si&1.61887&18&48.304&0.50&0.0554&19\\[4pt]
\hline\\[-4pt]
\end{tabular*}
\tablefoot{
\tablefoottext{a}{SB2: $M_1\,\sin^3i=0.1081~M_{\sun}$,
$M_2\,\sin^3i=0.0692~M_{\sun}$.}
\tablefoottext{b}{Visual binary: $M_1=1.99~M_{\sun}$,
$M_2=0.69~M_{\sun}$ (based on Hipparcos parallax).} 
\tablefoottext{c}{SB2: $M_1\,\sin^3i=1.772~M_{\sun}$,
$M_2\,\sin^3i=1.290~M_{\sun}$.} 
\tablefoottext{d}{Eclipsing binary.}}\\
\tablebib{
(1)~\citet{2016A&A...589A..47A};
(2)~\citet{2002A&A...394..151C};
(3)~\citet{2007A&A...475.1053A}; 
(4)~\citet{1998A&AS..130..223N};
(5)~\citet{1980AN....301...71M};
(6)~\citet{2003ApJ...585.1007D};
(7)~\citet{1980ApJS...42..421B};
(8)~\citet{1999A&A...343..273L};
(9)~\citet{1973ApJ...186..951W};
(10)~\citet{1973ApJS...25..137A};
(11)~\citet{2013MNRAS.431.1513F};
(12)~\citet{2007MNRAS.379..349H};
(13)~\citet{1991A&AS...89..121M}; 
(14)~\citet{1969MNRAS.142..543H}; 
(15)~\citet{2012AN....333...41K};
(16)~\citet{1980AN....301..317V};
(17)~\citet{1977A&AS...30...27F};
(18)~\citet{2000A&A...357..548A};
(19)~\citet{1979IBVS.1600....1O}.}
\end{table*}

Orbital elements were determined for 14 of the 22 binaries of
this paper. For three of these binaries (HD~29578, HD~50169, and HD~61468),
an orbital solution was obtained for the first time. We also presented
revised solutions for five other systems: 
HD~9996, HD~12288, HD~55719, HD~94660, HD~116458 and HD~187474. For four
systems, 
either orbital elements based on our observations 
had already been published (HD~59435, HD~81009) or our new data did
not allow us to improve determinations that were already based on much
more extensive sets of measurements (HD~65339, HD~137909). The last
system for which (partial) orbital elements are available, HD~201601,
is a visual binary, for which we did not observe radial velocity
variations. The relevant references are quoted in the respective sections of
Appendix~\ref{sec:notes}. For seven other systems, constraints on the
orbital periods (mostly lower limits) have been obtained by
\citet{2002A&A...394..151C} for HD~2453, HD~18078, HD~116114, and
HD~142070; 
and in this paper for HD~144897, HD~165474, and HD~200311. For the last
spectroscopic binary, HD~208217, our data are insufficient to
characterise the orbit. The main properties of the remaining 21 stars
(rotation and orbital periods, or lower limits; eccentricity and mass
function) 
are summarised in Table~\ref{tab:rot_orb}. The reference given in the
last column is to the paper where the listed orbital parameters or
orbital period estimate were originally published. When they were
derived in the present study, there is no entry in this column. 

The shortest of the 21 orbital periods that are known or constrained
is 27 days long. This is remarkable. While the (almost) complete lack
of binaries containing an Ap star with orbital periods shorter than 3
days has long been known
\citep{1985A&A...146..341G,2002A&A...394..151C}, one can see in Fig.~8
of \citeauthor{2002A&A...394..151C} that for about one-third of the binaries of their sample with Si or
SrCrEu peculiarity type (similar to the stars studied here), the
logarithm of the orbital period is less than 1.4 (that is, the period
is shorter than 25 days). The difference in the behaviour of the
spectroscopic binaries of the present paper is highly significant: if the
\citeauthor{2002A&A...394..151C} sample and ours had similar properties, we would
expect our sample to include six or seven binaries with orbital periods
shorter than 25 days. Also, the number of spectroscopic
binaries considered here is about half the number of Si and SrCrEu
stars in the \citeauthor{2002A&A...394..151C} study. In other words, with an increased
number of binaries and an improved knowledge of their orbits,
those considerations strengthen
our suspicion of \citetalias{1997A&AS..123..353M}, that the subset of Ap stars with
resolved magnetically split lines shows an extreme deficiency of
binaries with short orbital periods as compared to the class of the
magnetic Ap stars as a whole. In this context, Ap star must
  be understood in a rather strict sense, that is, roughly limited
  to spectral type A: formally, our sample ranges from B8p (HD~70331,
  HD~144897, HDE~318107) to F1p (HD~213637). This should essentially
  coincide with the \citet{2002A&A...394..151C}
  SrCrEu and Si 
  subgroups, but exclude the He weak and (of course) HgMn stars of
  their study.\label{fn:trange}

On the other hand, recently published preliminary results of a
  systematic survey of magnetism in binaries 
  with short orbital periods ($\Porb<20$\,d) indicate that less than
  2\% of the components of spectral types O--F of such systems have
  large-scale organised magnetic fields
  \citep{2015IAUS..307..330A}. While this strongly suggests that the
  rate of occurrence of such fields is considerably lower in close
  O--F binaries than in single stars of similar spectral type, it is
  not inconsistent with the results of
  \citet{2002A&A...394..151C}. Indeed the sample of the latter study
  consists of stars that were initially selected for their chemical
  peculiarity, so that it comprises only a rather small fraction of
  the whole set of binary stars of spectral type A. Conversely, as can
  be seen in Fig.~1 of \citep{2015IAUS..307..330A}, only about 1/3 of
  the stars of their sample have spectral types between early F and
  late B, and the rate of occurrence of magnetism may not be uniform
  throughout the all range of spectral types covered by their sample.

In order to gain further insight into the nature and origin of
  the apparent deficiency of short-period spectroscopic binaries among
  the Ap stars with resolved magnetically split lines as compared to
  the \citet{2002A&A...394..151C} sample, we looked in the literature for 
other Ap stars in binaries for which both the orbital
and rotation periods are known. We critically assessed the reliability
of the published periods, checking the original papers in which they were
derived. To our surprise, we only identified 12 stars for which we felt
confident that the values of both periods were established firmly
enough to warrant consideration. Their properties are summarised in
Table~\ref{tab:rot_orb_litt}. Most columns should be
self-explanatory. The source of the rotation periods is identified by
the references in Col.~5; Col.~9 lists the papers in which the orbital
elements were originally published. 

Some caution must be exerted with respect to three of the stars in
Table~\ref{tab:rot_orb_litt}. Their HD numbers appear in italics and
we 
also use italics to identify the corresponding elements of
information that are subject to 
reservations. The listed value of the rotation period of HD~25267
satisfactorily represents the variations of its mean longitudinal
magnetic field observed by \citet{1980ApJS...42..421B}. However
\citet{1985A&A...144..251M} found this value of the period
inconsistent with their 
photometric observations. But the interpretation of these
observations, which 
indicate the presence of two periods of variation (1.2~d and 3.8~d),
has remained 
unclear to this day. On the other hand, the \citet{1999A&A...343..273L}
argument that the observed variations of the
longitudinal field, between $-400$~G and 0~G, are 
incompatible with a value close to 90\degr\ of the angle $i$ between
the rotation axis and line of sight, is weak. The non-reversal of
this field moment only requires that $i+\beta\leq90$\degr \ and the
example of HD~142070  (Appendix~\ref{sec:hd142070}) shows that $\Hz$
variation amplitudes of several hundred Gauss may be observed even
when either i or $\beta$ (the angle between the rotation and magnetic
axes) is only a few degrees. \citet{2002A&A...394..151C} cannot
definitely rule out a period alias very close to 1 day for the radial
velocity variations of HD~191654, which would then be due to
rotation. However, the 2121~d alternative corresponding to an orbital
motion seems more plausible, especially since the photometric period
of 1\fd85 \citep{1980AN....301..317V} appears robust and can very
plausibly be attributed to rotation.  Finally, the spectral subtype
of the Bp primary of HD~123335, and its temperature, may possibly be
above the upper limit of the range covered by the stars of our
sample, as specified on page \pageref{fn:trange}). 

Although \citet{1985A&A...146..341G} list several other cool Ap 
and Si stars as having known orbital and rotation periods, none of
these stars stand up to closer scrutiny. \citet{1995A&AS..114..253A} did
not detect any photometric variation of HD~15144, although
\citet{2008MNRAS.384.1588C} argue that synchronised rotation with the
orbital 
period (2\fd998; \citealt*{1997A&AS..121...71T,1999A&A...343..273L})
is consistent with the observed $\vsi$. Neither the 
rotation period \citep{2007A&A...475.1053A} nor the orbital period
\citep{1984A&AS...57...55S} of HD~68351 is unambiguously determined,
although it appears rather certain that the former is short (of the
order of a few days) and the latter long (probably 635 or 475
days). HD~77350 is actually a HgMn star \citep{1989MNRAS.239..487A},
not a Si star. The  
variability of the radial velocity of HD~90569 and HD~183056 is
questionable, while the orbital period of HD~170000 is ambiguous
\citep{2004A&A...424..727P}.  Furthermore, among the stars of the
\citet{1985A&A...146..341G} 
Tables~6b and 6c, for 
which the rotation period was unknown at the time, in the meantime  this period has
been determined only for HD~162588
\citep{1984A&AS...55..259N}. However, the 
orbital elements of this star are highly uncertain and it is even
questionable that it is a binary at all \citep{2004A&A...424..727P}.

The orbital periods of HD~5550, HD~25267, HD~25823 and HD~98088 are much
shorter than those of any binary  containing an Ap star with
magnetically resolved lines, and those periods are equal to the rotation
periods of their Ap component: these systems are synchronised. The other 
eight stars of Table~\ref{tab:rot_orb_litt} have orbital periods in
the same range as the stars studied in this paper. 

According to \citet{2016A&A...589A..47A}, HD~5550 has the weakest
magnetic field detected to this day in an Ap star with a dipole
strength of only 65~G.
\citet{2007A&A...475.1053A} also observed a weak field in
HD~8441 \citep[see also][]{2012AstL...38..721T}. Its mean
longitudinal component may reach a couple hundred 
Gauss at maximum, and its mean modulus is at most of the order of
1--1.5~kG. \citet{1998CoSka..27..470K} measured a longitudinal
field varying between $-300$~G and 500~G in the primary of the visual
binary HD~15089. It was confirmed by \citet{2007A&A...475.1053A},
who inferred a dipole field strength of the order of 2~kG. With
longitudinal fields varying, respectively, between 
$-0.4$ and $+1.3$~kG \citep{1973ApJ...186..951W},
$-1.2$ and $+1.0$~kG \citep{2000MNRAS.313..851W}, and $-1.8$ and
$+2.1$~kG \citep{1994A&AS..108..547M}, HD~25823, HD~98088 and
HD~125248 have 
magnetic fields whose mean strength must be at least of the order of the
weakest mean field moduli measured in stars with magnetically resolved
lines. As a matter of fact, the quadratic field of HD~125248 varies
between $\sim8$ and $\sim11$~kG (see Table~\ref{tab:hquad_rev}). The
longitudinal field 
of HD~184471 is only marginally weaker, with extrema of $-0.1$ and
$+0.9$ kG \citep{2012AN....333...41K}. We obtained one
AURELIE spectrum of this star in May 1993, where the line
\ion{Fe}{ii}~$\lambda\,6149.2$ shows incipient resolution.
This splitting is insufficient to measure $\Hm$, but it suggests that full
resolution of the magnetically split components may be achieved at
other phases; this is similar to other stars, such as HD~9996, where
the magnetic splitting is resolved only over
part of the rotation cycle. Unfortunately, this
eventuality was overlooked upon original inspection of the considered
spectrum, and no follow-up observations were taken. No definite detection
of the magnetic field of HD~216533 has 
been reported, besides the \citet{1958ApJS....3..141B}
eye estimate of a 
possible longitudinal component of the order of 1~kG. No magnetic
field was detected by \citet{1993A&A...269..355B} or by
\citet{2006MNRAS.372.1804K} in HD~219749. We did not find in the
literature 
any attempt to determine the magnetic field of HD~123335 or
HD~191654. We obtained a single high-resolution spectrum 
of the latter in April 1994 with the ESO CES (see
Table~1 of \citetalias{1997A&AS..123..353M}) as part of our search for Ap stars with
magnetically resolved lines. 
This spectrum does not show any evidence of a magnetic field. However the
possible presence of a field
of up to a few kG cannot be ruled out because of the significant
rotational broadening of the spectral lines. In summary, the stars of
Table~\ref{tab:rot_orb_litt} comprise a mixture of magnetic fields of
strength that is comparable to the stars with magnetically resolved lines and
of weaker fields.

\begin{figure}[t]
  \resizebox{\hsize}{!}{\includegraphics{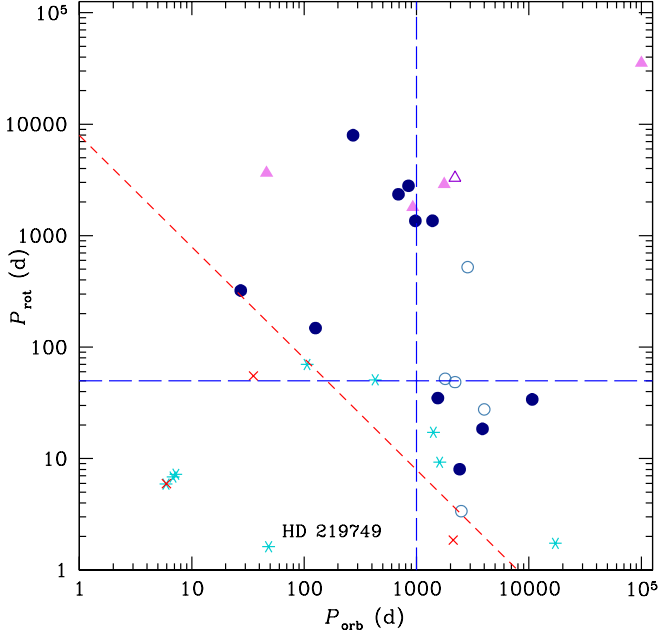}}
  \caption{Rotation period is plotted against the orbital period
    for the spectroscopic binaries for which these two periods are
    sufficiently constrained by the existing observations. 
      Asterisks and crosses represent stars not known
    to have magnetically 
    resolved lines, for which both periods were retrieved from the
    literature; crosses distinguish
    those for which uncertainties remain about some of the relevant
    parameters (see text). All other symbols correspond to Ap stars
    with 
    magnetically resolved lines studied in detail in this paper. 
      Dots identify those stars for which both periods are exactly
    known; open circles, stars with fully determined rotation
    periods for which only a lower limit of the orbital period has
    been obtained so far; filled triangles, stars with a known
    orbital period but only a lower limit of the rotation period; and
    open triangles, stars that have only been observed over a
    fraction of both their rotation and orbital periods. The short-dashed
    diagonal line represents the locus $\log\Prot=3.9-\log\Porb$. The
    long-dashed horizontal line corresponds to $\Prot=50$~d and the
    long-dashed vertical line to $\Porb=1000$~d.}
  \label{fig:Prot_vs_Porb}
\end{figure}

 For the systems appearing in Tables~\ref{tab:rot_orb} and
\ref{tab:rot_orb_litt} for which both the rotation and orbital periods
are constrained, we plot these two periods (or their lower limits)
against each other in Fig.~\ref{fig:Prot_vs_Porb}. Open symbols are used
for stars for which only a lower limit of the orbital period could be
obtained so far, while triangles identify lower limits of the rotation
period. Asterisks and crosses represent stars not known to show
magnetically 
resolved lines, for which the values of both periods were retrieved
from the literature (that is, the stars listed in
Table~\ref{tab:rot_orb_litt}); among them, crosses distinguish
the three stars discussed above, for which some of the relevant
parameters are subject to uncertainties. 

A striking feature of
Fig.~\ref{fig:Prot_vs_Porb} is that all the stars with magnetically
resolved lines are confined to the upper right region of the
figure. To help visualise 
this, we plotted a dashed diagonal line defined by
$\log\Prot=3.9-\log\Porb$, which is approximately the lower envelope
(both in $\Prot$ and in $\Porb$) of the representative points of the
stars studied in this paper. 

Five of the stars of
Table~\ref{tab:rot_orb_litt} also have representative points lying
above or very near this dividing line. The main exceptions are
HD~219749 and the four
synchronised systems, HD~5550, HD~25267, HD~25823, and HD~98088, which appear in 
isolation in the lower left corner of the figure; the
representative points of HD~25267 and HD~98088 are superimposed owing to the similarity of their periods.  

The separation between two of the 
stars below and to the left of the diagonal line, HD~123335 and
HD~191654, and the bulk of the sample, is much less clear cut. Indeed,
a different way to look at the result illustrated in
Fig.~\ref{fig:Prot_vs_Porb} is highlighted by the horizontal and
vertical long-dashed lines drawn in this figure. The former 
corresponds the value 50~d of the
  stellar rotation period, the 
latter corresponds to a 1000-d long orbital
period. Of the stars represented, only the four
synchronised systems and HD~219749 are found  in the
lower left quadrant defined by these two lines. But for
HD~219749, barring synchronisation,  
(relatively) short rotational, and orbital periods seem to be mutually 
exclusive. More specifically, in
Fig.~\ref{fig:Prot_vs_Porb}, 11 of the 12 non-synchronised
spectroscopic binaries with a rotation period shorter
than $\sim50$~d have orbital periods in excess of
1000~d. Conversely, only 1 of the 12 non-synchronised spectroscopic
binaries with an orbital period shorter than 1000~d has a rotation period
shorter than $\sim50$~d. On the other
hand, a few stars have both a long rotation period and a long orbital
period. Also keep in mind that stars represented by open symbols are
bound to move further right in the diagram, while those appearing as
triangles are due to raise further up when observations have been
obtained over full periods. 
 
\begin{figure}[t]
  \resizebox{\hsize}{!}{\includegraphics{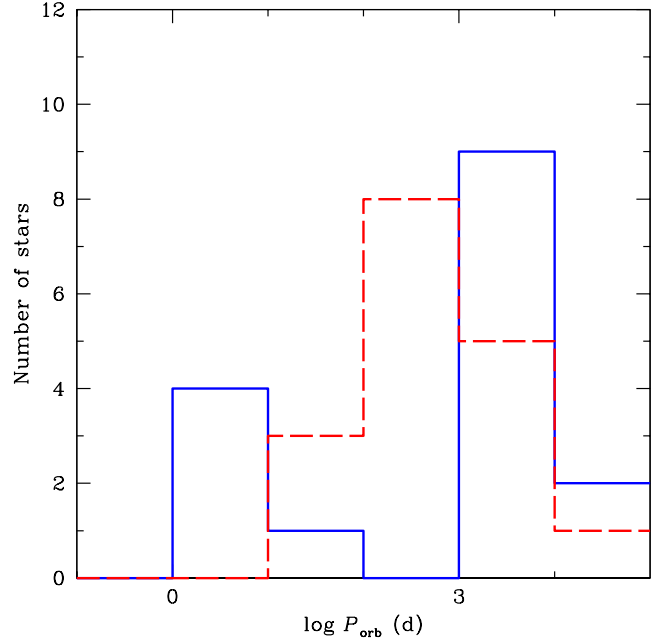}}
  \caption{Comparison between the distribution of the orbital periods
    of binaries containing an Ap star with a rotation period shorter
    (solid blue line) and greater (dashed red line)
    than 50 days.}
  \label{fig:Porb_hist}
\end{figure}

A different illustration of the systematics of the rotation and
orbital periods of the stars of Tables~\ref{tab:rot_orb} and
\ref{tab:rot_orb_litt} is shown in Fig.~\ref{fig:Porb_hist}. The solid
line histogram represents the distribution of the orbital periods of
systems in which the Ap component has a rotation period shorter than
50 days. It is visibly different from the dashed line histogram,
which corresponds to the orbital periods of the systems with an Ap
component whose rotation period exceeds 50 days. To quantify this
comparison, we have run a two-sided Kolmogorov-Smirnov test, which
shows that the distribution of $\Porb$ in the two subsamples under
consideration is different at the 89\% confidence level. If we leave
the three synchronised systems (the lower peak of the solid line
histogram) out of the $\Prot<50$\,d sample, the confidence level increases
to 99.6\%. 

We can further test that the inferred deficiency of non-synchronised
systems with both 
(relatively) short rotation and orbital periods is not
the result of a chance 
coincidence through consideration of the
projected equatorial velocity of the Ap component of systems with
$\Porb<1000$ days, in some of the cases for which the rotation period
of this component is not known. Such a test is meaningful only for
those stars for which accurate determinations of $\vsi$ based on
spectra of high enough resolution are available. Remember that for a
star whose radius is $2R_{\sun}$, and for which $\Prot=10$\,d,
$\vsi\leq10$\,\kms, its value cannot be accurately determined from the
analysis of low to moderate spectral resolution spectra and, hence, is not
accurately constrained by the $\vsi$ data available in most catalogues. 

In this context, the study of \citet{2002A&A...394..151C} contains
the most extensive and homogeneous set of suitable $\vsi$
determinations. We briefly discuss the stars of this study with
$\Porb<1000$\,d that do not have a known rotation period; these are HD22128,
54908, 56495, 73709, 105680, 138426, 188854, and 
200405.  

Four of these stars (HD~56495, 73709, 105680, and 188854) are
classified as Am stars in the \citet{2009A&A...498..961R}
catalogue.  \citet{2002A&A...394..151C} acknowledge the ambiguity
of their classification. \citet{2015NewA...38...55C} recently reported
that HD~188854 is an eclipsing binary whose primary is a $\gamma$~Dor
pulsating star; this makes it very unlikely that it is an Ap
star. \citet{1967JO.....50..425B} explicitly mentioned that they could
not confirm the anomalous strength of the Sr lines in HD~105680
reported in the HD catalogue, and that they found its spectrum
consistent with an Am classification. The Am nature of HD~73709 is 
confirmed by the detailed abundance analysis of
\citet{2007A&A...476..911F}, who also ruled out the presence of a
detectable 
magnetic field in this star (which was suggested by
\citeauthor{2002A&A...394..151C}). In another detailed analysis,
\citet{2013MNRAS.433.3336F}
confirmed that HD~56495 is not an Ap star; furthermore, they also
found both components of the SB2 system HD~22128 to be Am stars.
While \citet{2009A&A...498..961R} give a spectral type A0p
Si for HD~54908, we could not find this classification in the
\citet{1973AJ.....78..687B}
work to which \citet{2002A&A...394..151C} 
refer, and the star is assigned spectral type A1m in the Michigan
Catalogue \citep{1999mctd.book.....H}. 

For HD~138426 ($\Porb=11\fd34$),
$\vsi$ is small enough ($\la2$\,\kms) to be compatible with
a rotation period of the order of 50 days or longer. For the remaining
star, HD~200405 ($\Porb=1\fd63$), the
existing constraints are 
compatible with the synchronisation of the rotation of the Ap
component with the orbital period. Indeed, assuming that the Ap
component is rotating synchronously, hence that
$\Prot=\Porb=1.635$\,d and that its radius is of
the order of $2.4R_{\sun}$, which should be fairly representative for the
spectral type of interest, application of the relation
\begin{equation}
\Prot=50.6\,R/v\\
\end{equation}
yields $v=74$\,km\,s$^{-1}$. Then $\vsi=9.6$\,\kms 
implies that $i\simeq7.4$\degr, which is consistent with the conclusion by
\citeauthor{2002A&A...394..151C} \citeyearpar{2002A&A...394..151C} that the inclination of the orbital plane  
must be very small. The fact that the eccentricity of the orbit of
HD~200405 is not significantly different from 0
further supports the plausibility that this systems is
synchronous. 

Thus, for the systems
studied by \citet{2002A&A...394..151C} for which the Ap character
of one of the components is unambiguously established, we did not find
any definite indication that the relation between the rotation period
of the Ap component and the orbital period of the system may not fit
the pattern that we derived from consideration of the binaries
containing Ap stars with known rotation periods. 

This pattern is not questioned either by any of the spectroscopic
binaries listed as Ap (including B9p and F0p) in the
\citet{2004A&A...424..727P} 
catalogue. There are only four such stars that have not been already
discussed above; these are HD~108945, HD~134759, HD~147869, and
HD~196133. The binarity of HD~108945 cannot be regarded
as confirmed. Consideration of Fig.~8 of \citet{2002A&A...392..637S} and
of Fig.~24 of \citet{2007A&A...475.1053A} strongly suggests that the
distortion of its line profiles resulting from abundance
inhomogeneities on its surface must have mistakenly been
interpreted as an SB2 signature by
\citet{1999ApJ...521..682A}. The orbital period of HD~134759, 23.42\,y
\citep{1982A&AS...47..569H}, is much longer than 1000\,d. It appears
doubtful that HD~147869 and HD~196133 are actually Ap stars. Although
HD~147869 is listed as an A1p Sr star by \citet{2009A&A...498..961R},
it had been classified as A1 III by \citet{1995ApJS...99..135A};
\citet{2002A&A...383..558L} also regard it as an evolved star rather
than a (chemically peculiar) main-sequence
star. \citet{2009A&A...498..961R} note that HD~196133 is perhaps an Am
star rather than an Ap star, while \cite{1967JO.....50..425B}
explicitly state that it is an A2 star without any noticeable peculiarity.

In summary, we have established the existence, with a high
probability, of a connection between rotation and orbital periods for
Ap stars in binaries. Namely, barring synchronisation,
Ap components with short rotation periods (i.e. less than about
50 days) are found only in systems with very long orbital periods
(more than 1000 days or so). Conversely, whenever a binary containing
an Ap component has a short orbital period (that is, shorter than
1000 days), the rotation period of the Ap component is long (greater
than 50 days). So far, this mutual exclusion of short rotational and
orbital periods, as we shall refer to it from here on, suffers only
one exception, HD~219749, for which $\Prot=1.61$\,d and
$\Porb=48.30$\,d. HD~68351, if its suspected periods are
  confirmed, may formally represent another exception. However, this
  exception is at most marginal, since it would vanish if the
  cut-off line between short and long orbital periods, which
  involves a significant degree of arbitrariness, was shifted from the
  adopted value of 1000 days to 500--600 days. 

We do not regard the HD~219749
exception as sufficient to question the reality of the mutual short period
exclusion, since there is a realistic probability, in a sample of 29
systems, that one may have undergone some external disturbance
altering its evolution. We are strengthened in this position by the fact
that consideration of the available high-quality $\vsi$ information
for the Ap component of additional systems, in which the rotation period
of this component is unknown, did not indicate a
definite inconsistency for any of these systems with the mutual exclusion of short rotational
and orbital periods.

The lack of non-synchronised systems where both the rotational and
orbital periods are short is unlikely to result from an observational
bias. On 
the contrary, one might argue that long orbital periods are more
likely to be overlooked in faster rotating stars, in which the achievable
precision of radial velocity determinations is limited by the
rotational Doppler broadening of the spectral lines. 

Understanding the origin of the observed behaviour represents a
challenge. There is no obvious physical mechanism to slow down the
rotation of the Ap component of systems with orbital periods shorter
than 1000 days. Tidal braking should be practically insignificant in the majority, if not
all, of these systems because the shortest orbital period of the considered
systems is 27 days, and all of the orbital periods, except for three, are longer than 100 days.
As a matter of fact, in 9 of the 11 systems with
$\Porb<1000$\,d and $\Prot>50$\,d from Tables~\ref{tab:rot_orb} and
\ref{tab:rot_orb_litt}, the rotation period of the Ap component is
longer than the orbital period of the system: tidal interactions
cannot account for the slow rotation of the Ap star. 

With nothing in the present state of the systems apparently accounting
for the mutual exclusion of short rotational and orbital periods, the
explanation of the latter must be sought in the earlier evolution of
these systems. Recently, two groups have, independently from each
other, proposed a scenario for the origin of the magnetic fields of Ap
stars in which these stars result from the merging of two lower mass
stars or protostars
\citep{2009MNRAS.400L..71F,2010ARep...54..156T}. In this scenario, the
binaries that we observe 
now were triple systems earlier in their history. One can expect
such a dynamical event as the merger of two of the three components
of a triple system not only to determine the rotation rate of the
resulting merged star, but also to have a significant impact on the
orbit of the third component. Thus the merging scenario for the origin
of Ap stars  provides a plausible link between the rotational period of
the Ap component of binary sytems and their orbital
period. Conversely, the mere existence
of a connection between rotational 
and orbital periods in Ap binaries represents a very strong argument in
support of the merging scenario for the origin of Ap stars. The
details of the merging process are complex, and their study is beyond
the scope of this paper. The observational results reported here
should represent useful constraints for its theoretical modelling. 

The merging scenario is consistent with the recent identification of a
group of magnetic Herbig Ae/Be stars as the likely progenitors of the
main-sequence Ap stars \citep[e.g.][and references
therein]{2013MNRAS.429.1001A,2009A&A...502..283H} as long as the
merger is completed early enough at the pre-main-sequence stage for
its result to appear as a Herbig Ae/Be
star. \citet{2009MNRAS.400L..71F} explicitly stress that one (at
least) of the merging stars must be on the Henyey part of the
pre-main-sequence track towards the end of its contraction to
the main sequence. This makes it very plausible that the outcome
becomes observable as a Herbig Ae/Be star. Once the merger is
completed, the stellar field should then be essentially
indistinguishable from the fossil fields hypothesised by the more
traditional theories of the origin of Ap star magnetism. Furthermore,
since the angular momentum is shed out during the merging process, 
no further rotational braking is needed at either the
pre-main-sequence or main-sequence stage, which is consistent with the
observation that Herbig Ae/Be stars achieve slow rotation at an
extremely early phase \citep{2013MNRAS.429.1027A}.

How the synchronised binaries fit in the merging scenario is not
entirely clear. In such a binary, the initial rotation period of the
Ap component at the end of its formation must have been shorter than
the orbital period of the system (so that, with time, tidal braking
would synchronise it). By contrast, if we are correct in our interpretation that the
connection between rotational and orbital periods is due to the
formation of the Ap component through merging, the long
rotation period of the Ap component of binaries with short orbital
periods (that is, the systems in the upper left quadrant of
Fig.~\ref{fig:Prot_vs_Porb}) must have been defined at the time when
they formed from the merger of two lower mass stars. Thus we have two
very different populations of Ap stars in binaries: one with stars
that had initial rotation periods not exceeding a few days, and the other
whose members have had rotation periods longer than $\sim50$ days
since the time of their formation. 

Synchronised binaries may be systems whose components were originally
close enough to develop an interaction leading to the generation of a
magnetic field in one of the stars, but not to lead to
merging. The field then allowed chemical peculiarities to develop in
the magnetic component, which became an Ap star without
merging. Alternatively, it is very possible  
that Ap stars in synchronised binaries formed  
through a completely different channel. As stressed by
\citet{2010ARep...54..156T}, even if merging is the main channel of 
formation of Ap stars, this does not rule out the possibility that some are
formed through other processes.

Within this picture, there are various plausible paths leading to the
exception of HD~219749, the only non-synchronised binary known so far
with a 
(relatively) short orbital period (48 days) and an Ap component with a
short rotation period (1.6 day). HD~219749 may have formed through the merger
scenario as a single Ap star with a short rotation period, and it may have
acquired a companion later in its evolution through a chance
encounter. Or this binary may have originally been a triple system, of
which two components merged to form an Ap star with a distant
companion, with an 
orbital period greater than 1000 days. In such a binary, the Ap
component could have a short rotation period. This binary would then
have undergone some external perturbation later in its life,
leading to its components getting closer to each other with a
considerably shortened orbital period. A third possibility is that the
system formed through the synchronised binary channel, rather than
the merger channel, and that an external perturbation later in its
life led to its components moving further apart, with an increased
orbital period. Thus, as already mentioned, the exception of HD~219749
does not question the general conclusion that, as a rule, short orbital
and rotational periods are mutually exclusive in non-synchronised
systems. 

\section{Conclusions}
\label{sec:conclusion}
We carried out the most exhaustive statistical study to date of a
sample of Ap stars with magnetically resolved lines. It is based on
extensive sets of previously unpublished measurements of various
parameters characterising them:
\begin{itemize} 
\item Measurements of their mean magnetic field modulus at
  different epochs. These measurements complement those presented in
  \citetalias{1997A&AS..123..353M} for 40 of these stars, considerably extending the time span
  that they covered.
\item Multi-epoch determinations from spectropolarimetric observations
  of their lowest order field moments: mean longitudinal magnetic
  field, crossover, and mean quadratic field. For about a quarter of
  the studied stars, our data represent the first ever determinations
  of these field moments. For many others, only very few such
  measurements had been obtained before, in most cases sparsely
  sampling the rotation cycles of the considered stars, hence
  unsuitable to constrain their variability. Improved phase
  sampling and extended time span coverage result from the
  combination of the new measurements of this paper with those of our
  previous studies
  (\citealt{1994A&AS..108..547M,1995A&A...293..733M,1995A&A...293..746M,1997A&AS..124..475M};
  note also the revised quadratic field values of  
  Appendix~\ref{sec:hquad_rev}), allowing many variation curves to be
  defined for the first time.
\item Radial velocity determinations from almost all the observations
  of \citet{1991A&AS...89..121M}, \citet{1997A&AS..124..475M}, \citetalias{1997A&AS..123..353M}, and
  the present study.
\end{itemize}

Whenever appropriate, we complemented the measurements detailed above
with similar data from the literature. We exerted great care in the
selection of the latter, to ensure as much as possible that they were
of similar (or better) quality and reliability as our own data. There
was little concern in that respect for mean magnetic field modulus
determinations from other authors. But our critical evaluation of the
published determinations of other parameters revealed a much wider
range in their suitability for our purposes. In particular,
many ambiguities were found with respect to the peculiarity
class (Ap or Am, or even non-peculiar) of a number of stars, to their
rotation and orbital periods, or even to their binarity. 

Analysis of these data enabled us to confirm and refine the results
established in \citetalias{1997A&AS..123..353M} with increased significance, reveal the
existence of additional statistical trends, and present supporting
evidence for the occurrence of physical processes that were little
studied until now. Hereafter, we summarise these results, we consider
possible ways of submitting them to additional tests, and we discuss
their implications for our general understanding of the Ap stars and
their magnetic fields.

Neither our new mean magnetic field modulus data, nor the much less
extensive measurements of this field moment in the large number of
additional Ap stars with resolved magnetically split lines whose
discovery was reported in the literature in recent years, call into question the
existence of a sharp discontinuity at the low end of the $\Hm$
distribution. The location of this discontinuity, about 2.8\,kG (in
terms of average of the field modulus over the stellar rotation cycle),
is confirmed. The observations presented in this paper also strengthen
the arguments developed in \citetalias{1997A&AS..123..353M} against
the possibility that 
this discontinuity is only the apparent consequence of observational
limitations. 

A number of detections of magnetic fields below the magnetic
resolution threshold through modelling of differential broadening of
lines of different magnetic sensitivities were published in the
meantime, all yielding field strengths not exceeding 1.5\,kG. Thus,
while based on the results presented in
\citetalias{1997A&AS..123..353M}, one could speculate 
the existence of a cut-off at the low end of the field intensity
distribution, a much more intriguing picture now seems to emerge, in
which the distribution of the magnetic field strengths of the slowly
rotating Ap stars is
bimodal, with a narrow peak below $\sim2$\,kG and a broad, spread
out distribution above approximately 2.8\,kG, separated from each
other by a gap in which no star is found. 

The reality of the existence of this gap should be further tested. Its
boundaries should be better established, especially on the low field
side. The rate of occurrence and distribution of field strengths below
2\,kG should be studied. Whether the bimodal distribution of the field
strengths is specific to the slowly rotating Ap stars or whether it
extends to their faster rotating counterparts also needs to be
investigated. 

Those stars that show (or
may show) magnetically resolved lines only for part of their rotation
cycle are especially interesting, and they deserve to be followed
throughout their whole period. This is needed in particular to
confirm our suspicion that in stars where the line
\ion{Fe}{ii}~$\lambda\,6149.2$ is magnetically resolved at some
phases, the average over the rotation cycle of the mean field modulus
is always greater than or equal to $\null\sim2.8$\,kG. 

It is almost
impossible for it to be significantly lower in HD~9996, as $\Hm$
reaches $\null\sim5.1$\,kG at its maximum and the observed shape of
its variation around the latter (Fig.~\ref{fig:hd9996}) indicates that
it must exceed 3\,kG over nearly half the rotation period of the star. In
HD~18078, the phase interval over which the
\ion{Fe}{ii}~$\lambda\,6149.2$ line is not resolved is too narrow for 
the value $\Hav=3.45$\,kG, which is derived by fitting the
$\Hm$ measurements over the rest of the cycle to be much greater than
the actual average value of the mean field modulus over the whole
rotation period. The star that may be best suited
for a critical test of the $\Hav$ lower limit is
HD~184471, provided that our suspicion that
\ion{Fe}{ii}~$\lambda\,6149.2$ can be resolved into its split
components over part of its rotation cycle (see
Sect.~\ref{sec:binarity}) is correct. Furthermore, its fairly short
rotation period, $50\fd8$, lends itself well to achieving good phase
coverage of its field modulus variation curve in a reasonable time. 

On the low side of the conjectured $\Hm$ distribution gap, as part of
our systematic search for Ap stars with magnetically resolved lines,
we identified a significant number of (very) low $\vsi$ Ap stars
with unresolved lines. Systematic determination of their magnetic
fields, or upper limits thereof, from the analysis of differential
broadening of lines of different magnetic sensitivities, will allow us
to probe the low end of the $\Hm$ distribution. However, the rotation
period of the vast majority of these stars is unknown. It is likely
long for most of them and its determination may not be
straightforward. Photometry has proved poorly suited to
  studying variations on timescales of months or years. Spectral line
  profile variations in natural light may be small and difficult to
  analyse as magnetic broadening is not the dominant contributor to
  the overall line width. Studying the variability of the mean
  longitudinal magnetic field appears to be the most effective
  approach to constrain the rotation periods of the sharp-lined Ap
  stars whose lines are not resolved into their magnetically split
  components.
In any event, characterisation of the distribution of the magnetic
field strengths of Ap stars below 1.5--2.0\,kG will require a
significant, long-term effort. 

Here we confirm, refine, and complement our previously established dependencies of the magnetic field strength and structure on the
rotation period (\citetalias{1997A&AS..123..353M};
\citealp{2000A&A...359..213L}; \citealp{2002A&A...394.1023B}). 

The conclusion from
\citetalias{1997A&AS..123..353M} 
that very strong magnetic fields ($\Hav>7.5$\,kG) occur only in stars
with rotation periods of less than $\null\sim150$ days was
strengthened by the larger size of the sample of stars on
which it is based (in part thanks to the inclusion of additional
stars from the literature) as well as by the increase in significance
of the values of $\Hav$ resulting from the improved phase coverage of
the measurements for many of the studied stars. In order to further
test it, one should focus first on determining the rotation periods of those
stars most likely to have $\Hav>7.5$\,kG for which their values have
not yet been derived and for which $\vsi$ is small; these stars are HD~47103 and
HD~66318 (see Sect.~\ref{sec:Hmdisc}).

Continued observation
of the stars for which the data available so far do not fully
constrain the mean field modulus variation curve will make the set of
$\Hav$ values increasingly representative of the actual stellar field
strengths. This pending improvement will be particularly important for
the stars whose rotation period is longer than the time frame spanned
by the measurements performed up to now. For the majority of these
stars, the actual value of $\Hav$ may be somewhat greater than the
current estimate based on data sampling only a fraction of the
period, if the tendency of the $\Hm$ variation curves of the slowest
rotating stars to have broad, nearly flat bottoms advocated in
Sect.~\ref{sec:Hmdisc} is confirmed. This putative systematic increase
of $\Hav$ for the slowest rotators is however
very unlikely to be sufficient to question or significantly alter
the above-mentioned conclusion about the lack of very strong magnetic
fields in the long-period Ap stars. However, it will be interesting to
confirm if the apparent tendency for very slowly rotating stars to
show in general higher values of the ratio $q$ between the extrema of the
field modulus (Fig.~\ref{fig:q_vs_P}) is strengthened as the fraction
of their period over which $\Hm$ measurements were obtained
increases. An open question is whether $q$ may in some cases be much
greater than 2. 

\citet{2000A&A...359..213L} and \citet{2002A&A...394.1023B}
mostly used the values of the magnetic moments presented here for
those stars for which full coverage of the rotation cycle is achieved
to reach the conclusion that the angle between the rotation and
magnetic axes is generally large in stars with rotation periods
shorter than 1--3 months, while these axes tend to be aligned in slower
rotating stars. The longest period stars are omitted from those
studies. Nevertheless, even though the data presented here for
  the slowest rotators do not completely sample the field variations,
they are useful to complement the results of the
above-mentioned works and to amend their conclusions. Their indication
that for $\Prot\ga1000$\,d, large $\beta$ angles become predominant
again appears convincing. 

The point that remains more debatable at
this stage is whether near alignment of the magnetic and rotation axes
actually occurs in an intermediate period range (approximately
100--1000\,d), or whether this apparent trend is just coincidental,
resulting from the insufficient population of the sample of stars that
could be studied in that range. Hopefully, that population will grow
once the rotation periods of the stars with magnetically resolved
lines that have so far only been observed at few epochs (that is,
primarily, most stars of Table~\ref{tab:stars_litt}) have been
determined. At present, there are two stars with periods in that range
for which the existing data are insufficient to constrain the magnetic
field geometry; these are HD~110274 (which has magnetically resolved lines) and
HD~221568 (which does not). Systematic measurements of their mean
longitudinal magnetic fields, and of the mean magnetic field modulus
of HD~110274, throughout their rotation cycles, will valuably
complement our current knowledge of the field properties in the
100--1000\,d $\Prot$ range. 

One of the particularly interesting outcomes of this study is the
identification of a small number of stars showing considerable
deviations with respect to the statistical trends that apply to the
bulk of the studied sample and, hence, presumably to the majority of Ap
stars with moderate or slow rotation. Remarkably, the exceptions
comprise some of the most famous and best-studied Ap stars, i.e. HD~65339,
HD~137909, HD~215441. On the one hand, it is very fortunate that
these stars have been extensively studied for several decades, since
the consideration of exceptions and extreme cases often provide
especially valuable clues towards the understanding of the physical
processes that are at play in the type of objects to which they 
belong. This represents a strong incentive to continue to study them
in ever greater detail. On the other hand, these outliers should not be regarded as representative or typical of
the common properties or general behaviour of magnetic Ap stars: one
should be cautious not to generalise the conclusions drawn about them
to the class as a whole -- this is a mistake that has occasionally
been made in the past. 

Other, less studied stars that prove to be atypical in some respects
include HD~18078, HD~51684, HD~93507, HD~94660, and HD~126515. All
deserve additional investigation, as understanding what makes their
magnetic fields different from the majority may also provide important
clues about the physical processes at play for the latter. 

The better quality and coverage of the new spectropolarimetric
observations analysed here, compared to those of our previous studies,
and the resulting improvement in the precision of the magnetic field
moments that we derived from them, have enabled us to achieve significant
measurements of rotational crossover in longer period stars. That is not
unexpected, but a more intriguing result is the apparent detection of
crossover in stars whose rotation is definitely too slow to generate
the Doppler line shifts required to account for that effect. Those
detections are, however, close to the limit of formal significance,
and while we have presented various arguments to support their
reality, they beg for further confirmation. 

The most straightforward way to obtain such confirmation will be to
try to reproduce the detections achieved here for new, similar
observations of the same stars. The higher resolution, broader
spectral coverage and better stability of modern spectropolarimeters
such as ESPADONS or HARPSpol 
should ensure that much more precise determinations of the crossover
should be achievable. With lower uncertainties, crossover values of
the order reported here should be measurable at a sufficient level of
significance to dispel any doubt about their reality. 

If that endeavour is successful, the next step will be to test the
interpretation that we proposed, that pulsation is responsible for
that new type of crossover. In that case, the crossover should vary
with the pulsation frequency; this  can be probed by acquiring
spectropolarimetric observations with a time resolution that is
shorter than 
the pulsation period, similar to the high time and spectral resolution
observations that have been used to study pulsation-induced line
profile variations \citep[e.g.][]{2008CoSka..38..317E}. 

Admittedly,
among the stars in which we detect a formally significant crossover
that does not appear to be of rotational origin, only HD~116114 is
known to be pulsating. This makes it a prime candidate for the
considered observational test, all the more since it has one of the
longest pulsation periods known among roAp stars
\citep{2005MNRAS.358..665E}, such that it is particularly well suited
to obtaining pulsation-phase-resolved spectropolarimetric
observations; it is also relatively bright. 

However that should not
detract one from attempting a 
similar experiment for some of the other stars in which crossover that
cannot be accounted for by rotation has apparently been detected (see
Sect.~\ref{sec:xdisc}). It should be reasonable to assume that, if
they pulsate at all, they do so at frequencies similar to those of
the known roAp stars, and to obtain observations with a matching time
resolution. Those observations would then probe the occurrence of
pulsations that cannot be detected photometrically or via
studies of radial velocity and line profile variations, possibly
because they correspond to non-radial modes of high 
degree $\ell$. Such success would open the door to a whole new region of
the parameter space in asteroseismological studies of Ap stars. 

Our long-term monitoring of the magnetic variations of Ap stars with
resolved magnetically split lines has allowed us to set new
constraints on many rotation periods. A significant number of stars proved to rotate so slowly that even
observations spanning time intervals of several years -- in some
cases, more than a decade -- fall very short of covering a full cycle. The
resulting increase in the number of Ap stars known to have extremely
long periods led to the realisation of the existence of a considerable
population of Ap stars that rotate extremely slowly. Stars with
periods in excess of one year must represent several percent of all Ap
stars. 

The long-period tail of the distribution of the rotation
periods of Ap stars is now sufficiently populated to allow statistical
inferences to be drawn. The most definite one at this stage is the
first concrete evidence that some of the rotation periods must be as
long as 300 years. The existence of such long periods makes it
absolutely essential to continue to monitor the variations of the
corresponding stars. It makes us responsible to future generations
of researchers; any gaps that we may leave today in the phase coverage
of the variations of those super-slow rotators
may not be recoverable for decades or centuries. Such gaps may
heavily hamper scientific progress by making it impossible to fully
characterise the properties of the longest period stars and to
identify possible differences between them and faster (or less slow)
rotators. 

The impact is not restricted to the knowledge of the slowest rotators;
it also extends to our general understanding of the Ap
phenomenon. The 5 to 6 orders of magnitude spanned by their rotation
periods represent a unique property that distinguishes
them from all other stellar types. Since the evolution of their
rotation rates during their lifetimes on the main sequence are
mostly consistent with conservation of the angular momentum, the
definition of those rates must be a fundamental part of the stellar
formation and early evolution processes. Knowledge of the longest
periods and the other stellar properties with which they are
correlated must therefore be one of 
the keys to the unravelling of those processes.  

On the other hand, the knowledge acquired about the long-period
tail of the distribution of the rotation periods of Ap stars  so far  is mostly
restricted to the strongly magnetic members of that class (that is, in
practice, stars with $\Hav\ga2.8$\,kG). As mentioned in
Sect.~\ref{sec:rotation}, of the 36 stars with $\Prot>30$\,d currently
known, only 4 do not show magnetically resolved lines. But there must
also exist a sizeable population of weakly magnetic stars that rotate
very slowly. We need to identify and to study them to obtain
a complete picture of super-slow rotation in Ap stars and to gain
further insight into the processes at play during their formation and
early evolution that lead to the huge spread in their rotation
velocities by the time when they arrive on the main
sequence. Observationally, this undertaking overlaps with the
above-mentioned study of the distribution of the magnetic field
strengths of Ap stars below 1.5--2.0\,kG.

The significance of rotation for the understanding of the Ap
phenomenon is further emphasised by the unexpected discovery of
correlations between the rotation periods of Ap components of binaries
and the orbital periods of those systems. It is highly significant that the orbital period is not shorter than 27 days for any of the
binaries in which the Ap component has magnetically resolved lines. It
is a very solid result because the probability that a
short-period spectroscopic binary may have been overlooked by our
intensive monitoring is vanishingly small. This could have happened
only for very rare combinations of an orbital plane seen very nearly
face-on and a very low mass secondary. Even if a couple of such
systems had been overlooked, the general conclusion that there is a
deficiency of binaries with short orbital periods among systems
containing an Ap star with resolved magnetically split lines would
still remain valid. 

Actually, among Ap stars in binaries that do not show
magnetically resolved lines for which both the rotation and the
orbital periods are known (see Table~\ref{tab:rot_orb_litt}), there is
also a deficiency of orbital periods shorter than 27 days: only the three
synchronised systems have such periods. All others, except for HD~219749, have
either $\Porb>1000$\,d, or $\Prot>50$\,d in combination with a
magnetic field too weak for their spectral lines to be observationally
resolved into their split components. 

 That lack of known non-synchronised
systems in which both the rotation and orbital periods are short
for Ap stars that do not show magnetically resolved lines as well as
for those that do show them led us to the conclusion that, for Ap
stars in binaries, barring synchronism, short rotation periods and short orbital
periods are mutually exclusive. This result begs further
confirmation, even though it is established at a high level of
statistical significance from the most complete sample that we could
build by combining our new data with all the relevant information of
sufficient quality that we could find in the literature. 

The acid test consists of trying to find Ap stars with rotation
periods shorter than 50 days in non-synchronised binaries with orbital
periods shorter than 1000 days. To this effect, it is critical to
confirm the peculiarity of the presumed Ap component of those
binaries. Indeed, we found in Sect. ~\ref{sec:binarity} that in a
surprisingly large number of the systems that have in the past been
considered as Ap binaries, that classification appears to be definitely or very
probably mistaken. Accordingly, there remain only very few binaries
that have a reliably known orbital period shorter than 1000 days and
contain an Ap component whose rotation period has not been
determined. We have been able to identify only two such systems; these are
HD~138426 and HD~200405. An observational effort should be
undertaken to study the variability of their Ap components. 

More generally, taking into account the classification inaccuracies
that appear to have plagued previous studies, it is worth
reconsidering the distribution of the orbital periods of
the binaries containing Ap components. The exhaustive BinaMIcS
project is already revisiting its
short-period tail in a systematic manner and the (still partial)
results obtained until now unambiguously show it to be 
significantly less populated than previously thought
\citep{2015IAUS..307..330A}. On the other 
hand, while the nature of our study of the magnetic fields of the Ap
stars with resolved magnetically split lines has also made it particularly
appropriate for the discovery and characterisation of long-period
spectroscopic binaries, complementing the knowledge of that population
with systems containing faster-rotating Ap components may prove much
more elusive. 

We argued in Sect.~\ref{sec:binarity} that the existence of a
connection between the rotation and orbital periods of Ap stars in
binaries represents a strong argument in favour of a merger scenario
for their formation. Many of the other results presented in this
paper, including those about the distributions of the rotation periods
of the Ap stars and of the strengths of their magnetic fields, and about the
correlation between rotation and magnetic properties, are products of
the evolution 
process that they have undergone. Taken together, those various
elements must guide the theoretical developments aimed at explaining
the origin of magnetism in a fraction of the main-sequence stars of
spectral type A and possibly also of hotter stars. This remains one of the
least well-understood aspects of stellar physics, in which hopefully
our contribution will enable significant progress to be achieved. As
discussed in the present section, this contribution does not represent
the final word on the observational side, as some of its conclusions
beg for further confirmation, and even more, as it raises a number of
additional, sometimes unexpected, questions. We have highlighted some
of these questions and proposed ways to address them, tracing a
possible path for continued investigation of some of the most
intriguing and potentially enlightening aspects of the magnetism of Ap
stars.

\begin{acknowledgements}
A large portion of this study has
been carried out during stays of the author in the Department Physics and
Astronomy of the University of Western Ontario (London, Ontario,
Canada). I thank John Landstreet for giving me the
opportunity of these visits and for providing partial funding to
support them. Financial support received from the ESO
Director General's Discretionary Fund (DGDF) also contributed to make
these stays possible. I am grateful to Swetlana Hubrig, John Landstreet, Jean
Manfroid, and Gregg Wade, for their participation in the excution of
some of the observations, and to Steven Arenburg, Jean Manfroid, and
Erich Wenderoth for their contributions to their initial reduction;
Wenderoth's participation in the project was also
funded by the ESO DGDF.
This research has made use of the SIMBAD database,
operated at the CDS, Strasbourg, France
\end{acknowledgements}

\appendix 

\section{Notes on individual stars}
\label{sec:notes}
\begin{figure*}[t]
\resizebox{12cm}{!}{\includegraphics[angle=270]{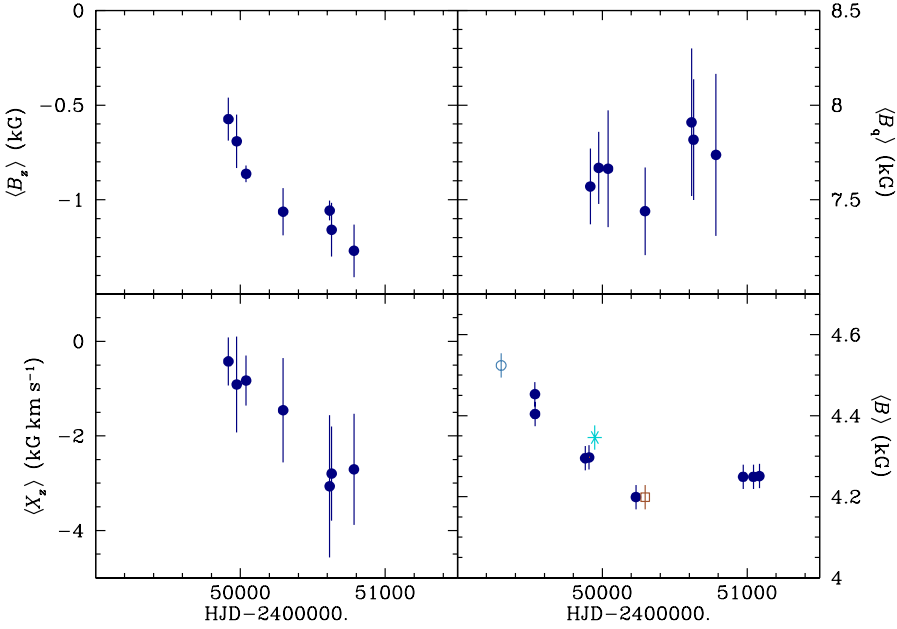}}
\parbox[t]{55mm}{
\caption{Mean longitudinal magnetic field ({\it top left\/}),
crossover ({\it bottom left\/}), 
mean quadratic magnetic field ({\it top right\/}),
and mean magnetic field modulus ({\it bottom right\/}) 
of the star HD~965,
against heliocentric Julian date. The symbols are as described at the
beginning of Appendix~\ref{sec:notes}.}
\label{fig:hd965}}
\end{figure*}

\begin{figure*}[t]
\resizebox{12cm}{!}{\includegraphics[angle=270]{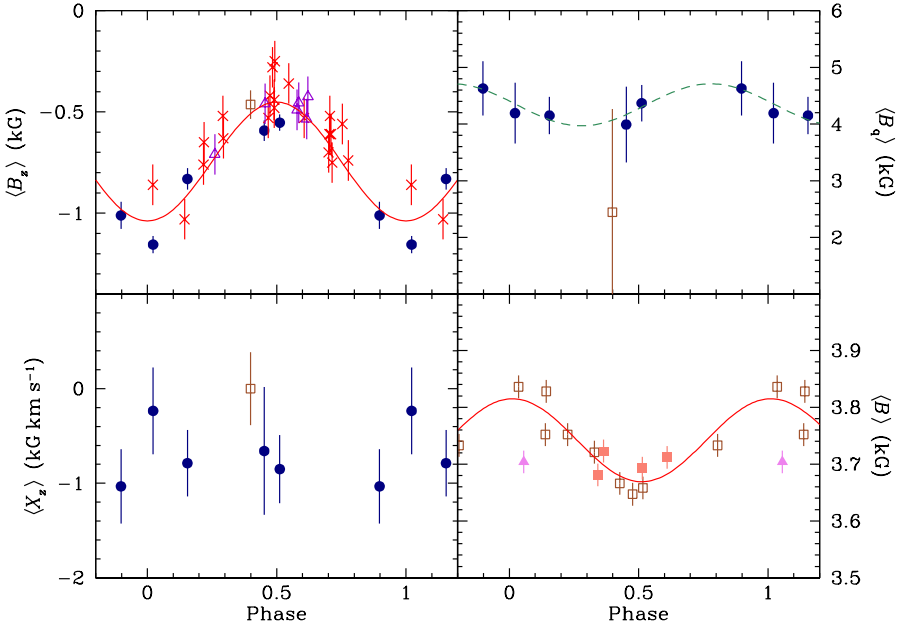}}
\parbox[t]{55mm}{
\caption{Mean longitudinal magnetic field ({\it top left\/}),
crossover ({\it bottom left\/}), 
mean quadratic magnetic field ({\it top right\/}),
and mean magnetic field modulus ({\it bottom right\/}) 
of the star HD~2453,
against rotation phase. 
For the longitudinal field, measurements of \citeauthor
{1958ApJS....3..141B} (\citeyear {1958ApJS....3..141B}; 
open triangles) and \citeauthor{1975ApJ...202..127W} (\citeyear{1975ApJ...202..127W}; 
crosses) are also 
shown. The other symbols are as described at the
beginning of Appendix~\ref{sec:notes}.} 
\label{fig:hd2453}}
\end{figure*}

In this Appendix, we discuss the measurements that we obtained for
each star individually and their implications for
our knowledge of its properties. This discussion is illustrated by
figures showing the curves of variation of the mean magnetic field modulus,
and when available, the mean longitudinal magnetic field, the
crossover, and the quadratic field. Those variations are plotted
against the rotation phase, when the period is known, or against the
time of observation. In the former case, the fits of the variation
curves computed with the parameters listed in Tables~\ref{tab:mfit}
to \ref{tab:qfit} are also shown. The different 
symbols used to represent the $\Hm$ data of
\citetalias{1997A&AS..123..353M} and of this paper  distinguish the
various instrumental configurations with which those measurements have
been obtained, as specified in Table~\ref{tab:plot_sym}. In the plots
of $\Hz$, $\xover,$ and $\Hq$, filled dots (navy blue) correspond to
the new determinations reported in this paper. For our older data, we
used the same symbols as \citet{1997A&AS..124..475M}, namely open
squares (brown) for the measurements that these authors obtained with
the long camera of CASPEC, filled triangles (violet) for those that they
dervived from CASPEC short camera spectra, and filled squares (salmon)
for the earlier data of
\citet{1994A&AS..108..547M,1995A&A...293..733M,1995A&A...293..746M}. 
In a few cases, values of some of the magnetic field moments that were
obtained by other authors are also included in the figures; the
symbols used for that purpose are specified in the individual captions
of the relevant figures. Solid
curves (red) represent fits for which all parameters are
formally significant, long-dashed curves (navy blue) represent fits
including a first harmonic that is not formally significant, and
short-dashed curves (green) represent non-significant fits by the
fundamental only. 

For a number of stars for which the diagnostic lines used for the new
determinations of $\Hz$, $\xover,$ and $\Hq$ of this paper show
significant intensity variations, figures showing the equivalent widths of
those lines against rotation phase are included. 

The orbital solutions that we derived for several spectroscopic
binaries and the radial velocity measurements for other stars whose
radial velocity is variable are also illustrated. The symbols of
Table~\ref{tab:plot_sym} were used to represent the determinations
based on the high-resolution spectra recorded in natural light that
were analysed in \citetalias{1997A&AS..123..353M} and in this
paper. The individual figure captions specify the symbols used for all
other radial velocity measurements, whether based on the CASPEC
spectropolarimetric observations of \citet{1991A&AS...89..121M},
\citet{1997A&AS..124..475M}, and this paper, or published by
other authors. 

\subsection{HD~965}
\label{sec:hd965}
% hall
While it had hardly been studied prior to the discovery of resolved
magnetically split lines in its spectrum \citepalias{1997A&AS..123..353M}, HD~965 has since
turned out to be a very unusual and intriguing object. Interest in
this star originally arose from our realisation, as part of the
present analysis of its CASPEC spectra, of the many apparent
similarities between its spectrum and that of HD~101065 (Przybylski's
star), the most peculiar Ap star known. The resemblance was
fully confirmed from consideration of higher resolution spectra by
\citet{2002ASPC..279..365H}, who further characterised HD~965 with
preliminary quantitative abundance determinations. Like for HD~101065,
the hydrogen Balmer lines of HD~965 show a marked core-wing anomaly; both
stars were among the first in which this anomaly was identified
\citep{2001A&A...367..939C}. It reflects the fact that the
atmospheric structure of the star is not normal. More remarkably
still, the strongest evidence yet for the presence of promethium (an
element with no stable isotope) in the spectrum of an Ap star was
found in HD~965 \citep{2004A&A...419.1087C}.  However, rather
intriguingly, despite its many features in common with Przybylski's
star and other roAp stars, all attempts to detect pulsation in HD~965
so far have yielded negative results \citep{2005MNRAS.358.1100E}. 

The richness of the spectrum of HD~965 makes it difficult to find
lines sufficiently free from blends to be usable for determination of
the longitudinal field, the crossover, and the quadratic field. As a
result, the line set from which these field moments were derived is,
together with that of HD~47103, the smallest of all the stars analysed
here. All diagnostic lines used are \ion{Fe}{i} lines, except for one
\ion{Fe}{ii} line. No significant variation of the equivalent width of
these lines was detected, so that there is no indication of large
inhomogeneities in the distribution of the element on the stellar
surface. 

The magnetic measurements of HD~965 are plotted against time in
Fig.~\ref{fig:hd965}. After its monotonic decrease observed over the
whole time span of the data published in \citetalias{1997A&AS..123..353M}, the mean field
modulus apparently went through a minimum between August 1996 and June
1998 (unfortunately no high-resolution spectrum of the star was
obtained between these two epochs). The position in the plot of the
last three measurements with respect to the previous points suggests
that the slope of the post-minimum increase of $\Hm$ is
significantly less steep than that of its pre-minimum decrease. This
is fully confirmed by the more recent data shown in Fig. 3 of
\citet{2005MNRAS.358.1100E}, which indicates a broad, almost flat minimum over
approximately seven years with possibly a shallow secondary maximum
around the middle of this time interval. With this feature, the field
modulus variation curve is reminiscent of those of other stars
discussed here, such as HD~187474 (Fig.~\ref{fig:hd187474}). But the
duration of the minimum in HD~965 is exceptionally long.

\citet{2005MNRAS.358.1100E} and \citet{2014AstBu..69..427R} also 
reported new mean longitudinal magnetic 
field measurements obtained after those presented here. Although
systematic differences between $\Hz$ determinations performed with
different instruments are not unusual, the combination of the two
data sets unambiguously indicates that the longitudinal field of HD~965
went through a negative extremum sometime during the broad minimum of
its field modulus (possibly close to the shallow secondary
maximum). The difference of slope in the variation of $\Hz$ before and
after the time of negative extremum suggests that the field is not
symmetric about an axis passing through the centre of the star.
The new $\Hz$ data of \citet{2005MNRAS.358.1100E} and
\citet{2014AstBu..69..427R} also reveal
that the field reverses its polarity along the stellar rotation cycle.
This cycle has not yet been fully covered by all the observations
obtained until now; the rotation period must significantly exceed 13
years. 

The quadratic field does not show any definite variation over the
time span covered by our observations. The ratio between it and the
field modulus is 
remarkably high, reaching of the order of 1.8. This suggests that the
field must be unusually inhomogeneous over the observed part of the
star. 

With the very long rotation period, no significant crossover was
expected. Consistently with this, none of the individual
determinations of this quantity yield values significantly different
from zero (in the best case, a $2.8\sigma$ value is obtained). But
the facts that all measurements are negative, and that they seem to
show a monotonic trend with time towards increasingly negative
values, could be regarded as rather convincing indications that
crossover is actually detected. This is further discussed in
Sect.~\ref{sec:xdisc}.

\subsection{HD~2453}\label{sec:hd2453}
% hzc11 xhc1 hqc1 hmc1 hallk
For determination of all magnetic field moments but the mean field
modulus in HD~2453, a set of lines of \ion{Fe}{ii} was analysed. No
variability of the equivalent widths of these lines was observed. 

Our five new determinations of the mean longitudinal magnetic field were
combined with previous measurements of \citet{1958ApJS....3..141B},
\citet{1975ApJ...202..127W}, and \citet{1997A&AS..124..475M} to
redetermine the 
rotation period of the star. Some uncertainty arises from the
possibility that our determinations may yield values that are systematically
100--200\,G more negative than those of Babcock and Wolff. Taking
that uncertainty
into account, we confirm the best value of the period already adopted
in \citetalias{1997A&AS..123..353M}: $\Prot=(521\pm2)$\,d. Using this period, we derive a
revised phase origin coinciding with the minimum of $\Hz$, 
${\rm HJD}_0=2442288.0$. These values of $\Prot$ and HJD$_0$ are used
to compute the phases against which the various magnetic field moments
are plotted in Fig.~\ref{fig:hd2453}. 

The inclusion of four new measurements of $\Hm$ into the plot of this field
moment against phase increases the scatter about a smooth curve with 
respect to Fig.~6 of \citetalias{1997A&AS..123..353M}, to the extent that one may be inclined 
to question the reality of the detection of actual variations. This
doubt is also fuelled by the comparatively low value (0.77) of
the multiple correlation coefficient $R$ corresponding to the fit of
the $\Hm$ data with a cosine wave. It should be noted, however, that 
a large part of 
the impression of scatter is due to the point corresponding to the
single observation of this star with 
the CFHT. It has been seen in \citetalias{1997A&AS..123..353M}, and is
shown again elsewhere in 
the present paper, that for a number of stars, differences
exist between $\Hm$ values based on AURELIE observations and on
observations obtained with other instruments; these differences are
systematic for a given star but vary from star to star for reasons
that are not fully understood. Although such differences have not been
identified previously for CFHT/Gecko data (of which, between
\citetalias{1997A&AS..123..353M} and 
this work, we have considerably less than AURELIE data), it is not
implausible that they may exist for some stars: HD~2453 may be one of
these stars. This view is supported by the fact that, if the CFHT point is
removed from the data set, the multiple correlation coefficient is
significantly increased for a fit of the remaining data by a cosine
wave: 
$R=0.90$. The fit coefficients appearing in Table~\ref{tab:mfit} and
the fitted curve shown in Fig.~\ref{fig:hd2453} correspond to this
case. The excellent coincidence of the phase of maximum of the mean
magnetic field modulus and of the phase of minimum (i.e. largest
negative value) of the mean longitudinal magnetic field represents a
further indication that we very probably do detect the actual
variations of $\Hm$.  

\begin{figure}
\resizebox{\hsize}{!}{\includegraphics[angle=270]{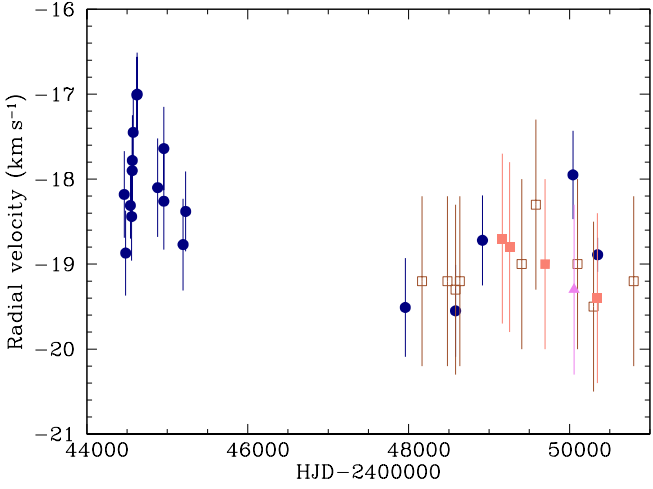}}
\caption{Our radial velocity measurements for HD~2453 (from
  high-resolution spectra recorded in natural light)  are plotted
  together with those of \citeauthor{2002A&A...394..151C} (\citeyear{2002A&A...394..151C};  dots)
  against heliocentric Julian date. The symbols are as described at the
beginning of Appendix~\ref{sec:notes}.}
\label{fig:hd2453_rv}
\end{figure}

By contrast, no definite variation is observed for the quadratic
field. The five measurements of this paper, which are much more
precise than the determination of \citet{1997A&AS..124..475M}, line up 
nicely along a cosine curve. But this may be coincidental as the fitted
amplitude is only at the $2.5\sigma$ level. 

None of the individual values derived for the crossover have
significance levels above $2.6\sigma$, so that no definite detection
is achieved from any of them separately. It is, however, surprising
that all of them are negative. This is further discussed in
Sect.~\ref{sec:xdisc}. 

\begin{figure}[t]
\resizebox{\hsize}{!}{\includegraphics[angle=270]{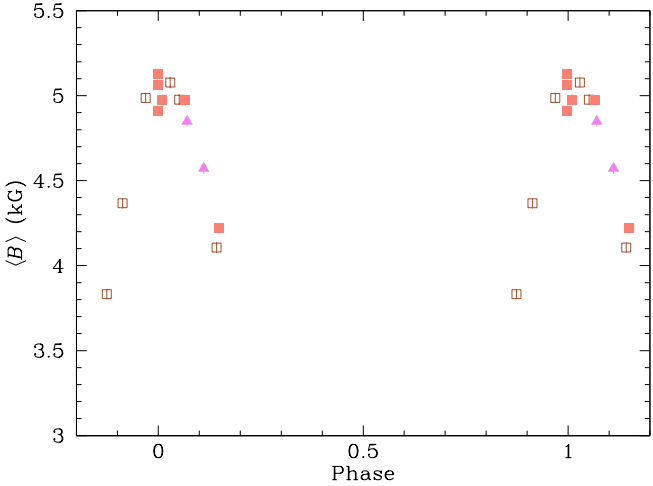}}
\caption{Mean magnetic field modulus of the star HD~9996,
against rotation phase. The symbols are as described at the
beginning of Appendix~\ref{sec:notes}.}
\label{fig:hd9996}
\end{figure}

\begin{figure}
\resizebox{\hsize}{!}{\includegraphics{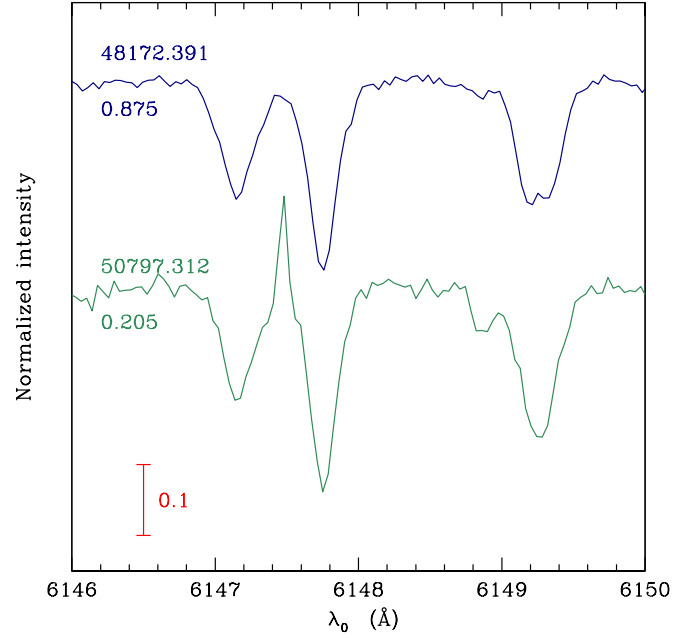}}
\caption{Portion of the spectrum of HD~9996 recorded at two different
  epochs (resp. rotation phases), identified above (resp. below) each
  tracing (the epochs are given as ${\rm HJD}-2400000$), 
  showing the lines \ion{Cr}{ii}\,$\lambda\,6147.1$,
  \ion{Fe}{ii}\,$\lambda\,6147.7$, and
  \ion{Fe}{ii}\,$\lambda\,6149.2$. At the second epoch, the blue wing
  of the latter is blended with an unidentified rare earth line. That
  spectrum is affected by a cosmic ray hit between the lines
  \ion{Cr}{ii}\,$\lambda\,6147.1$ and
  \ion{Fe}{ii}\,$\lambda\,6147.7$.}
\label{fig:hd9996_6149}
\end{figure}
 
\begin{figure}
\resizebox{\hsize}{!}{\includegraphics{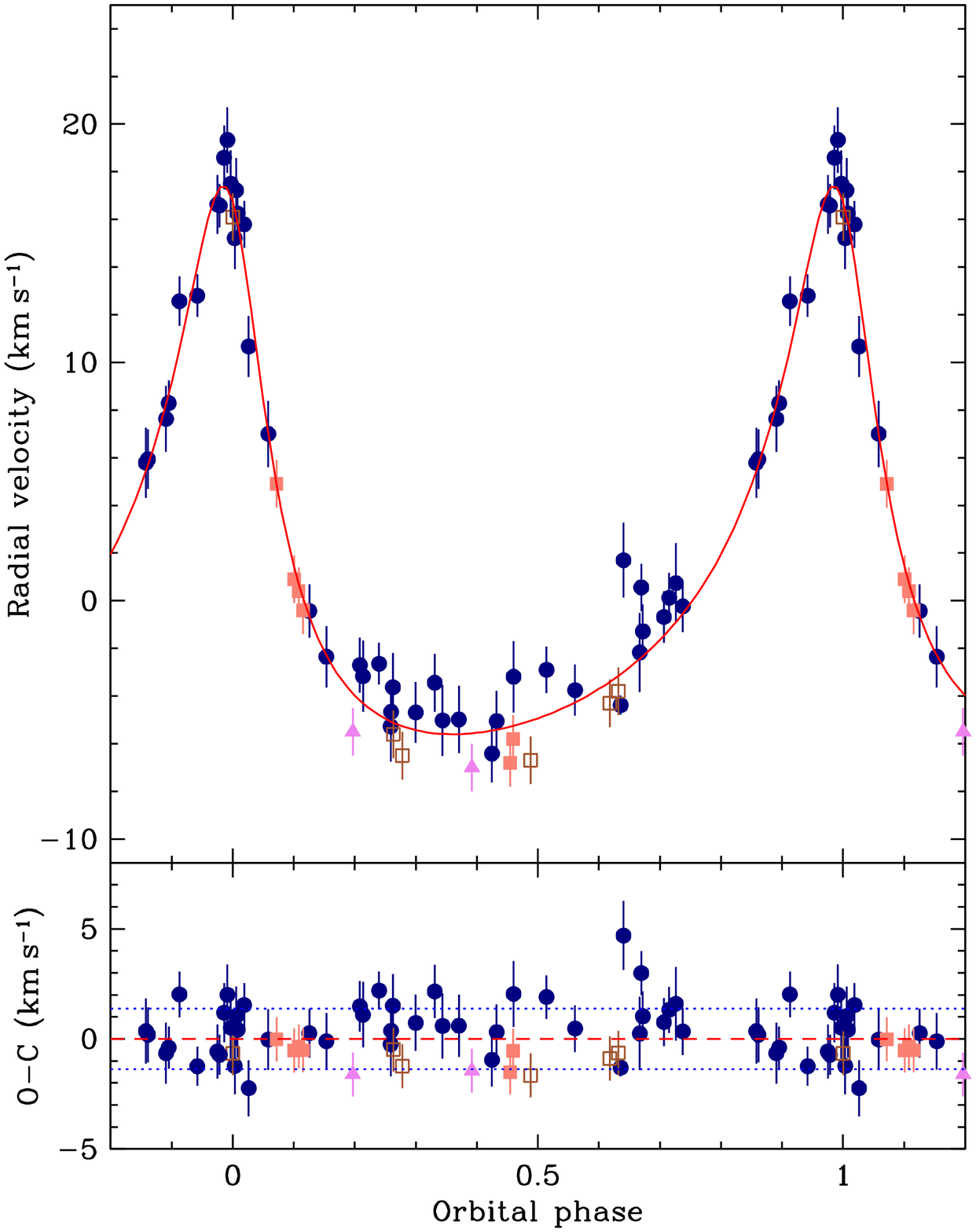}}
\caption{{\it Upper panel\/}: Our radial velocity measurements for
  HD~9996 are plotted 
  together with those of \citet{2002A&A...394..151C} against
  orbital phase. The  solid curve  
  corresponds to the orbital solution given in
  Table~\ref{tab:orbits}. The time $T_0$ of periastron passage is
  adopted as phase origin. {\it Bottom panel:\/} Plot of the
  differences ${\rm O}-{\rm C}$ between the observed values of the
  radial velocity and the predicted values computed from the orbital
  solution. The  dotted lines correspond to $\pm1$ rms
  deviation of the observational data about the orbital solution 
    (dashed line).  Dots represent data from \citeauthor{2002A&A...394..151C}; all
  other symbols refer to our high-resolution spectra 
  obtained with various instrumental configurations, as indicated in
  Table~\ref{tab:plot_sym}.}  
\label{fig:hd9996_rv}
\end{figure}

\citet{2002A&A...394..151C} noted the occurrence of significant
radial velocity variations in HD~2453. As can be seen in
Fig.~\ref{fig:hd2453_rv}, their measurements cluster in two separate
time intervals: 14 from HJD~2444464 to HJD~2445228, whose mean is
$(-18.00\pm0.56)$\,km\,s$^{-1}$, and 5 from HJD~2447958 to HJD~2450349,
corresponding to a mean value of $(-18.92\pm0.59)$\,km\,s$^{-1}$. In
each interval, the standard deviation of the
measurements is consistent with their errors. Our
measurements are mostly contemporaneous with those of the second group
of \citeauthor{2002A&A...394..151C}.  Combining the latter with those
derived from our 13 high-resolution 
observations of the star (obtained between HJD~2448166 and
HJD~2450797), the resulting mean radial velocity is
$(-19.04\pm0.40)$\,km\,s$^{-1}$, which is fully consistent with the value
computed from data from \citeauthor{2002A&A...394..151C} alone. In other words, we confirm
that no variations are detected during each of the two considered time
intervals. But the $\sim1$\,km\,s$^{-1}$ difference in the mean
value of the radial velocity between them, though small, is
significant. Thus HD\,2453 appears to be a spectroscopic binary. Even
in combination with those of \citet{2002A&A...394..151C}, our data
are insufficient to characterise its orbit. But the orbital period
must be (much) longer than the time interval HJD~2447958--2450797,
$\sim2850$\,d (or close to 8 years). The poorer accuracy of the radial velocity
  determinations achieved from the spectropolarimetric CASPEC
  observations makes them unsuitable for consideration in this
  discussion, given the low amplitude of the radial velocity
  variations observed in HD\,2453. We have also omitted them
  from Fig.~\ref{fig:hd2453_rv}.

\subsection{HD~9996}
\label{sec:hd9996}
% hallk
HD~9996 is located too far north to be observable from La
Silla. Accordingly, we have not obtained any spectropolarimetric observations 
for this star. But the variations of its mean longitudinal magnetic
field have been monitored over a long time span
by the Bychkov group at the Special Astrophysical Observatory
\citep{1997smf..proc..204B,2012AcA....62..297B,2014AstBu..69..315M}. 

The three new measurements of the mean magnetic field
modulus that are presented here fully confirm that $\Hm$ went through
its maximum at the end of 1993 or the beginning of 1994, as suspected
in \citetalias{1997A&AS..123..353M}. This coincides roughly
with the negative extremum of the longitudinal field, which is taken
as the phase origin in the ephemeris of \cite{2014AstBu..69..315M}
used to plot the variation of the field modulus
in Fig.~\ref{fig:hd9996}. In an additional spectrum  
taken with AURELIE in December 1997 (HJD~2450797.312), the line
\ion{Fe}{ii}\,$\lambda\,6149$ is no longer resolved, as shown in
Fig.~\ref{fig:hd9996_6149}. The 
interval of approximately seven years separating the two spectra appearing
in this figure, which correspond to our first \citep{1992A&A...256..169M}
and last observations of the star, represents about one-third of its
rotation period \citep{2014AstBu..69..315M}. The field modulus in December 1997 must
have been significantly lower than 3.5\,kG. Because of
the unusually large width of the resolved line components in
HD\,9996, pointed out in \citetalias{1997A&AS..123..353M}, the limit of resolution of the
\ion{Fe}{ii}\,$\lambda\,6149$ line in this star is considerably higher
than the 1.7\,kG limit estimated in \citetalias{1997A&AS..123..353M}
for the majority of slowly 
rotating Ap stars. 

The weak line appearing as a blend in the blue wing
of \ion{Fe}{ii}\,$\lambda\,6149$ in the more recent spectrum is seen,
with different intensities, in many Ap stars. It is not
definitely identified but undoubtedly pertains to a rare earth
element (REE). The fact that it was not seen in our previous spectra
but starts to become visible at the end of our sequence of
observations is consistent with the fact that the part of the stellar
disk where REEs are concentrated is progressively coming into view as
a result of stellar rotation, in the transition from REE minimum and
magnetic maximum to REE maximum \citep{1988A&A...199..299R} and
magnetic minimum. 

 The similarity 
of the line shapes in the two spectra shown in
Fig.~\ref{fig:hd9996_6149} (note in particular the unusual triangular
shape of \ion{Cr}{ii}\,$\lambda\,6147.1$, already noted in
\citetalias{1997A&AS..123..353M}), which were obtained before and
after the mean field modulus maximum (see Fig.~\ref{fig:hd9996}),
suggests that the 
structure of the magnetic field on the  
surface of the star, while unusually inhomogeneous, is to some extent
symmetrical about the plane defined by the rotation and magnetic
axes. On the other hand, a fit of the existing $\Hm$ measurements
by a function of the form of Eq.~(\ref{eq:fit1}) (that is, by a single
sinusoid) leads to a negative minimum of the mean field
modulus (of the order of $-1.1$\,kG), which is of course
non-physical. This clearly indicates that the actual curve of
variation of this field moment must have a high degree of
anharmonicity, presumably with a broad, almost flat minimum, similar
to other very long period stars (e.g. HD~965, as discussed in
Sect.~\ref{sec:hd965}). 

HD~9996 is a spectroscopic binary \citep{1970ApJ...160.1071P} whose
orbital period, $\Porb=273$\,d, is much shorter than the rotation
period of the Ap primary. Combining our radial velocity
measurements with those of \citet{2002A&A...394..151C}, we computed a
revised orbital solution. Its parameters, which are reported in
Table~\ref{tab:orbits}, do not significantly differ from those derived
by \citeauthor{2002A&A...394..151C}, but their standard deviations are
somewhat smaller. The fitted orbit is shown in Fig.~\ref{fig:hd9996_rv}.

\begin{figure}[t]
\resizebox{\hsize}{!}{\includegraphics[angle=270]{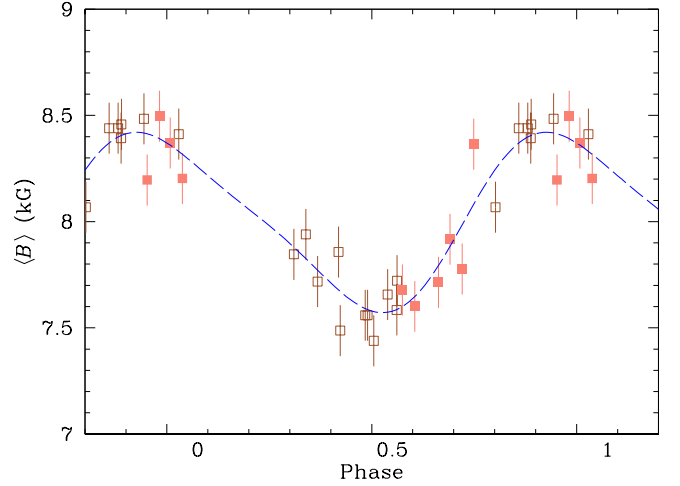}}
\caption{Mean magnetic field modulus of the star HD~12288,
against rotation phase.}
\label{fig:hd12288}
\end{figure}

\begin{figure}
\resizebox{\hsize}{!}{\includegraphics{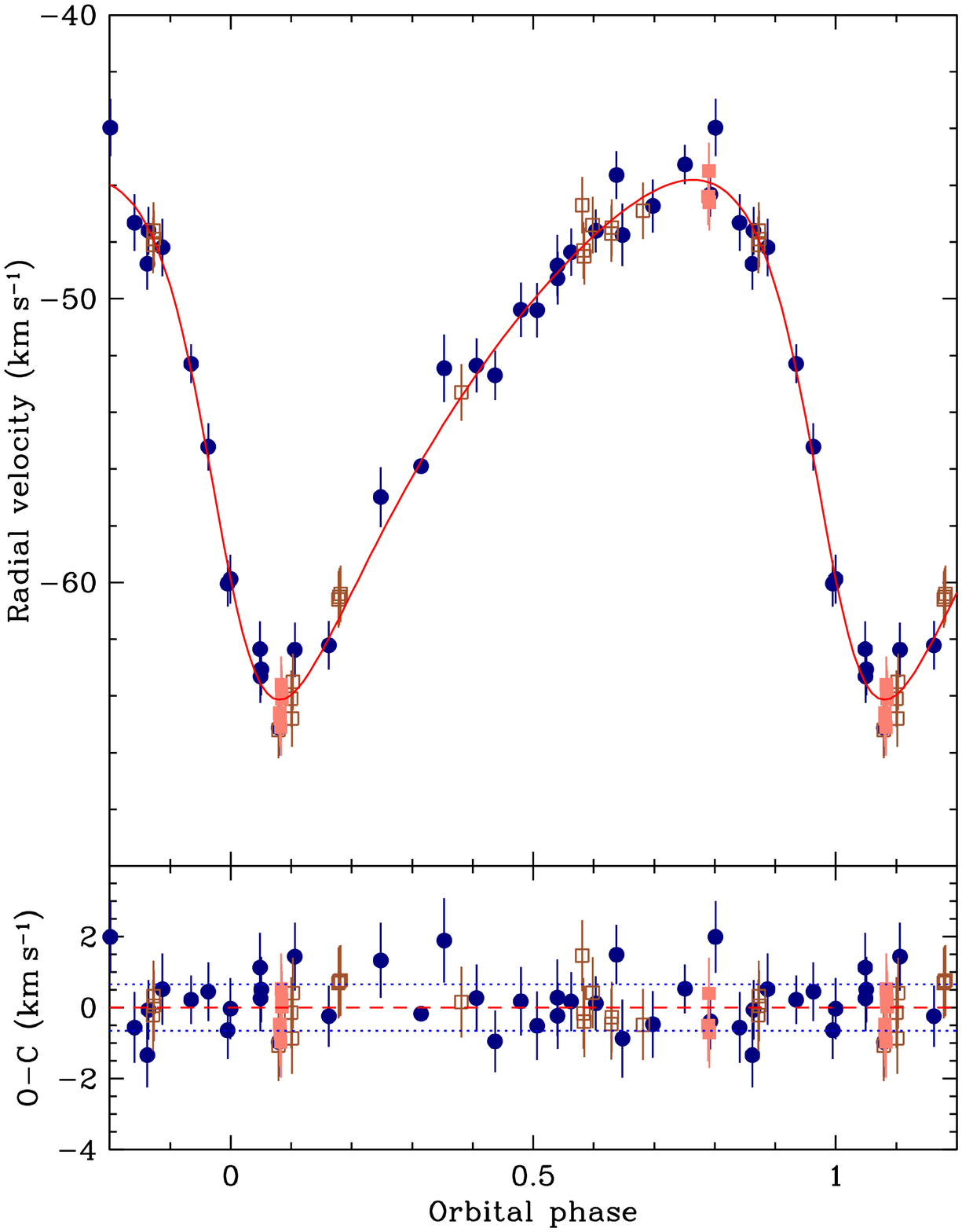}}
\caption{{\it Upper panel\/}: Our radial velocity measurements for
  HD~12288 are plotted 
  together with those of \citet{2002A&A...394..151C} against
  orbital phase. The  solid curve  
  corresponds to the orbital solution given in
  Table~\ref{tab:orbits}. The time $T_0$ of periastron passage is
  adopted as phase origin. {\it Bottom panel:\/} Plot of the
  differences ${\rm O}-{\rm C}$ between the observed values of the
  radial velocity and the predicted values computed from the orbital
  solution. The  dotted lines correspond to $\pm1$ rms
  deviation of the observational data about the orbital solution 
    (dashed line).  Dots represent the data from \citeauthor{2002A&A...394..151C}; all
  other symbols refer to our high-resolution spectra 
  obtained with various instrumental configurations, as indicated in
  Table~\ref{tab:plot_sym}.}  
\label{fig:hd12288_rv}
\end{figure}

\begin{figure}
\resizebox{\hsize}{!}{\includegraphics[angle=270]{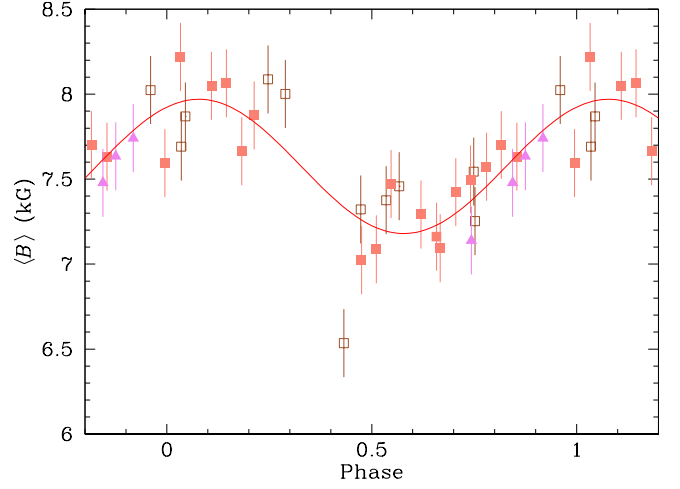}}
\caption{Mean magnetic field modulus of the star HD~14437,
against rotation phase. The symbols are as described at the
beginning of Appendix~\ref{sec:notes}.}
\label{fig:hd14437}
\end{figure}

\subsection{HD~12288}
\label{sec:hd12288}
The eight new $\Hm$ measurements presented in this paper, combined with
the data of \citetalias{1997A&AS..123..353M}, were used by \citet{2000A&A...355.1080W} to
obtain a refined value of the rotation period of HD~12288:
$\Prot=(34.9\pm0.2)$\,d. The corresponding phase diagram is shown in
Fig.~\ref{fig:hd12288}; the variation of the field modulus seems to
have a slightly anharmonic character, although it is not formally
significant (see Table~\ref{tab:mfit}). Its maximum coincides
roughly, but possibly not exactly, with the minimum of the
longitudinal field (i.e. its largest absolute value), according to the measurements of
\citet{2000A&A...355.1080W} 
of the latter. We could not obtain additional data on that
field moment, nor on the crossover or the quadratic field, owing to the
northern declination of the star. 

The variability of the radial velocity of HD~12288 was announced in
\citetalias{1997A&AS..123..353M}. Since then, \citet{2002A&A...394..151C} determined the
orbital elements of this binary from a separate set
of data, contamporaneous with ours. Here we combined both sets to
obtain the revised orbital solution presented in
Table~\ref{tab:orbits} and illustrated in
Fig.~\ref{fig:hd12288_rv}. It is fully consistent with that 
of \citeauthor{2002A&A...394..151C}, but significantly more precise. 

\begin{figure}[t]
\resizebox{\hsize}{!}{\includegraphics[angle=270]{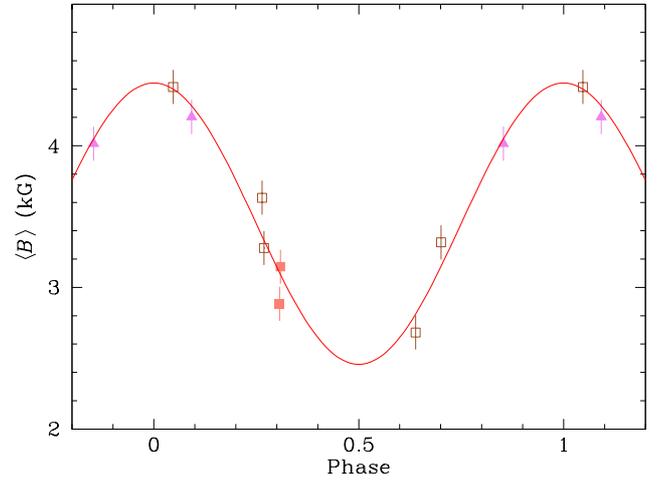}}
\caption{Mean magnetic field modulus of the star HD~18078,
against rotation phase. The symbols are as described at the
beginning of Appendix~\ref{sec:notes}.}
\label{fig:hd18078}
\end{figure}

\begin{figure}
\resizebox{\hsize}{!}{\includegraphics[angle=270]{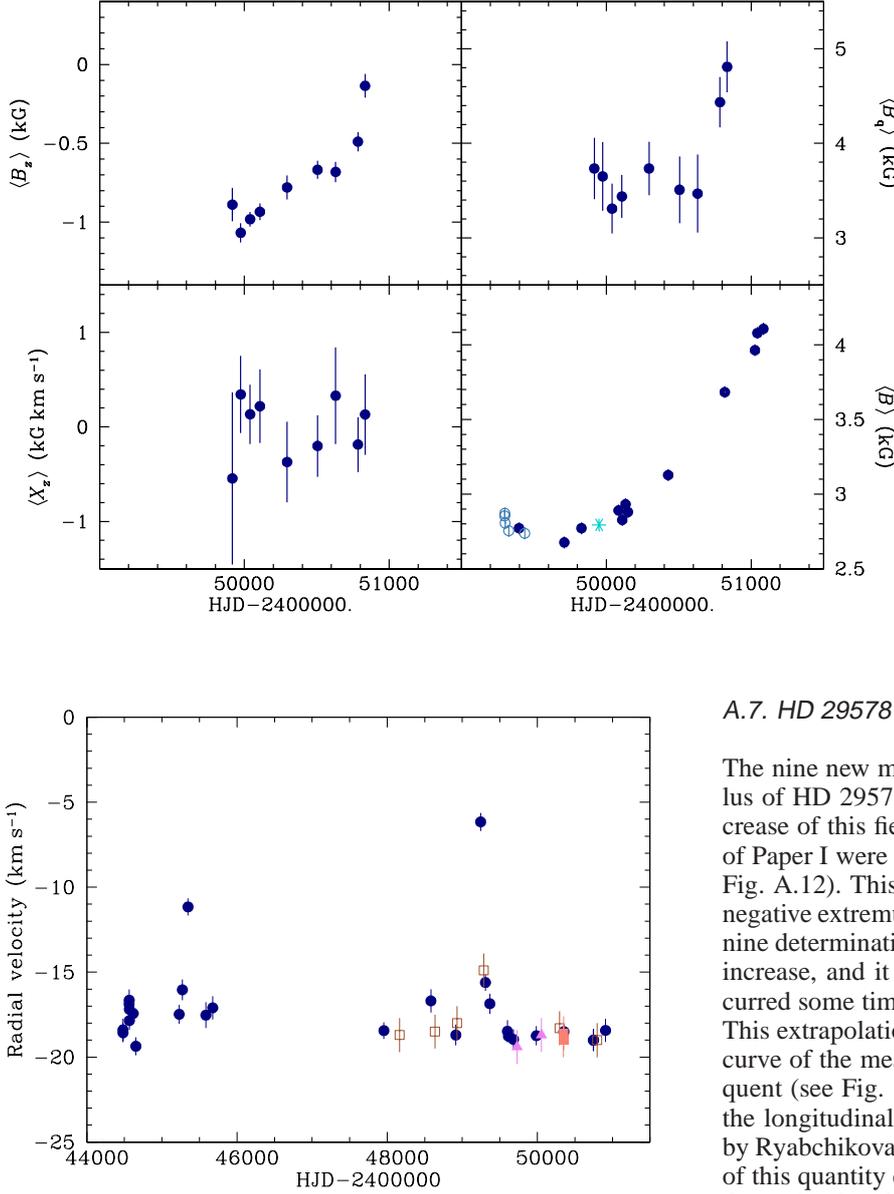}}
\caption{Our radial velocity measurements for HD~18078 are plotted
  together with those of \citeauthor{2002A&A...394..151C}
  (\citeyear{2002A&A...394..151C};  dots) 
  against heliocentric Julian date. The symbols are as described at the
beginning of Appendix~\ref{sec:notes}.}
\label{fig:hd18078_rv}
\end{figure}

\subsection{HD~14437}
\label{sec:hd14437}
Like for HD~12288, we could not obtain spectropolarimetric
observations of HD~14437 owing to its northern
declination. \citet{2000A&A...355.1080W}  published a set of mean
longitudinal magnetic 
field measurements of this star, from which they established the value
of its rotation period, $\Prot=26\fd87$. They also obtained a similar,
but less accurate value from analysis of the field modulus data of
\citetalias{1997A&AS..123..353M}, complemented by the 15 new measurements presented here. The
variation of $\Hm$ with this period shows no significant anharmonicity
(see Fig.~\ref{fig:hd14437}). 

\begin{figure*}[!ht]
\resizebox{12cm}{!}{\includegraphics[angle=270]{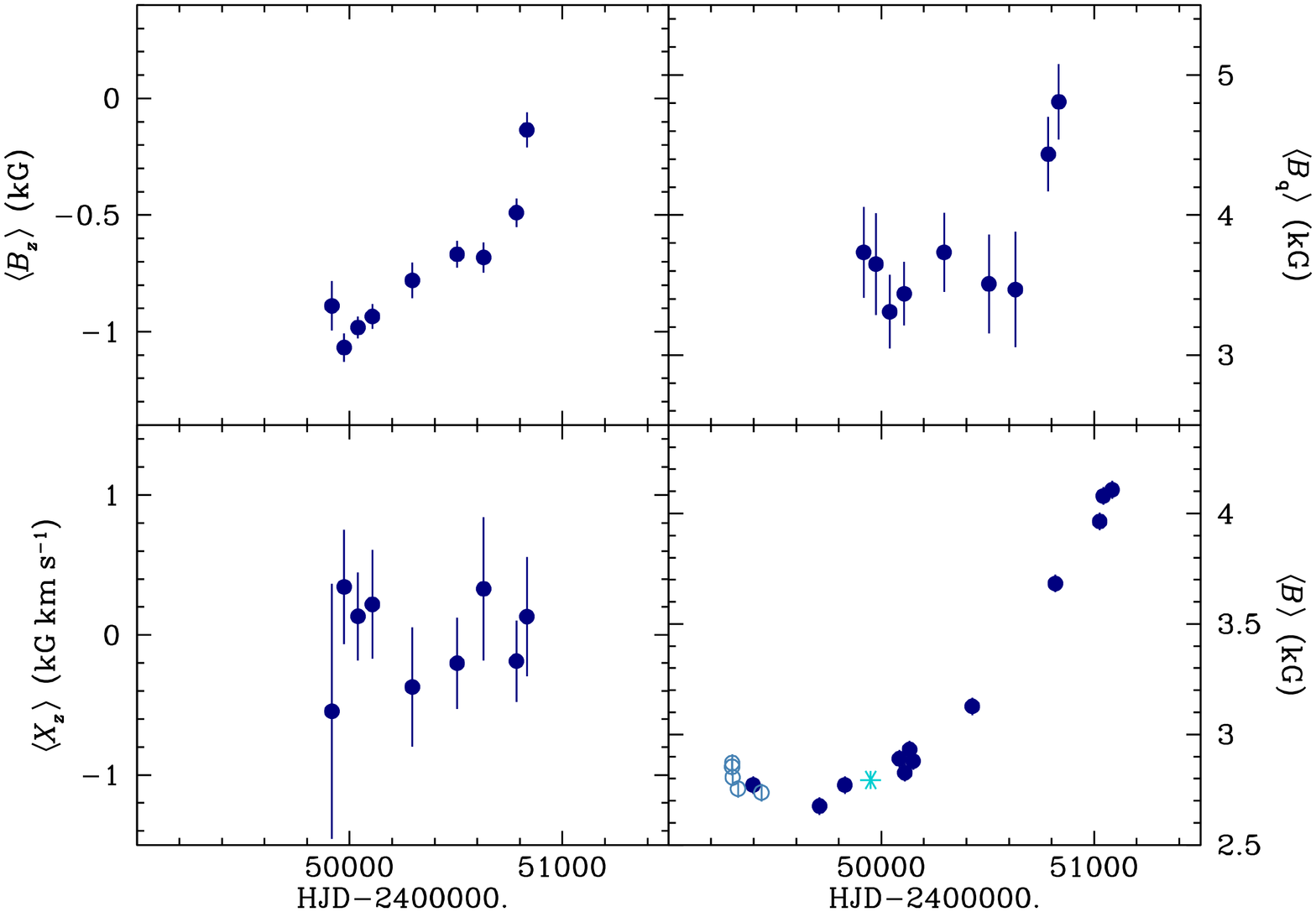}}
\parbox[t]{55mm}{
\caption{Mean longitudinal magnetic field ({\it top left\/}),
crossover ({\it bottom left\/}), 
mean quadratic magnetic field ({\it top right\/}),
and mean magnetic field modulus ({\it bottom right\/}) 
of the star HD~29578,
against heliocentric Julian date. The symbols are as described at the
beginning of Appendix~\ref{sec:notes}.}
\label{fig:hd29578}}
\end{figure*}

 The longitudinal field is negative throughout the rotation period. The
near coincidence of the phase at which it reaches its most negative
value with the phase of the minimum of the field modulus
is not the expected behaviour for a centred
dipole, from which the actual structure of
the magnetic field must significantly depart.

\subsection{HD~18078}
\label{sec:hd18078}
% hallk
The rotation period of HD~18078, $\Prot=1358$\,d, was recently derived
\citep{2016A&A...586A..85M}
by combining
our magnetic field modulus measurements (including five new
determinations that did not appear in
\citetalias{1997A&AS..123..353M}) with longitudinal field data
obtained at the Special Astrophysical Observatory. This makes it one
of only five Ap stars for which a period longer than 1000\,d has been
exactly determined. 

Like in HD~9996, the \ion{Fe}{ii}\,$\lambda\,6149$ line in HD~18078 is
magnetically resolved only over part of the rotation cycle, and the
resolution limit, close to 2.6\,kG, is considerably higher than we
would expect in the best cases ($\sim1.7$\,kG), owing to the fact that 
the split components are exceptionally broad. This indicates that the
geometrical structure of the magnetic field must significantly depart
from a simple dipole, which is a conclusion that is further strengthened by the
consideration of the variations of the field moments (see
Tables~\ref{tab:mfit} and \ref{tab:zfit}). The combination 
of an anharmonic $\Hz$ curve with a $\Hm$ curve that does not
significantly depart from a sinusoid (Fig.~\ref{fig:hd18078}) is
highly unusual, as is the 
large phase shift between their respective extrema. The very high
value of the ratio between the maxium and minimum of the mean field
modulus, $\sim1.8$, is another manifestation of considerable
departures from a dipolar structure. 

\citet{2002A&A...394..151C} infer that HD\,18078 may be a
spectroscopic binary with an orbital period of the order of 978~d and
a very excentric orbit, from the fact that 
two of their radial velocity measurements yield values much higher
than all others. Our data, which are contemporaneous with theirs,
lends some support to this suggestion. In particular, in
Fig.~\ref{fig:hd18078_rv} we note how our point
obtained on HJD~2449286 is higher than the others, which is consistent with the
\citeauthor{2002A&A...394..151C} measurements around the same epoch
indicating the 
probable presence of a narrow peak in the radial velocity curve.

\subsection{HD~29578}
\label{sec:hd29578}
% hall
The nine new measurements of the mean magnetic field modulus of HD~29578
presented in this paper show a monotonic increase of this field moment
and indicate that all the data points of \citetalias{1997A&AS..123..353M} were concentrated
around its phase of minimum (see Fig.~\ref{fig:hd29578}). This phase
likely coincides approximately 
with the negative extremum of the mean longitudinal magnetic
field. Our nine determinations of this field moment also show a monotonic
increase, and it is almost certain that a change of polarity occurred
some time after our last spectropolarimetric
observation. This extrapolation is borne out by consideration
  of the variation curve of the mean field modulus, assuming that, as
  is most frequent (see Fig.~\ref{fig:phi_zm}), its extrema roughly
  coincide with those of the longitudinal field curve. Indeed, later
measurements of $\Hm$ by \citet{2004A&A...423..705R} show that
the monotonic increase of this quantity continued for more
  than two years after our last
determination, leading these authors to suggest that the rotation
period should exceed 12 years. We do not necessarily concur with that
conclusion, which is based on the implicit assumption that the
ascending and descending slopes in the variation curve of the field
modulus are similar. Several counterexamples in the present work show
that this cannot be taken for granted. But in any case the period of
HD~29578 must be significantly longer than the seven year time span
between our first observation and the most recent
point of Ryabchikova et al. As also pointed out by Ryabchikova et al.,
the ratio between 
the highest value of $\Hm$ obtained so far and the minimum of this
field moment, of the order of 2.0 or greater, is exceptionally large. 

\begin{figure}
\resizebox{\hsize}{!}{\includegraphics{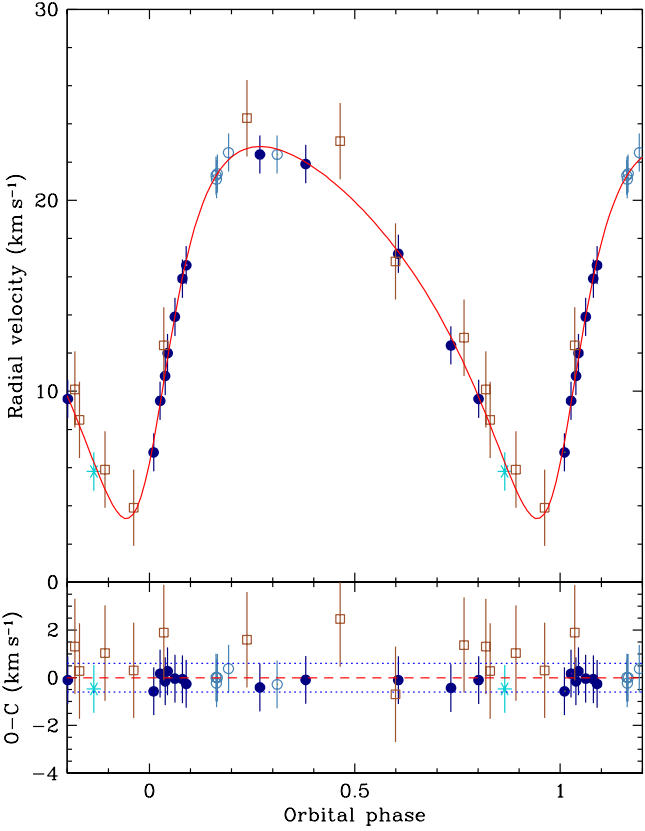}}
\caption{{\it Upper panel\/}: Our radial velocity measurements for
  HD~29578 are plotted against orbital phase. The  solid curve 
  corresponds to the orbital solution given in
  Table~\ref{tab:orbits}. The time $T_0$ of periastron passage is
  adopted as phase origin. {\it Bottom panel:\/} Plot of the
  differences ${\rm O}-{\rm C}$ between the observed values of the
  radial velocity and the predicted values computed from the orbital
  solution. The  dotted lines correspond to $\pm1$ rms
  deviation of the observational data about the orbital solution
    (dashed line).  Open squares represent our CASPEC
  observations; all other symbols refer to our high-resolution spectra
  obtained with various instrumental configurations, as indicated in
  Table~\ref{tab:plot_sym}.}  
\label{fig:hd29578_rv}
\end{figure}

\begin{figure}
\resizebox{\hsize}{!}{\includegraphics[angle=270]{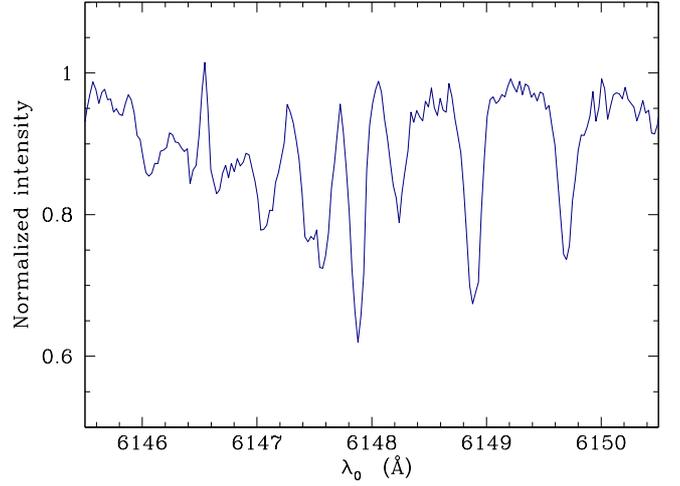}}
\caption{Portion of the spectrum of HD~47103 recorded on HJD~2450427.689, 
  showing the lines \ion{Cr}{ii}\,$\lambda\,6147.1$ (affected by a
  cosmic ray hit),
  \ion{Fe}{ii}\,$\lambda\,6147.7$, and
  \ion{Fe}{ii}\,$\lambda\,6149.2$.}
\label{fig:hd47103_6149}
\end{figure}

The quadratic field, after remaining approximately constant since our
first point, seems to show a sudden increase in our last two
measurements. We would discard this very unusual and surprising
behaviour as spurious, if a somewhat similar feature did not also seem to
be present for the longitudinal field. Indeed, the determinations of the
longitudinal field and the quadratic field are fundamentally
different (wavelength shift between lines recorded in opposite
circular polarisations against line broadening in natural light), so
that it is not very plausible that both are affected by the same
errors. If the effect is real, it suggests that the magnetic field of
HD~29578 must have a very peculiar structure. Yet the ratio of the
quadratic field to the field modulus is similar to that of most stars
considered here.

No crossover is detected, which is consistent with the expectation for a star
with such a long rotation period. The diagnosis of $\Hz$, $\xover$, and $\Hq$ is based on the analysis
of lines of \ion{Fe}{i}, which show no significant variation of
equivalent width with time. No radial velocity variation is detected
either. 

As announced in \citetalias{1997A&AS..123..353M}, HD~29578 is a
spectroscopic binary; its
orbital elements, which appear in Table~\ref{tab:orbits}, are
determined here for the first time. The fitted orbit is shown in
Fig.~\ref{fig:hd29578_rv}. The orbital period, though long
($\Porb=927$\,d) is much shorter than the rotation period of the Ap
primary.

\subsection{HD~47103}
\label{sec:hd47103}
\begin{figure*}
\resizebox{12cm}{!}{\includegraphics[angle=270]{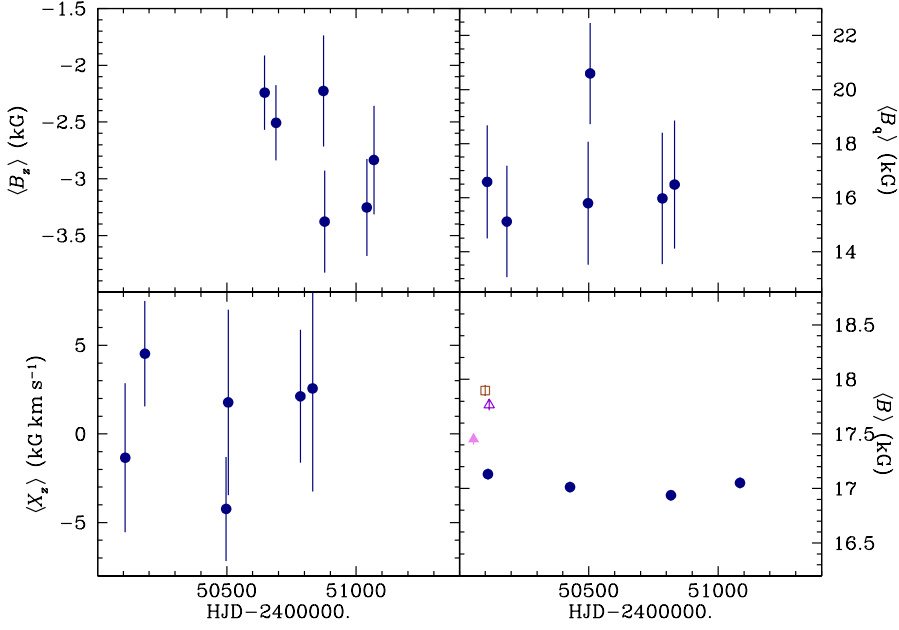}}
\parbox[t]{55mm}{
\caption{Mean longitudinal magnetic field ({\it top left\/}),
crossover ({\it bottom left\/}), 
mean quadratic magnetic field ({\it top right\/}),
and mean magnetic field modulus ({\it bottom right\/}) 
of the star HD~47103,
against heliocentric Julian date. The symbols are as described at the
beginning of Appendix~\ref{sec:notes}.}
\label{fig:hd47103}}
\end{figure*}

% hall
\citetalias{1997A&AS..123..353M} did not include any measurement of HD~47103, in which the
presence of resolved magnetically split lines had been
announced only shortly before publication \citep{1995A&A...303L...5B}. The region
of the spectrum around the line \ion{Fe}{ii}\,$\lambda\,6149.2$ is
shown in Fig.~\ref{fig:hd47103_6149}. Four of our
determinations of the mean magnetic field modulus are based on ESO-CES
spectra; the other three come from different instruments; these data
are plotted in Fig.~\ref{fig:hd47103}. The CES measurements are very
similar; their standard deviation is only 70\,G, which is very small
compared to the magnetic field strength ($\sim17$\,kG). The other
three values of $\Hm$ are several hundred Gauss greater. This
difference seems unlikely to reflect the occurrence of actual variations of the
stellar field, as this would imply that all four CES observations were
obtained at either of two phases at which the field modulus had the
same value (assuming a typical single-wave variation of this field
moment). The probability of such a chance coincidence is very
low. Thus we are inclined to suspect that the observed differences
between our measurements obtained with different instruments are caused
by systematic instrumental effects, whose nature cannot be further
assessed because of the small number of observations. 

The assumption that
the field modulus of HD~47103 remained constant over the time interval
covered by our observations (nearly three years), is further supported
by the fact that, within the limits of the uncertainties related to
the use of different instruments and a different diagnostic method,
our $\Hm$ data are also consistent with the seven measurements published
by \citet{1997A&A...325..195B}. The latter show no variations over a
time range of $\sim1.5$~years, partly overlapping with the time range
during which our observations were obtained. By contrast,
\citet{2008A&A...480..811R} measured $\Hm=16.3$\,kG from a single ESO
UVES 
spectrum obtained about three years after our latest observation. This
value of the field modulus is significantly smaller than those we
determined from our CES spectra, and barring systematic differences
between the two instruments, it may be suggestive of a slow
decrease of $\Hm$ over the considered time span, hence of a long period
(or the order of years). 

At present, we adopt 70\,G as the estimated uncertainty of our $\Hm$
measurements of HD~47103, bearing in mind the fact that, like for
HD~192678 \citepalias{1997A&AS..123..353M}, this value corresponds to
random measurement 
uncertainties, but ignores possible systematic instrumental errors. 

 HD~47103 has one of the strongest magnetic fields known. As a result,
even at moderate spectral resolutions, 
many of its spectral lines are resolved into their Zeeman components,
or at least are strongly broadened. This makes it difficult to find
lines sufficiently free from blends for magnetic field diagnosis from
the CASPEC spectra. Accordingly the set of lines used for this purpose
is smaller than for any other star studied in this paper but
HD~965. As a consequence, it was not possible to use the
multi-parameter fit described in Sect.~\ref{sec:moments} for determination of the
mean quadratic magnetic field. Instead the simpler, less accurate
form of the fit function given in Eq.~\ref{eq:Hqold} was 
adopted. This is not expected to have a major impact in the present
case, since the magnetic field is by far the dominant broadening
factor and the accuracy of its determination should not be strongly
affected by approximate treatment of the other contributions to the
line width. All the lines used for determination of $\Hz$, $\xover$,
and $\Hq$ are \ion{Fe}{i} lines, except for one \ion{Fe}{ii}
line. They show no significant equivalent width variations.

\begin{figure*}
\resizebox{12cm}{!}{\includegraphics[angle=270]{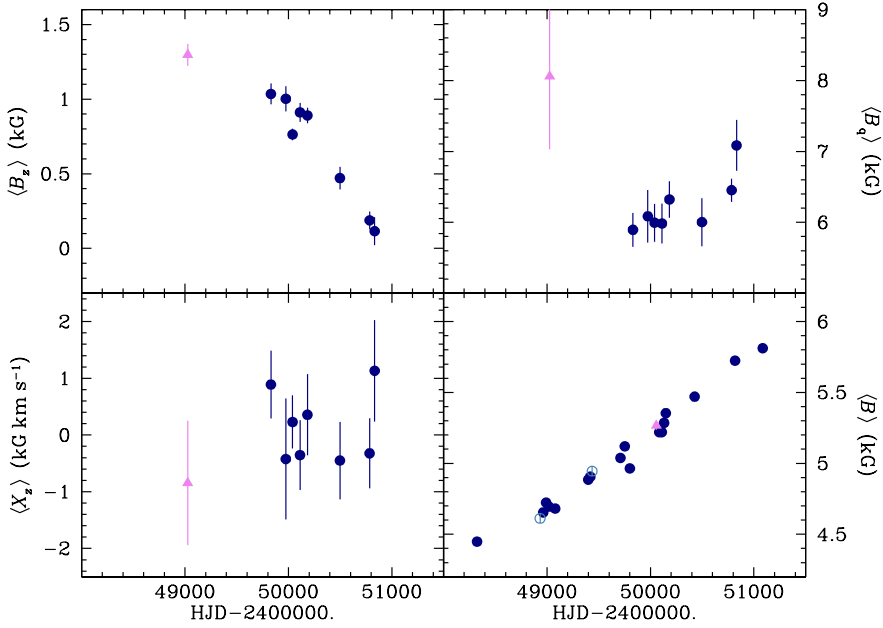}}
\parbox[t]{55mm}{
\caption{Mean longitudinal magnetic field ({\it top left\/}),
crossover ({\it bottom left\/}), 
mean quadratic magnetic field ({\it top right\/}),
and mean magnetic field modulus ({\it bottom right\/}) 
of the star HD~50169,
against heliocentric Julian date. The symbols are as described at the
beginning of Appendix~\ref{sec:notes}.}
\label{fig:hd50169}}
\end{figure*}

 Variation of the longitudinal field is not definitely detected in our data,
although there is some hint in Fig.~\ref{fig:hd47103} that it may have
undergone a slow, monotonic variation from less negative to more negative
values over the time span covered by our observations. However, this
apparent trend is probably only coincidental. It vanishes when
combining our data with eight additional $\Hz$ measurements obtained
by \citet{1997smf..proc..106E} over a period slightly
shorter than two months, which includes the date of our first CASPEC
observation of HD~47103. These measurements are more scattered than
ours. A period search combining both data sets remained inconclusive.

No definite
variation of the quadratic field is seen either. The ratio of 
$\Hq$ to $\Hm$ is marginally smaller than one. Formally, this is
physically meaningless and may at first sight suggest that the
quadratic field has been systematically underestimated in this
star. But taking the measurement uncertainties into account, the
discrepancy vanishes; considering the high value of the field modulus,
compared to the longitudinal field, it is not unexpected that $\Hq$
hardly differs from $\Hm$. On the other hand, from better
observational data, \citet{2006A&A...453..699M} obtained from analysis
of a set of \ion{Fe}{i} 
lines $\Hq=19.3\pm0.5$\,kG on HJD~2450115.615 (that is, one week after
the first measurement reported here). This value is more consistent
with the field modulus, and the difference between it and most
measurements of this paper, though somewhat marginal, is formally
significant. Thus we cannot fully rule out the possibility that the
$\Hq$ values reported here for this star are systematically
underestimated, plausibly because of the above-mentioned application
of the simplified Eq.~(\ref{eq:Hqold}) to derive them.

No crossover is detected, but the error bars are among the largest for
any star in this paper.

\begin{figure}[t]
\resizebox{\hsize}{!}{\includegraphics[angle=270]{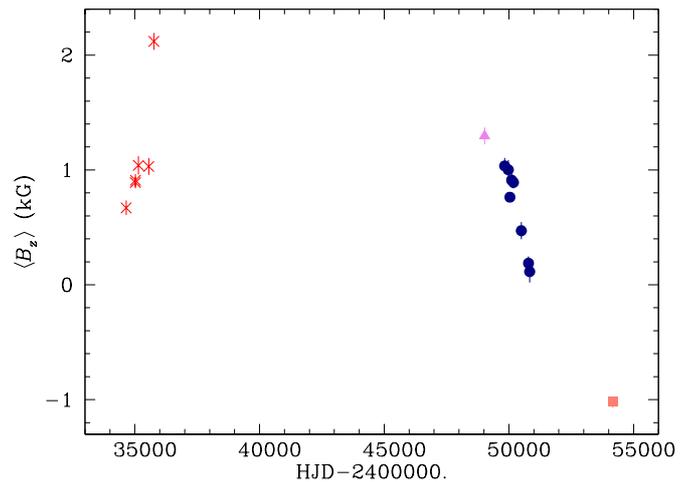}}
\caption{Mean longitudinal magnetic field of the star HD~50169,
against heliocentric Julian date.  Crosses represent 
the measurements of \citet{1958ApJS....3..141B}, and the filled square
is the point obtained by \cite{2014AstBu..69..427R}. The other symbols
are as described at the 
beginning of Appendix~\ref{sec:notes}.}
\label{fig:hd50169long}
\end{figure}

\begin{figure}[!ht]
\resizebox{\hsize}{!}{\includegraphics{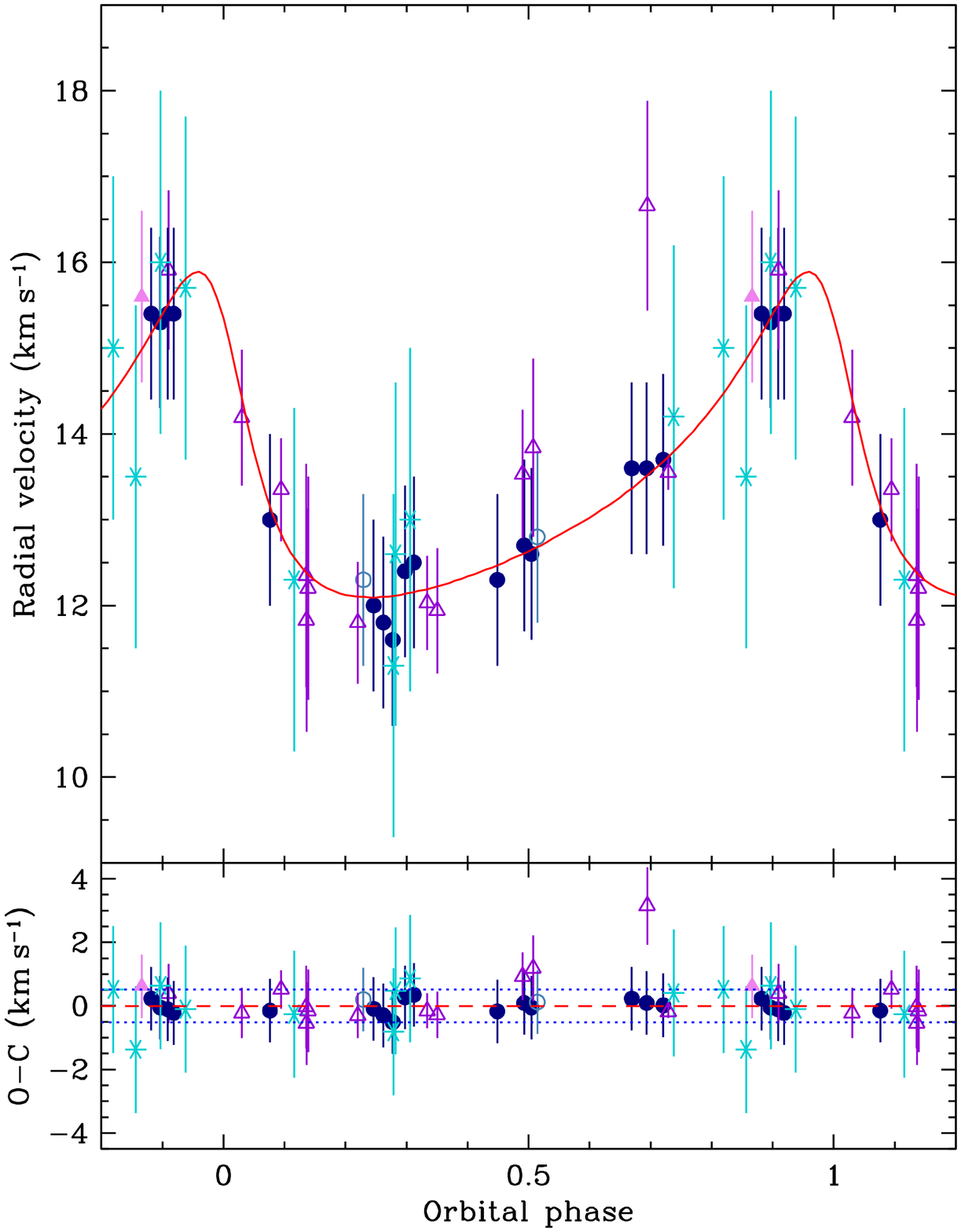}}
\caption{{\it Upper panel\/}: Our radial velocity measurements for
  HD~50169 are plotted against orbital phase. The  solid curve 
  corresponds to the orbital solution given in
  Table~\ref{tab:orbits}. The time $T_0$ of periastron passage is
  adopted as phase origin. {\it Bottom panel:\/} Plot of the
  differences ${\rm O}-{\rm C}$ between the observed values of the
  radial velocity and the predicted values computed from the orbital
  solution. The  dotted lines correspond to $\pm1$ rms
  deviation of the observational data about the orbital solution 
    (dashed line).  Open triangles represent data from \citeauthor{2002A&A...394..151C} and  asterisks our CASPEC
  observations; all other symbols refer to our high-resolution spectra
  obtained with various instrumental configurations, as indicated in
  Table~\ref{tab:plot_sym}.}  
\label{fig:hd50169_rv}
\end{figure}

\begin{figure}[!h]
\resizebox{\hsize}{!}{\includegraphics[angle=270]{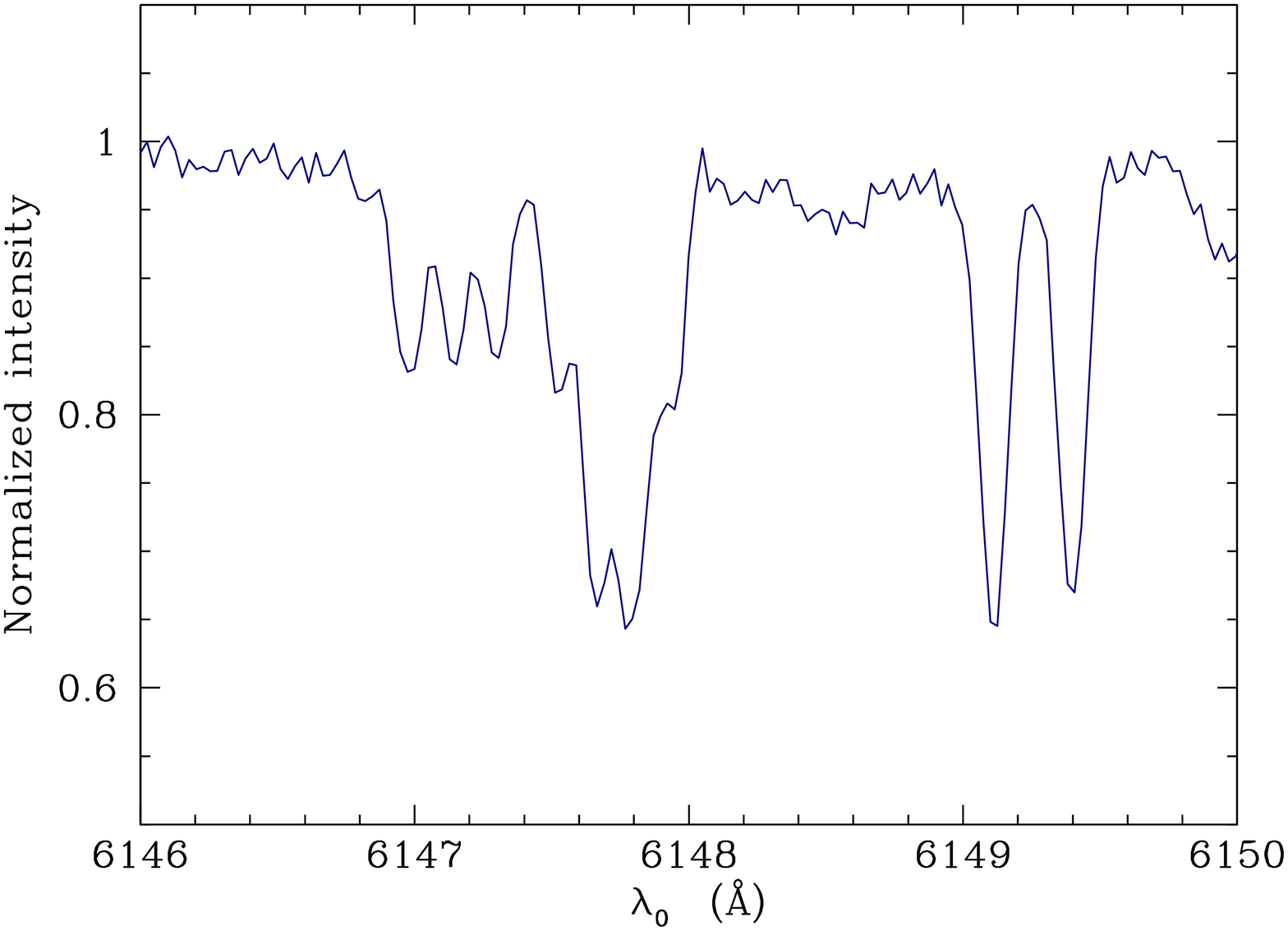}}
\caption{Portion of the spectrum of HD~51684 recorded on HJD~2450427.623, 
  showing the lines \ion{Cr}{ii}\,$\lambda\,6147.1$,
  \ion{Fe}{ii}\,$\lambda\,6147.7$, and
  \ion{Fe}{ii}\,$\lambda\,6149.2$.}
\label{fig:hd51684_6149}
\end{figure}

The lack of definite variations of any of the observables that we
considered implies that either HD~47103 has an extremely long period,
or that for this star, one at least of the angles $i$ (between the
rotation axis and the line of sight) or $\beta$ (between the magnetic
and rotation axes) is too close to zero for any detectable variability
to occur. Longer term follow-up of the star is needed to solve
this ambiguity. This is an especially important issue, as a very long
period, if confirmed, would make 
HD~47103 an exception to the conclusion that fields stronger than
7.5\,kG occur only in stars with rotation periods shorter than
150\,d (see Sect.~\ref{sec:Hmdisc}). 

\citet{1997A&A...325..195B} and \citet{2002A&A...394..151C}
report a possible variability of the radial velocity in HD~47103.
The scatter of our radial velocity measurements supports this
possibility, but the data currently available are insufficient to set
significant constraints on the orbital period.

\subsection{HD~50169}
\label{sec:hd50169}
% hall
The monotonic increase of the mean magnetic field modulus of HD~50169
noted in \citetalias{1997A&AS..123..353M}
continues throughout the time span covered by the new observations
presented here, indicating that the rotation period of the star must
be much longer than the 7.5 years separating our oldest determination
of this quantity from the most recent one (see Fig.~\ref{fig:hd50169}). 

Our measurements of the other field moments, combined with the single
one of \citet{1997A&AS..124..475M}, are spread over five years. During
this interval the longitudinal field showed a monotonic
decrease. There is some hint in both the $\Hm$ and $\Hz$ curves
that an extremum may have taken place not too long before our first
observations, suggesting that phases of minimum of the field modulus
and of positive extremum of the longitudinal field may roughly
coincide. The monotonic increase of $\Hz$ observed by 
\citet{1958ApJS....3..141B} over a three-year time span is approximately
mirror symmetric 
with the decrease that we recorded over a similar time range in our
observations (Fig.~\ref{fig:hd50169long}), but the more than 40 years
of separation between the epochs of the Babkock observations and ours
does not allow us to be sure 
that both pertain to the same rotation cycle. Actually it seems
somewhat more likely that at least one full rotation cycle has been
completed in between. This is consistent with the more recent
measurement of \citet{2014AstBu..69..427R}, also shown in  
Fig.~\ref{fig:hd50169long}, which is the only one until now to yield a
negative value of $\Hz$ and establishes that both poles of the
star come into view in alternation. 

Over the three years covered by the new spectropolarimetric CASPEC data
analysed here, the quadratic field shows a slow, but definite
monotonic increase, roughly following that of the field modulus. The
single measurement of \citet{1997A&AS..124..475M} has a 
large formal uncertainty, and its deviation from the general trend
more probably reflects its low quality rather than the actual
behaviour of the star. No significant crossover is detected.

The determinations of $\Hz$, $\xover,$ and $\Hq$ are based on the
analysis of a set of lines of \ion{Fe}{ii}, which show no equivalent
width variations.

The variability of the radial velocity of HD~50169 was reported for
the first time in \citetalias{1997A&AS..123..353M}. Combining our
radial velocity measurements with those of
\citet{2002A&A...394..151C}, we achieved the first determination of
its orbital parameters, which are presented in
Table~\ref{tab:orbits}. The fitted orbit is shown in
Fig.~\ref{fig:hd50169_rv}. That the orbit could be determined despite
the low  amplitude of the radial velocity variations of this star
(less than 4\,\kms\ peak-to-peak) and its long orbital period
(almost five years long -- still much shorter than the rotation period 
of the Ap primary) is testament to the precision and the stability of
our radial velocity 
determinations.

\subsection{HD~51684}
\label{sec:hd51684}
The presence of resolved magnetically split lines in HD~51684 was
discovered as part of this project after \citetalias{1997A&AS..123..353M} was written, and it
is reported here for the first time. We are not aware of any attempt
to study the magnetic field of this star prior to this discovery. A
portion of a high-resolution spectrum around the line
\ion{Fe}{ii}\,$\lambda\,6149.2$ is shown in
Fig.~\ref{fig:hd51684_6149}. The sharpness of the split components of
this line and the fact that the spectrum goes back almost all the way
up to the continuum between them reflect the narrow spread of the
magnetic field strengths over the stellar disk. 

\begin{figure*}
\resizebox{12cm}{!}{\includegraphics[angle=270]{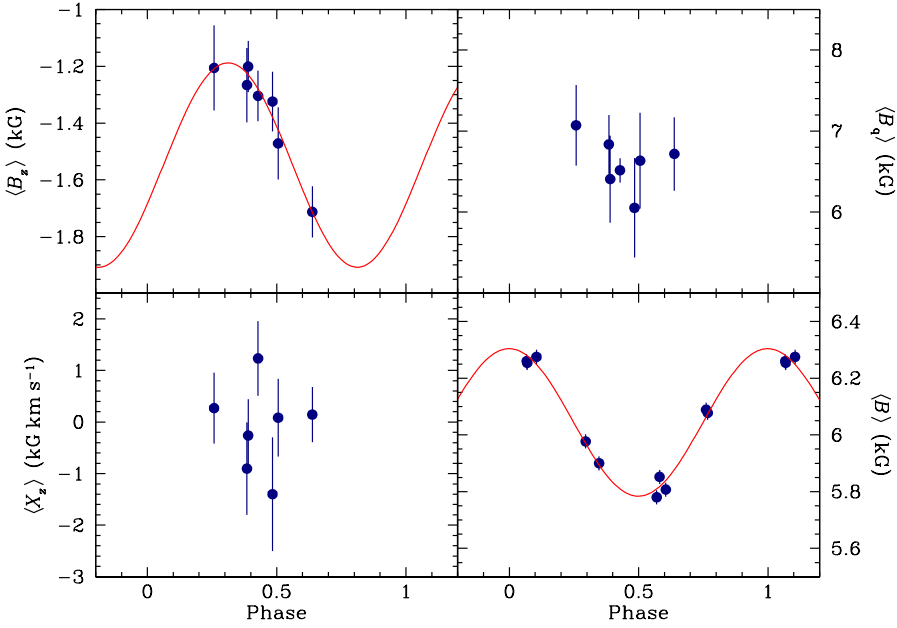}}
\parbox[t]{55mm}{
\caption{Mean longitudinal magnetic field ({\it top left\/}),
crossover ({\it bottom left\/}), 
mean quadratic magnetic field ({\it top right\/}),
and mean magnetic field modulus ({\it bottom right\/}) 
of the star HD~51684, against rotation phase. The symbols are as described at the
beginning of Appendix~\ref{sec:notes}.} 
\label{fig:hd51684}}
\end{figure*}

Only the mean magnetic field modulus measurements lend themselves to
determination of the rotation period, for which the value
$\Prot=(371\pm6)$\,d was obtained. Our data are plotted against this
period in Fig.~\ref{fig:hd51684}. Their standard deviation about the
best-fit curve of the form given in Eq.~(\ref{eq:fit1}) is
25\,G. We adopted this value for the uncertainties of
our $\Hm$ measurements of this star.

Because the rotation period of HD~51684 is
unfortunately close to one~year, the distribution across the rotation
cycle of our spectropolarimetric observations, which are less numerous
than our high-resolution spectra in natural light, is very
uneven. Nevertheless, the longitudinal field data unambiguously rule
out a much shorter period, of the order of 10\fd5, which looks
plausible from consideration of the $\Hm$ data alone. While both $\Hm$
and $\Hz$ undergo definite variations, the shape of their variation
curves is not strongly constrained (especially in the case of $\Hz$)
by the available data. But the existence of a considerable phase shift
between their respective extrema seems inescapable since the variation of
$\Hz$ is very steep around the phase of minimum of $\Hm$.

Out-of-phase variations of the longitudinal field and the field
modulus represent an
unusual feature among the stars studied here, which is suggestive of a
magnetic field structure significantly different from a simple
dipole. This is rather surprising, since sizeable departures from a
centred dipole tend to spread the distribution of the field strengths
across the stellar surface, which manifests itself observationally by
a broadening of the resolved line components, such as observed in
for example HD~9996 (see Fig.~\ref{fig:hd9996_6149}) or HD~18078 \citep[see
Fig.~1 of][]{2016A&A...586A..85M}. As explained above, this effect is
definitely not apparent in HD~51684. Further investigation is
warranted to identify a magnetic field structure that can account for
both the phase shift between the variations of $\Hm$ and $\Hz$ and the
very clean profiles of the resolved split lines. 

On the other hand, it appears unlikely that $\Hz$ reverses
its sign over the rotation cycle. If the field of HD~51684 bears any
resemblance to a dipole, it is probable that only its negative pole is
observed. The quadratic field shows no definite variation over the covered phase
range, and no significant detection of the crossover is achieved,
which is consistent with the length of the rotation period.

The determination of $\Hz$, $\xover,$ and $\Hq$ is based on the
analysis of a mixed set of \ion{Fe}{i} and \ion{Fe}{ii} lines. These
lines are found to undergo equivalent width variations with an
amplitude of the order of 10\%; the lines appear weaker close to
phase 0.5, that is, close to the phase of minimum of the field modulus.
The exact phases of extrema are somewhat uncertain, however, owing to
the poor sampling of the rotation cycle by our
observations. The variability of the equivalent widths means that the
distribution of Fe on the surface of HD~51684 is less uniform than in
most Ap stars. As the amplitude of the variations is
rather small, the inhomogeneities are not extreme, and their impact on
the derived values of the field moments should be limited. Nevertheless
one should keep in mind in the interpretation of those moments that they
represent the convolution of the actual structure of the magnetic
field with the distribution of Fe over the stellar surface.

\begin{figure*}
\resizebox{12cm}{!}{\includegraphics[angle=270]{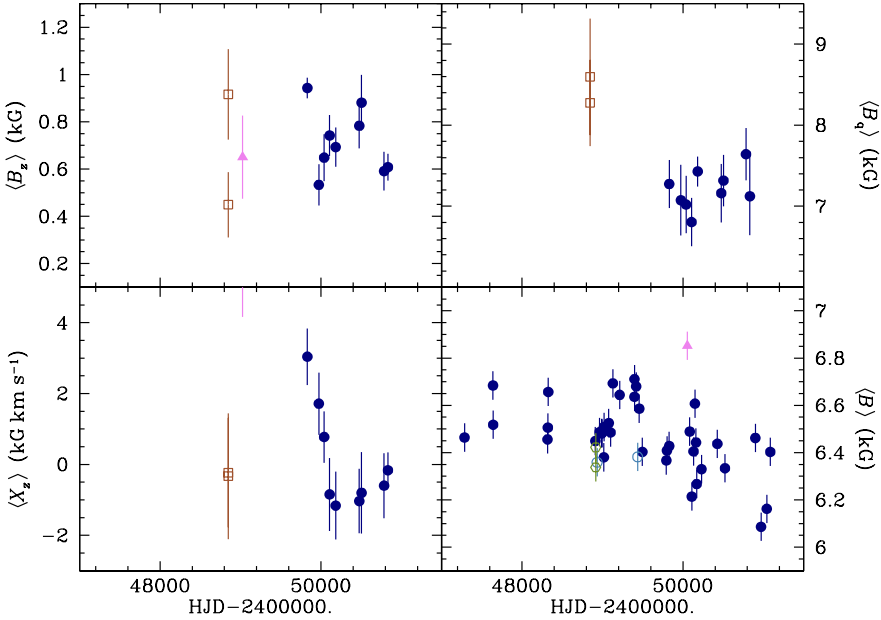}}
\parbox[t]{55mm}{
\caption{Mean longitudinal magnetic field ({\it top left\/}),
crossover ({\it bottom left\/}), 
mean quadratic magnetic field ({\it top right\/}),
and mean magnetic field modulus ({\it bottom right\/}) 
of the star HD~55719,
against heliocentric Julian date. The symbols are as described at the
beginning of Appendix~\ref{sec:notes}.}
\label{fig:hd55719}}
\end{figure*}

\subsection{HD~55719}
\label{sec:hd55719}
% hall
The 14 new determinations of the mean magnetic field modulus of
HD~55719 presented here definitely rule out the tentative values of
the period proposed in \citetalias{1997A&AS..123..353M} (847\,d or 775\,d). Instead, one can
see in Fig.~\ref{fig:hd55719} that a very slow decrease of $\Hm$ is
becoming apparent. The unexpectedly large scatter of the data points
about this long-term trend had previously misled us into believing
that we had observed shorter timescale actual variations, but it now
clearly appears that the rotation period of the star must vastly
exceed the time span covered by our data, which is
$\sim3800$~days. The origin of the 
scatter of the measurements about this slow variation is
unclear. The alternative of a very short period, for example close to
  1 day, has been ruled out in \citetalias{1997A&AS..123..353M}. The
star is bright, so that virtually all spectra that we have recorded
have signal-to-noise ratios amongst the highest of all spectra
analysed in this project. The blend of the blue component of
\ion{Fe}{ii}\,$\lambda\,6149$ by an unidentified REE line, which
complicates its measurement in many of the studied stars, increasing
its uncertainty, has not been seen over the part of the rotation 
cycle observed so far in HD~55719. The peculiar shape of the split
components of \ion{Fe}{ii}\,$\lambda\,6149$, attributed in \citetalias{1997A&AS..123..353M}
to an unusual structure of the magnetic field, does not introduce
particular complications in the measurement of their wavelengths,
since they are very well separated (see Fig.~2 of \citetalias{1997A&AS..123..353M}). As already
discussed at length in previous papers
\citep{1990A&A...232..151M,1992A&A...256..169M}, while the star is a
double-lined 
spectroscopic binary (SB2), no feature of the secondary is
readily seen in the region around 6150\,\AA. Admittedly, some
contamination of the $\Hm$ measurements by contribution of the
secondary that escapes visual inspection might be present. Considering the very slow variation of the primary, however, we would then
expect the orbital period to come out in a frequency analysis of the
$\Hm$ data, but it definitely does not. Furthermore, the quality of the
radial velocity data derived from the same measurements of the
\ion{Fe}{ii}\,$\lambda\,6149$ line performed for determination of the
field modulus (see below) does not suggest that they are in the least
polluted by the secondary spectrum. In summary, for now we do not know
what causes the large scatter of the $\Hm$ data about their long-term
trend. Further investigation is needed to identify this
  cause. But whatever it is, we are confident that it does not
  question the conclusion that the rotation period of the star must be
  longer than a decade.

Like the mean magnetic field modulus, the longitudinal field and
crossover (mostly null) data show somewhat more scatter than would be
expected from their estimated uncertainty. However, for these latter
two field moments, a plausible explanation exists. Namely, the set of
diagnostic lines (of \ion{Fe}{ii}) used for their determination
contains a fraction of 
lines from regions of the spectrum affected by telluric water vapour
absorption lines that is considerably higher than for most other stars
of the analysed sample. Although the generally very dry conditions on
La Silla make these telluric lines much less disruptive than in most
other observatory sites, they are variable with the time and with the
airmass of the observations, and they can have some effect on the
measurements, especially on the most humid nights. These effects, from
the point of view of the stellar properties, come in as essentially
random noise. This interpretation is all the more plausible since
we observed HD~55719 through clouds more often than most other stars
because of its brightness. 
However, it is not entirely supported by
consideration of the mean quadratic magnetic field, which appears
constant, with little scatter, for the measurements of this paper. The
marginal discrepancy of the older two data points of
\citet{1997A&AS..124..475M} can most probably, as in other cases, be
attributed to 
the shortcomings of the quadratic field measurements of this reference.
The equivalent widths of the lines used for determination of the field
moments from the CASPEC spectra also show some scatter, which is plausibly
related to the scatter in the magnetic field data.
 
\begin{figure}
\resizebox{\hsize}{!}{\includegraphics{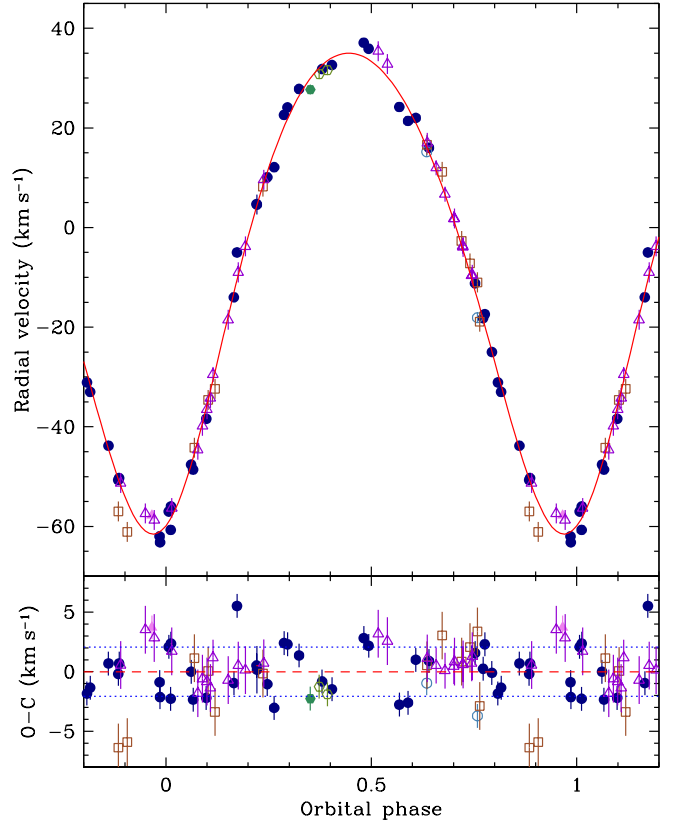}}
\caption{{\it Upper panel\/}: Our radial velocity measurements for
  HD~55719 are plotted 
  together with those of \citet{1976ApJ...209..160B}  against orbital
  phase. The  solid curve  
  corresponds to the orbital solution given in
  Table~\ref{tab:orbits}. The time $T_0$ of periastron passage is
  adopted as phase origin. {\it Bottom panel:\/} Plot of the
  differences ${\rm O}-{\rm C}$ between the observed values of the
  radial velocity and the predicted values computed from the orbital
  solution. The  dotted lines correspond to $\pm1$ rms
  deviation of the observational data about the orbital solution 
    (dashed line).  Open triangles represent data from Bonsack and  open squares our CASPEC
  observations; all other symbols refer to our high-resolution spectra
  obtained with various instrumental configurations, as indicated in
  Table~\ref{tab:plot_sym}.}  
\label{fig:hd55719_rv}
\end{figure}

We combined our radial velocity measurements 
to those of \citet{1976ApJ...209..160B} to refine the orbital elements
originally derived by this author
(see Table~\ref{tab:orbits} and Fig.~\ref{fig:hd55719_rv}). The orbital period,
$\Porb=46\fd31810$, is much shorter than the
rotation period of the Ap component.

%\clearpage

% \clearpage

%\clearpage

\subsection{HD~59435}
\label{sec:hd59435}
\begin{figure}
\resizebox{\hsize}{!}{\includegraphics[angle=270]{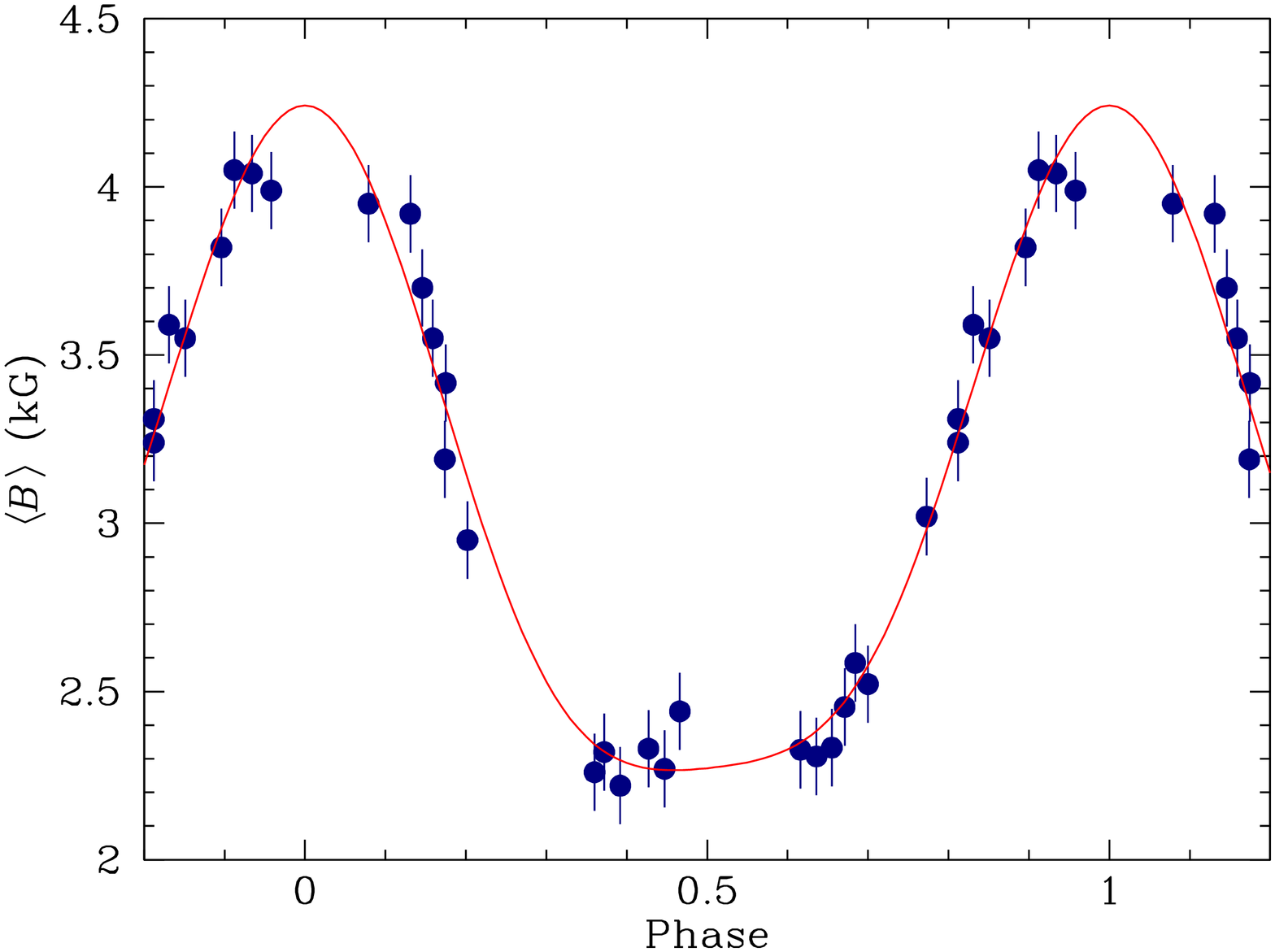}}
\caption{Mean magnetic field modulus of the star HD~59435,
against rotation phase. The symbols are as described at the
beginning of Appendix~\ref{sec:notes}.}
\label{fig:hd59435}
\end{figure}

A complete study of HD~59435, a SB2 with an orbital period
$\Porb=1386\fd1$, has been published elsewhere
\citep{1999A&A...347..164W}. It includes the new field modulus
measurements 
obtained since \citetalias{1997A&AS..123..353M}, which are repeated here for completeness and
for comparison with the other studied stars. The
rotation period of the Ap component, $\Prot=1360$\,d, is only slightly
shorter than the orbital period; Wade et al. argue that this
similarity is coincidental. The curve of variation of $\Hm$
(Fig.~\ref{fig:hd59435}) is
definitely anharmonic, with a fairly broad, almost flat minimum,
that is reminiscent of the behaviour observed in other stars, such as HD~965
(see Appendix~\ref{sec:hd965}) or HD~187474 (see
Appendix~\ref{sec:hd187474}); the maximum also shows some hint of
flattening, albeit less pronounced. It is also worth noting that even though
the field modulus was only of the order of 2.3\,kG around minimum, the
components of the Zeeman doublet \ion{Fe}{ii}\,$\lambda\,6149$ were
clearly resolved and their separation could be measured without major
uncertainty. This clearly demonstrates the feasibility of the
determination of mean field moduli of that order or even somewhat
smaller. 

 Based on our previous experience of the intricacies of untangling the
contributions of the two components of the binary in our high-resolution spectra recorded in natural light
\citep{1996A&A...314..491W}, we gave up on attempting  
to observe the star in spectropolarimetry, so as to save the available
telescope time for
objects for which we were more confident that it would actually be
possible to exploit the data. 

\begin{figure*}
\resizebox{12cm}{!}{\includegraphics[angle=270]{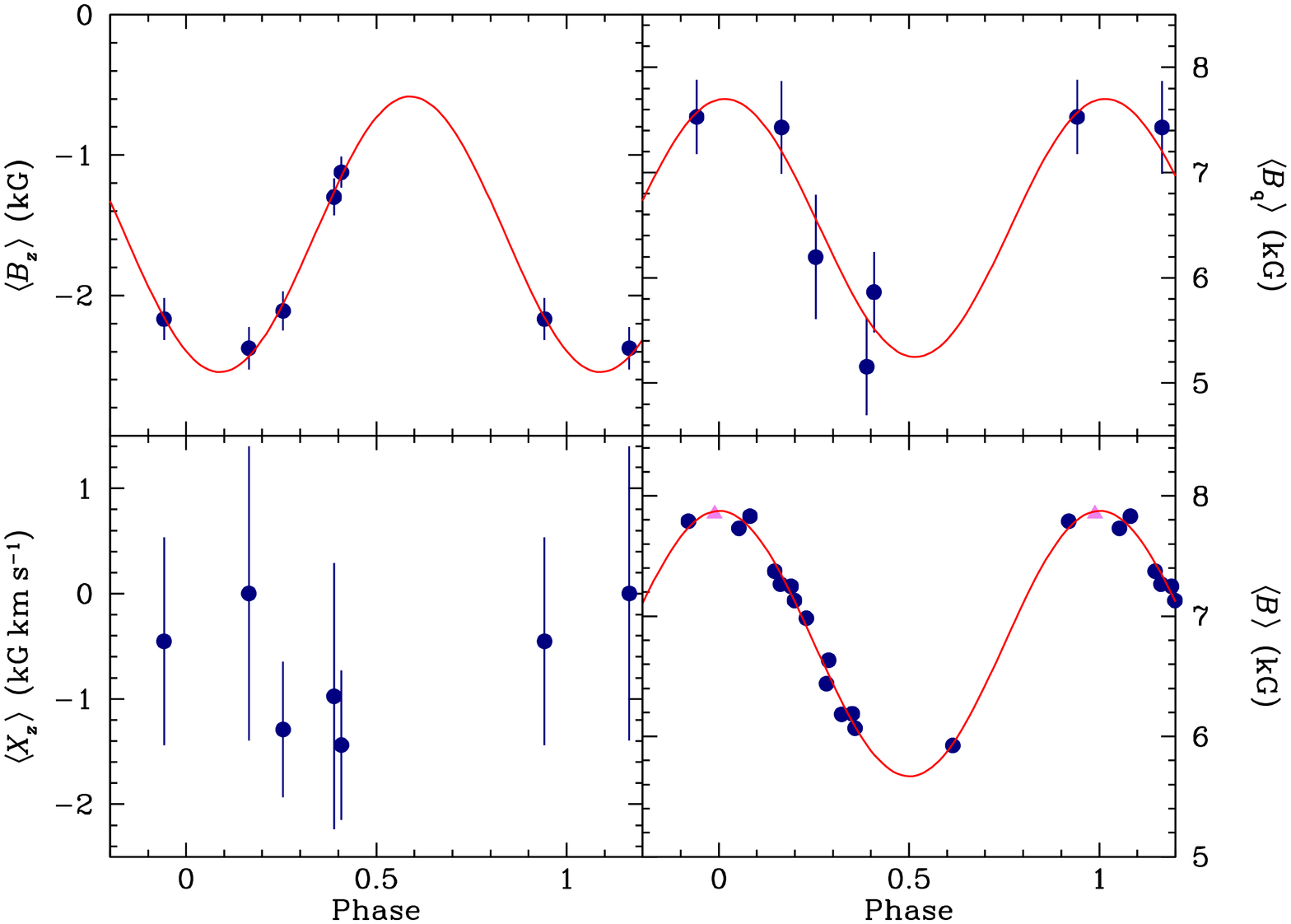}}
\parbox[t]{55mm}{
\caption{Mean longitudinal magnetic field ({\it top left\/}),
crossover ({\it bottom left\/}), 
mean quadratic magnetic field ({\it top right\/}),
and mean magnetic field modulus ({\it bottom right\/}) 
of the star HD~61468, against rotation phase. The symbols are as described at the
beginning of Appendix~\ref{sec:notes}.} 
\label{fig:hd61468}}
\end{figure*}

\subsection{HD~61468}
\label{sec:hd61468}

 The 11 new determinations of the mean magnetic field modulus presented
here, combined with the four data points of \citetalias{1997A&AS..123..353M}, unambiguously define
the rotation period of HD~61468: $\Prot=(322\pm3)$\,d. This period
also matches the variations of the longitudinal
field and the quadratic field (see Fig.~\ref{fig:hd61468}). The
variations of all three field moments are well represented by cosine
waves, but the variation curves are only loosely constrained for $\Hz$
and $\Hq$ because of the small number of measurements
obtained. Accordingly, the phase shift between the extrema of $\Hm$
and $\Hq$, on the one hand, and of $\Hz$, on the other hand, cannot be
regarded as significant. To the precision achieved, the largest
negative value of the longitudinal field coincides with the
maxima of the quadratic field and field modulus. The
longitudinal field probably does not reverse its polarity as the star
rotates, but this cannot be definitely ascertained because of incomplete
phase coverage of the observations. The ratio between the extrema of
$\Hm$, close to 1.4, exceeds the limit value for a centred dipole,
1.25, from which its structure must show some
departure. The quadratic field also shows an unusually large amplitude
of variation. The latter is certainly real, even though, in absolute
terms, the fact that $\Hq$ is systematically marginally smaller than
$\Hm$ is non-physical and suggests that it may be somewhat
underestimated. However, taking their uncertainties into account, the
difference between the two field moments is not significant.

 As can be expected from the long rotation period, no significant
crossover is detected: no derived value exceeds $2\sigma$. But it is
somewhat intriguing that all measurements (but one, which is null)
yield negative values. 

The equivalent widths of the \ion{Fe}{ii} lines, which are used to
determine $\Hz$, $\xover$, and $\Hq$, vary with a full amplitude of
the order of 15\%. The lines appear to be stronger close to the phase of
maximum of the field modulus and weaker around its minimum. This may
have a small effect on the measured field moments, which should be
kept in mind in their interpretation.

\begin{figure}
\resizebox{\hsize}{!}{\includegraphics{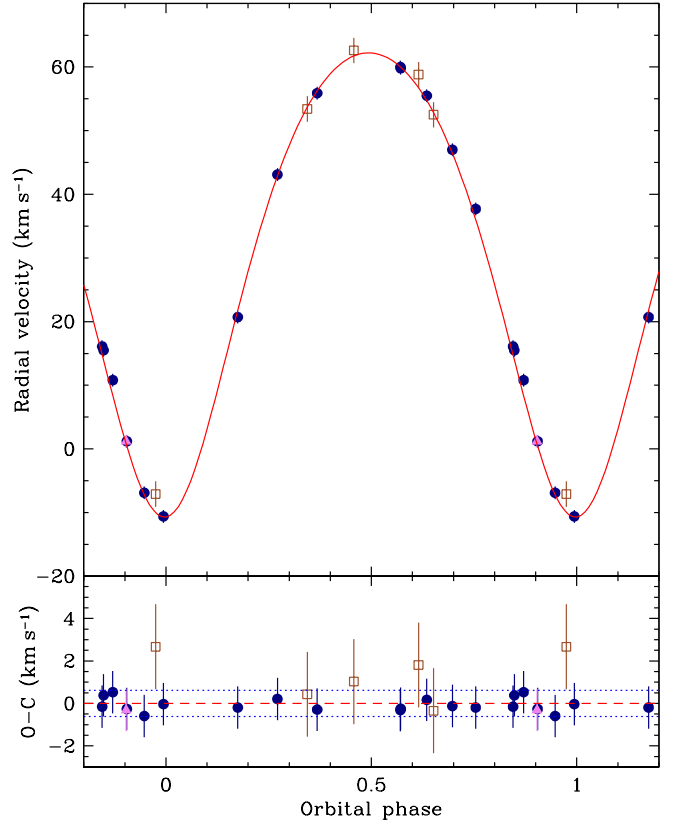}}
\caption{{\it Upper panel\/}: Our radial velocity measurements for
  HD~61468 are plotted against orbital phase. The  solid curve  
  corresponds to the orbital solution given in
  Table~\ref{tab:orbits}. The time $T_0$ of periastron passage is
  adopted as phase origin. {\it Bottom panel:\/} Plot of the
  differences ${\rm O}-{\rm C}$ between the observed values of the
  radial velocity and the predicted values computed from the orbital
  solution. The  dotted lines correspond to $\pm1$ rms
  deviation of the observational data about the orbital solution 
    (dashed line).  Open squares represent our CASPEC
  observations; all other symbols refer to our high-resolution spectra
  obtained with various instrumental configurations, as indicated in
  Table~\ref{tab:plot_sym}.}  
\label{fig:hd61468_rv}
\end{figure}

The variability of the radial velocity of the star was pointed out in
\citetalias{1997A&AS..123..353M}. Here we determine an orbital
solution for the first time; the parameters of this orbital solution are given in
Table~\ref{tab:orbits} and the resulting fit appears in
Fig.~\ref{fig:hd61468_rv}. At 27\fd3, the orbital period of HD~61468,
which is considerably shorter than the rotation period of its Ap component,
is the shortest one for the known binaries containing an Ap star
with resolved magnetically split lines. It is very remarkable 
that none of these binaries have shorter $\Porb$.

% \clearpage

\subsection{HD~65339}
\label{sec:hd65339}
HD\,65339 is one of the Ap stars whose magnetic field has been best
studied. In particular, the \citet{2004A&A...414..613K}
model of its 
field is one of the most detailed and precise models of an Ap star magnetic
field obtained so far. The primary reason why the star is included in
the present study is to allow its consideration in the statistical
discussion of the properties of the Ap stars with resolved
magnetically split lines by characterising it with data of the same
nature as for the other members of the class. 

\begin{figure}
\resizebox{\hsize}{!}{\includegraphics[angle=270]{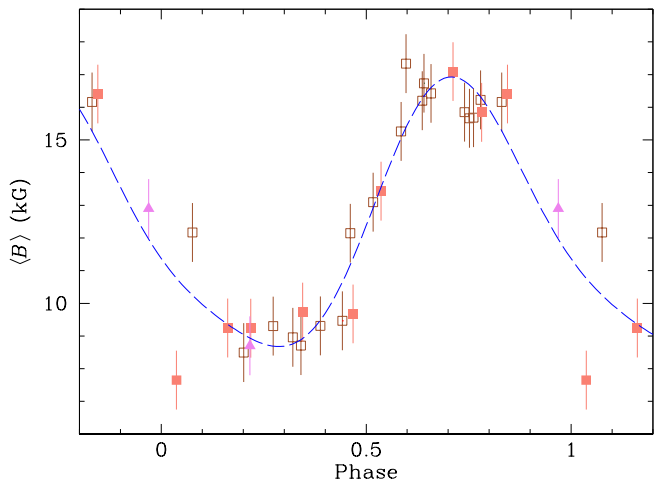}}
\caption{Mean magnetic field modulus of the star HD~65339,
against rotation phase. The symbols are as described at the
beginning of Appendix~\ref{sec:notes}.}
\label{fig:hd65339}
\end{figure}

The 14 new measurements of the mean magnetic field modulus reported
here are plotted together with the data of \citetalias{1997A&AS..123..353M} in
Fig.~\ref{fig:hd65339}; the phases are based on the 
very accurate value of the rotation period of
\citet{1998MNRAS.297..236H}. To the eye, the variations are strongly
anharmonic, which is not 
surprising considering the complexity of the field structure revealed
by the \citet{2004A&A...414..613K}
analysis. The variation curve shown
in the figure is the best fit by the
superposition of a cosine wave and of its first harmonic, but the
harmonic term is not formally significant (see
Table~\ref{tab:mfit}). With the complex structure of the field, such a
superposition is not likely 
to provide a good fit to the data. But we did not try to improve on
it, for example by adding higher harmonics, because the points
on the descending branch of the variation curve are scarce and the $\Hm$
determinations 
involve unusually large uncertainties, owing to the distortion, strong
blending, and marginal resolution of
the \ion{Fe}{ii}\,$\lambda\,6149$ line over part of the rotation cycle
 (see \citetalias{1997A&AS..123..353M}). Actually
considering the 
difficulty of the measurements, the fairly small dispersion of the
data points about a smooth variation curve is rather surprising.

The northern declination of the star did not allow us to obtain
spectropolarimetric observations, but there are many published
longitudinal field measurements of HD~65339; see
\citet{2012AN....333...41K} and references therein.  
 
\citet{2002A&A...394..151C} determined the orbital parameters of
HD~65339 by fitting radial velocity data, speckle
measurements, and the Hipparcos parallax of the star simultaneously. Our radial
velocity measurements do not add anything to their solution.

\subsection{HD~70331}
\label{sec:hd70331}
\begin{figure*}
\resizebox{12cm}{!}{\includegraphics[angle=270]{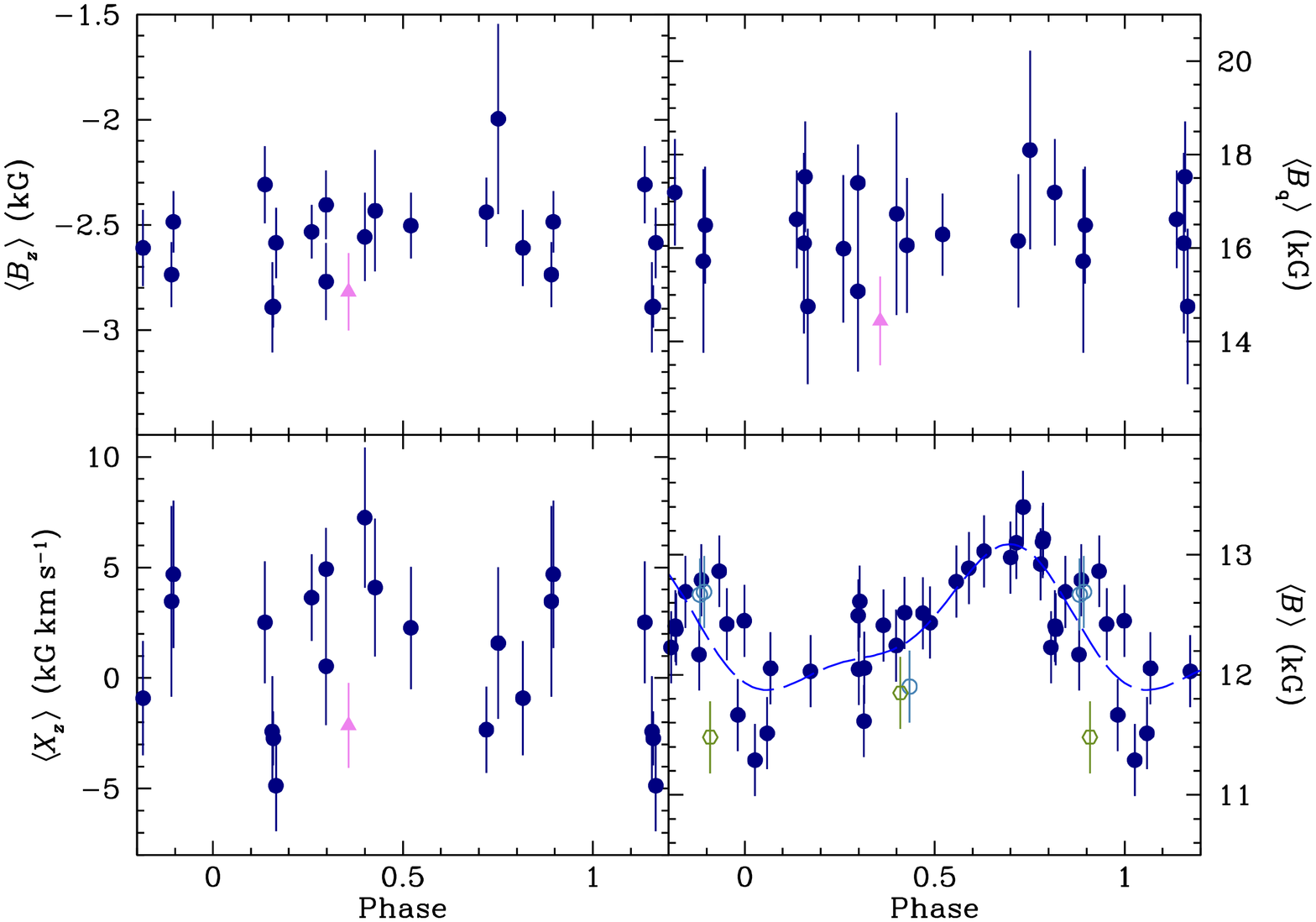}}
\parbox[t]{55mm}{
\caption{Mean longitudinal magnetic field ({\it top left\/}),
crossover ({\it bottom left\/}), 
mean quadratic magnetic field ({\it top right\/}),
and mean magnetic field modulus ({\it bottom right\/}) 
of the star HD~70331, against rotation phase, computed assuming
that the rotation period is 1\fd9989. The symbols are as described at the
beginning of Appendix~\ref{sec:notes}.} 
\label{fig:hd70331j}}
\end{figure*}

\begin{figure*}
\resizebox{12cm}{!}{\includegraphics[angle=270]{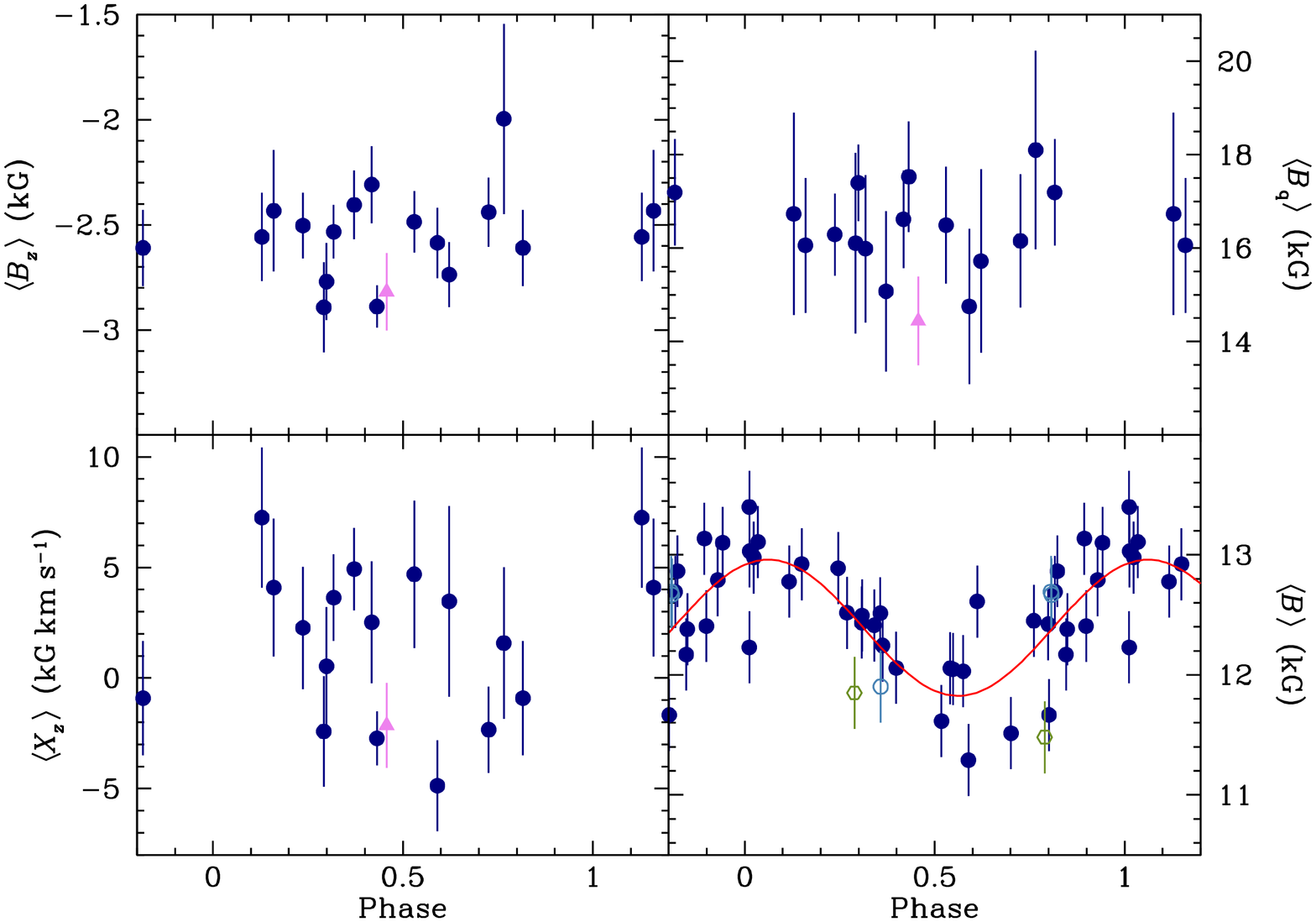}}
\parbox[t]{55mm}{
\caption{Mean longitudinal magnetic field ({\it top left\/}),
crossover ({\it bottom left\/}), 
mean quadratic magnetic field ({\it top right\/}),
and mean magnetic field modulus ({\it bottom right\/}) 
of the star HD~70331, against rotation phase, computed assuming
that the rotation period is 1\fd9909. The symbols are as described at the
beginning of Appendix~\ref{sec:notes}.} 
\label{fig:hd70331k}}
\end{figure*}

The seven new measurements of the mean magnetic field modulus of HD~70331
definitely rule out the tentative value of the rotation period
$\Prot=3\fd0308$
suggested in \citetalias{1997A&AS..123..353M}. Period determination is
hampered by strong 
aliasing, even with the larger set of $\Hm$ data of the present
study. The most probable values now are 
$\Prot=(1.9989\pm0.0008)$\,d or $\Prot=(1.9909\pm0.0008)$\,d. The
proximity of these values to two days may raise some concern about their
reality. However visual inspection of the spectra reveals the
occurrence of considerable
variations from one night to the next, and significant
distortion  of the resolved components of
\ion{Fe}{ii}\,$\lambda\,6149$ by the combination of different amounts
of Doppler and Zeeman effects on different parts of the stellar
surface. All the field moments are plotted against phases computed
with these two periods in Figs.~\ref{fig:hd70331j} and
\ref{fig:hd70331k}. In the former, there may be some very marginal
hint of variation of $\Hz$ with the 1\fd9989 period. However,
the scatter of the individual measurements of this field moment is
consistent with its being constant, so that the absence of correlation
with phase seen in the other figure for $\Prot=1\fd9909$ does not
represent an argument against the reality of that
period. A formally significant non-zero rms value of the crossover is
computed by combining all individual measurements, but none of these measurements
represent definite detections in themselves (in the best case, a $2.6\sigma$
value is obtained). Neither the quadratic
field nor the
equivalent widths of the \ion{Fe}{ii} lines used to diagnose the
magnetic field from the CASPEC spectra show any definite variability. 

% \clearpage

 In any event, there is no question that the rotation period of
HD~70331 must be of the order of days. While our data do not lend
themselves well to the determination of its $\vsi$, the presence of
well-resolved lines in its spectrum -- much better resolved than in
HD~65339, which has a field of comparable strength and
$\vsi=12.5$\,km\,s$^{-1}$ \citep{2004A&A...414..613K} -- indicates
that it cannot exceed 10\,km\,s$^{-1}$. For a star of this
temperature, the stellar radius cannot be significantly smaller than
$2.5\,R_{\odot}$, so that the angle between the rotation axis and the
line of sight must be at most 10\degr. With such a low inclination,
the amplitude of variation of $\Hm$ is remarkably large. This suggests
the existence of a strongly magnetic spot close to the stellar limb,
which alternately comes in and out of view as the star rotates. This
is consistent with the absence of large variation of the longitudinal
field provided that the magnetic vector in this spot is mostly
transversal.  The likelihood of such a configuration, in combination
with the large magnetic field 
strength, makes HD~70331 a promising
candidate for observations of linear polarisation in spectral lines. 

\subsection{HD~75445}
\label{sec:hd75445}
\begin{figure*}
\resizebox{12cm}{!}{\includegraphics[angle=270]{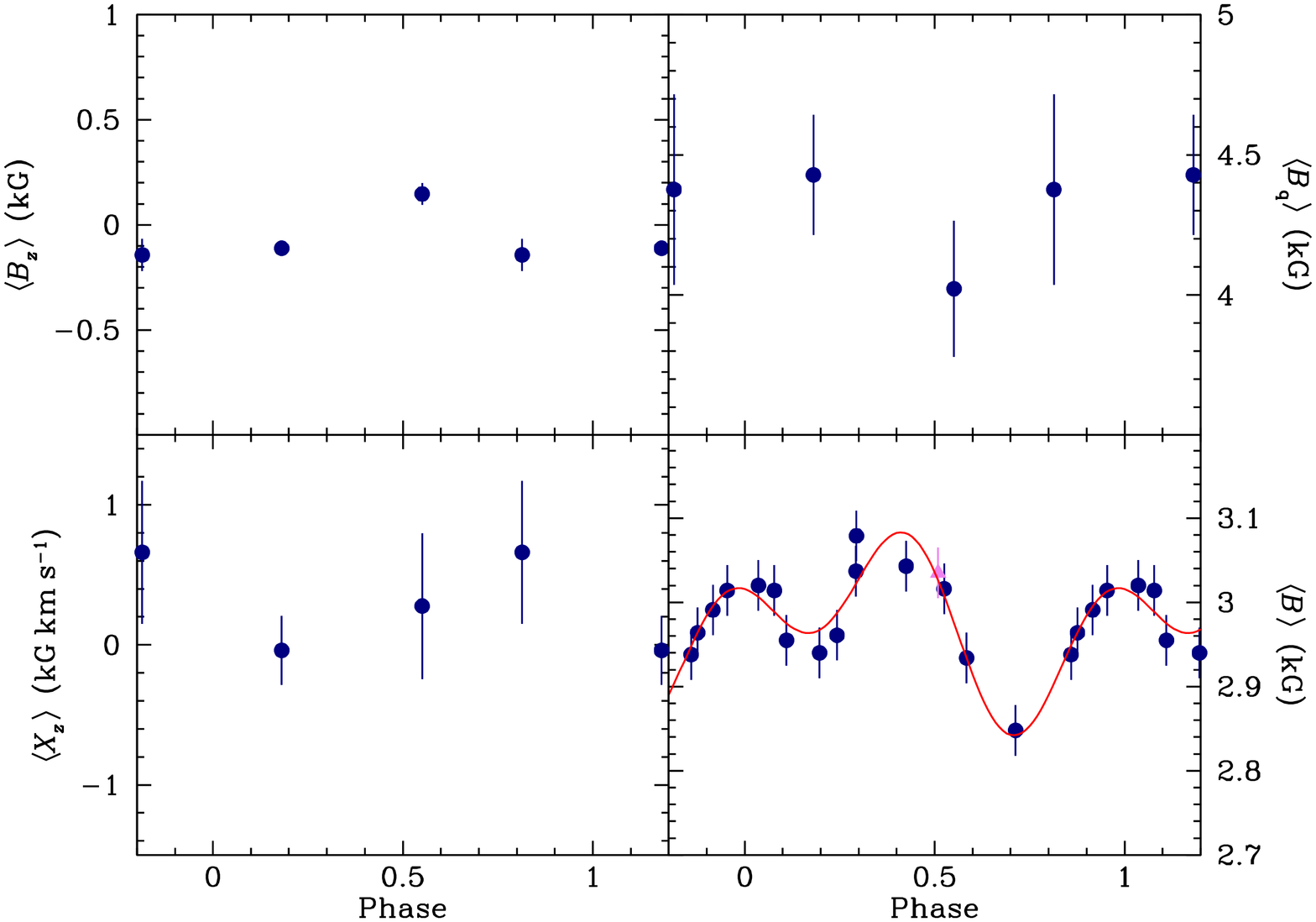}}
\parbox[t]{55mm}{
\caption{Mean longitudinal magnetic field ({\it top left\/}),
crossover ({\it bottom left\/}), 
mean quadratic magnetic field ({\it top right\/}),
and mean magnetic field modulus ({\it bottom right\/}) 
of the star HD~75445, against rotation phase. The symbols are as described at the
beginning of Appendix~\ref{sec:notes}.} 
\label{fig:hd75445}}
\end{figure*}

The clean, sharp profile of \ion{Fe}{ii}\,$\lambda\,6149$ in HD~75445
lends itself well to the precise determination of the mean magnetic field
modulus. Accordingly, there is little doubt that the range of the
derived values of $\Hm$ reflects actual variations of this field
moment. In a period analysis, one value stands out rather clearly:
$\Prot=(6.291\pm0.002)$\,d. When phased with this period, the $\Hm$ data
closely follow a smooth, though strongly anharmonic, variation curve
(Fig.~\ref{fig:hd75445}). Given the low amplitude of the variations, a
comparison of our data with field values of other authors may not be
meaningful, so that we shall restrict ourselves to note that
the three measurements of this star by \citet{2004A&A...423..705R} are not 
inconsistent with the proposed period (or a very close
value). Actually they are more consistent with the rather short period
proposed here than with the very long period suggested by these
authors. 

Only three spectropolarimetric observations of the star could be
secured, which are insufficient to characterise the variations of the
field moments derived from their analysis (based on a set of
\ion{Fe}{i} lines). Only one of the three values of $\Hz$
that are derived is formally non-null, albeit marginally so (at the
$3.0\,\sigma$ level), and no crossover is
detected. By contrast, the quadratic field comes out clearly, and is
about 1.4 times greater than the field modulus.

In any event, more observations are required to establish definitely
that the value of the period proposed here is correct.

%No point in trying to adopt a specific phase as origin, given
%remaining uncertainty on P and on variation curve shape.

\begin{figure*}
\resizebox{12cm}{!}{\includegraphics[angle=270]{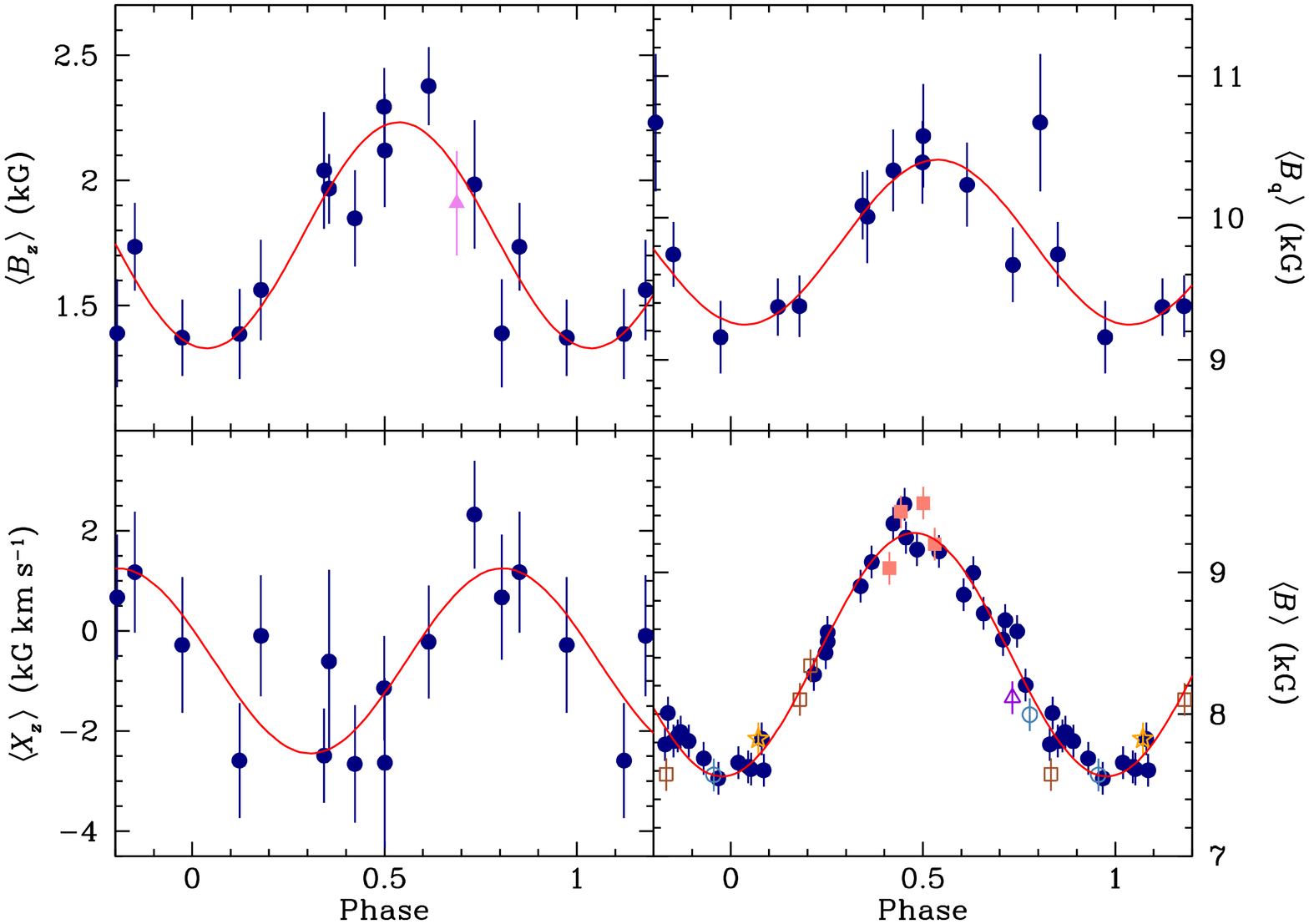}}
\parbox[t]{55mm}{
\caption{Mean longitudinal magnetic field ({\it top left\/}),
crossover ({\it bottom left\/}), 
mean quadratic magnetic field ({\it top right\/}),
and mean magnetic field modulus ({\it bottom right\/}) 
of the star HD~81009, against rotation phase. The symbols are as described at the
beginning of Appendix~\ref{sec:notes}.} 
\label{fig:hd81009}}
\end{figure*}

\subsection{HD~81009}
\label{sec:hd81009}
The visual and single-lined spectroscopic binary (SB1) HD~81009
was studied in 
detail by \citet{2000A&A...361..991W}. This work was based in part on
the measurements of the mean magnetic field modulus (only two new
determinations since \citetalias{1997A&AS..123..353M}), mean longitudinal magnetic field and
radial velocity that are reported in this
paper. 

Figure~\ref{fig:hd81009} shows the variations of the four field
moments considered here against the rotation phase determined using
the refined value of the rotation period derived from photometry by
\citet{1997PASP..109....9A} and confirmed by \citet{2006PASP..118...77A},
which is more 
accurate than the period obtained 
in \citetalias{1997A&AS..123..353M}. The variations of all four moments are well represented by
cosine waves. The values of the crossover
and quadratic field published by \citet{1997A&AS..124..475M}
for their single observation of HD~81009, which are of much poorer
quality than the data presented here, have been left out of the fits
and do not appear in the figure (where they would be out of scale).
The value of the quadratic field, which was obtained by
\citet{2006A&A...453..699M} from the analysis of a larger set of
\ion{Fe}{i} lines 
measured in a higher resolution spectrum covering a broader wavelength
range recorded at phase 0.733, $\Hq=9825\pm438$\,G, is in excellent
agreement with the measurements presented here.

The phases of maximum of the longitudinal field and the quadratic field coincide almost exactly; the maximum of the field
modulus seems to occur 
slightly earlier (about 0.06 rotation cycle), but this phase
difference is at the limit of significance. Although none of the
individual values of the crossover are formally significant (they are all
below $2.7\,\sigma$), the rms value obtained by combining them
indicates a definite detection and they show a formally significant
sinusoidal variation with rotation 
phase, with a fairly high value of the $R$ coefficient, 0.83; this is
also apparent visually in Fig.~\ref{fig:hd81009}. That we indeed
detect the actual variation of the 
crossover is made all the more plausible by the fact that the curve
fitted to the variations of $\xover$ is, within the errors, in phase
quadrature with the $\Hz$ fit; see Sect.~\ref{sec:xdisc} for a general
discussion of the phase relation between these two field 
moments.  

\begin{figure}[!h]
\resizebox{\hsize}{!}{\includegraphics[angle=270]{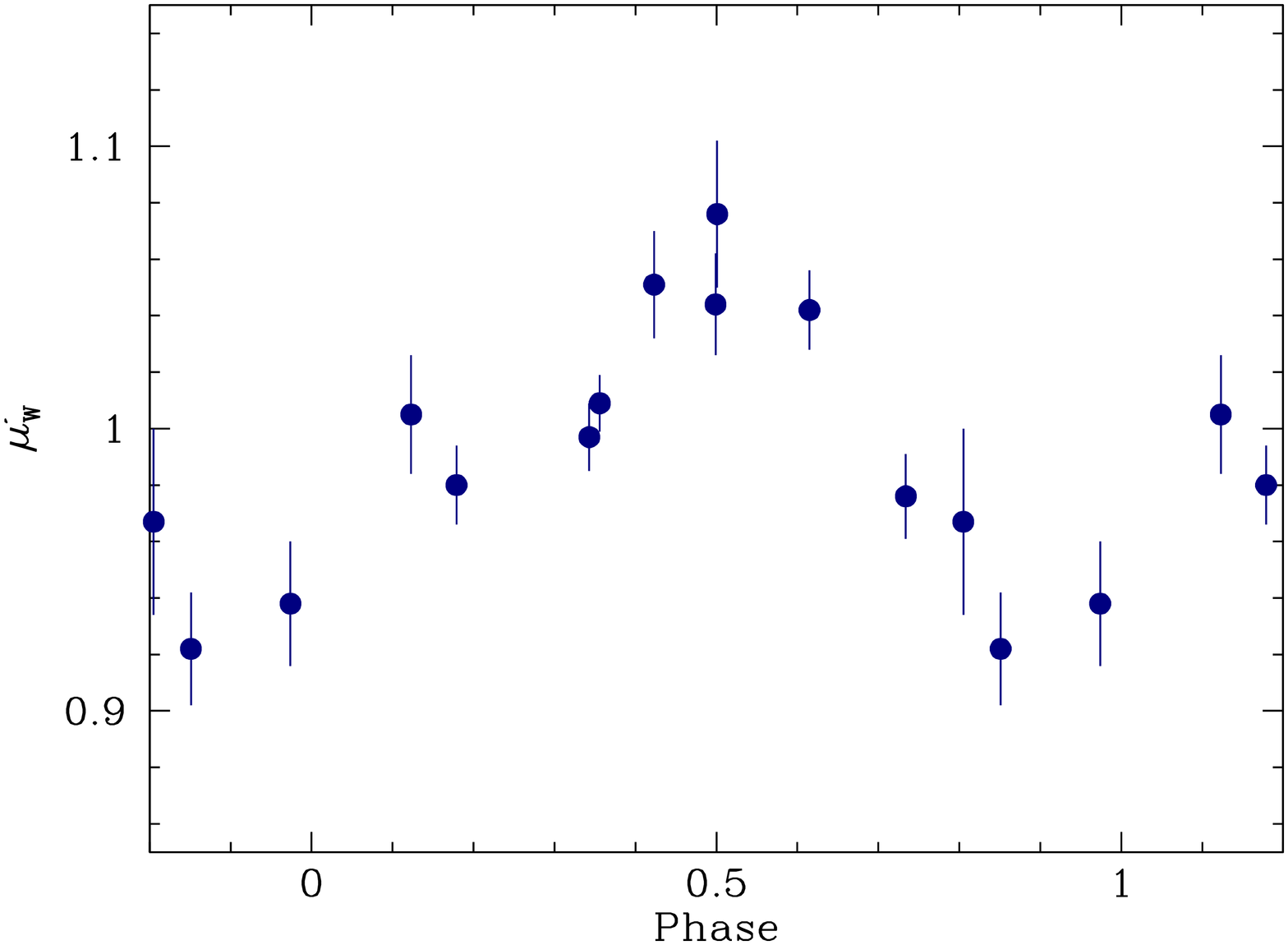}}
\caption{Variation with rotation phase of the average $\mu^\prime_W$
  of the normalised equivalent widths \citep{1994A&AS..108..547M} of the
  \ion{Fe}{i} and \ion{Fe}{ii} lines analysed in the CASPEC spectra of
  HD~81009. }
\label{fig:hd81009_ew}
\end{figure}

The variability of the \ion{Fe}{i} and \ion{Fe}{ii} lines from
which $\Hz$, $\xover,$ and $\Hq$ are determined is illustrated in
Fig.~\ref{fig:hd81009_ew}, where the average $\mu^\prime_W$ of their
normalised equivalent widths \citep[see][for definition
of this quantity]{1994A&AS..108..547M} is plotted against rotation
phase. The lines are 
stronger close to the phase of $\Hm$ maximum. As for other stars, the
corresponding differences of local line intensities are convolved into
the derived values of the considered field moments.

\subsection{HD~93507}
\label{sec:hd93507}
\begin{figure*}
\resizebox{12cm}{!}{\includegraphics[angle=270]{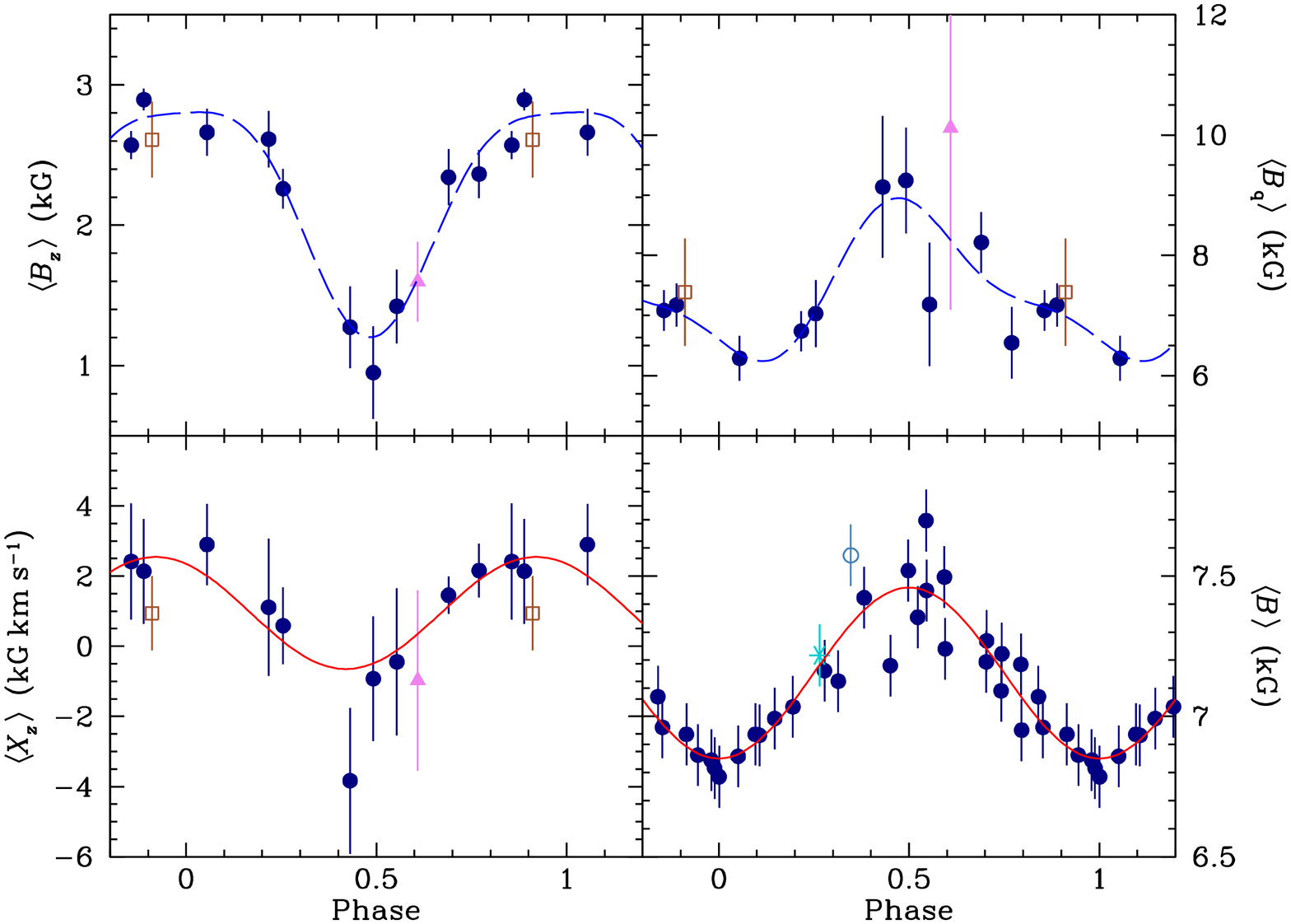}}
\parbox[t]{55mm}{
\caption{Mean longitudinal magnetic field ({\it top left\/}),
crossover ({\it bottom left\/}), 
mean quadratic magnetic field ({\it top right\/}),
and mean magnetic field modulus ({\it bottom right\/}) 
of the star HD~93507, against rotation phase. The symbols are as described at the
beginning of Appendix~\ref{sec:notes}.} 
\label{fig:hd93507}}
\end{figure*}

\begin{figure}
\resizebox{\hsize}{!}{\includegraphics[angle=270]{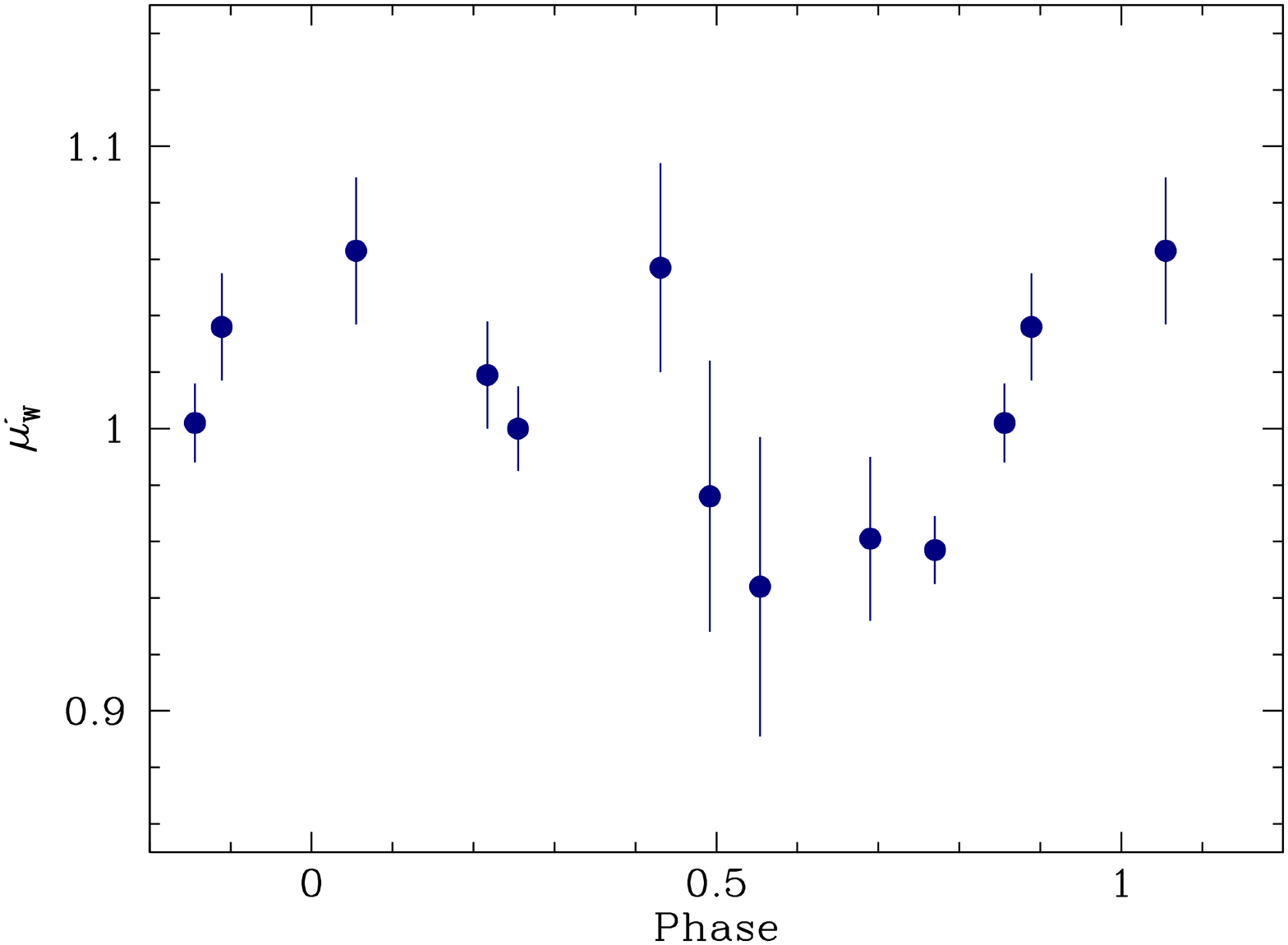}}
\caption{Variation with rotation phase of the average $\mu^\prime_W$
  of the normalised equivalent widths \citep{1994A&AS..108..547M} of the
  \ion{Fe}{ii} lines analysed in the CASPEC spectra of
  HD~93507. }
\label{fig:hd93507_ew}
\end{figure}

Only two new measurements of the mean magnetic field modulus of HD~93507
are reported here, and are plotted in Fig.~\ref{fig:hd93507} together
with our previous data against the phase computed using the value of
the rotation period derived in \citetalias{1997A&AS..123..353M}, $\Prot=556$\,d. This period
also matches the variations of the other field moments. The
longitudinal field is positive throughout the rotation cycle. The
phase of its minimum coincides with the phase of $\Hm$ maximum,
indicating that the field structure must significantly depart from a
centred dipole. There is a hint of anharmonicity in the variation of
$\Hz$, for which the superposition of a cosine wave and of its first
harmonic gives a somewhat better fit both visually and
mathematically (the reduced $\chi^2$ decreases from 1.7 to 1.4 when
adding the first harmonic), but is not formally significant. The
improvement resulting from 
inclusion of the first harmonic in the fit is less definite for the
quadratic field; we cannot decide at
present if the $\Hq$ variation curve actually departs significantly
from a single cosine wave, or if its apparent anharmonicity results
from some combination of the limited precision of the measurements and
of their uneven coverage of the rotation cycle. The fact that $\Hq$ is
marginally smaller than $\Hm$ over a sizable fraction of the stellar
rotation cycle indicates that it is probably somewhat
underestimated. This discrepancy is more pronounced around phases
0.1--0.2, while the error bars of all field moments are larger in the
phase interval 0.4--0.6 than over the rest of the rotation cycle. This
appears due primarily to the fact that, around the maximum of the
field modulus, the spectral lines show significant distortion,
suggesting that the distribution of the magnetic field strength and/or
orientation over the corresponding part of the star is much less
uniform than on the rest of its surface. The fact that \ion{Fe}{ii}
lines, from which the field is diagnosed, are weaker around phase 0.5
may also have some impact on the achievable measurement precision, but
contrary to what was conjectured in \citetalias{1997A&AS..123..353M}, it seems unlikely to be
the main cause of its degradation. Indeed, the amplitude of variation
of the equivalent widths is rather small (see
Fig.~\ref{fig:hd93507_ew}) -- smaller, in particular, than in other
stars where these variations have no noticeable effect on the magnetic
field determination. The distortion of the spectral lines around phase
0.5 may possibly account for the apparent variation of the crossover
around that phase, but they certainly cannot be invoked to explain
its consistently positive value over the rest of the period. Actually,
it rather looks as though a definite non-zero crossover is detected in this
star, which is somewhat unexpected because of its long rotation
period. This is further discussed in Sect.~\ref{sec:xdisc}.

 \subsection{HD~94660}
\label{sec:hd94660}
The eight new determinations of the mean magnetic field modulus of
HD~94660 presented here fully confirm the suspicion expressed in
\citetalias{1997A&AS..123..353M} that the time interval covered by the
data of that paper is 
almost identical to the rotation period of the star. The value of the
latter can now be determined from the whole set of $\Hm$ measurements:
$\Prot=(2800\pm200)$~d. This value is in good agreement with that
derived independently by \citet{2014A&A...572A.113L} from analysis of
FORS-1 $\Hz$ data spanning nearly one rotation cycle,
$\Prot=(2800\pm250)$\,d. 

\begin{figure*}
\resizebox{12cm}{!}{\includegraphics[angle=270]{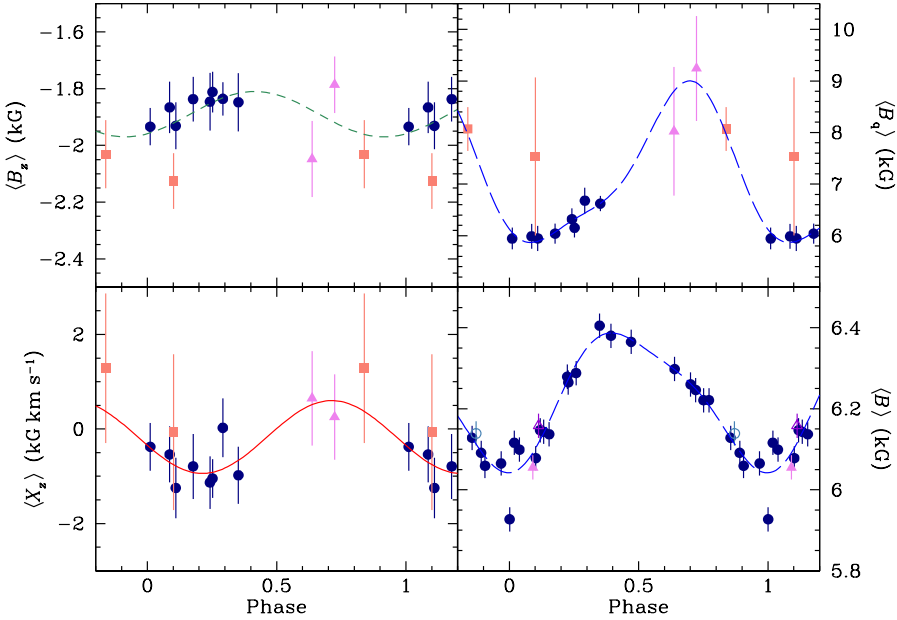}}
\parbox[t]{55mm}{
\caption{Mean longitudinal magnetic field ({\it top left\/}),
crossover ({\it bottom left\/}), 
mean quadratic magnetic field ({\it top right\/}),
and mean magnetic field modulus ({\it bottom right\/}) 
of the star HD~94660, against rotation phase. The symbols are as described at the
beginning of Appendix~\ref{sec:notes}.} 
\label{fig:hd94660}}
\end{figure*}

The rather large uncertainty affecting the value of the period
reflects the fact that, even with the new 
observations, the $\Hm$ data obtained so far cover only $\sim1.3$ rotation
cycle. For the same reason, and because of the apparently rather
complex shape of the variation of $\Hm$ around its minimum (see
Fig.~\ref{fig:hd94660}), the time of the latter, adopted as phase
origin, must be regarded as rather uncertain. 

The harmonic term of a
fit of the field modulus values by a function of the form of
Eq.~(\ref{eq:fit2}) is just below the threshold of formal
significance. But the anharmonicity of the curve of variation
of $\Hm$ appears definite. Its asymmetry indicates that
the magnetic field of HD~94660 is not symmetric about an axis passing
through its centre. 

The other field moments are diagnosed from analysis of a set of lines
of \ion{Fe}{ii}, whose equivalent widths appear constant over the
stellar rotation period. Based on the new measurements of the present
paper, 
variations of the quadratic field are undoubtedly observed. But it is
far from certain that the fitted curve describes their actual shape,
since its determination heavily depends on the older data of
\citet{1995A&A...293..746M} and  
\citet{1997A&AS..124..475M}, derived by application of Eq.~(\ref{eq:Hqold}),
which have much larger uncertainties than 
the new values reported here. We note that, close to phase 0, the values
of $\Hq$ are only marginally consistent with those of $\Hm$, which
suggests that they may be somewhat underestimated. However, they are
fully consistent with the value obtained by
\citet{2006A&A...453..699M} from the analysis of a better quality
spectrum recorded 
at phase 0.113.  

\begin{figure}[!h]
\resizebox{\hsize}{!}{\includegraphics{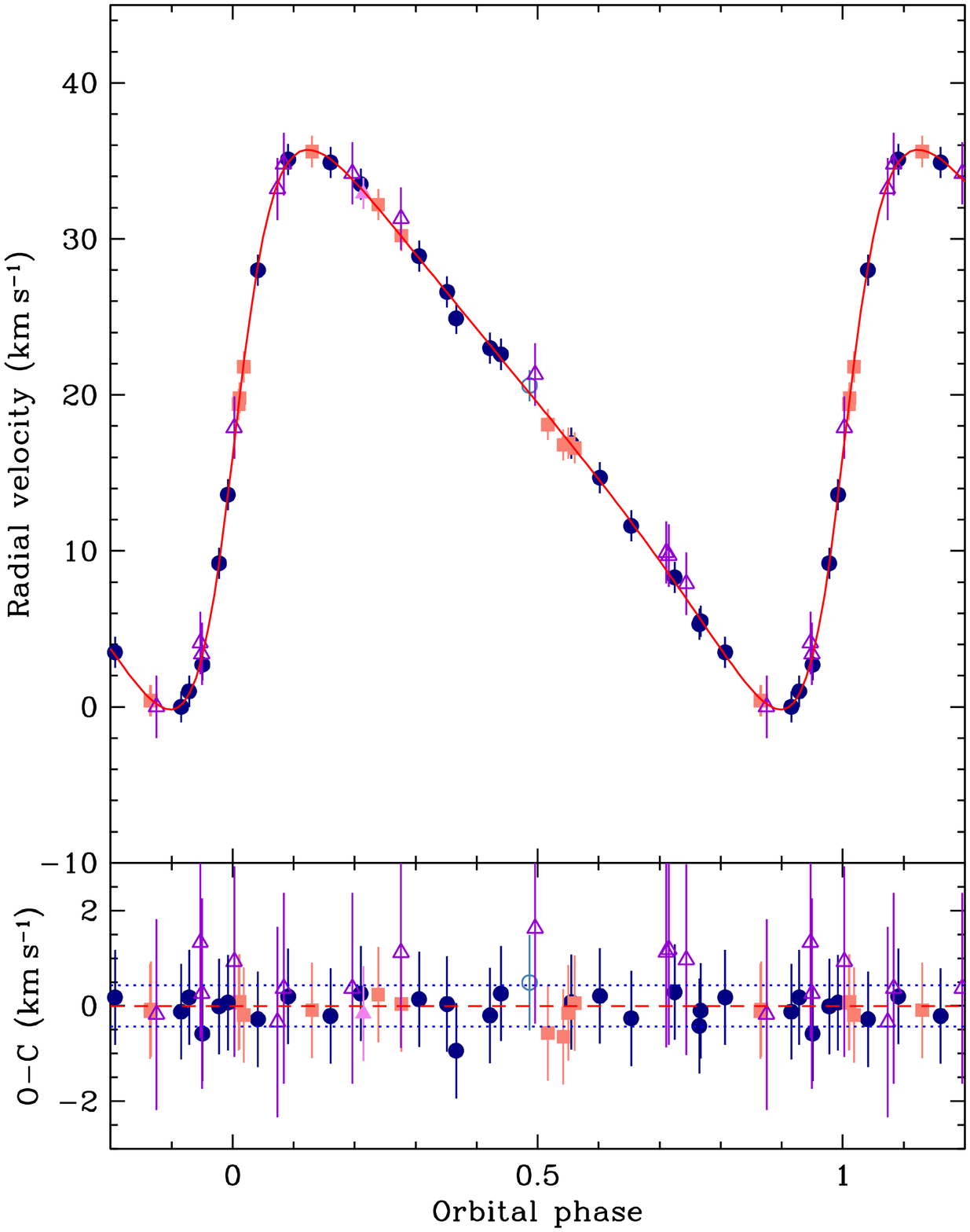}}
\caption{{\it Upper panel\/}: Our radial velocity measurements for
  HD~94660 are plotted together with those of
  \citet{2015A&A...575A.115B} against orbital phase. The  solid
    curve   
  corresponds to the orbital solution given in
  Table~\ref{tab:orbits}. The time $T_0$ of periastron passage is
  adopted as phase origin. {\it Bottom panel:\/} Plot of the
  differences ${\rm O}-{\rm C}$ between the observed values of the
  radial velocity and the predicted values computed from the orbital
  solution. The  dotted lines correspond to $\pm1$ rms
  deviation of the observational data about the orbital solution 
    (dashed line).  Open triangles represent our CASPEC
  observations; all other symbols refer to our high-resolution spectra
  obtained with various instrumental configurations, as indicated in
  Table~\ref{tab:plot_sym}.}  
\label{fig:hd94660_rv}
\end{figure}

\begin{figure}[h]
\resizebox{\hsize}{!}{\includegraphics[angle=270]{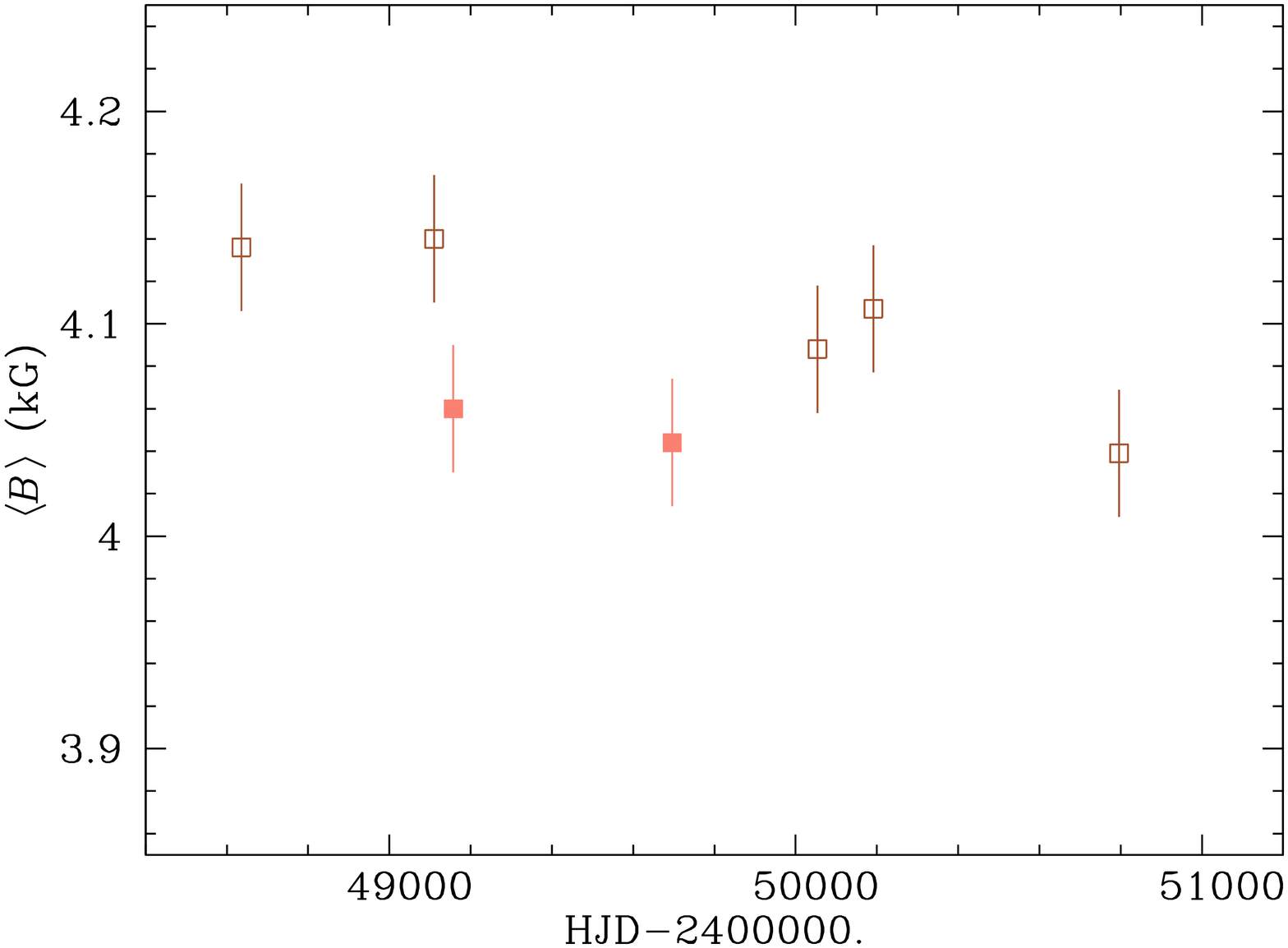}}
\caption{Mean magnetic field modulus of the star HD~110066,
against heliocentric Julian date. The symbols are as described at the
beginning of Appendix~\ref{sec:notes}.}
\label{fig:hd110066}
\end{figure}

 The new $\Hz$ data of this paper seem to to indicate a slow
increase (from more to less negative values) over the time range that
they cover. But neither in isolation nor combined with our previous
measurements do they undisputedly establish the occurrence of
variations of the longitudinal field. The fitted variation
curve is not formally significant, and it must therefore be regarded
as tentative at best. It is 
intriguing, though, that within the limits of the uncertainties, this
curve is in phase quadrature with the fitted $\xover$ curve. However,
the latter must be taken with even more caution, since it is 
primarily defined by the difference between the present set of
measurements and the old, lower quality data of \citet{1995A&A...293..733M}
and \citet{1997A&AS..124..475M}, since none of the crossover
determinations (old or new), taken individually, are formally
significant (they never exceed $2.6\,\sigma$), and since crossover is
not expected to occur for rotation periods of several years;  see
Sect.~\ref{sec:xdisc}, however. 

That the fitted $\Hz$ curve reflects real variations is furthermore
supported by the near coincidence of the phase of the largest negative
value of the longitudinal field with that of the minimum of the mean
field modulus. While such a phase relation between $\Hz$ and $\Hm$
indicates that the field structure departs significantly from a centred
dipole, it is observed in about half of the stars studied in this
paper for which the variation curves of the two field moments are
defined (see Sect.~\ref{sec:Hzdisc}).

On the other hand, we announced in \citetalias{1997A&AS..123..353M}
that the radial velocity of HD~94660 is variable. Preliminary orbital
elements were derived by \citet{2015A&A...575A.115B}. They are fully
consistent with those obtained from analysis of our data, which sample
the orbit more densely. We combined both data sets to determine the
refined 
orbital elements reported in Table~\ref{tab:orbits}. The fitted orbit is
shown in Fig.~\ref{fig:hd94660_rv}. The orbital period, 849\,d, is
shorter than the rotation period of the Ap component.

\subsection{HD~110066}
\label{sec:hd110066}
The three new determinations of the mean magnetic field modulus of
HD~110066 reported here marginally suggest that we may start to see a
very slow decrease of this field moment (see
Fig.~\ref{fig:hd110066} -- as for other stars, there is probably a
systematic difference of instrumental origin between the $\Hm$ values
derived from AURELIE observations and from KPNO coud\'e spectrograph
data).  However, considering that the time interval
between our first \citepalias{1997A&AS..123..353M} and our most recent
measurements of $\Hm$ 
in this star corresponds to almost half of the value of the rotation
period proposed by \citet{1981A&AS...44..265A}, $\Prot=4900$~d, if
this rotation period is correct, the peak-to-peak amplitude of
variation of the field modulus 
is unlikely to be much greater than 100~G. 

No significant detection of the longitudinal field or of the crossover
is achieved. It is not clear either that old measurements of $\Hz$ by
\citet{1958ApJS....3..141B}, yielding values of up to 300~G, are actually
significant: this critically depends on the correct evaluation of the
uncertainties, which have proven to be underestimated for several
stars (but definitely not for all) in the works of Babcock. The $\Hz$
values obtained by \citet{2014AstBu..69..427R} from two more recent
observations do not significantly differ from zero either. 

On the other
hand, the three values of the quadratic field that we derive
are formally significant, but physically meaningless, since they are
considerably smaller than the field modulus. They are obviously
spurious. This is not entirely surprising as previous practical
experience as well as numerical experiments have
shown that for comparatively weak fields, determination of the
quadratic field 
becomes impossible, or at least unreliable (with an apparently
systematic tendency to underestimate the actual field). In summary,
the spectropolarimetric observations do not add any meaningful element
to our knowledge of the magnetic field of HD~110066.

\subsection{HD~116114}
\label{sec:hd116114}
\begin{figure*}
\resizebox{12cm}{!}{\includegraphics[angle=270]{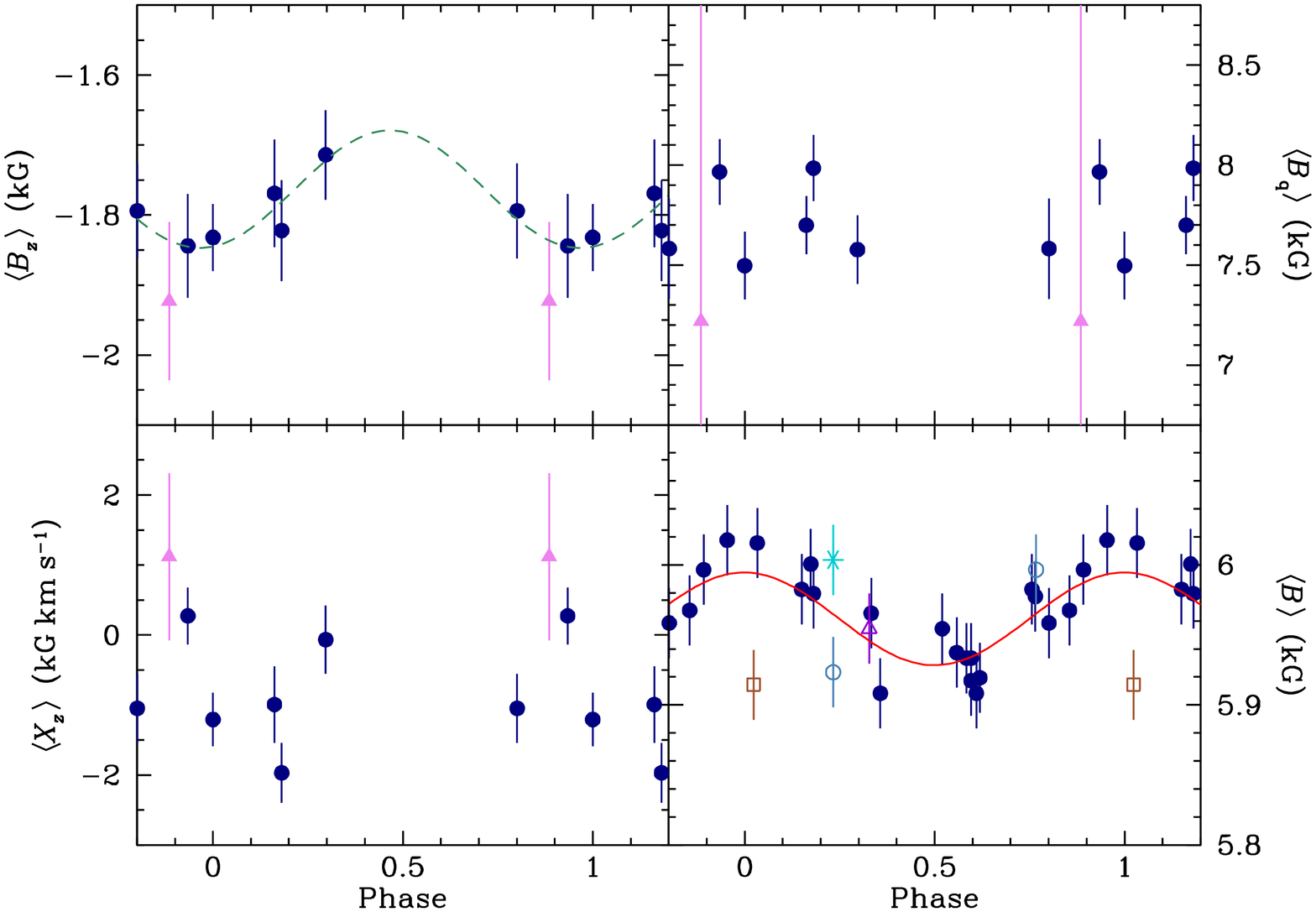}}
\parbox[t]{55mm}{
\caption{Mean longitudinal magnetic field ({\it top left\/}),
crossover ({\it bottom left\/}), 
mean quadratic magnetic field ({\it top right\/}),
and mean magnetic field modulus ({\it bottom right\/}) 
of the star HD~116114, against rotation phase. The symbols are as described at the
beginning of Appendix~\ref{sec:notes}.} 
\label{fig:hd116114}}
\end{figure*}

After addition of six new determinations of the mean magnetic field
modulus to the measurements of \citetalias{1997A&AS..123..353M}, we do
not confirm the 
suspicion expressed in this paper of a slow increase of this field
moment. Instead, leaving aside the only AURELIE data point, which may
be affected by a systematic shift of instrumental origin, a value
$\Prot=(27.61\pm0.08)$\,d 
stands out convincingly from the period analysis. With this period,
the $\Hm$ and $\Hz$ data show a dependence on rotation phase that is
very well fitted by a cosine wave (see Fig.~\ref{fig:hd116114}, however the amplitude of the cosine fit for the longitudinal
field is somewhat below the
level of formal significance: $Z_1/\sigma(Z_1)=2.5$). The
fact that the period derived from consideration of the field modulus
alone appears to match the variations of the longitudinal field
adds to its credibility. But even more convincingly, the coincidence
between the phases of minimum (largest negative value) of $\Hz$ and of
maximum of $\Hm$, within their uncertainties, strengthens our
confidence that, in spite of the very low amplitude of the
variations, we have correctly identified the rotation period of the
star. 

\citet{2014AstBu..69..427R} state that their $\Hz$ data are
incompatible with $\Prot=27\fd6$. However, all their measurements but
the most recent three have uncertainties about twice as large as ours, and
also twice as large as the amplitude of the sinusoid that we fit to
the longitudinal field values that we obtained. Since that amplitude
is below the level of formal significance, we are led to suspect that
the data of \citet{2014AstBu..69..427R} are not precise enough to
detect the variation. This suspicion is also borne out by consideration of
their Fig.~4; the period shown in the latter, $\Prot=4\fd1156$ is
definitely inconsistent with our determinations of the mean magnetic
field modulus. On the other hand, the $\Hz$ measurements of
\citet{2014AstBu..69..427R} appear systematically shifted by about
$-200$\,G with respect to ours; such minor differences are not unusual
between longitudinal field determinations from observations obtained
with different telescopes and instruments. 

The value $\Prot=5.3832$\,d proposed by
\citet{2012MNRAS.420..757W} on the basis of photometric data obtained
with the 
STEREO satellites is not supported by our magnetic field
data either. 

 The quadratic field shows no definite variation, which is not
surprising, considering that the uncertainty of its determinations is
large compared with the amplitudes observed for the variations of the
longitudinal field and the field modulus. The value of $\Hq$
obtained by \citet{2006A&A...453..699M} from the analysis of a
sample of \ion{Fe}{i} lines in a spectrum of HD~116114 recorded over a
broader wavelength range and at higher resolution than our CASPEC
spectra, $(7158\pm190)$\,G, is somewhat smaller than any of the values
determined here, but is consistent with most of them within their
respective uncertainties. Two of the six crossover values are
significant at a level $>3\,\sigma$. While it is plausible that
crossover due to stellar rotation is detected for a 27\fd6 period, the
fact that all values of $\xover$ derived here are negative (with one
exception, which is not formally significant), albeit with a coverage
gap between phases 0.3 and 0.8, and that no clear
variation with the rotation period is observed, raises questions about
the nature of the measured effect. This is further discussed in
Sect.~\ref{sec:xdisc}. 

 The field moments determined from the CASPEC spectra are based on the
analysis of a set of \ion{Fe}{i} lines that show no equivalent width
variation.

\begin{figure}[!h]
\resizebox{\hsize}{!}{\includegraphics[angle=270]{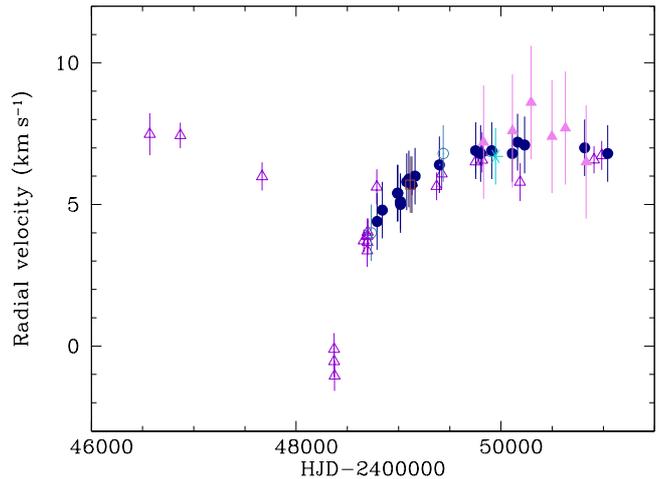}}
\caption{Our radial velocity measurements for HD~116114 are plotted
  together with those of \citet{2002A&A...394..151C}  against
  heliocentric Julian date.  Open triangles correspond to the
  \citeauthor{2002A&A...394..151C} data and  filled triangles to our CASPEC
  observations; all 
  other symbols refer to our high-resolution spectra obtained with
  various instrumental configurations, as indicated
  in Table~\ref{tab:plot_sym}.}
\label{fig:hd116114_rv}
\end{figure}

The variability of the radial velocity of HD~116114 was announced in
\citetalias{1997A&AS..123..353M} and confirmed by \citet{2002A&A...394..151C}, who suggested
that the orbital period of this spectroscopic binary is of the order
of 4000~days. This suggestion appears founded from inspection 
of Fig.~\ref{fig:hd116114_rv}, where their data and ours are plotted
together. This figure is a good illustration of the precision of the
radial velocity measurements obtained from our high-resolution
spectra, and of their consistency with those of
\citeauthor{2002A&A...394..151C}. 

A few years ago, \citet{2005MNRAS.358.1100E} announced the discovery of
oscillations with a period of 21\,min in HD~116114, making it the
rapidly oscillating Ap (roAp) star with the longest pulsation period
known at the time, and the most evolved one.

\subsection{HD~116458}
\label{sec:hd116458}
\begin{figure*}
\resizebox{12cm}{!}{\includegraphics[angle=270]{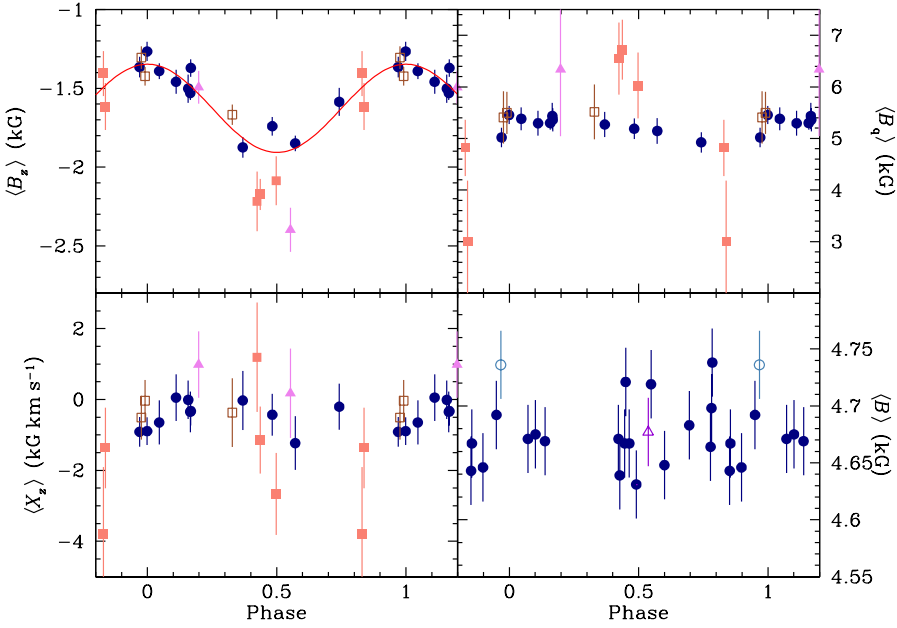}}
\parbox[t]{55mm}{
\caption{Mean longitudinal magnetic field ({\it top left\/}),
crossover ({\it bottom left\/}), 
mean quadratic magnetic field ({\it top right\/}),
and mean magnetic field modulus ({\it bottom right\/}) 
of the star HD~116458, against rotation phase. The symbols are as described at the
beginning of Appendix~\ref{sec:notes}.} 
\label{fig:hd116458}}
\end{figure*}

HD~116458 has already been extensively studied in our previous papers,
both from high-resolution spectroscopy in natural light
\citep[\citetalias{1997A&AS..123..353M};][]{1990A&A...232..151M,1992A&A...256..169M}, and from 
spectropolarimetric observations
\citep{1994A&AS..108..547M,1995A&A...293..733M,1995A&A...293..746M,1997A&AS..124..475M}. Combining the
data from those 
papers with the 11 new measurements of the mean longitudinal magnetic
field presented here, we refined the determination of the value of the
rotation period, $\Prot=(148.39\pm0.33)$\,d, which is used to compute
the phases against which the measurements are plotted in
Fig.~\ref{fig:hd116458}. The new $\Hz$ data have considerably smaller
error bars than the old data, and they nicely line up along a cosine
wave, with little scatter. The amplitude of the variation appears
smaller than could have been suspected from the old, less precise
measurements.

The mean magnetic field modulus and the mean quadratic field do not
show any significant variability. The standard deviation of all our
$\Hm$ data about 
their average, 4677\,G, is only 31\,G; in our sample, the field modulus measurements only show less
scatter for HD~137949
and HD~177765. For the quadratic field, considering only the new
measurements of this paper (which are considerably better than the old
ones), the standard deviation of the data is 169\,G, which is smaller
than the average uncertainty of the individual determinations
(211\,G). The average of the new determinations of $\Hq$, 5252\,G, is
however somewhat larger than the value of 4716\,G obtained by
\citet{2006A&A...453..699M} from analysis of a sample of 83
\ion{Fe}{ii} 
lines observed at higher spectral resolution. But the
\citet{2006A&A...453..699M} value may
somewhat underestimate the actual field strength, as can be inferred
from the fact that it is not significantly larger than the field
modulus. 

As in other stars, the best-determined value of the crossover does not
reach the threshold of significance, at $2.2\sigma$. However, one
notes that among the 11 new determinations of this field moment, all
but one (which does not significantly differ from 0) are
negative. Furthermore, these 11 data points appear to define a curve of
variation of the crossover with the rotation period that is in phase
quadrature with the variation of the longitudinal field. This is
further discussed in Sect.~\ref{sec:xdisc}.

 No variability is observed for the equivalent widths of the
\ion{Fe}{ii} lines from which $\Hz$, $\xover,$ and $\Hq$ are
diagnosed. 

\begin{figure}
\resizebox{\hsize}{!}{\includegraphics{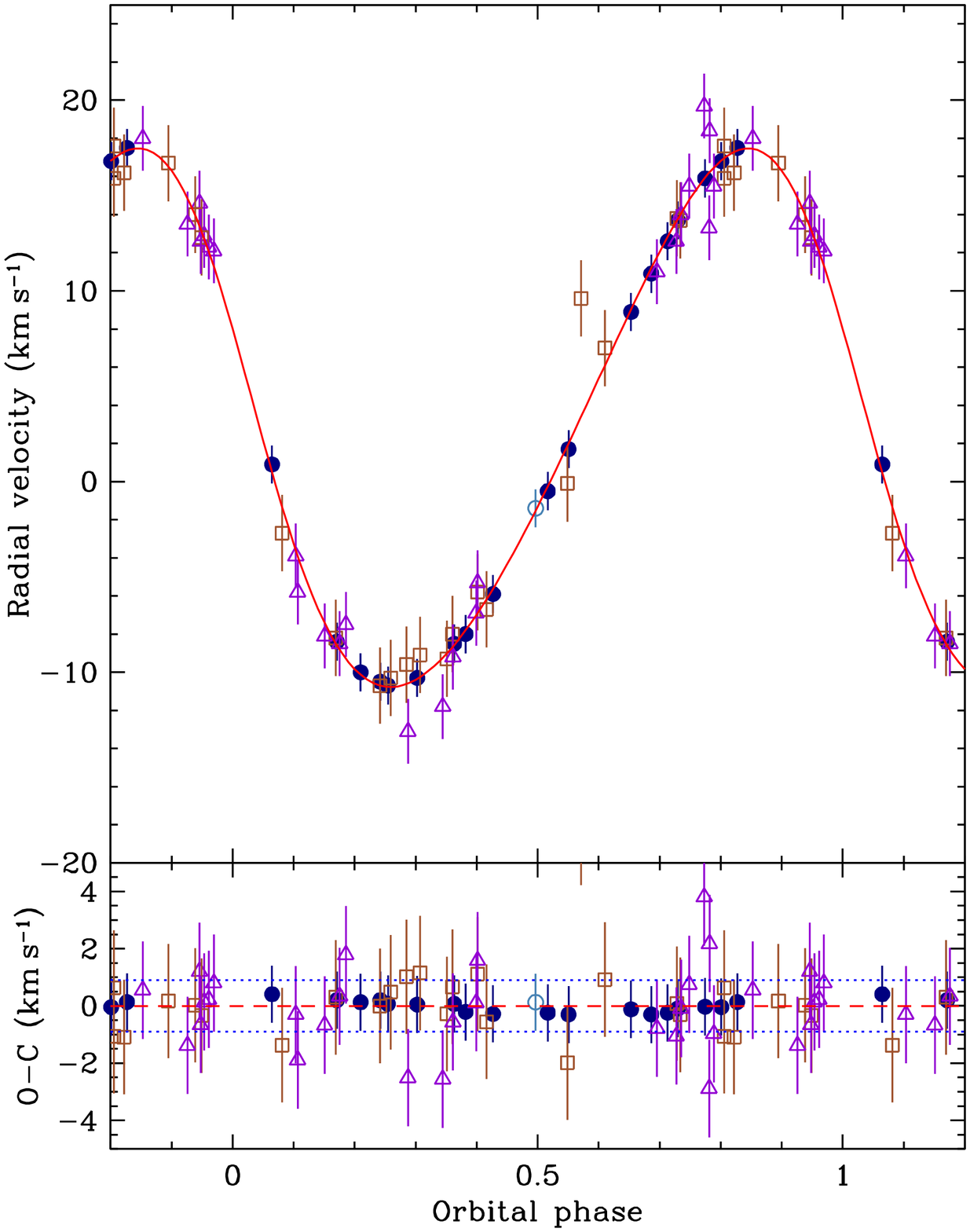}}
\caption{{\it Upper panel\/}: Our radial velocity measurements for
  HD~116458 are plotted 
  together with those of \citet{1982Obs...102..138D} against orbital
  phase. The  solid curve   
  corresponds to the orbital solution given in
  Table~\ref{tab:orbits}. The time $T_0$ of periastron passage is
  adopted as phase origin. {\it Bottom panel:\/} Plot of the
  differences ${\rm O}-{\rm C}$ between the observed values of the
  radial velocity and the predicted values computed from the orbital
  solution. The  dotted lines correspond to $\pm1$ rms
  deviation of the observational data about the orbital solution 
   (dashed line).  Open triangles represent Dworetsky's data
  and  open squares our CASPEC 
  observations; all other symbols refer to our high-resolution spectra
  obtained with various instrumental configurations, as indicated in
  Table~\ref{tab:plot_sym}.}  
\label{fig:hd116458_rv}
\end{figure}

HD~116458 is an SB1 whose orbit has first been determined by
\citet{1982Obs...102..138D}. Our first attempt to combine our radial
velocity 
measurements with those of this author led to a revised orbital
solution about which the residuals \OC\ for the former still showed a
low-amplitude variation with the orbital period. This appears 
to be due to the fact that the measurement errors of \citeauthor{1982Obs...102..138D} were
underestimated, so that his data were given too large a weight in the
combined data set. \citeauthor{1982Obs...102..138D} himself noted that the rms error derived
from his orbital solution was ``nearly three times the mean internal
error'' of his individual measurements. Adopting the former
(1.7\,\kms) for the uncertainty of all his radial velocity values, rather
than the (smaller) estimates quoted for each determination in Table~I
of his paper, we obtain an orbital solution in which the residuals no
longer show any significant periodic variation. The elements given in
Table~\ref{tab:orbits} and the orbital curve shown in
Fig.~\ref{fig:hd116458_rv} correspond to this solution. Not surprisingly,
given that consideration of our data together with those of
\citet{1982Obs...102..138D} more than doubles the time
interval covered by the  
measurements, our determination of the orbit represents a significant
improvement, albeit fully compatible with the original solution by
\citeauthor{1982Obs...102..138D} within the errors.   

\subsection{HD~119027}
\label{sec:hd119027}
\begin{figure}
\resizebox{\hsize}{!}{\includegraphics[angle=270]{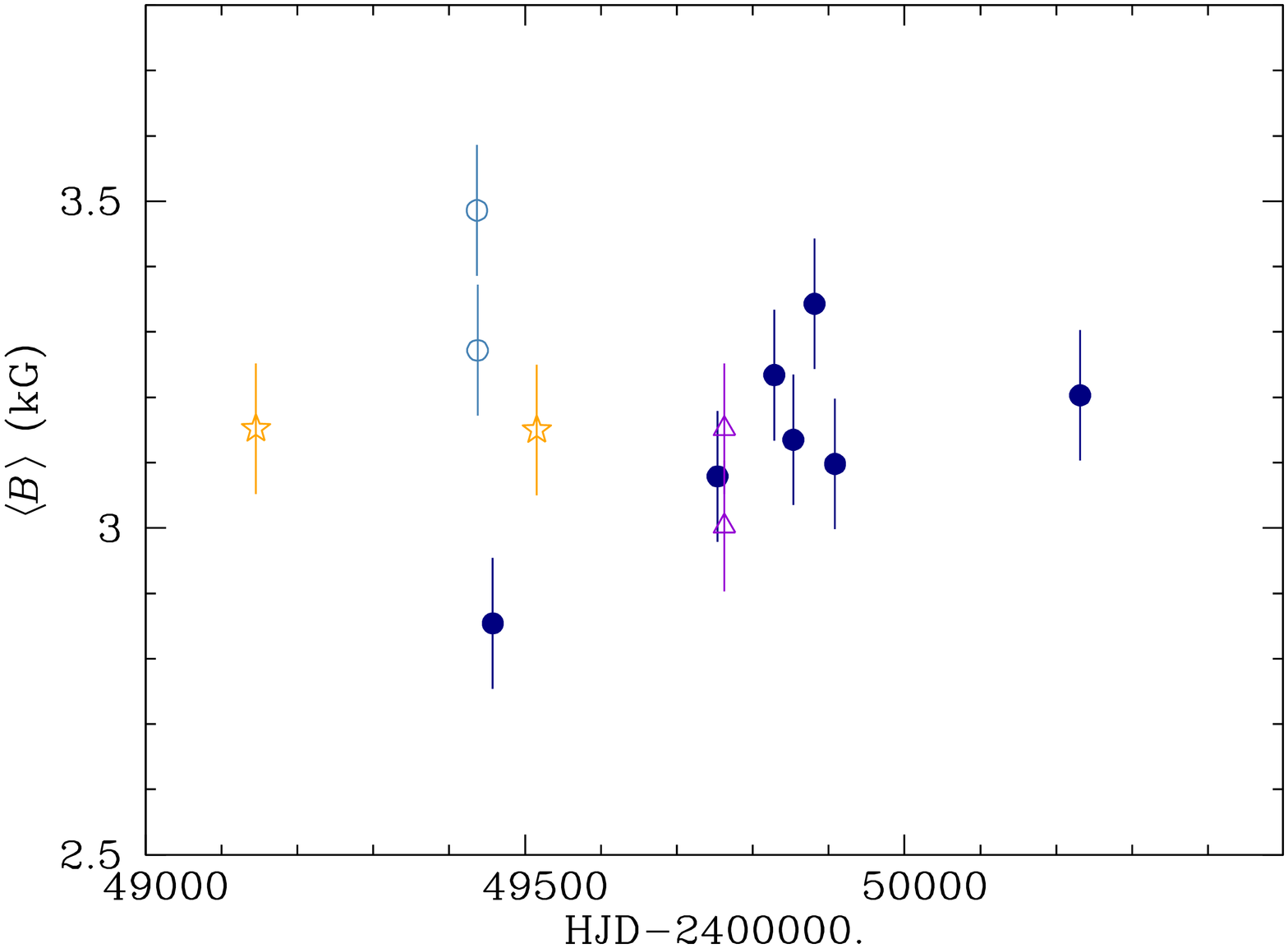}}
\caption{Mean magnetic field modulus of the star HD~119027,
against heliocentric Julian date. The symbols are as described at the
beginning of Appendix~\ref{sec:notes}.}
\label{fig:hd119027}
\end{figure}

In \citetalias{1997A&AS..123..353M}, it was argued that the rotation period of the roAp star
HD~119027 could plausibly be of the order of a few weeks, but its
value remained elusive. The addition of a single
magnetic field modulus measurement to the data of that paper (plotted
with them in Fig.~\ref{fig:hd119027}) is
insufficient to make any progress in its determination. 

We present here the first measurements of the longitudinal field and
the quadratic field of the star, which were obtained from
analysis of a set of 
\ion{Fe}{i} lines; both are well above the significance
threshold. No significant crossover was detected in our single
spectropolarimetric observation. 

\citet{2014AstBu..69..427R} also obtained a single measurement of
the longitudinal field in this star. Both the value that they
derived, $\Hz=930$\,G, and our determination, $\Hz=510$\,G, are
positive. But we do not have enough information to decide whether the
difference between those two values, which is formally very
significant, reflects the variation that the stellar field underwent
between the two observations, which took place 11 years apart, or
systematic effects resulting from the usage of different telescope and
instrument combinations. 

\subsection{HD~126515}
\label{sec:hd126515}
\begin{figure*}
\resizebox{12cm}{!}{\includegraphics[angle=270]{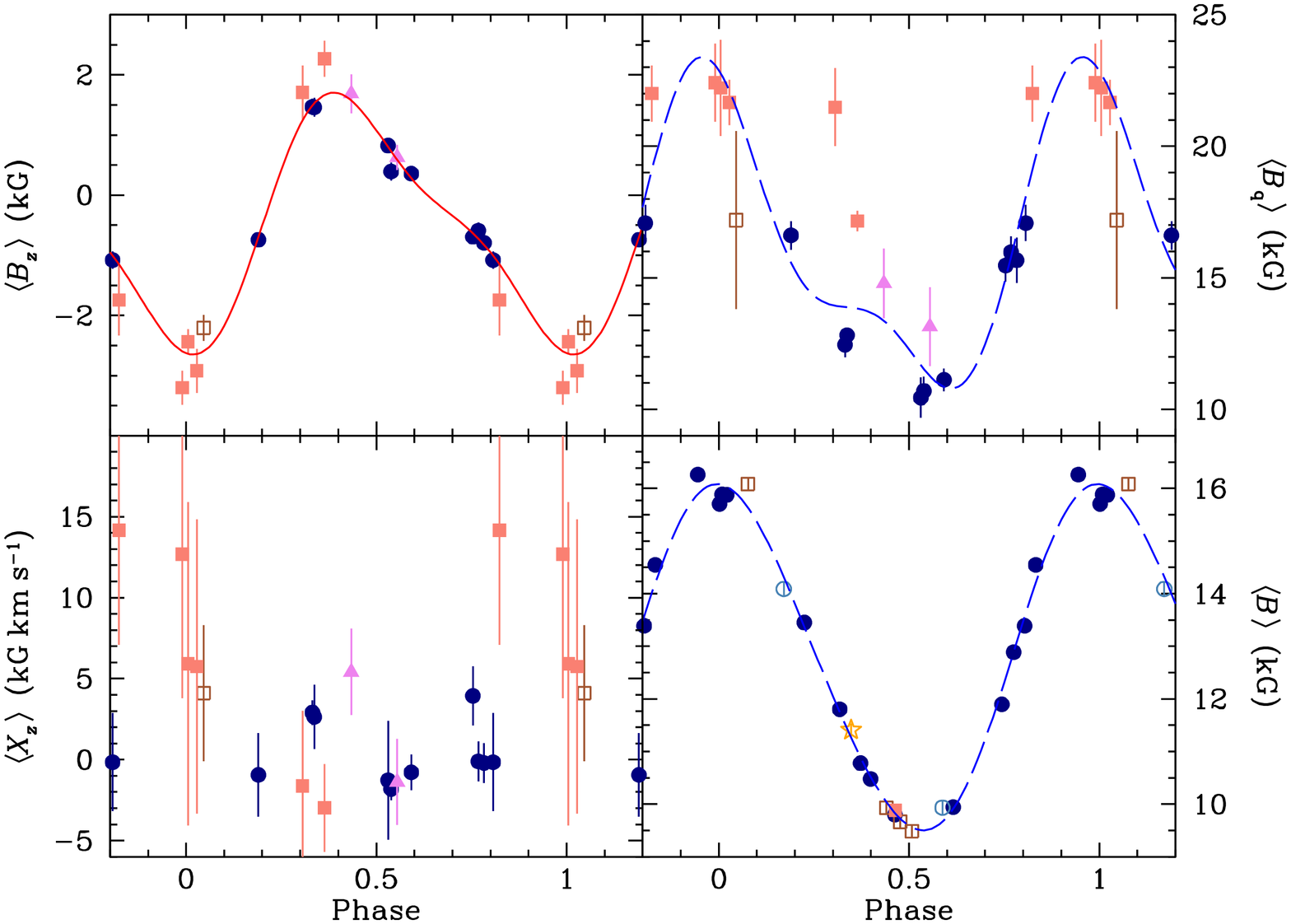}}
\parbox[t]{55mm}{
\caption{Mean longitudinal magnetic field ({\it top left\/}),
crossover ({\it bottom left\/}), 
mean quadratic magnetic field ({\it top right\/}),
and mean magnetic field modulus ({\it bottom right\/}) 
of the star HD~126515, against rotation phase. The symbols are as described at the
beginning of Appendix~\ref{sec:notes}.} 
\label{fig:hd126515}}
\end{figure*}

The new determinations of the mean magnetic field modulus and the
mean longitudinal magnetic field of HD~126515 (Preston's star) agree
well with the refined value of the rotation period derived in \citetalias{1997A&AS..123..353M},
$\Prot=129\fd95$ (see Fig.~\ref{fig:hd126515}). Their combination with
our older data for these field moments allow their variation curves to
be more precisely characterised. The strong anharmonicity of the $\Hz$
curve, mentioned in \citetalias{1997A&AS..123..353M}, is fully
confirmed. The departure of the 
$\Hm$ curve from a single cosine wave is much less pronounced; the
contribution of the harmonic is not visually compelling in
Fig.~\ref{fig:hd126515}, but at $2.9\sigma$, the second-order term
of the fit is just 
below the threshold of formal significance. That is
very unusual because, in general, the longitudinal field variation curves are
very nearly sinusoidal even in those stars in which a strong anharmonicity
of the field modulus variations is observed. The
negative extremum of $\Hz$ coincides in phase with the maximum of
$\Hm$. 

The shape of the quadratic field variations is less well
characterised, in particular because from phase 0.81 to phase 0.19, it
is only constrained by older determinations, which are considerably
less precise than those published in this paper. The anharmonicity of
the $\Hq$ curve critically depends on the precision of the old
measurements of this field moment; it cannot be regarded as definitely
established. 

We almost certainly detect crossover in this star; see
the detailed discussion of Sect.~\ref{sec:xdisc}. However, its
variation with rotation phase is ill-defined, in part because over a
large portion of the cycle (from phase $\sim0.8$ to phase $\sim
0.2$), it is
constrained only by the old, less precise measurements of
\cite{1995A&A...293..733M} and \cite{1997A&AS..124..475M}. 

 The large ratio between the extrema of $\Hm$ (1.7) indicates significant
departure of the field geometry from a centred dipole. The asymmetry
of the $\Hz$ variation curve reflects the lack of symmetry of the
field about an axis passing through the centre of the star.

\begin{figure}
\resizebox{\hsize}{!}{\includegraphics[angle=270]{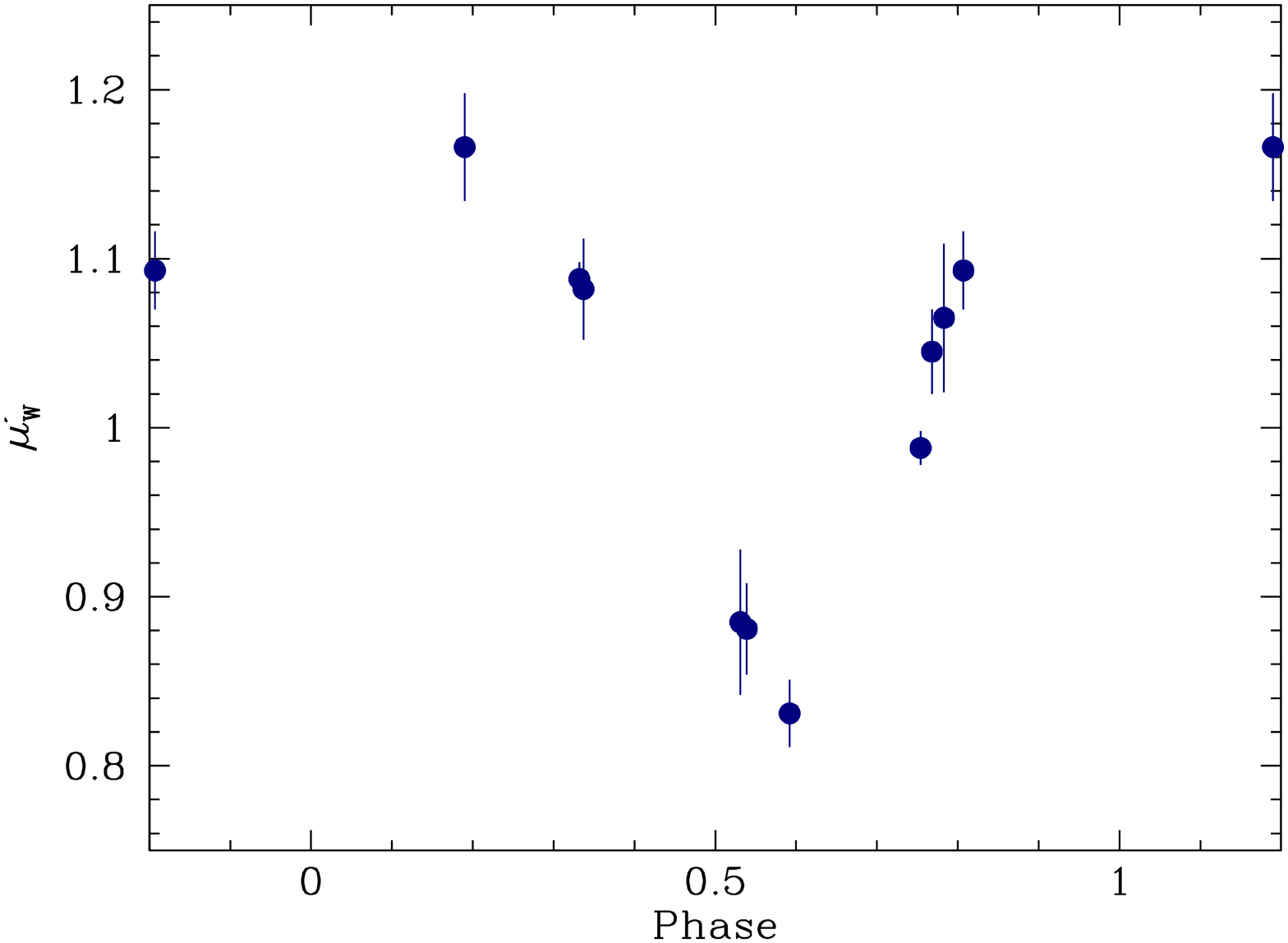}}
\caption{Variation with rotation phase of the average $\mu^\prime_W$
  of the normalised equivalent widths \citep{1994A&AS..108..547M} of the
  \ion{Fe}{ii} lines analysed in the CASPEC spectra of
  HD~126515. }
\label{fig:hd126515_ew}
\end{figure}

\citet{1970ApJ...160.1059P} had already reported the variability of the
equivalent widths of the Fe lines in HD~126515. This variability is
confirmed and more precisely characterised from consideration of the
\ion{Fe}{ii} lines from which $\Hz$, $\xover$, and $\Hq$ are
determined. As can be seen in Fig.~\ref{fig:hd126515_ew}, the
amplitude of variation of the equivalent widths of these lines is
larger than in most other stars discussed in this paper. This has to
be kept in mind in the interpretation of the magnetic field
measurements. The lines are stronger close to the negative pole of the
star (minimum of the field modulus). 

\subsection{HD~134214}
\label{sec:hd134214}
\begin{figure*}
\resizebox{12cm}{!}{\includegraphics[angle=270]{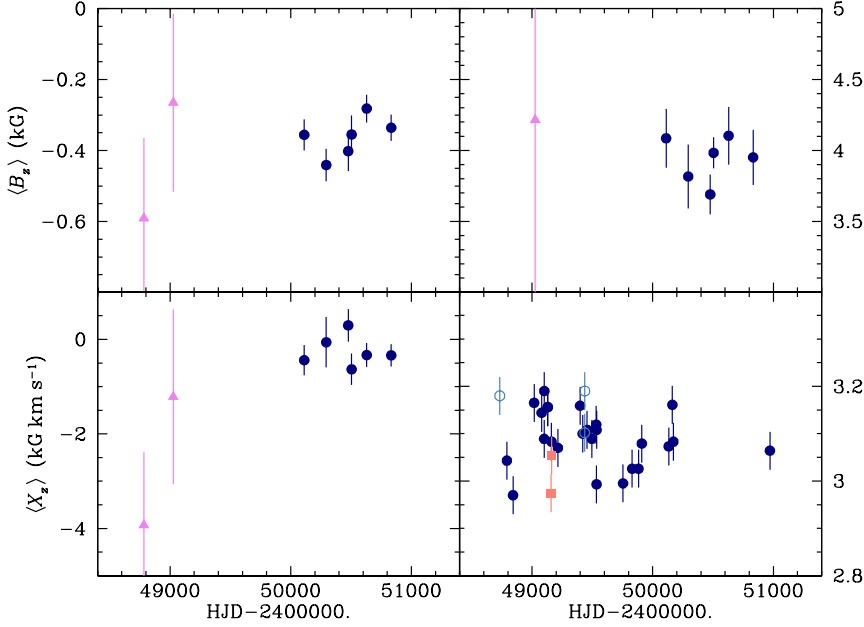}}
\parbox[t]{55mm}{
\caption{Mean longitudinal magnetic field ({\it top left\/}),
crossover ({\it bottom left\/}), 
mean quadratic magnetic field ({\it top right\/}),
and mean magnetic field modulus ({\it bottom right\/}) 
of the star HD~134214,
against heliocentric Julian date. The symbols are as described at the
beginning of Appendix~\ref{sec:notes}.}
\label{fig:hd134214}}
\end{figure*}

The four new measurements of the mean magnetic field modulus of the roAp
star HD~134214 presented here do not enable us to confirm or rule
out the tentative value $\Prot=4\fd1456$ of its rotation period
proposed in \citetalias{1997A&AS..123..353M}. However, several other
values of the period seem 
equally probable. This is why we prefer to plot our measurements
against the Julian date in Fig.~\ref{fig:hd134214}. The standard deviation
of our $\Hm$ determinations, 63\,G, is larger than the
precision that we expect for our measurements (compare e.g. with the
standard deviations for HD~137949 or HD~177765). This suggests that
the scatter of our data reflects actual variations, but until now their
periodicity eludes determination. On the other hand,
\citet{2000A&AS..146...13A} confirmed that no photometric
variations could 
be detected in this star, besides those caused by its pulsation.

 The
magnetic field moments derived from our spectropolarimetric
observations do not show any variability either; the equivalent widths
of the \ion{Fe}{i} lines on which their diagnosis is based
appear constant as well. HD~134214 is one of the two stars of the sample
for which the best precision is consistently achieved in the
determination of $\Hz$. With error bars about five times
smaller than those of \citet{1997A&AS..124..475M}, we
independently 
confirm the definite detection of a moderate longitudinal field
in HD~134214 reported by \cite{2014AstBu..69..427R}. Our measurements,
obtained between Januray 1996 and January 1998, show little, if any,
variations of $\Hz$ around a mean value of $\sim-360$\,G. The
difference with the values obtained between 1999 and 2007 by
\cite{2014AstBu..69..427R}, which range from $-700$ to $-900$\,G, is
probably of instrumental origin. 

The quadratic field of HD~134214 is about 1.3 times larger than the
field modulus. No crossover is detected. 

\subsection{HD~137909}
\label{sec:hd137909}
Since the publication of \citetalias{1997A&AS..123..353M}, a large number of measurements of
the mean longitudinal magnetic field modulus of HD~137909
($\beta$~CrB) have been published by various groups, i.e. 
\citet{2000AN....321..115H}, 
\citet{2000MNRAS.313..851W}, and \citet{2001A&A...365..118L}. The
latter authors, combining published data of various groups
accumulated over more than 40 years, confirm the adequacy of
the 
value of the rotation period derived by
  \citet{1989MNRAS.238..261K}, $\Prot=18\fd4868$. In  
particular, the revised value proposed by \citet{2000A&A...358..929B}  
does not represent a significant improvement over the latter.

\begin{figure*}
\resizebox{12cm}{!}{\includegraphics[angle=270]{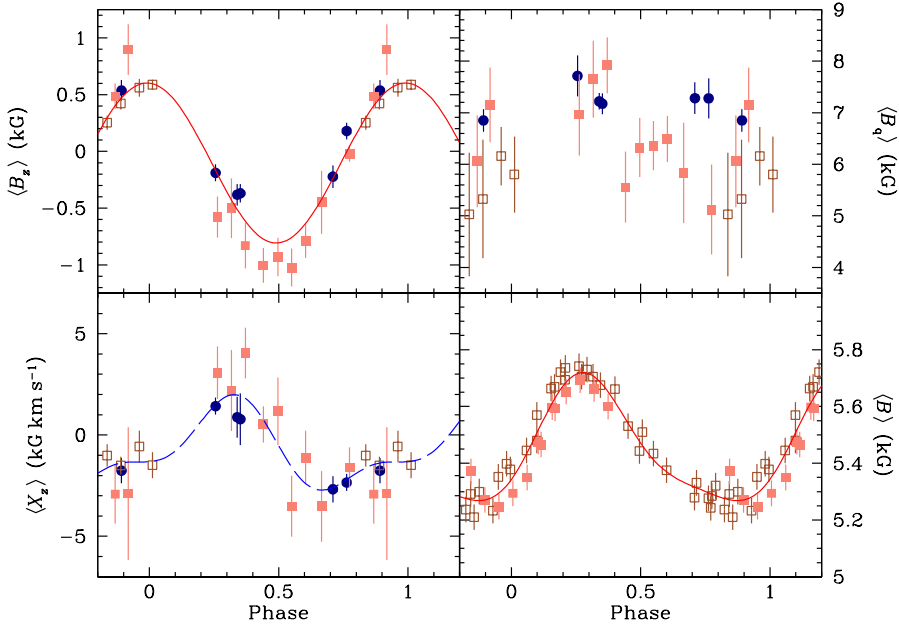}}
\parbox[t]{55mm}{
\caption{Mean longitudinal magnetic field ({\it top left\/}),
crossover ({\it bottom left\/}), 
mean quadratic magnetic field ({\it top right\/}),
and mean magnetic field modulus ({\it bottom right\/}) 
of the star HD~137909, against rotation phase. The symbols are as described at the
beginning of Appendix~\ref{sec:notes}.} 
\label{fig:hd137909}}
\end{figure*}

All our data are plotted in Fig.~\ref{fig:hd137909} against the phases
computed using Kurtz's ephemeris. The fairly large numbers of new
measurements of the mean magnetic field modulus obtained with both
AURELIE and the KPNO coud\'e spectrograph (8 and 6, respectively) 
allow the systematic differences between the values of $\Hm$ obtained
at the two sites to be better visualised. Unfortunately, a gap remains
in the coverage of the KPNO data between phases 0.37 and 0.84, which
leaves some ambiguity as to their exact nature. It
seems inescapable, however, that as already suspected in
\citetalias{1997A&AS..123..353M}, a 
somewhat different shape of the $\Hm$ variation curve is defined by
the measurements from the two instruments. This indicates that caution
is in order when interpreting the shapes of the variation curves in
terms of the actual structure of the stellar magnetic fields; this is the case not
only in HD~137909, but more generally in any star. 

However, it appears unquestionable that the variation of the field
modulus in HD~137909 has a significant degree of anharmonicity. Since
we do not know how to handle the inconsistency between the OHP
and KPNO data better, we fitted all of them together with a cosine wave and
its first harmonic. By contrast, $\Hz$, for which we
present here six new measurements, has a variation
curve that shows no significant departure from a cosine wave. The
introduction of the first harmonic in the $\Hz$ fit, driven by the
behaviour of $\Hm$, does not in the least improve it. The
anharmonic character of the crossover variation curve is more
convincing to the eye but not formally significant, and one should
note that it is, to a large extent, 
defined by the less precise old data points; the new determinations
are not conclusive in this respect. The new quadratic field
measurements do not show any significant variability (but they do not
rule it out either); the old data points were reasonably represented
by a cosine wave alone, but this cosine wave is inconsistent with the more precise
determinations of this paper. 

The new values of $\Hz$, $\xover,$ and $\Hq$ presented here were
derived from the analysis of a set of \ion{Fe}{i} and \ion{Fe}{ii} lines,
for which no significant equivalent width variation is observed.

\begin{figure}
\resizebox{\hsize}{!}{\includegraphics[angle=270]{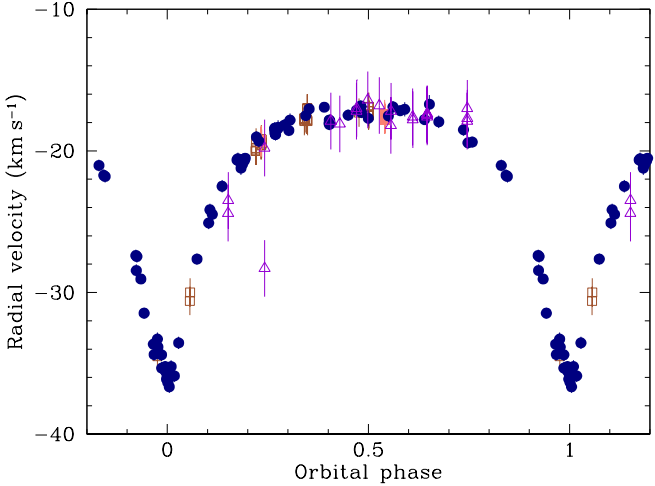}}
\caption{Our radial velocity measurements for HD~137909 are plotted
  together with those of \citet{1998A&AS..130..223N} against
  the orbital phase computed using the (final) orbital elements of
  these authors.  Dots correspond to 
  the North et al. data and  open triangles to our CASPEC
  observations; all 
  other symbols refer to our high-resolution spectra obtained with
  various instrumental configurations, as indicated
  in Table~\ref{tab:plot_sym}.}
\label{fig:hd137909_rv}
\end{figure}

\citet{1998A&AS..130..223N} refined the determination of the orbital
elements of HD~137909, which is an SB1, by combining their own radial
velocity measurements with those of previous authors. Our radial
velocity data are contemporaneous with theirs, and agree well with
them (see Fig.~\ref{fig:hd137909_rv}), but they do not set any
additional constraint on the orbit. In addition, radial velocity
oscillations, with a period of 
16.21\,min, were discovered in this star
\citep{2004MNRAS.351..663H}. 

% \clearpage

% \clearpage

% \clearpage

% \clearpage

\subsection{HD~137949}
\label{sec:hd137949}
\begin{figure*}[t]
\resizebox{12cm}{!}{\includegraphics[angle=270]{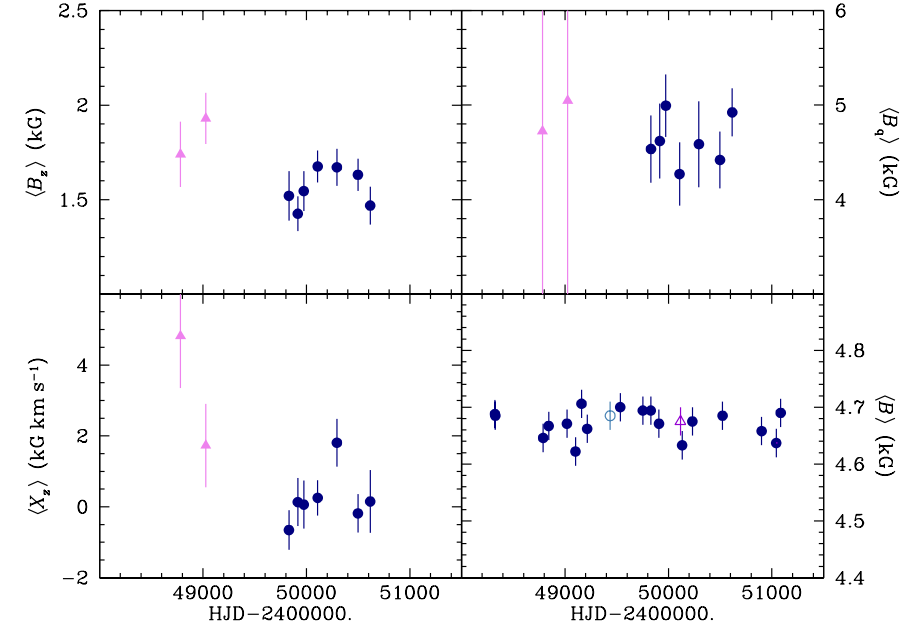}}
\parbox[t]{55mm}{
\caption{Mean longitudinal magnetic field ({\it top left\/}),
crossover ({\it bottom left\/}), 
mean quadratic magnetic field ({\it top right\/}),
and mean magnetic field modulus ({\it bottom right\/}) 
of the star HD~137949,
against heliocentric Julian date. The symbols are as described at the
beginning of Appendix~\ref{sec:notes}.}
\label{fig:hd137949}}
\end{figure*}

The standard deviation of all our mean magnetic field modulus
measurements of the roAp star HD~137949 (13 points from
\citetalias{1997A&AS..123..353M}, and 7 
new determinations reported here; see Fig.~\ref{fig:hd137949}) is
only 23\,G. Thus $\Hm$ has remained remarkably constant over the
time span covered by our 
observations, 7.5 years. This indicates that either the field modulus
of this star does not vary at all, or its rotation period must be much
longer than 7.5 years. 

 Consideration of the other observables for which we have obtained
measurements supports a long period. The equivalent widths of the
\ion{Fe}{i} lines used to diagnose the magnetic field in our CASPEC
spectra do not show any definite variation. No crossover is detected;
the formally most significant determination of this field moment,
reaches the $2.7\sigma$ level, but none of the other exceed
$1.2\sigma$. The longitudinal field and quadratic field both
appear constant over the time interval of more than 2 years separating
our first and most recent observations. As already mentioned in
\citetalias{1997A&AS..123..353M}, however, different values of $\Hz$ have been
obtained by various authors at different epochs
\citep{1958ApJS....3..141B,1971A&A....11..461V,1975ApJ...202..127W,1997A&AS..124..475M,2004A&A...415..661H}. We
gave arguments 
in \citetalias{1997A&AS..123..353M} supporting the view that no significant variation of $\Hz$
is definitely observed 
within any of those data sets taken in isolation. On the other
hand, comparison between longitudinal field values determined by
application of different diagnostic techniques to observations
obtained with different telescope and instrument combinations is always
subject to ambiguity owing to the possible existence of different
systematic errors. Assuming that they can be neglected, we suggested
in \citetalias{1997A&AS..123..353M} that $\Hz$ had been
increasing monotonically since the time of
the discovery of the magnetic field of HD~137949 by \citet{1958ApJS....3..141B}.

A recent analysis of FORS longitudinal field data by
\citet{2014A&A...572A.113L} seems to support that
interpretation. However, this conclusion rests on the 
assumption that there are no major systematic differences between the
$\Hz$ values derived from the low spectral resolution FORS observations and
those of 
\citetalias{1997A&AS..123..353M} and of other studies based on
high-resolution spectropolarimetry. The validity of
this assumption is challenged by comparison of the measurements of
\citet{2014A&A...572A.113L} with those of \citet{2014AstBu..69..427R},
which were published at almost the same time. In particular,
\citet{2014AstBu..69..427R} determined $\Hz=(1415\pm70)$\,G on
HJD\,2452333, while \citet{2014A&A...572A.113L} obtained
$\Hz=(2689\pm70)$\,G and $\Hz=(2843\pm64)$\,G from two observations
obtained with slightly different instrumental configurations on
HJD\,2452383. The difference of a factor of $\sim2$ between those
measurements performed by the two groups only 50 days apart
represents a major inconsistency if the stellar rotation period is
indeed of the order of decades. Furthermore, the value of
$\Hz=(2146\pm55)$\,G 
derived by \citet{2004A&A...415..661H} through a separate analysis of
the same FORS data strengthens the suspicion that, for HD~137949, FORS
yields systematically higher longitudinal field values than other
instruments. 

\begin{figure}
\resizebox{\hsize}{!}{\includegraphics[angle=270]{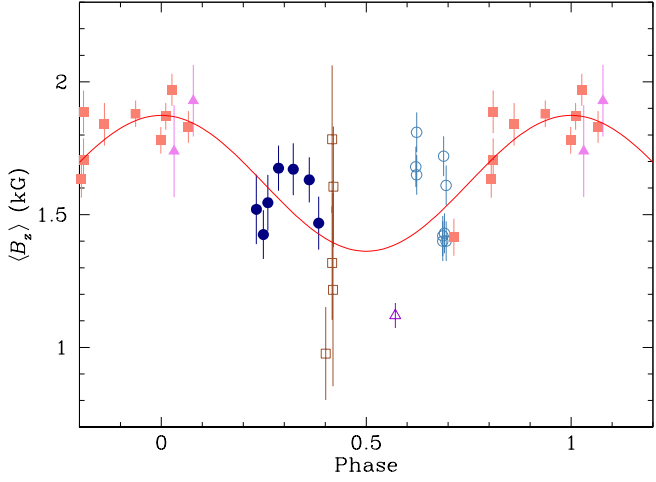}}
\caption{Mean longitudinal magnetic field of the star HD~137949
  against rotation phase assuming $\Prot=5195$\,d (see text).
The various symbols correspond to
measurements by different authors:  the open triangle indicates
\citet{1958ApJS....3..141B};   open squares indicate \citet{1971A&A....11..461V}; 
 open circles indicate  \citet{1975ApJ...202..127W};  filled triangles indicate
\citet{1997A&AS..124..475M};  dots indicate this paper; and  filled
  squares indicate \citet{2014AstBu..69..427R}.} 
\label{fig:hd137949_bz}
\end{figure}

By contrast, when combining the data of \citet{2014AstBu..69..427R}
with those of \citet{1997A&AS..124..475M}
and of this paper, a coherent picture
emerges in which the longitudinal field has passed through a minimum
between 1998 and 2002, and which hints at a period of the order of
5300 days. To seek confirmation, we carried out a period search
including the above-mentioned data sets and the earlier measurements of 
\citet{1958ApJS....3..141B}, \citet{1971A&A....11..461V}, and
\citet{1975ApJ...202..127W}. A peek stands out rather clearly in the
periodogram, corresponding to $\Prot=5195$\,d. An error bar cannot be
associated with that value given the uncertainty of the possible
existence of systematic differences between $\Hz$ measurements
obtained with different instrumental configurations (in particular,
between the determinations obtained since 1992 using CCD observations
and those from 1975 and before that relied on photographic
plates). But consideration of Fig.~\ref{fig:hd137949_bz}, where the
longitudinal field variations are plotted as a function of the
rotation phase computed using the proposed value of the period,
strengthens our confidence that this value is essentially correct,
especially in view of its consistency with all the data obtained since
1992. 

% \clearpage

Despite their overall consistency with the measurements of
\citet{2014AstBu..69..427R}, our data are difficult to reconcile with
their suggested short period, $\Prot=7\fd0187$. In particular, with
that value of the period, the
lowest value of the longitudinal field 
that we derive, $\Hz=1425$\,G, on HJD\,2449916, would be found close
to the phase of maximum of $\Hz$. The measurements of
\citet{1997A&AS..124..475M} would also be significantly
discrepant. Overall, the long period $\Prot=5195$\,d (or 14.2\,y)
represents a much 
better match to the combination of our data and of those of
\citet{2014AstBu..69..427R}.

The derived values of $\Hq$ are marginally smaller than those of $\Hm$
(by about 50\,G, in average). This non-physical difference is not
formally significant, but it probably indicates that our quadratic
field determinations for this star are slightly underestimated. They
are marginally smaller than the value obtained by
\citet{2006A&A...453..699M} from analysis of \ion{Fe}{i} line profiles
in a 
contemporaneous spectrum of higher resolution covering a broader
wavelength range.

\begin{figure}
\resizebox{\hsize}{!}{\includegraphics[angle=270]{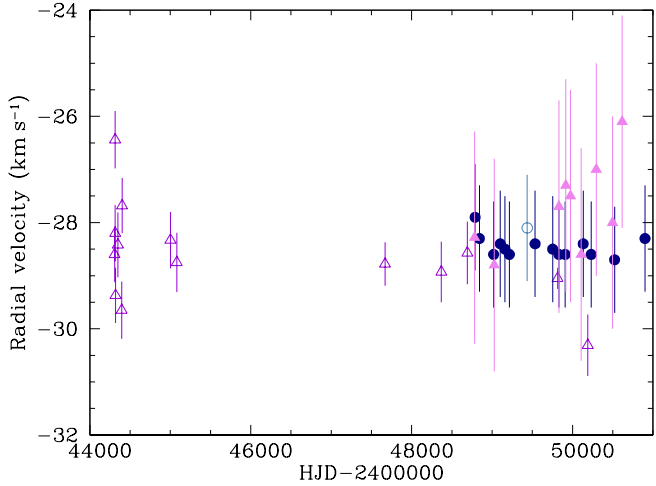}}
\caption{Our radial velocity measurements for HD~137949 are plotted
  together with those of \citet{2002A&A...394..151C}  against
  heliocentric Julian date.  Open triangles correspond to the
  \citeauthor{2002A&A...394..151C} data and  filled triangles to our CASPEC
  observations; all 
  other symbols refer to our high-resolution spectra obtained with
  various instrumental configurations, as indicated
  in Table~\ref{tab:plot_sym}.}
\label{fig:hd137949_rv}
\end{figure}

Our radial velocity data and those of \citet{2002A&A...394..151C}
are shown against Julian date in Fig.~\ref{fig:hd137949_rv}. The
measurements that we obtained from our high-resolution spectra
recorded in natural light (all with the CES) definitely do not
show the scatter reported by \citeauthor{2002A&A...394..151C}; their
standard deviation is only 0.22\,km\,s$^{-1}$. Our CASPEC data show
somewhat more scatter, but not more than expected from their more
limited accuracy. The main difference between our determinations and
those of \citeauthor{2002A&A...394..151C} is that ours rely exclusively on Fe lines,
while theirs use lines of a large number of chemical elements. This is
consistent with 
their tentative interpretation that the radial velocity variations
that they detect are due to spots coming in out and view as the star
rotates, since typically Fe is fairly uniformly distributed on the
surface of Ap stars, while other elements tend to show larger
inhomogeneities. But this interpretation cannot be reconciled with the
length of the rotation period that is inferred from the magnetic field
variations. Also, the non-negligible apparent $\vsi$
mentioned by \citeauthor{2002A&A...394..151C} is more probably
reflecting the magnetic 
broadening of the spectral lines. An alternative explanation to the
different behaviour of our radial velocity measurements and those of
\citeauthor{2002A&A...394..151C} is that the scatter of the latter (and also the
variability of the width of the correlation dip) is caused by
pulsation. This is made plausible by the fact that, in this star (as
in many roAp stars), the amplitude of the radial velocity variations
due to pulsation is considerably smaller for Fe than for a number of
other elements \citep{2005MNRAS.358L...6K}. However the magnitude of
the scatter of the \citeauthor{2002A&A...394..151C} data seems too large with respect to
the amplitudes observed by Kurtz et al. in individual lines,
especially if one takes into account the fact that, in the Carrier et
al. case, the signal is diluted by the lines that show lower
amplitude variations, or no variation at all, and (probably) by the
fact that the integration time represents a non-negligible fraction of
the period of the stellar pulsation. One would have to assume that the
pulsation amplitude varies with time, which is actually observed in
other stars and, hence, cannot be ruled out. But it will need to be
established by future observations.

\subsection{HD~142070}
\label{sec:hd142070}
The value of the rotation period of HD~142070 published in \citetalias{1997A&AS..123..353M} was
erroneous because the Julian dates of two of the magnetic field
modulus measurements were wrong by one day. The dates appearing in
Table~4 of \citetalias{1997A&AS..123..353M} are correct, but the correction took place after
the period was determined. With the correct dates, and the two new
$\Hm$ determinations presented here, a revised value of the period is
obtained, $\Prot=(3.3718\pm0.0011)$\,d. This value is in excellent
agreement with the value derived by \citet{2001A&A...368..225A} from the
analysis of photometric data, $\Prot=(3.37189\pm0.00007)$\,d. Although
the latter is formally more accurate, in what follows we use the
value of the period that we obtained because the variation of the
magnetic field is much better defined; that is, its amplitude is much
larger with respect to the individual measurement errors. 

\cite{2014AstBu..69..427R} also report that their longitudinal field
measurements are compatible with $\Prot=3\fd3719$. The $\Hz$ values
that they derive appear to be systematically shifted by $\sim-150$\,G
with respect to ours, as a result of the usage of different
instruments and line samples for their determination. 

\begin{figure*}
\resizebox{12cm}{!}{\includegraphics[angle=270]{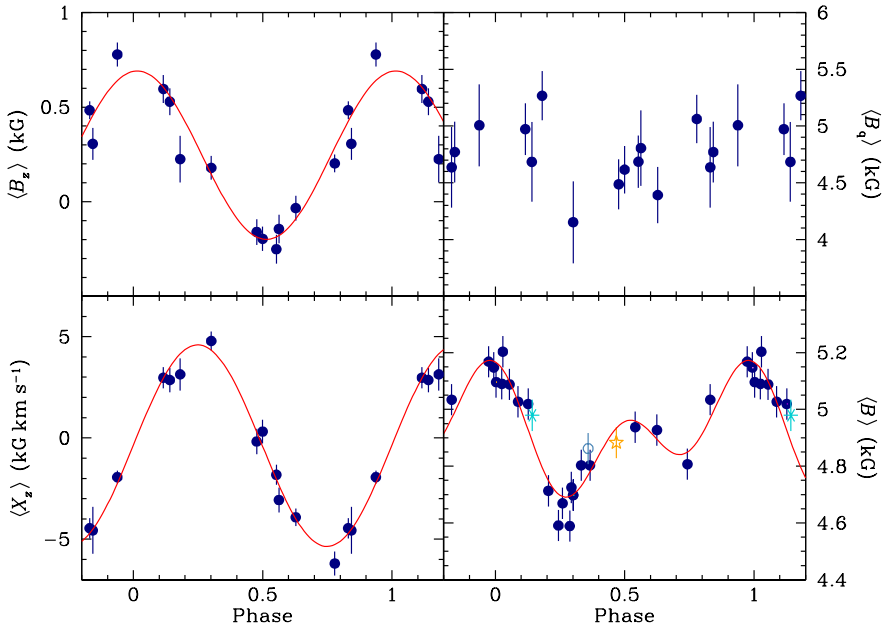}}
\parbox[t]{55mm}{
\caption{Mean longitudinal magnetic field ({\it top left\/}),
crossover ({\it bottom left\/}), 
mean quadratic magnetic field ({\it top right\/}),
and mean magnetic field modulus ({\it bottom right\/}) 
of the star HD~142070, against rotation phase. The symbols are as described at the
beginning of Appendix~\ref{sec:notes}.} 
\label{fig:hd142070}}
\end{figure*}

In \citetalias{1997A&AS..123..353M} we inferred from consideration of
the fairly short rotation 
period of HD~142070 and of its low $\vsi$ 
that the inclination of its rotation axis over the line of sight
cannot exceed 8\degr. It is particularly remarkable that, in spite of
this low inclination, all field moments but the quadratic field show
very well-defined variations with significant amplitude (see
Fig.~\ref{fig:hd142070}). The 
longitudinal field reverses its sign: as the star rotates, both
magnetic poles come alternatively into view. This implies that the
magnetic axis must be very nearly perpendicular to the rotation
axis. The crossover is large as well. The variation curve of $\Hm$ is
strongly anharmonic, although the scarcity of measurements between
phases 0.6 and 0.0 implies that its shape is not fully
constrained. The variation curves of $\Hz$ and $\xover$ do not
significantly depart from cosine waves. They are in phase quadrature
with each other, within the limits of the precision of the phase origin
determinations. The positive extremum of $\Hz$ coincides in phase with
the primary maximum of $\Hm$; to the extent that the phase of the
secondary maximum of $\Hm$ is constrained, it seems to occur close to
the negative extremum of $\Hz$. Variations of the quadratic field are
not definitely detected. As in several other stars, this field moment
appears to be somewhat underestimated, since it tends at most phases
to be marginally smaller than the field modulus.

The longitudinal field, the crossover and quadratic field were
determined through analysis of a set of \ion{Fe}{ii} lines. The
equivalent widths of these lines do not appear to be variable.

\begin{figure}
\resizebox{\hsize}{!}{\includegraphics[angle=270]{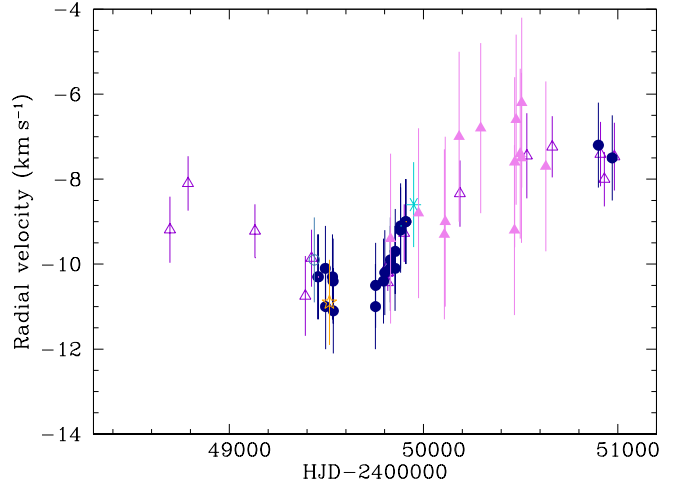}}
\caption{Our radial velocity measurements for HD~142070 are plotted
  together with those of \citet{2002A&A...394..151C}  against
  heliocentric Julian date.  Open triangles correspond to the
  \citeauthor{2002A&A...394..151C} data and  filled triangles to our CASPEC
  observations; all 
  other symbols refer to our high-resolution spectra obtained with
  various instrumental configurations, as indicated
  in Table~\ref{tab:plot_sym}.}
\label{fig:hd142070_rv}
\end{figure}

In \citetalias{1997A&AS..123..353M}, we had found that the radial velocity of HD~142070
is variable. \citet{2002A&A...394..151C} confirmed this
discovery. Their measurements are plotted together with ours in
Fig.~\ref{fig:hd142070_rv}. The two sets are contemporaneous and there
is good agreement between them. Our data bracket the radial velocity
minimum more narrowly than the data of \citeauthor{2002A&A...394..151C}, and accordingly our data
define somewhat better the shape of the radial velocity curve
around that phase. But our data do not provide new constraints about the
orbital period, and the conclusion by \citeauthor{2002A&A...394..151C} that it must be
of the order of 2500\,d or longer remains valid. 

\subsection{HD~144897}
\label{sec:hd144897}
As for HD~142070, the value of the rotation period derived in
\citetalias{1997A&AS..123..353M} for HD~144987 is affected by a one-day error in the Julian dates
of two of the mean magnetic field modulus measurements. The dates
appearing in Table~4 of \citetalias{1997A&AS..123..353M} are correct, but the correction took
place after the period was determined. After correction of these two
dates, and addition of the two new $\Hm$ measurements reported in this
paper, a revised value of the period is obtained,
$\Prot=(48.57\pm0.15)$\,d. 

\begin{figure*}
\resizebox{12cm}{!}{\includegraphics[angle=270]{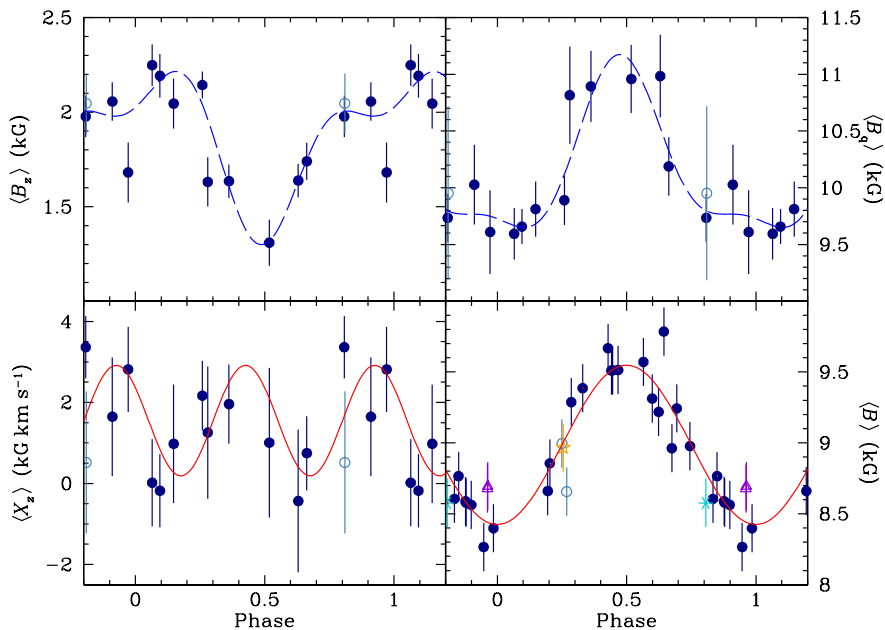}}
\parbox[t]{55mm}{
\caption{Mean longitudinal magnetic field ({\it top left\/}),
crossover ({\it bottom left\/}), 
mean quadratic magnetic field ({\it top right\/}),
and mean magnetic field modulus ({\it bottom right\/}) 
of the star HD~144897, against rotation phase. The symbols are as described at the
beginning of Appendix~\ref{sec:notes}.} 
\label{fig:hd144897}}
\end{figure*}

This value also matches the variations of
the other field moments, as can be seen in Fig.~\ref{fig:hd144897},
where it is used to compute the phases. While a cosine wave represents
an excellent approximation to the behaviour of the field modulus, the
addition of the first harmonic considerably improves the quality of
the fits to the variation curves of the longitudinal field and quadratic field (for which
the data set includes one observation of \citealt{1997A&AS..124..475M}),
both visually and from consideration of $\chi^2/\nu$, even though the
fitted amplitudes of the harmonic are just below the limit of formal
significance. 

Only 1 of the 13 determinations of the
crossover is significantly non-zero, but it is intriguing that all but 2
yield positive values. Furthermore if the measurement of
\citet{1997A&AS..124..475M}, which may be of lower
quality as 
in a number of other stars, is discarded, a fit by a cosine wave with
twice the rotation frequency (that is, a first harmonic alone) appears
significant with a computed variation amplitude at the
$4.4\sigma$ level. The parameters appearing in Table~\ref{tab:xfit}
and the $\xover$ curve plotted in Fig.~\ref{fig:hd144897} correspond
to this case. This is further discussed in Sect.~\ref{sec:xdisc}.

The longitudinal field is always positive. The fact that its primary
minimum very nearly coincides in phase with the maxima of the field
modulus and of the quadratic field indicates that its structure must
depart significantly from a centred dipole, as already suspected from
the probable anharmonicity of the variation curves of several field
moments.

The lines of \ion{Fe}{ii}, which were used to diagnose the magnetic
field from the CASPEC spectra, do not show any significant equivalent
width variation over the stellar rotation period.

\begin{figure}
\resizebox{\hsize}{!}{\includegraphics[angle=270]{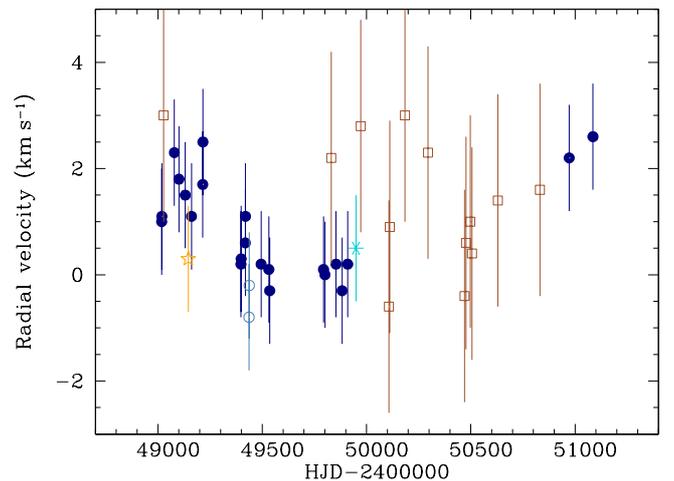}}
\caption{Our radial velocity measurements for HD~144897 are plotted
  against heliocentric Julian date.  Open squares
  correspond to our CASPEC observations; all 
  other symbols refer to our high-resolution spectra obtained with
  various instrumental configurations, as indicated
  in Table~\ref{tab:plot_sym}.}
\label{fig:hd144897_rv}
\end{figure}

Figure~\ref{fig:hd144897_rv} shows that
the radial velocity of HD~144897 has been slowly decreasing from the
time of our first observation of the star until the seasonal
observability gap of end of 1994 and early 1995; since then until
September 1998, it appears to have been increasing monotonically (even
though the larger uncertainties of the CASPEC measurements tend to
confuse the picture). The overall amplitude of the variations observed
until now is of the order of 3\,km\,s$^{-1}$: the star appears to be a
spectroscopic binary with an orbital period in excess of six 
years. To the best of our knowledge, the binarity of HD~144897 is
reported here for the first time.

% \clearpage

% \clearpage

\begin{figure}
\resizebox{\hsize}{!}{\includegraphics[angle=270]{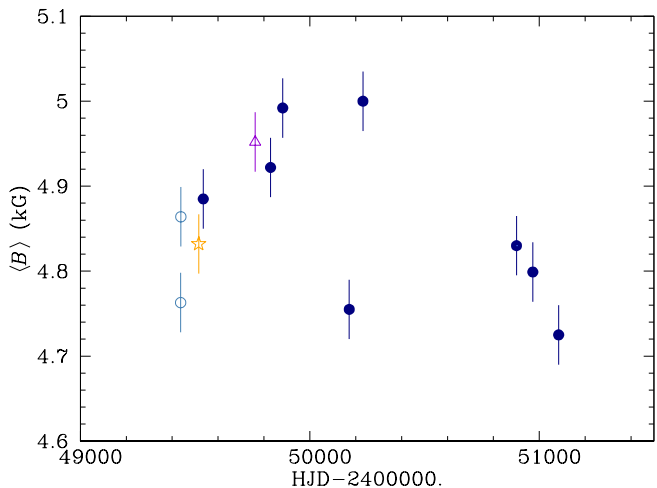}}
\caption{Mean magnetic field modulus of the star HD~150562,
against heliocentric Julian date. The symbols are as described at the
beginning of Appendix~\ref{sec:notes}.}
\label{fig:hd150562}
\end{figure}

% \clearpage

\subsection{HD~150562}
\label{sec:hd150562}
The presence of magnetically resolved split lines in the spectrum of
the little-studied roAp star HD~150562 was announced in \citetalias{1997A&AS..123..353M}. The
seven $\Hm$ measurements of this paper are plotted together with five
new determinations in Fig.~\ref{fig:hd150562}. All data points but one
convincingly indicate that the rotation period must be longer than the
time range covered by our observations, which is 4.5 years. Most likely, the
deviating measurement, obtained on HJD 2450171.802, is wrong and it
should be ignored, although we cannot clearly identify a reason for
its bad quality; an unrecognised cosmic ray event in one of the components
of the line \ion{Fe}{ii}\,$\lambda\,6149$ seems the most plausible
source of error. 

The maximum of the field modulus of HD~150562 appears close to
5.0~kG. Only one spectropolarimetric observation of the star could be
obtained, revealing a positive longitudinal field and a quadratic
field that must have been of the order of 10\% greater than the field
modulus around the same phase. No crossover is detected, which is consistent
with a rotation period of several years. The available observations
are insufficient to draw definite conclusions about the possible
variability of the radial velocity, or of the equivalent widths of the
\ion{Fe}{i} lines that were used for determination of the longitudinal
field, the crossover, and the quadratic field.

\begin{figure*}
\resizebox{12cm}{!}{\includegraphics[angle=270]{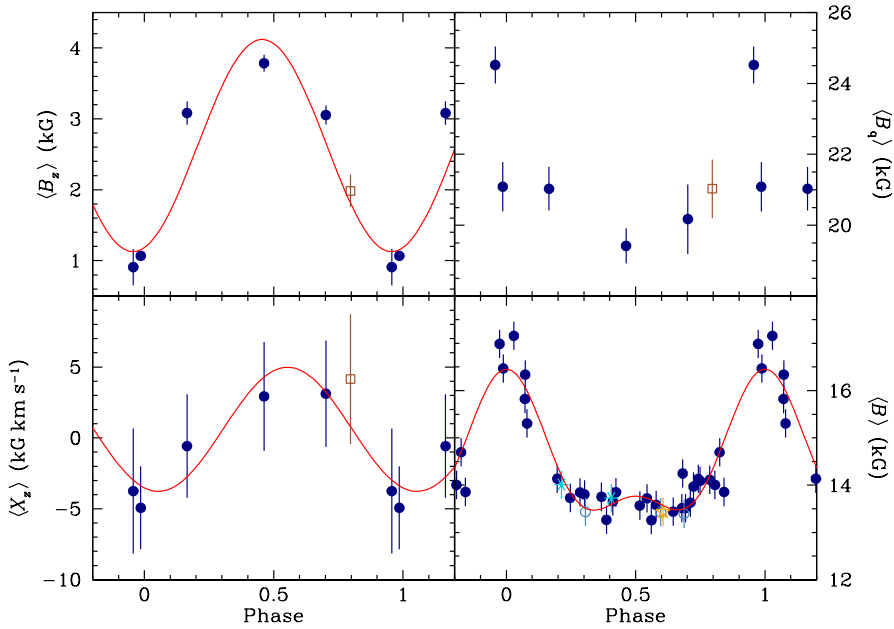}}
\parbox[t]{55mm}{
\caption{Mean longitudinal magnetic field ({\it top left\/}),
crossover ({\it bottom left\/}), 
mean quadratic magnetic field ({\it top right\/}),
and mean magnetic field modulus ({\it bottom right\/}) 
of the star HDE~318107, against rotation phase. The symbols are as described at the
beginning of Appendix~\ref{sec:notes}.} 
\label{fig:hd318107}}
\end{figure*}

\begin{figure*}
\resizebox{12cm}{!}{\includegraphics[angle=270]{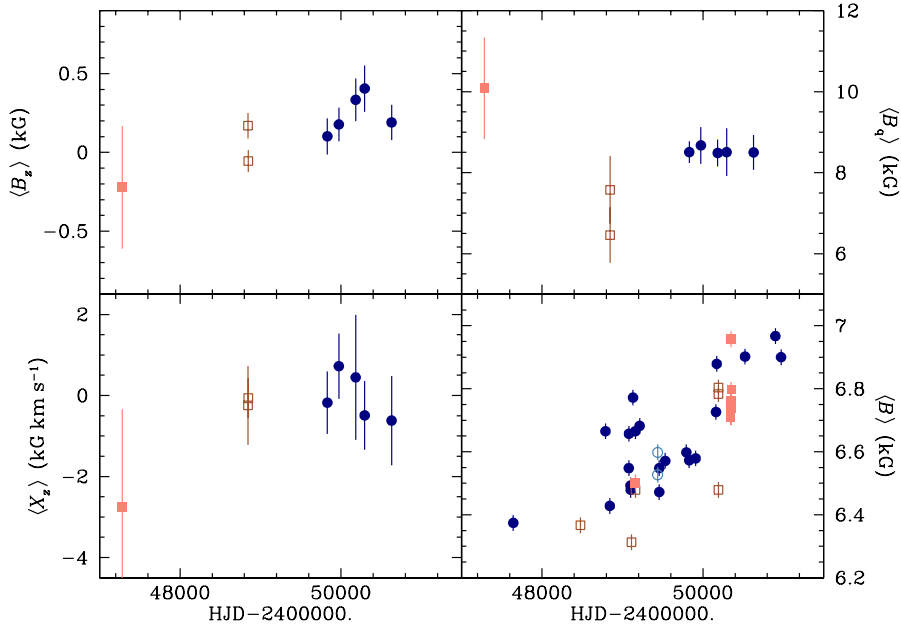}}
\parbox[t]{55mm}{
\caption{Mean longitudinal magnetic field ({\it top left\/}),
crossover ({\it bottom left\/}), 
mean quadratic magnetic field ({\it top right\/}),
and mean magnetic field modulus ({\it bottom right\/}) 
of the star HD~165474,
against heliocentric Julian date. The symbols are as described at the
beginning of Appendix~\ref{sec:notes}.}
\label{fig:hd165474}}
\end{figure*}

\subsection{HDE~318107}
\label{sec:hd318107}
A detailed study of the variations of HDE~318107 was carried out by
\citet{2000A&A...364..689M}, who succeeded in establishing the
value of the rotation period of the star,
$\Prot=(9.7085\pm0.0021)$\,d. This value was subsequently refined to
$\Prot=(9.7088\pm0.0007)$\,d by \citet{2011A&A...535A..25B}, from
consideration of additional magnetic field measurements. This improved
value is used here. 

The study of \citet{2000A&A...364..689M} was based in part on a revision
of the values of the mean magnetic field modulus derived in \citetalias{1997A&AS..123..353M},
as well as on the four new measurements presented here. The revision of
part of the \citetalias{1997A&AS..123..353M} data was carried out in an attempt to measure the line
\ion{Fe}{ii}\,$\lambda\,6149$ in a more uniform manner at all phases,
which is an endeavour complicated by its variable strong blending, as well as for some observations, by the 
fairly low S/N ratio of the spectra. The revised measurements are
included in Table~\ref{tab:hm}. 

All the measurements of the mean magnetic field modulus, as well as
the values of the other field moments derived from spectropolarimetric
observations (including one set from \citealt{1997A&AS..124..475M}) are
plotted against rotation phase in Fig.~\ref{fig:hd318107}. The
variation curve of $\Hm$ is definitely anharmonic with a broad,
almost flat minimum over about half of the rotation period. The shapes
of the variation curves of the other field moments are not well
defined owing to the small number of data points. The longitudinal field
appears to be always positive with a phase of minimum close to the
phase of maximum of the field modulus. As already pointed out for
other stars, this is indicative of a field structure significantly
departing from a centred dipole. 

None of the individual crossover
determinations are significant (there is only one above the $1\sigma$
level), but when plotted against phase they seem to line up along a
cosine wave, and the amplitude coefficient of a fit by such a wave is 
formally significant, just at the $3\sigma$ level. The multiple
correlation coefficient is also rather high, $R=0.87$.

The quadratic
field is nearly constant over the rotation period with the exception
of one measurement being considerably higher than the others. The most
likely explanation is that this determination is spurious (without
obvious reason, though), but the proximity of this point to the phase
of maximum of the field modulus, and the steep variation of the latter
around this phase do not allow one to rule out the reality of the
considered $\Hq$ value. 

The lines of \ion{Fe}{ii}, from which $\Hz$, $\xover,$ and $\Hq$ are
determined do not show any definite equivalent width variations.

\begin{figure}
\resizebox{\hsize}{!}{\includegraphics[angle=270]{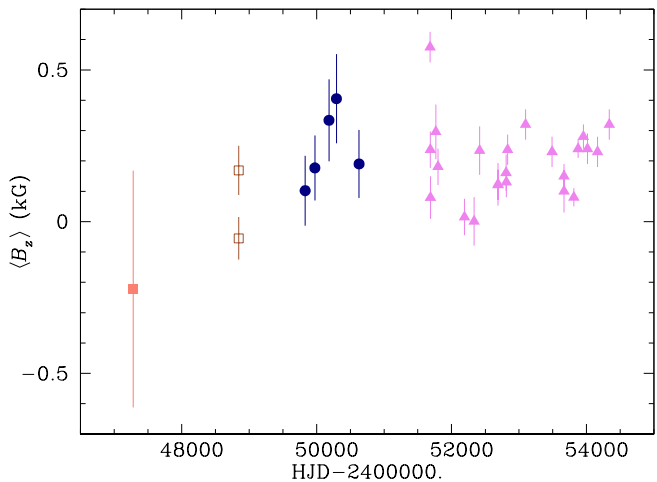}}
\caption{Mean longitudinal magnetic field of the star HD~165474 against
  heliocentric Julian date. The different symbols correspond to
  measurements by different authors: the filled square indicates
  \citet{1994A&AS..108..547M}; open squares indicate
  \citet{1997A&AS..124..475M}; dots indicate this paper; and filled triangles indicate
  \citet{2014AstBu..69..427R}.} 
\label{fig:hd165474_bz}
\end{figure}

\begin{figure}
\resizebox{\hsize}{!}{\includegraphics[angle=270]{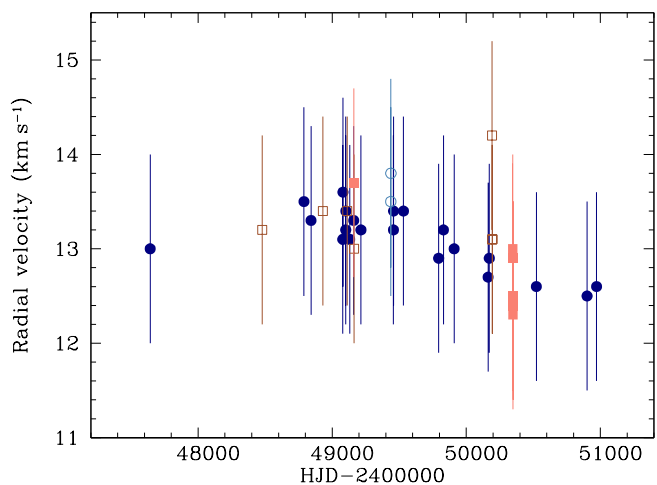}}
\caption{Our radial velocity measurements for HD~165474 are plotted
  against the heliocentric Julian date. All
  symbols refer to our high-resolution spectra obtained with
  various instrumental configurations, as indicated
  in Table~\ref{tab:plot_sym}.}
\label{fig:hd165474_rv}
\end{figure}

\subsection{HD~165474}
\label{sec:hd165474}
The tentative value of 2\fd54065\,d proposed in \citetalias{1997A&AS..123..353M} for the
rotation period of HD~165474, which more recently seemed borne out by 
independent measurements of \citet{2002A&A...395..549N}, is not
confirmed by the 14 new measurements of its mean magnetic field
modulus that are reported here, and no other short period
  emerges from the analysis of the whole set of $\Hm$ data. When all
our measurements of this 
field moment are plotted against the Julian date
(Fig.~\ref{fig:hd165474}), a slow monotonic increase is clearly seen,
from our first point to our last one, indicative of a period
considerably longer than the time interval separating them, of nine
years. The evidence for a long period becomes even stronger if one
allows for the existence of a systematic difference of the $\Hm$ 
values derived from AURELIE data with respect to those measured in
spectra recorded with other instruments, as observed in a number of
stars. 

Even after taking that shift into account, the scatter of the individual
measurements about a smooth variation curve is somewhat higher than would be
expected from their adopted uncertainty. There is no obvious reason to
suspect that the latter has been underestimated: the star is fairly
bright, so that almost all spectra have high S/N ratio; the
components of the \ion{Fe}{ii}\,$\lambda\,6149$ line are well
resolved, to such an extent that the line goes back up all the way to
the continuum between them (HD~165474 is one of the very few stars
where this is observed); their profiles are very regular and clean; and, in
particular, there is no hint at all of the blend that hampers the
measurement of the blue component in many other stars. The case is
somewhat similar to that of HD~55719. For the time being we have no
explanation for the higher than expected scatter of the individual
$\Hm$ determinations about a smooth variation curve. This, however,
does not question the reality of the variation. From the first to the
last observation, the field modulus has increased by approximately
500\,G, while once the AURELIE measurements have been shifted to make
them consistent with the data obtained with other instruments, the rms
of the points about a smooth variation trend is only of the order of 100\,G.

\begin{figure*}
\resizebox{12cm}{!}{\includegraphics[angle=270]{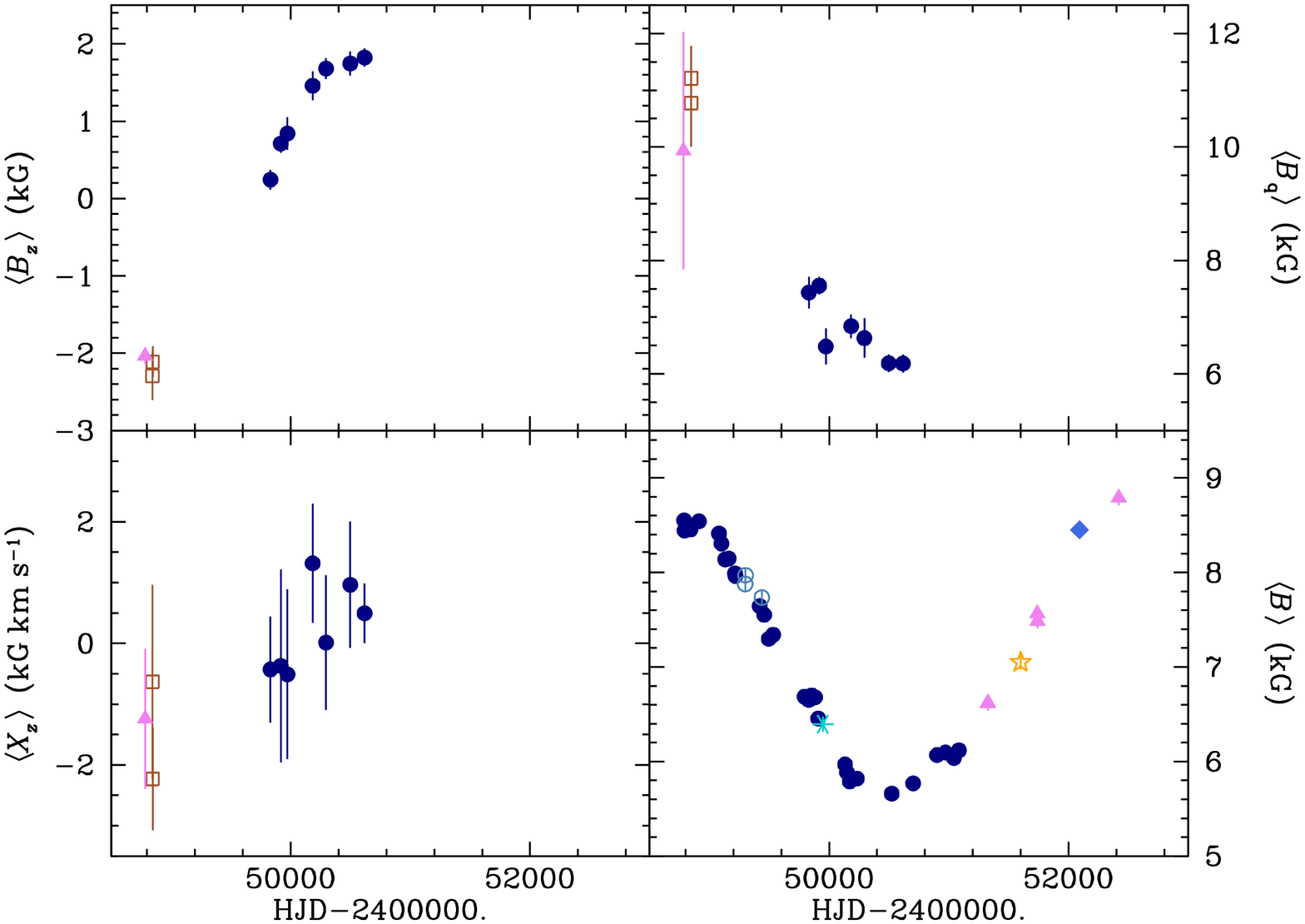}}
\parbox[t]{55mm}{
\caption{Mean longitudinal magnetic field ({\it top left\/}),
crossover ({\it bottom left\/}), 
mean quadratic magnetic field ({\it top right\/}),
and mean magnetic field modulus ({\it bottom right\/}) 
of the star HD~166473,
against heliocentric Julian date. A diamond is used to
represent the measurement of \citet{2007MNRAS.380..181M} obtained
with the UVES spectrograph at Unit Telescope 2 of ESO's Very Large
Telescope. All other additional measurements of these authors are
based on observations performed with instrumental configurations
similar to those used for the present study; the same symbols are used
to identify them as in the rest of this paper (see beginning of
Appendix~\ref{sec:notes}.} 
\label{fig:hd166473}}
\end{figure*}

 Several other arguments support the view that the rotation period of
HD~165474 is long. With the variation trend observed in the data
discussed here, it is not implausible that,  at a later time, the field modulus may have
reached a value close to the 7.2\,kG value determined
by \citet{1971ApJ...164..309P} from the spectrum from which he discovered
resolved magnetically split lines in the
star. \citet{2002A&A...395..549N} did not publish their individual
$\Hm$ measurements,  
from spectra recorded with the SOFIN spectrograph at the Nordic
Optical Telescope, but
one can see from their Fig.~8  that they are all concentrated around a
mean value of 6.6\,kG. They were all obtained in a ten night interval
centred on HJD\,2451769, that is, a little more than two years after our
last measurement. If there is no systematic shift between the data from Nielsen \&
Wahlgren and our data (which should not be taken for granted),
their combination suggests that the field modulus of HD~165474 may
have been through a maximum during 1999, and from then started to
decrease with a significantly steeper slope than it had increased
between 1989 and 1998. However, this will have to be confirmed because of the uncertain consistency of measurements obtained with
different instruments.

None of the individual determinations of $\Hz$ presented here
yield formally significant non-zero values. The older measurements of
\citet{1994A&AS..108..547M} and \citet{1997A&AS..124..475M} did not yield
any definite detection either. The corresponding $3\sigma$ upper
limits of the longitudinal field range from 240\,G to 450\,G (except
for the measurement performed by \citet{1994A&AS..108..547M} of a
specrtrum obtained in 1988, which is less precise). The more recent
$\Hz$ measurements of \citet{2014AstBu..69..427R} have smaller error
bars, and for 13 of them (i.e. a little more than half), a formally
significant detection (at the $3\sigma$ level) is achieved. 

The tentative value of the period proposed by
\citet{2014AstBu..69..427R}, $\Prot=24\fd38$, is definitely ruled out
by our mean field modulus data. As a matter of fact, the scatter of
the $\Hz$ values obtained by those authors is mostly consistent with
the error bars; the occurrence of variations cannot be firmly
established from consideration of their data set alone. We plotted
their measurements together with ours against observation dates in
Fig.~\ref{fig:hd165474_bz}. Even allowing for possible systematic differences between
$\Hz$ values derived from observations with different telescopes and
instruments, the trend seen in this figure is consistent with the slow
variability of the field over a very long period that we inferred from
consideration of the field modulus.

Moreover, \citet{1958ApJS....3..141B} had measured a longitudinal field of
900\,G in a spectrum recorded in 1957. The difference between this
value and the more recent data shown in Fig.~\ref{fig:hd165474_bz}
strengthens the 
argument favouring a long
period. 

This argument is also consistent with the fact that no crossover
is detected and with the behaviour of the quadratic field. The
determinations of the quadratic field presented in this paper, which are much
more reliable than those of previous work, show it to be remarkably
constant over the time interval April 1995--June 1997. The ratio
between it and contemporaneous values of $\Hm$ is of the order of
1.25. This suggests that, unless the field structure is unusually
tangled, a sizeable mean longitudinal field must be observable
over a significant fraction of the rotation period. This in turn
represents additional evidence that the rotation period of HD~165474
is very long, probably of the order of several decades. If, as
hypothesised above, the value reported by \citet{1971ApJ...164..309P}
was close to the maximum of $\Hm$ and the star went through another
maximum in 1999, it is not implausible that these two maxima may
correspond to consecutive cycles, so that the rotation period may be of
the order of the $\sim30$ years elapsed between them. 

Not surprisingly, if the period is very long, no variations of
equivalent widths of the \ion{Fe}{i} and \ion{Fe}{ii} lines used for
diagnosis of the magnetic field from CASPEC data are observed
over the 2.2 y time interval covered by them. This does not set any
strong constraints on the distribution of the element on the surface of
HD~165474. 

\begin{figure}
\resizebox{\hsize}{!}{\includegraphics[angle=270]{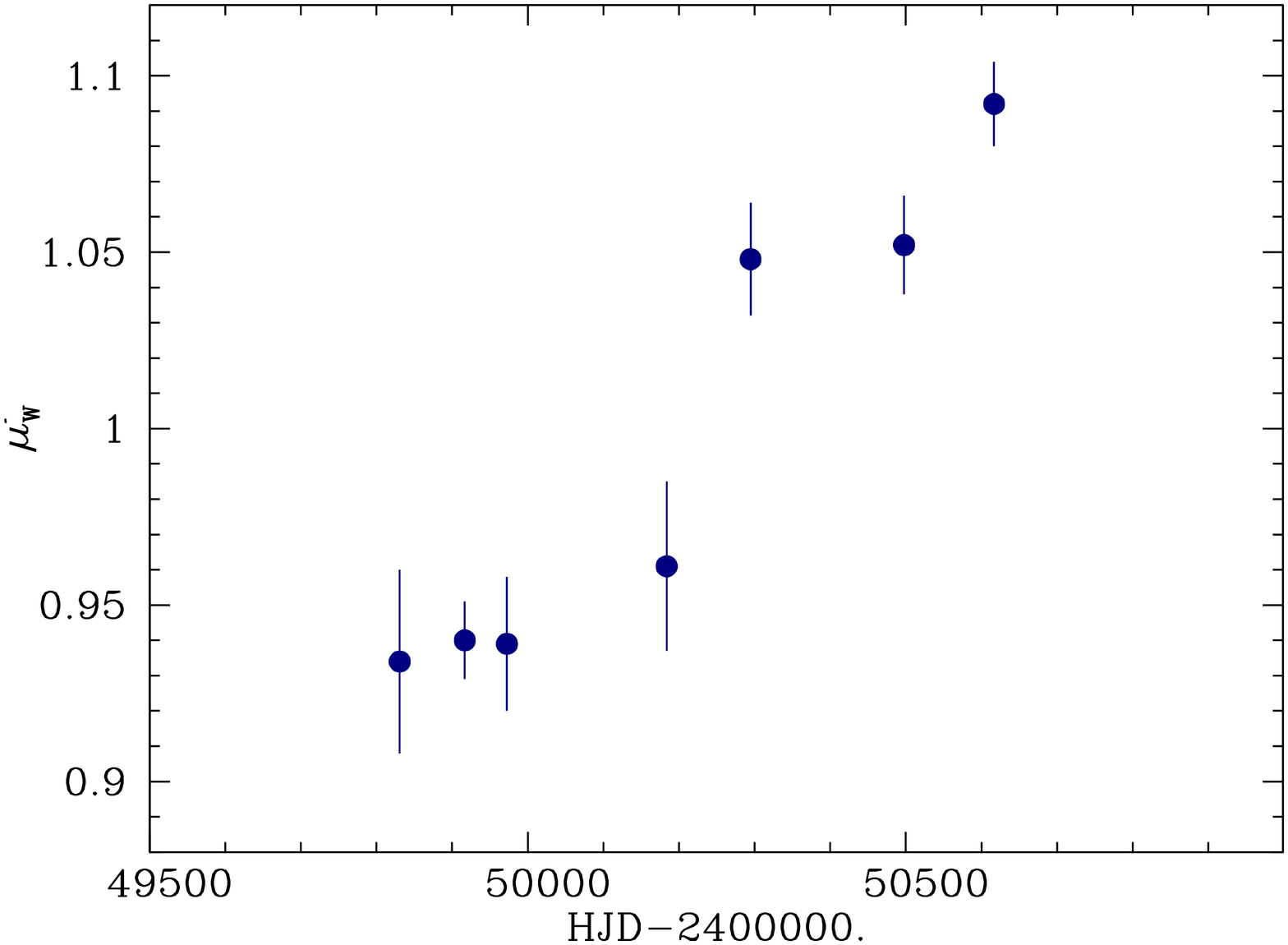}}
\caption{Variation with time of the average $\mu^\prime_W$
  of the normalised equivalent widths \citep{1994A&AS..108..547M} of the
  \ion{Fe}{i} lines analysed in the CASPEC spectra of
  HD~166473. }
\label{fig:hd166473_ew}
\end{figure}

 Figure~\ref{fig:hd165474_rv} shows that the radial velocity of
HD~165474 has been decreasing very slowly from June 1992 (third data
point from the left) to June 1998 (last point on the right). Over this
six year time interval, the total amplitude of the radial velocity
variation has barely reached 1\,km\,s$^{-1}$. The figure only shows
the measurements performed on the high-resolution spectra recorded in
natural light. The CASPEC data are not included because their larger
uncertainties tend to blur the picture on visual inspection; but they
are fully consistent with the above-described variation. The fact that
such a minute progressive variation can be followed over a time span
of six years, from analysis of data obtained with various instrumental
configurations at different telescopes, demonstrates a remarkable
stability and mutual consistency of these systems. In particular, the
exceedingly small scatter about a smooth trend of the measurements
obtained with the ESO Coud\'e Auxiliary Telescope and the Long Camera
of the Coud\'e Echelle Spectrograph is a testimony to the exquisite
performance of this instrumental configuration. However, the two data
points obtained before June 1992 are insufficient to characterise the
behaviour of HD~165474 before that time. It is not implausible that
its radial velocity went through a maximum shortly before or around
June 1992, but one cannot rule out either the possibility that the
slow decrease seen later had started much earlier, and that the two
pre-June 1992 measurements are just unfortunate outliers. In any
event, it appears unquestionable that the star is a spectroscopic
binary with an orbital period considerably longer than six
years. HD~165474 has
a visual companion, HD~165475, which is a fast rotating, superficially normal A
star, located approximately $5"$ apart. It is unclear if they are
physically related, and if so, whether the observed radial velocity
variation corresponds to the orbit of the visual binary, or if it is
caused by
another, unseen component, which had not been detected before. 

\subsection{HD~166473}
\label{sec:hd166473}
Ten new measurements of the mean magnetic field modulus of the roAp
star HD~166473 are presented here, showing that it went through a
minimum, of the order of 5.7\,kG, between March and September
1997. This can be seen in Fig.~\ref{fig:hd166473}, where more recent
measurements \citep{2007MNRAS.380..181M} are also plotted, suggesting
that the rotation period may be close to ten years, although this
cannot be regarded as definitely established. In particular, the most
recent value is 250\,G higher than the first value, which is a formally
significant difference.

In addition, seven determinations of $\Hz$, $\xover,$ and $\Hq$ were obtained, which complement earlier measurements of
\citet{1997A&AS..124..475M}, showing in particular that the
longitudinal 
field reverses its polarity, hence that both magnetic poles of the
star come into view as it rotates. 

\citet{2007MNRAS.380..181M} have discussed in detail the variations
of all four field moments. We  point out a couple of
remarkable features here. The
amplitude of variation of $\Hm$ is larger 
than in most stars with a ratio of 1.55 between the highest and lowest values determined from the observations obtained until now. On
the other hand, the ratio between the quadratic field and the field
modulus is found to be of the order of 1.00 around minimum and 1.25
around maximum. The latter value, however, may be affected by a large
uncertainty, owing to the lower quality of the quadratic field
determinations of \citet{1997A&AS..124..475M}. 

The equivalent widths of the \ion{Fe}{i} lines, from which the
longitudinal field, the crossover, and the quadratic field are
determined, are
definitely variable. This is illustrated in
Fig.~\ref{fig:hd166473_ew}, where the average $\mu^\prime_W$ is
plotted against Julian date. The number of observations is
insufficient to characterise the shape of the variation, but its
extrema do not seem to coincide with those of the field modulus. In
any case, one
should bear in mind that the magnetic field moments involve a
weighting by the inhomogeneous distribution of \ion{Fe}{i} over the
stellar surface.

\begin{figure}
\resizebox{\hsize}{!}{\includegraphics[angle=270]{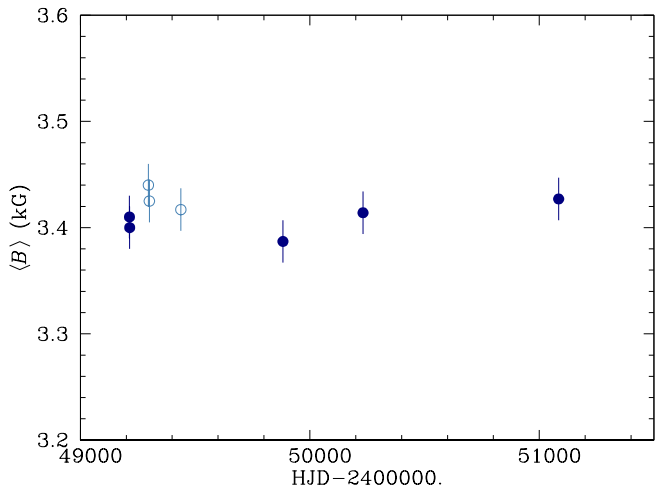}}
\caption{Mean magnetic field modulus of the star HD~177765,
against heliocentric Julian date. The symbols are as described at the
beginning of Appendix~\ref{sec:notes}.}
\label{fig:hd177765}
\end{figure}

% \clearpage

% \clearpage

\subsection{HD~177765}
\label{sec:hd177765}

\begin{figure*}
\resizebox{12cm}{!}{\includegraphics[angle=270]{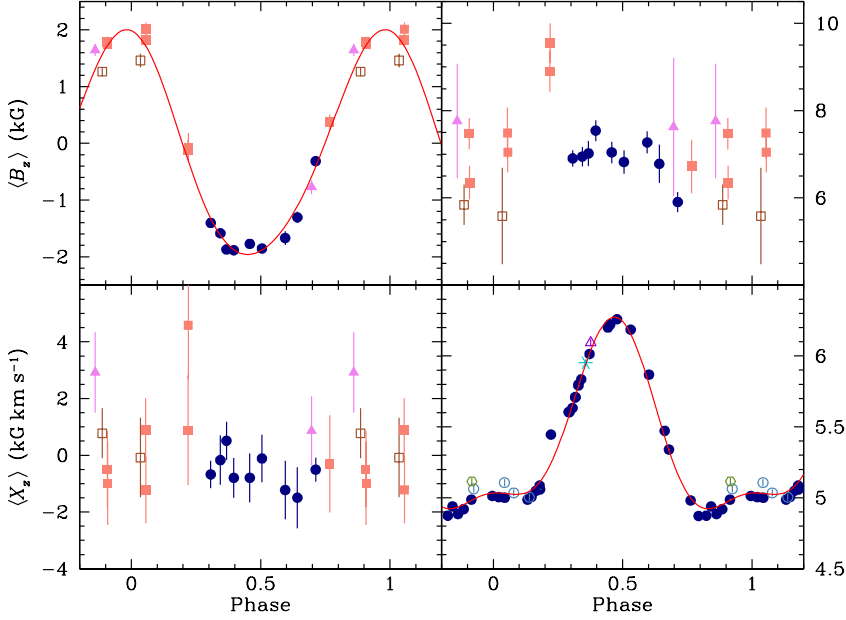}}
\parbox[t]{55mm}{
\caption{Mean longitudinal magnetic field ({\it top left\/}),
crossover ({\it bottom left\/}), 
mean quadratic magnetic field ({\it top right\/}),
and mean magnetic field modulus ({\it bottom right\/}) 
of the star HD~187474, against rotation phase. The symbols are as described at the
beginning of Appendix~\ref{sec:notes}.} 
\label{fig:hd187474}}
\end{figure*}

The two new values of the mean magnetic field modulus of HD~177765
presented here do not significantly differ from those of \citetalias{1997A&AS..123..353M}. We
have now obtained a total of eight measurements of this star, between
August 1993 and September 1998 (see Fig.~\ref{fig:hd177765}), and
their standard deviation about 
their mean is only 17\,G. This is the smallest standard deviation
derived for any of the stars studied in this paper. It must be very
close to the ultimate precision achievable in our determinations of
$\Hm$. As far as the star itself is concerned, the total lack of
detectable variations most likely indicates that its rotation period
is much longer than the time span over which it has been observed so
far, which is five years. The recent measurement of a significantly higher mean
field modulus, $\Hm=3550$\,G, in a spectrum obtained in June 2010 by 
\citet{2012MNRAS.421L..82A} fully supports this view. It is very possible that HD~177765 may have a rotation period exceeding the time
interval, 17 years, between our first $\Hm$ measurement and the recent observation by 
\citeauthor{2012MNRAS.421L..82A}  These authors also
reported the detection 
of rapid oscillations in this star with the longest pulsation period
known so far for an roAp star. 

We did not obtain any spectropolarimetric observation of HD~177765,
for which no measurement of the longitudinal field was found in the
literature. 

\subsection{HD~187474}
\label{sec:hd187474}
The shape of the curve of variation of the mean magnetic
field modulus of HD~187474 between phases of 0.36 and 0.86, which was
left mostly unconstrained (but for one data point) in
\citetalias{1997A&AS..123..353M}, is now 
fully characterised thanks to 12 new measurements
(Fig.~\ref{fig:hd187474}). This field moment reaches its maximum
($\sim6.3$\,kG) close to phase 0.47. This very nearly coincides with
the phase of negative extremum of $\Hz$. The negative part of the
variation curve of the latter is now fully characterised by our
measurements, that is, by a set of data that are all obtained with the same
instrument (although with different configurations). This
  curve shows a small, 
but formally significant (at the $3\,\sigma$ level), degree of
anharmonicity with a variation 
that is somewhat steeper about the positive extremum than about the negative
extremum, but no significant departure from mirror symmetry about the
phases of the two extrema. This behaviour contrasts with that of the
field modulus, which has a broad, almost flat minimum around the phase
of $\Hz$ positive extremum, and a sharp maximum close to the phase of
$\Hz$ negative extremum. The $\Hm$ variation is not as symmetric
as that of $\Hz$, but the departure from symmetry is small, so
that contrary to other stars, the magnetic structure may plausibly be
(almost) symmetric about an axis passing through the stellar centre.

The behaviour of
the quadratic field is ill-defined because earlier
determinations (revised values of the measurements of
\citealt{1995A&A...293..746M}, and especially data from
\citealt{1997A&AS..124..475M}, 
obtained by application of Eq.~(\ref{eq:Hqold}) rather 
than the more realistic Eq.~(\ref{eq:Hq})) are considerably less
accurate than those presented 
here. The latter hardly show any hint of variation, but this is mostly
inconclusive because of their limited phase coverage. The ratio
between $\Hq$ and $\Hm$ is of the order of 1.2 or greater, while the
ratio between the maximum and minimum of the field modulus, 1.27,
is close to the expected value for a centred dipole. The shape of the
field modulus variation however indicates that the actual field
structure does show non-negligible deviations from the latter. 

None of the
individual values of the crossover depart significantly from 0, as
can be expected for a rotation period of 2345~d. Yet, it is intriguing
that all but one of the nine new determinations of this paper yield
marginally negative values; the larger uncertainties of older
measurements \citep{1995A&A...293..733M,1997A&AS..124..475M} make
them unsuitable to test whether this effect is real, or purely
coincidental. This is further discussed in Sect.~\ref{sec:xdisc}

\begin{figure}
\resizebox{\hsize}{!}{\includegraphics[angle=270]{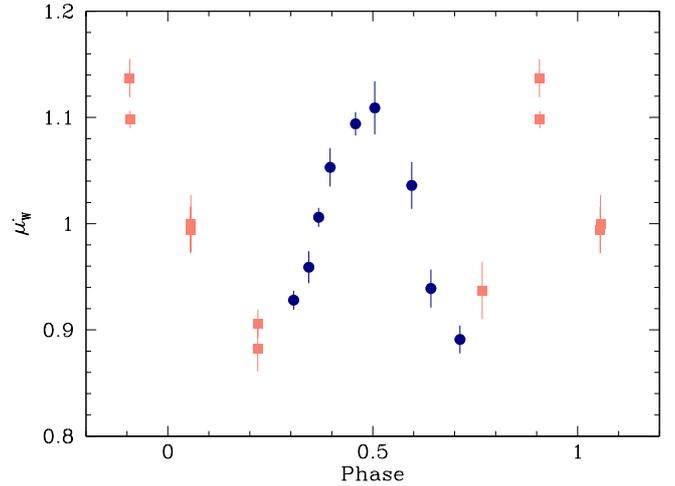}}
\caption{Variation with rotation phase of the average $\mu^\prime_W$
  of the normalised equivalent widths \citep{1994A&AS..108..547M} of the
  \ion{Fe}{ii} lines analysed in the CASPEC spectra of HD~187474. 
    Filled squares correspond to the observations of
  \citet{1991A&AS...89..121M}; filled dots correspond to the spectra
  analysed in the 
  present paper.}
\label{fig:hd187474_ew}
\end{figure}

In the interpretation of the behaviour of the various field moments,
one should keep in mind that the lines of \ion{Fe}{ii}, from which
they are derived, show significant equivalent width variations over
the stellar rotation period. This is illustrated in
Fig.~\ref{fig:hd187474_ew}, where the equivalent width parameter
$\mu^\prime_W$ is plotted against phase, for the spectropolarimetric
observations of \citet[open circles]{1991A&AS...89..121M} and of the
present paper (dots). The two samples are distinguished because the
definition of $\mu^\prime_W$ is such that its values are not a priori
consistent between sets of observations covering different parts of
the rotation cycle. Fortuitously, in the present case, the phase
distributions of the two subsamples of observations under
consideration are such that the sets of values of $\mu^\prime_W$
defined for each of these subsamples happen to be fairly
consistent with each other. Figure~\ref{fig:hd187474_ew} show a double
wave variation 
of the equivalent widths of the \ion{Fe}{ii} lines in HD~187474 over
its rotation period with maxima close to phases 0 and 0.5 and minima
near phases 0.25 and 0.75. This is consistent with the results of
 the \citet{2001A&A...378..153S}
study of the abundance 
distribution over the surface of the star, which shows some
concentration of Fe around both magnetic poles.

\begin{figure}
\resizebox{\hsize}{!}{\includegraphics{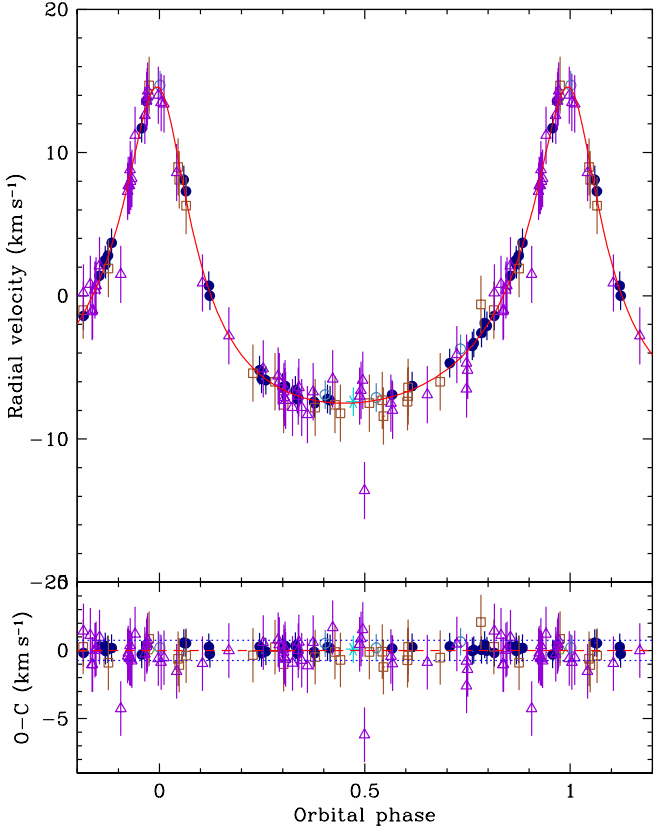}}
\caption{{\it Upper panel\/}: Our radial velocity measurements for
  HD~187474 are plotted together with those of Babcock (private
  communication)  against orbital phase. The solid curve   
  corresponds to the orbital solution given in
  Table~\ref{tab:orbits}. The time $T_0$ of periastron passage is
  adopted as phase origin. {\it Bottom panel:\/} Plot of the
  differences ${\rm O}-{\rm C}$ between the observed values of the
  radial velocity and the predicted values computed from the orbital
  solution. The dotted lines correspond to $\pm1$ rms
  deviation of the observational data about the orbital solution 
    (dashed line). Open triangles represent Babcock's data
  and  open squares our CASPEC 
  observations; all other symbols refer to our high-resolution spectra
  obtained with various instrumental configurations, as indicated in
  Table~\ref{tab:plot_sym}.}  
\label{fig:hd187474_rv}
\end{figure}

HD~187474 is a spectroscopic binary for which \citet{1964MNSSA..23....6L}
derived an approximate value of the orbital period, $\Porb=690$\,d
(shorter than the rotation period of the Ap component) and
quoted preliminary orbital elements derived by Miss Sylvia Burd from
the analysis of 33 spectra obtained by Babcock. Here we combine 
50 unpublished radial velocity
measurements of Babcock (private communication) with our data
to determine a first complete orbital solution, based on data spanning
a time interval of 42 years. Its parameters are
reported in Table~\ref{tab:orbits} and the corresponding fit is shown
in Fig.~\ref{fig:hd187474_rv}. 

\subsection{HD~188041}
\label{sec:hd188041} 
\begin{figure*}
\resizebox{12cm}{!}{\includegraphics[angle=270]{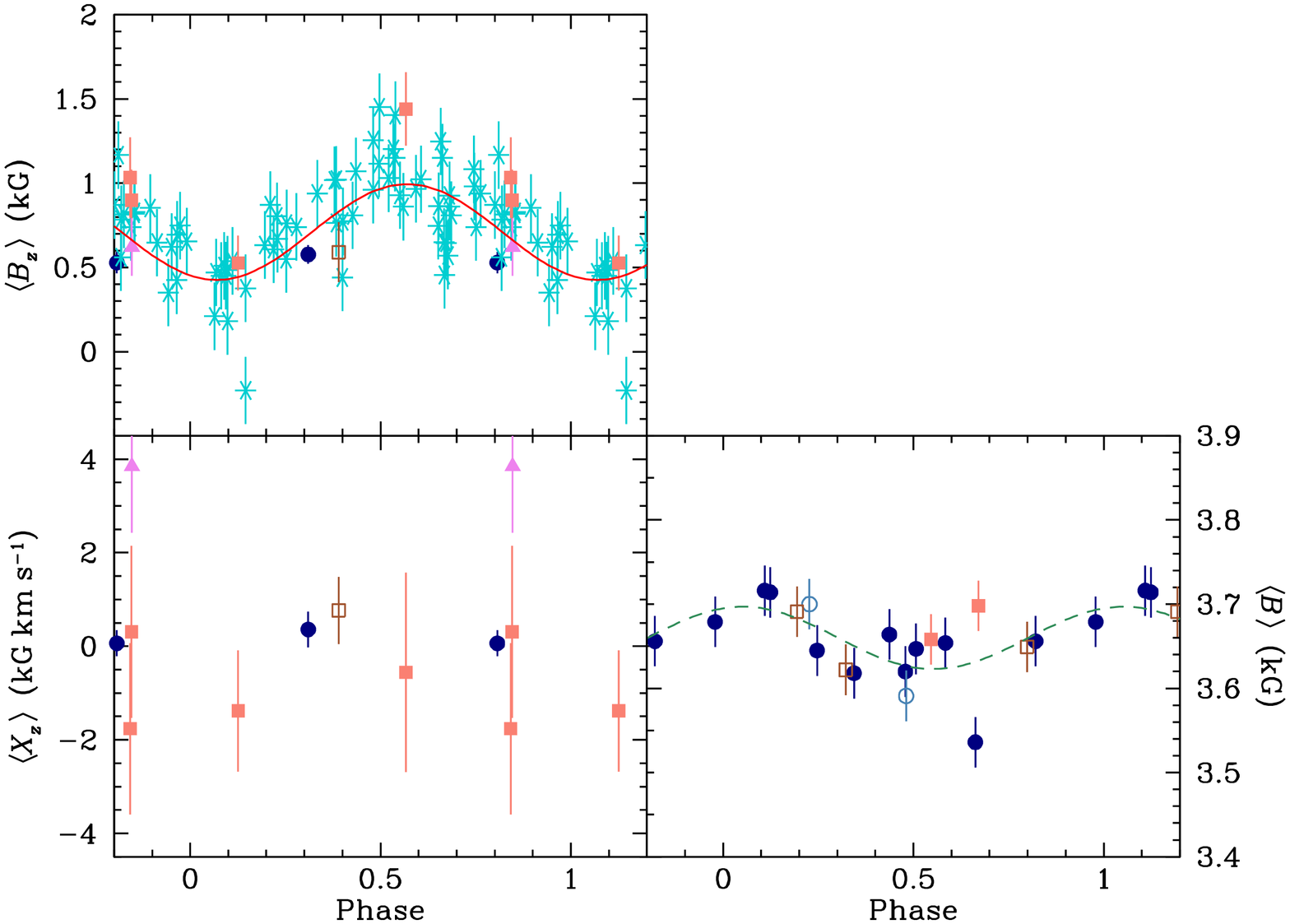}}
\parbox[t]{55mm}{
\caption{Mean longitudinal magnetic field ({\it top left\/}),
crossover ({\it bottom left\/}), 
and mean magnetic field modulus ({\it bottom right\/}) 
of the star HD~188041, against rotation phase. The longitudinal field
measurements of the present paper and those of
\citet{1994A&AS..108..547M} and
\citet{1997A&AS..124..475M} have been arbitrarily shifted by
$-400$~G for consistency with the more numerous data of
\citet{1954ApJ...120...66B,1958ApJS....3..141B}
(asterisks). Except for the latter, the symbols are as described at the
beginning of Appendix~\ref{sec:notes}. The mean quadratic
field could not be determined in this star ({\it see text for
details\/}).}
\label{fig:hd188041}}
\end{figure*}

Among the stars discussed in this paper, HD~188041 is one of those
showing the highest line density in the spectral range covered by the
CASPEC spectra. Accordingly, virtually all lines that could be used
for determination of the longitudinal field, the crossover, and the
quadratic field, show some degree of blending. This makes their
measurements difficult and increases the uncertainty of the derived
field moments. However, the combination of the two new $\Hz$ values
obtained here with the earlier data of \citet{1994A&AS..108..547M} and
\citet{1997A&AS..124..475M} definitely confirms the existence of a
systematic difference between the longitudinal field measurements of
this star from CASPEC spectra and the earlier determinations of
\citet{1954ApJ...120...66B,1958ApJS....3..141B}, as already suggested by Fig.~35 of
\citet{1991A&AS...89..121M}. In order to bring both data sets in
coincidence, a shift 
of $-400$\,G has been applied to the CASPEC values, which are less
numerous. The resulting 
combined data set is shown in the upper left panel of
Fig.~\ref{fig:hd188041}, where it is plotted against the rotation
phase corresponding to the value $\Prot=(223.78\pm0.10)$\,d of the
period. This value, which we derived from the same data set,
is in excellent agreement with the values obtained by
\citet{2003CoSka..33...29M} from photometric data of various origins 
($\Prot=223.826\pm0.040$\,d) and from combination of the magnetic data
of \citet{1954ApJ...120...66B,1958ApJS....3..141B}, \citet{1969ApJ...158.1231W}, and
\citet{1991A&AS...89..121M} ($\Prot=223.78\pm0.30$\,d). We did not
include the \citet{1969ApJ...158.1231W} 
data in our analysis because of their rather large scatter and because the $\Hz$ determinations of the Hawaii group for many
stars show unexplained discrepancies with the measurements by Babcock (and ours, which are mostly consistent with those of Babcock). Moreover, the
values published by \citet{1991A&AS...89..121M} are superseded by the results
of the revised analysis of \citet{1994A&AS..108..547M}. The fit parameters
appearing in Table~\ref{tab:zfit} have also been computed for the
whole data set shown in Fig.~\ref{fig:hd188041}. However, one
of the measurements of \citet{1997A&AS..124..475M}, on JD~2448782,
has been discarded, on account of its abnormally large discrepancy with the
other points. We have adopted the same phase origin as
\citet{1969ApJ...158.1231W}, even though it is now seen to be slightly
shifted with 
respect to the minimum of $\Hz$.

The diagnostic lines measured in our CASPEC spectra, which belong to
\ion{Fe}{i} and \ion{Fe}{ii}, are stronger close to the
phase of maximum of the longitudinal field, and weaker around its
phase of minimum, as already pointed out by \citet{1969ApJ...157..253W}.
The $\Hz$ data therefore represent the convolution of the actual
distribution of the line-of-sight component of the magnetic field on
the stellar surface with the Fe inhomogeneities, which however do not
appear very large. 

No significant crossover is detected. The quadratic field is also
below the detection limit: the values derived from the two CASPEC
spectra analysed here do not significantly differ from 0, with values
of $\sigma_{\rm q}$ of 1.9 and 2.1\,kG. In our previous
  studies \citep{1995A&A...293..746M,1997A&AS..124..475M}, application
  of Eq.~(\ref{eq:Hqold}) to our observations of this star yielded
  non-physical, negative values of $a_2$. This is the only star of the
  present sample for which the quadratic field could not be determined
  at any phase because it is too small. 

This is consistent with the fairly
weak mean magnetic field modulus observed in HD~188041. The variations
of the latter have a very small amplitude: the fitted curve has an
amplitude below 
the threshold of formal significance, even though it almost certainly
reflects real variations. In particular, we note the fairly good phase
coincidence between the phases of extrema of $\Hz$ and $\Hm$; if the
variations of the latter are real, the quasi-coincidence of its
maximum with the minimum of $\Hz$ is indicative of departure of the
field structure from a centred dipole. The three new measurements of
the field modulus presented here 
are consistent with those of \citetalias{1997A&AS..123..353M}, but do
not otherwise bring much progress in the knowledge of the magnetic
field of HD~188041.

Our measurements of the radial velocity of the star show no
significant variability, and they are in excellent agreement with
those of \citet{2002A&A...394..151C}.

% \clearpage

%\clearpage

\begin{figure}
\resizebox{\hsize}{!}{\includegraphics[angle=270]{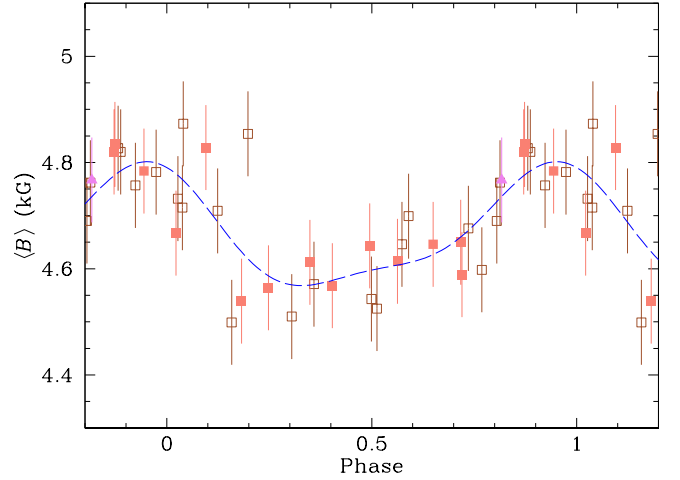}}
\caption{Mean magnetic field modulus of the star HD~192678,
against rotation phase. The AURELIE data have been shifted by
$-142$~G. The symbols are as described at the
beginning of Appendix~\ref{sec:notes}.}
\label{fig:hd192678}
\end{figure}

\subsection{HD~192678}
\label{sec:hd192678}
The magnetic field of HD~192678 has been studied in detail by
\citet{1996A&A...313..209W}. The value of the rotation period adopted
by these 
authors ($\Prot=(6.4186\pm0.002)$\,d) has recently been refined by
\citet{2006PASP..118...77A} using photometric data. This new value,
$\Prot=(6.4193\pm0.0003)$\,d, is used here to phase our observations. 

As
mentioned in \citetalias{1997A&AS..123..353M}, the values of the mean
magnetic field modulus derived for this star from AURELIE measurements
are systematically shifted with respect to those obtained from Kitt
Peak observations. In \citetalias{1997A&AS..123..353M}, we computed
this shift by comparing the 
mean values of all the measurements obtained with both
instruments. Here, instead, we compute separate least-squares fits of
the variations of $\Hm$ for the AURELIE data, on the one hand, and for
the Kitt Peak data (as well as one new measurement from a CFHT Gecko
spectrum), on the other hand, and we consider the difference between
the independent terms of these two fits as the systematic shift
between the respective data sets. As a result, we applied a correction
of $-142$\,G to the AURELIE data to bring them into coincidence with
the KPNO and CFHT measurements for plotting them in Fig.~\ref{fig:hd192678}
and for computation of the fit parameters given in
Table~\ref{tab:mfit}. Although the fit coefficient $M_2$ is below the
threshold of formal significance, the anharmonic character of the
$\Hm$ variation curve appears definite upon consideration of
Fig.~\ref{fig:hd192678}. The curve shows a hint of asymmetry,
suggesting that the magnetic field of HD~192678 may not be symmetric
about an axis passing through the centre of the star.

The northern declination of HD~192678 did not allow it to be observed
with CASPEC, but $\Hz$ determinations have been published by
\citet{1996A&A...313..209W}. According to these measurements, the
longitudinal 
field does not show any significant variation as the star rotates.  

Our determinations of the radial velocity do not show any significant
variation either and they are fully consistent with the data
of \citet{2002A&A...394..151C}.

\begin{figure*}
\resizebox{12cm}{!}{\includegraphics[angle=270]{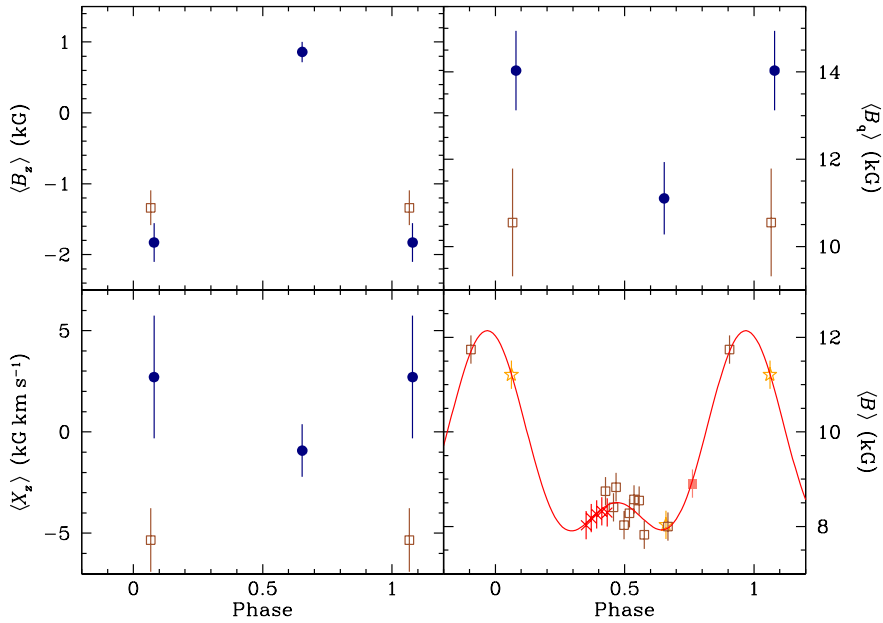}}
\parbox[t]{55mm}{
\caption{Mean longitudinal magnetic field ({\it top left\/}),
crossover ({\it bottom left\/}), 
mean quadratic magnetic field ({\it top right\/}),
and mean magnetic field modulus ({\it bottom right\/}) 
of the star HDE~335238, against rotation phase. The symbols are as described at the
beginning of Appendix~\ref{sec:notes}.} 
\label{fig:hd335238}}
\end{figure*}

\subsection{HDE~335238}
\label{sec:hd335238}
In \citetalias{1997A&AS..123..353M}, we had argued that the rotation
period of HDE~335238 must 
undoubtedly be between 40 and 50 days, but we had been unable to
establish its value definitely, owing to the very
unfortunate phase distribution of our observations. With all plausible
values, the $\Hm$ data were concentrated in two narrow
phase intervals, in which the mean field modulus measurements
clustered around two values, 8.5\,kG and 11.5\,kG. We had ruled out
the possibility that the higher of these two values, obtained in only
2 out of 16 observations, was spurious. Unfortunately, the 2 new $\Hm$
determinations reported here again happen to be close to 8.5\,kG,
so that they hardly add any new constraints on the shape of the
variation curve. They do, however, enable us to
rule out the preferred value of the rotation period tentatively
proposed in \citetalias{1997A&AS..123..353M},
$\Prot=(44.0\pm0.3)$\,d. Instead, they favour a 
revised value $\Prot=(48.7\pm0.1)$\,d, which also seems consistent
with Hipparcos photometric measurements of the star. This value is
used to compute the phases against which the magnetic field
data are plotted in Fig.~\ref{fig:hd335238}. From its
consideration, it appears
likely that the $\Hm$ curves shows a broad, fairly flat minimum and a
sharper-peaked maximum, as in a number of other stars considered in
this study. The uncertainty of the rotation period, and the very
incomplete phase coverage of the observations, do not allow the ratio
between the extrema of the field modulus to be exactly determined. But
it cannot be smaller than 1.35. This provides an additional indication
that the field structure in HDE~335238 must show some significant
departure from a centred dipole. 

Magnetic field determinations based on two new spectropolarimetric
observations are also presented here. The field moments are diagnosed
from lines of \ion{Fe}{ii} and of \ion{Cr}{ii}. With only three CASPEC
spectra available (the two considered here, and one from
\citealt{1997A&AS..124..475M}), we cannot definitely characterise the
possible 
variability of those lines. Blending and, to some extent, systematic
instrument-to-instrument differences, do not allow us to
constrain this variability from consideration of our
high-resolution spectra in 
natural light either.  

With only 3 $\Hz$ measurements available, the variations of this field
moment are not fully defined. But it definitely reverses its sign
(hence if the field bears any resemblance to a dipole, both poles come
into view as the star rotates). If the value of the rotation period
proposed above is correct, the negative extremum of the longitudinal
field may coincide roughly with the maximum of the field modulus, and
conversely, the $\Hz$ positive extremum may occur during the broad
minimum of $\Hm$, or possibly close to its little pronounced secondary
maximum. The two new measurements of the quadratic field presented
here are consistent with the field modulus variation, with the point
close to phase 0 about 3\,kG higher than the one close to phase
0.5. This difference is actually clearly seen in a visual comparison
of the 2 CASPEC spectra with all the lines appearing much broader in
the former than in the latter. In this picture, the $\Hq$
determination of \citet{1997A&AS..124..475M} is discrepant; this
can almost certainly be assigned to its comparatively low quality, which was
already mentioned for several other stars. This 
interpretation also applies to the discrepant crossover measurement of
Mathys \& Hubrig; the possible detection reported in that paper is
almost certainly spurious. There is no definite evidence of radial velocity variations in our
observations of HDE~335238.

% \clearpage

\subsection{HD~200311}
\label{sec:hd200311}
\begin{figure}
\resizebox{\hsize}{!}{\includegraphics[angle=270]{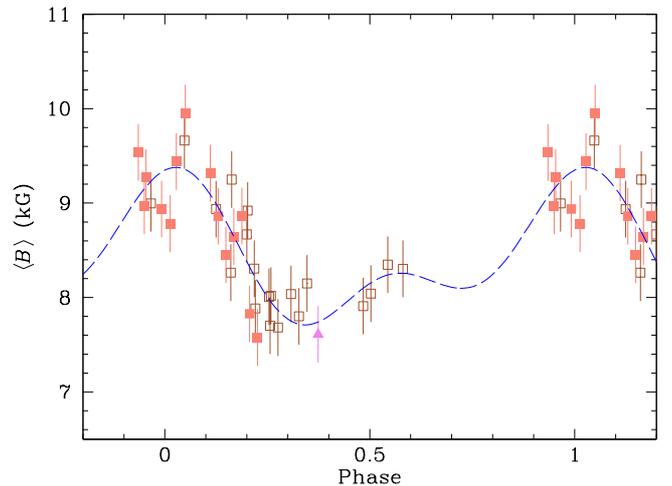}}
\caption{Mean magnetic field modulus of the star HD~200311,
against rotation phase. The symbols are as described at the
beginning of Appendix~\ref{sec:notes}.}
\label{fig:hd200311}
\end{figure}

The seven new determinations of the mean magnetic field modulus of
HD~200311 unfortunately do not fill the gap of about one-third of the
rotation cycle that was left in the phase coverage achieved in
\citetalias{1997A&AS..123..353M}. This gap is clearly seen between phases 0.58 and 0.93 in
Fig.~\ref{fig:hd200311}, in which the phases were computed using the
value of the rotation period derived by \citet{1997MNRAS.292..748W},
$\Prot=52\fd0084$. The best-fit curve to our $\Hm$ data suggests the
existence of a secondary maximum around phase 0.6. However, owing to the
above-mentioned phase gap, this curve is not strongly constrained. Its
marginal anarhmonicity (the $M_2$ coefficient of the fit is determined
at the $2.3\sigma$ level) is probably real but
its apparent asymmetry seems difficult to reconcile with
the almost perfect double-wave variation observed in photometry
\citep{1997MNRAS.292..748W}. The ratio between the extrema of $\Hm$ is
not 
well defined either, but it cannot be signficantly smaller than 1.2,
hence it is close to the upper limit for a centred dipole. 

The northern declination of the star did not allow us to observe it
with CASPEC. But $\Hz$ measurements have been
published by \citet{1997MNRAS.292..748W} and \citet{2000A&A...355..315L}. The longitudinal field reverses its sign
as the 
star rotates, and to the achieved precision, its variation is
sinusoidal. The phase relation between its extrema and those of the
field modulus is not accurately defined owing to the pretty large
uncertainties affecting the $\Hz$ determinations and to the gap in the
$\Hm$ phase coverage, but in first approximation, the positive
extremum of $\Hz$ seems to occur close to the phase of maximum of
$\Hm$. 

\begin{figure}
\resizebox{\hsize}{!}{\includegraphics[angle=270]{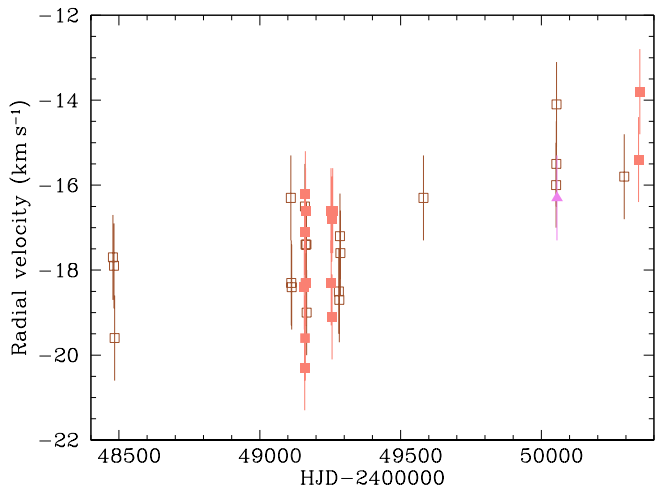}}
\caption{Radial velocity of HD~200311,  against heliocentric Julian
  date. The symbols are as described at the
beginning of Appendix~\ref{sec:notes}.}
\label{fig:hd200311_rv}
\end{figure}

Consideration of Fig.~\ref{fig:hd200311_rv} suggests that the
radial velocity of HD~200311 has been varying monotonically over the
time interval covered by our observations. The standard deviation of our
measurements about a straight-line fit, 1.1\,km\,s$^{-1}$, is
consistent with their uncertainty. While a linear variation has no
physical meaning, it may plausibly approximate well the actual shape
of a fraction of the radial velocity curve of a spectroscopic binary
with an orbital period much longer than the time separating our first
and last observations (more than five years). A shorter period, of the
order of 320 days, cannot be definitely ruled out, but this is most
likely coincidental.

\subsection{HD~201601}
\label{sec:hd201601}
HD~201601 is the prototype of the Ap stars with extremely long
rotation periods. Its mean longitudinal magnetic field had been
monotonically decreasing since the time of the measurements by
\citet{1958ApJS....3..141B}, about 60 years ago (which may have been close to the
positive extremum), until the latest determinations of
\citet{1997A&AS..124..475M}, in 1993 \citep[see e.g. Fig.~37
of][]{1991A&AS...89..121M}. These authors reported a hint of
flattening of the $\Hz$ 
curve, which raised suspicion that the star was approaching the phase
of negative extremum of this field moment. Similarly, but on a shorter
time span, the mean magnetic field modulus had been monotonically
increasing from the time of its first measurement in 1988
\citep{1990A&A...232..151M} to the latest determination of \citetalias{1997A&AS..123..353M}, in
1995. We had 
suggested at the time that the phase of $\Hm$ maximum might be 
near. 

\begin{figure*}
\resizebox{12cm}{!}{\includegraphics[angle=270]{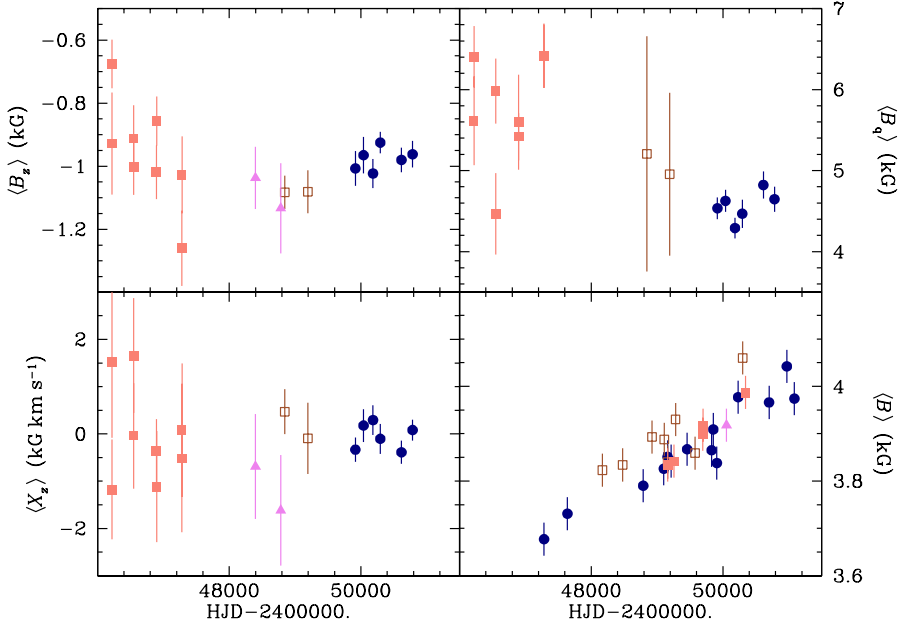}}
\parbox[t]{55mm}{
\caption{Mean longitudinal magnetic field ({\it top left\/}),
crossover ({\it bottom left\/}), 
mean quadratic magnetic field ({\it top right\/}),
and mean magnetic field modulus ({\it bottom right\/}) 
of the star HD~201601,
against heliocentric Julian date. The symbols are as described at the
beginning of Appendix~\ref{sec:notes}.}
\label{fig:hd201601}}
\end{figure*}

Both suspicions are fully borne out by the six new $\Hz$
determinations and the seven new $\Hm$ measurements presented here. These
show a slight, but definite, increase of the longitudinal field, as
well as more pronounced progressive flattening of the field modulus
variation curve (see Fig.~\ref{fig:hd201601}). 

The negative extremum
of $\Hz$ must have taken place very close to the time of the
observations of \citet{1997A&AS..124..475M}. The passage of the
longitudinal field through that extremum is fully confirmed by more
recent measurements of \citet{2006MNRAS.365..585B},
\citet{2014psce.conf..386S}, and \citet{2016MNRAS.455.2567B}. 

The rotation period of HD~201601 must be considerably longer than the
67 years time interval 
separating the earliest $\Hz$ measurement by \citet{1958ApJS....3..141B}
from the latest measurement from 
\citet{2016MNRAS.455.2567B}. The once preferred value of 77 years  
\citep{1994A&A...284..174L} now appears to be a very
conservative lower limit. The value of 97.16~y recently proposed by
\citet{2016MNRAS.455.2567B} may be more realistic, but it must still
be regarded with caution as it is the result of an extrapolation based
on the assumption that the longitudinal field variation curve does not
significantly depart from a sinusoid. That this cannot be taken for
granted is illustrated by the anharmonicity of the $\Hz$ variations
found for various other stars discussed in this paper. It may well be
that the rotation period of HD~201601 is actually even longer than
suggested by \citet{2016MNRAS.455.2567B}.

On the other hand, the view that the mean magnetic field modulus was
close to its maximum at the time of the new observations discussed
here (1996--1998) is supported by the more recent determination of
\citet{2006A&A...448.1165S}, on JD~2452920.6,
which is achieved by application of
the measurement technique and with one of the instrumental configurations
used in the present paper. Accordingly, the fact that the derived
value of $\Hm=(3879\pm25)$\,G is about 200\,G smaller than our most
recent values is a highly significant indication that the field
modulus is decreasing, after passing through its maximum. Thus the
phases of extremum of $\Hm$ and $\Hz$ appear to be in rough
coincidence, which is not unusual; whether there is some small phase
shift between them will have to be established by additional future
measurements.

\begin{figure*}
\resizebox{12cm}{!}{\includegraphics[angle=270]{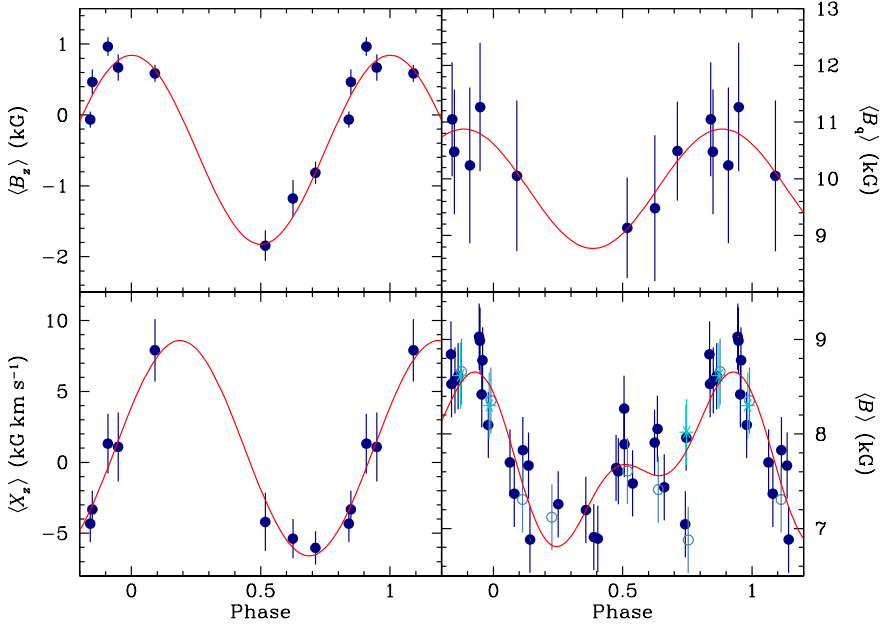}}
\parbox[t]{55mm}{
\caption{Mean longitudinal magnetic field ({\it top left\/}),
crossover ({\it bottom left\/}), 
mean quadratic magnetic field ({\it top right\/}),
and mean magnetic field modulus ({\it bottom right\/}) 
of the star HD~208217, against rotation phase. The symbols are as described at the
beginning of Appendix~\ref{sec:notes}.} 
\label{fig:hd208217}}
\end{figure*}

 The present determinations of the quadratic field of HD~201601 have on
average smaller uncertainties than for any other star considered
here. This must be related to the fact that the number of diagnostic
lines analysed in this star is greater than in any other, which is consistent
with the conclusion drawn by \citet{2006A&A...453..699M} that the  achievable
uncertainty in quadratic field determinations scales like the inverse
of the square root of the number of analysed lines. The apparent
decrease of $\Hq$ seen in Fig.~\ref{fig:hd201601} is more likely to
reflect primarily the lower quality of the measurements of
\citet{1995A&A...293..746M} and \citet{1997A&AS..124..475M} than to
be 
real. If we compare the quadratic field values of 1996--1998 with
contemporaneous field modulus data, the ratio between the two moments
appears to be of the order of 1.15, showing reasonable mutual
consistency. 

The measurements of the crossover in HD~201601 presented here also
have smaller errors than those of all other stars of the studied
sample. Thus the fact that no crossover is detected is highly
significant, and fully in line with the very long rotation period of
the star.

Our observations cover too small a fraction of the rotation cycle of
HD~201601 to assess the presence of variations of intensity of
the \ion{Fe}{i} and \ion{Fe}{ii} lines analysed in the CASPEC
spectra reliably. On the other hand, the absence of variations of the radial
velocity of the star is convincingly established by the fact that the
standard deviation of the values determined from the high-resolution
spectra in natural light is a mere 0.3\,km\,s$^{-1}$. Our measurements
are also very consistent with those of \citet{2002A&A...394..151C}
with which they are mostly contemporaneous. Both 
data sets are also contemporaneous with the radial velocity
determinations of \citet{2000AN....321..115H}, indicating
without ambiguity that the discrepant values reported by these authors
must be spurious. However, HD~201601 is a visual binary, for which
\citet{2011A&A...529A..29S} recently determined a preliminary
orbital solution. The long orbital period that they derived, 274.5
years, is compatible with the lack of significant radial velocity
variations over the time interval covered by our data and those of
\citet{2002A&A...394..151C}. 

\begin{figure}
\resizebox{\hsize}{!}{\includegraphics[angle=270]{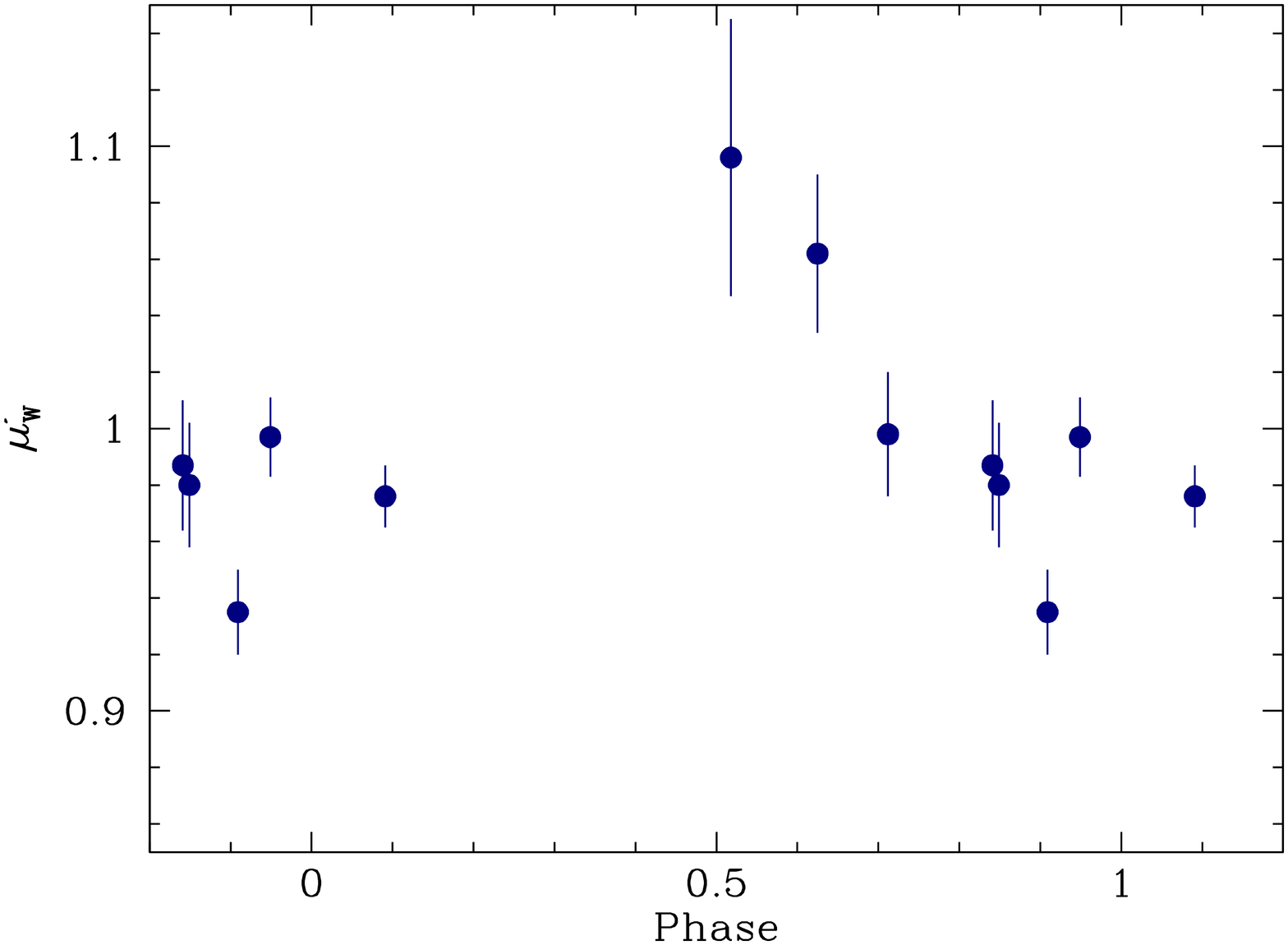}}
\caption{Variation with time of the average $\mu^\prime_W$
  of the normalised equivalent widths \citep{1994A&AS..108..547M} of the
  \ion{Fe}{ii} lines analysed in the CASPEC spectra of
  HD~208217. }
\label{fig:hd208217_ew}
\end{figure}

\subsection{HD~208217}
\label{sec:hd208217}
The seven new measurements of the mean longitudinal field magnetic field
of HD~208217 presented here combine well with the data of \citetalias{1997A&AS..123..353M}
using the value of the rotation period derived by
\citet{1997A&A...320..497M}, $\Prot=8\fd44475$
(Fig.~\ref{fig:hd208217}). They
improve the definition of the shape of the $\Hm$ variation curve,
which is significantly anharmonic, but with a secondary maximum
that is considerably lower than the primary maximum, so that contrary to our
speculation of \citetalias{1997A&AS..123..353M}, the field structure is different from a
centred dipole. The ratio between the highest and lowest $\Hm$ values
is fairly large, 1.27.

 By contrast, the variations of the other field moments, determined
through analysis of \ion{Fe}{ii} lines in eight CASPEC spectra, do not
significantly depart from sinusoids. The mean longitudinal magnetic
field, which is measured here for the first time, reverses its sign as
the star rotates, showing that both poles alternatively come into
view. The phase coverage of the observations is unfortunate, as all
concentrate in a little more than half a cycle. But within the limits
of the achieved precision in its determination, the phase of maximum of
$\Hz$ roughly coincides with that or the primary $\Hm$ maximum. 

The
difference between the phases of $\Hz$ positive extremum and of $\Hq$
maximum is somewhat below the formal significance threshold (see
Tables~\ref{tab:zfit} and \ref{tab:qfit}); as a matter of fact, the
phases of the extrema of the quadratic field variation curve are rather
poorly constrained. The ratio between the quadratic field and the
field modulus varies slightly with rotation phase around a value of
about 1.25. 

Crossover is definitely detected, not surprisingly since
the $\vsi$ estimate that is obtained as a by-product of the quadratic
field determination, $15\pm5$\,km\,s$^{-1}$, is
significant. The rotational Doppler effect also shows in the distortion
of the split components of the line \ion{Fe}{ii}~$\lambda\,6149.2$,
which as pointed out in \citetalias{1997A&AS..123..353M} complicates
the determination of 
$\Hm$. The crossover variations lag in phase behind those of the
longitudinal field by an amount that is not significantly different
from a quarter of the rotation period, and its average over the
rotation period is zero within the achieved accuracy; this behaviour
is fully consistent with that generally observed (and expected) for
this field moment \citep{1995A&A...293..733M}. 

\begin{figure}[!t]
\resizebox{\hsize}{!}{\includegraphics[angle=270]{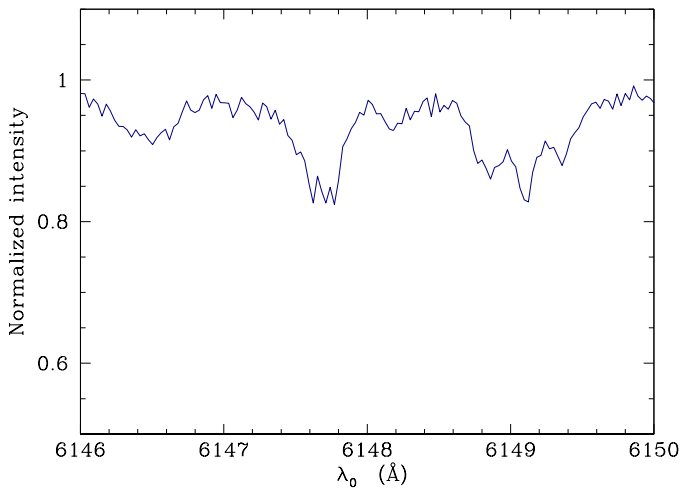}}
\caption{Portion of the spectrum of HD~213637 recorded on HJD~2451085.613, 
  showing the lines
  \ion{Fe}{ii}\,$\lambda\,6147.7$ and
  \ion{Fe}{ii}\,$\lambda\,6149.2$. The blue wing of the latter is
  heavily blended by a line from an unidentified rare earth. 
The \ion{Cr}{ii}\,$\lambda\,6147.1$ line is not observed in this star
at the considered epoch.}
\label{fig:hd213637_6149}
\end{figure}

\begin{figure}
\resizebox{\hsize}{!}{\includegraphics[angle=270]{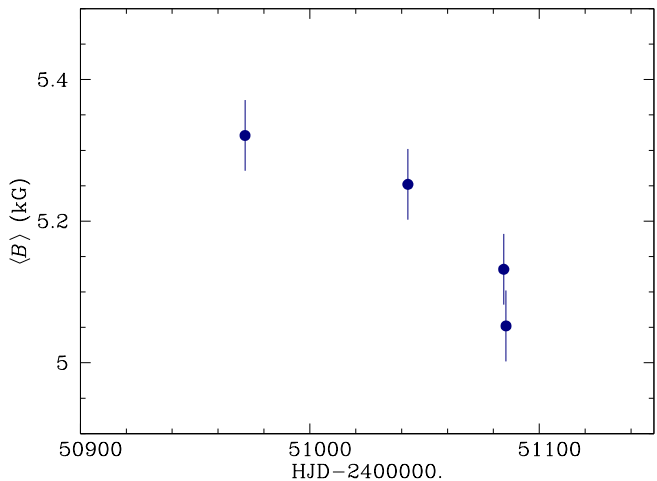}}
\caption{Mean magnetic field modulus of the star HD~213637,
against heliocentric Julian date.}
\label{fig:hd213637}
\end{figure}

 The equivalent widths of the \ion{Fe}{ii} lines used to diagnose the
field moments discussed in the previous paragraph show some
variability with rotation phase. As can be seen in
Fig.~\ref{fig:hd208217_ew}, they are stronger close to the positive
extremum of $\Hz$ and weaker close to its negative extremum,
suggesting a concentration of iron around the positive magnetic
pole. Accordingly, the measured magnetic field values represent a
convolution of the actual structure of the field with the
inhomogeneous distribution of Fe over the stellar surface.

\begin{figure*}
\resizebox{12cm}{!}{\includegraphics[angle=270]{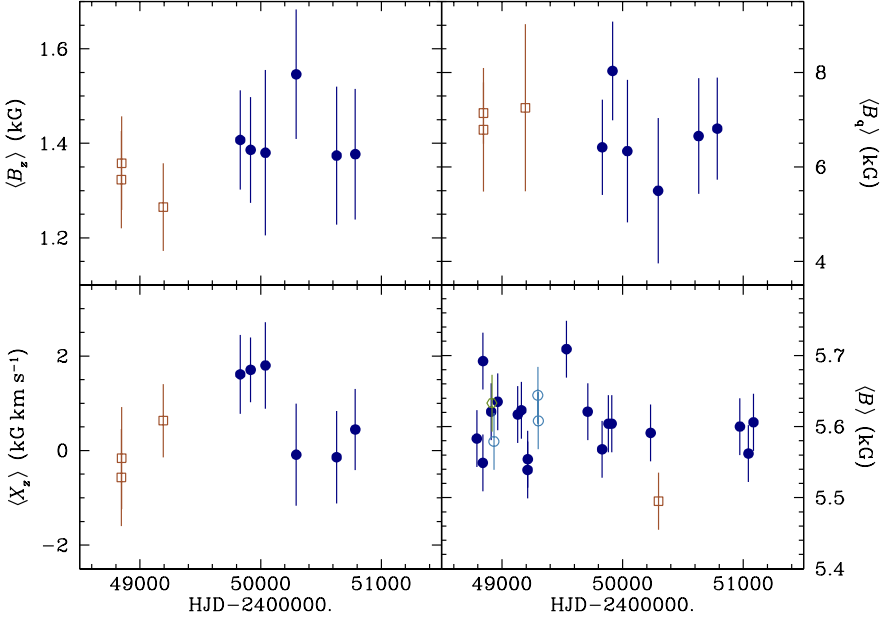}}
\parbox[t]{55mm}{
\caption{Mean longitudinal magnetic field ({\it top left\/}),
crossover ({\it bottom left\/}), 
mean quadratic magnetic field ({\it top right\/}),
and mean magnetic field modulus ({\it bottom right\/}) 
of the star HD~216018,
against heliocentric Julian date. The symbols are as described at the
beginning of Appendix~\ref{sec:notes}.}
\label{fig:hd216018}}
\end{figure*}

 As already reported in \citetalias{1997A&AS..123..353M}, the radial
 velocity of HD~208217 is 
variable. Our attempts to determine an orbital period were
unsuccessful owing to very strong aliasing. The difficulty is almost
certainly further compounded by the rather large uncertainty affecting
the radial velocity determinations in our high-resolution spectra, owing
to the strong blend affecting the blue component of the line
\ion{Fe}{ii}~$\lambda\,6149.2$ and to its distortion by the
combination of the Zeeman effect at different locations on the stellar
surface with rotational Doppler effect. The latter furthermore
contributes to some extent to the radial velocity variations. However,
the rotation period does not clearly appear in a frequency analysis of
the radial velocity measurements. In any case, the amplitude of the
variations (more than 
20\,km\,s$^{-1}$ peak-to-peak) is sufficient so that there is no
question that they are real and that they cannot be fully accounted
for by rotation, hence that the star must be a spectroscopic
binary. It is actually an astrometric binary
\citep{2005AJ....129.2420M}; it is impossible to determine
at this stage if 
the astrometric companion is the same as the spectroscopic one, or if
HD~208217 is part of a multiple system.

%\clearpage

%\clearpage

\subsection{HD~213637}
\label{sec:hd213637}
Resolved magnetically split lines were discovered in the little known
roAp star HD~213637 \citep{1997IBVS.4507....1M} on our
high-resolution spectrum obtained on November 11, 1997.
We announced this
  finding in an oral presentation 
  at the conference on ``Magnetic fields across the
  Hertzsprung-Russell diagram'' held in Santiago in January 2001
  \citep{2001ASPC..248..267M}.  Seven $\Hm$ measurements were reported
  at the conference on ``Magnetic fields in O, B, and A stars'' held in
  Mmabatho in November 2002 \citep{2003ASPC..305...65M}. An
  independent discovery report, based on a spectrum taken on April 17,
  2002, was published by \citet{2003A&A...404..669K}. This
  observation is posterior to all measurements presented here.

The resolution of the \ion{Fe}{ii}\,$\lambda\,6149.2$ line is
illustrated in the portion of one of our 
high-resolution spectra 
that is shown in Fig.~\ref{fig:hd213637_6149}.
The four determinations of the mean magnetic field modulus that we
obtained are consistent with a monotonic decrease over the 115~day
time interval that they span (see Fig.~\ref{fig:hd213637}), but they
do not rule out a shorter rotation period. The distortion of the lines
appearing in Fig.~\ref{fig:hd213637_6149} may plausibly be caused by
the rotational Doppler effect. On the other hand, the
value $\Hm=(5.5\pm0.1)$\,kG measured by \citet{2003A&A...404..669K}
indicates that the star has passed through at least one, and possibly
several minima of its field modulus between 1998 and 2002.

To estimate the uncertainties affecting our determinations of the mean
magnetic field modulus of this star, we computed their standard
deviation about a least-squares fit by a straight line, obtaining a
value of 40\,G. Admittedly, this is meaningful only if the rotation
period is longer than the time interval over which those data have
been obtained, so that the estimated error must be regarded with
caution. 

All our CASPEC observing nights took place before the discovery of
magnetically resolved lines in HD~213637. By chance, a single
spectropolarimetric observation of it had been obtained prior to this
discovery. Based on it, we report the measurement of a longitudinal
field slightly above the threshold of formal significance, at the
$3.6\sigma$ level. We also determined a value of the quadratic
  field. The latter cannot be 
directly compared to the field modulus data, since the measurements of
the two field moments were obtained at different epochs, at unknown
but presumably different rotation phases; but the orders of
magnitude are not incompatible. In contrast, the value of
$\Hz=(230\pm63)$\,G 
obtained here is significantly different from the one published
by \citet{2004A&A...415..685H}, $740\pm50$\,G, from low-resolution
spectropolarimetric observations performed with FORS-1 at the
VLT. Even allowing for possible discrepancies resulting from the usage
of different field determination techniques, the difference between
the two measurements most likely reflects actual variations of $\Hz$
(not surprisingly, considering the variability of $\Hm$). 
No crossover was detected, which tends
to favour a long period.

\subsection{HD~216018}
\label{sec:hd216018}
Our suspicion in \citetalias{1997A&AS..123..353M} that the mean
magnetic field modulus was 
starting to increase towards the end of our series of observations,
after remaining constant for about two years, is not supported by the five new
$\Hm$ measurements presented here. Actually, it was found that the last
three $\Hm$ determinations of \citetalias{1997A&AS..123..353M} were
incorrect. The reason is not 
definitely established, although it can be speculated that the strong
blend affecting the blue component of the line
\ion{Fe}{ii}~$\lambda\,6149.2$ had not been properly
dealt with. Those three measurements have been redone, and their
outcome appears in Table~\ref{tab:hm} and in
Fig.~\ref{fig:hd216018}. Following that revision, it appears that the
field modulus of HD~216018 has shown no significant variation over the
time interval covered by our entire
data set, which is slightly longer than six years. The standard
deviation of all our measurements, 47\,G, is 
compatible with a constant field over the considered period, if one
takes into account the difficulty to measure the heavily blended blue component of
\ion{Fe}{ii}~$\lambda\,6149.2$ with ultimate precision. 

The spectrum of HD~216018 is extremely dense, so that it
was impossible to find enough lines of a single ion, or
even of a single chemical element, that were sufficiently free from blends to
be used for magnetic field diagnosis. Like the mean field modulus, the
longitudinal field and the 
quadratic field, which were determined 
from the analysis of a sample of lines of \ion{Fe}{i}, \ion{Fe}{ii},
and \ion{Cr}{ii},
do not show any variability over the time span of more than five
years covered by our measurements and the earlier measurements of
\citet{1997A&AS..124..475M}. The ratio between the quadratic field and
the 
field modulus is close to 1.2. No crossover is detected.

The data described above indicate either that HD~216018 is not
variable (possibly seen rotation-pole on) or that its rotation period
is much longer than six years. 

The standard deviation of all the radial velocity measurements based
on our high-resolution spectra recorded in natural light is only
0.3\,km\,s$^{-1}$: we do not detect any significant variability. As
for the field modulus, the claim in \citetalias{1997A&AS..123..353M}
that the measurements of 
the last observing season under consideration were starting to show a
slow decrease of the radial velocity was the result of incorrect
determinations.

%\clearpage

\section{Revised mean quadratic magnetic field determinations}
\label{sec:hquad_rev}

\begin{table*}
\caption{Revised mean quadratic magnetic field measurements.}
\label{tab:hquad_rev}
\small{
\begin{tabular*}{\textwidth}[]{@{}@{\extracolsep{\fill}}ccc
@{\extracolsep{0pt}}@{}}
\parbox[t]{5.6cm}{
\centering 
\begin{tabular}[t]
{@{}@{\extracolsep{-2.5pt}}lrrr@{\extracolsep{0pt}}@{}}
\hline
\hline\\[-4pt]
HJD&$\Hq$&$\sigma_{\rm q}$&$n$\\
$\null-2400000.$&(G)&(G)\\[2pt]
\hline
\hline\\[-6pt]
\multicolumn{4}{c}{HD 83368 (HR~3831)}\\[2pt]
\hline\\[-6pt]
46218.483& 8178&1963&12\\
46547.516& 8504&2465&11\\
46548.505& 9853&1655&10\\
46548.687&10063&2265& 9\\
46549.526& 9950&2979&10\\
46894.481& 7171&1720&16\\
46895.531& 7351&2204&15\\
46895.707& 2728&4421&15\\
46896.620& 6237&2826&13\\
46897.765& 5508&2611&13\\
47280.518& 7711&1281&14\\
47281.548&11408&1771&14\\[2pt]
\hline\\[-6pt]
\multicolumn{4}{c}{HD 96446 (CPD $-$59 3038)}\\[2pt]
\hline\\[-6pt]
46547.585& 6489& 435& 9\\
46548.568& 6717& 830&10\\
46549.604& 5385&1211&10\\
46895.600& 4705&1126& 5\\
46896.702& 6977&1074& 6\\
46897.784& 8339& 931& 6\\
47279.618& 8774&1499& 6\\
47280.546& 8666& 906& 5\\
47281.571& 9450&1058& 6\\[2pt]
\hline\\[-6pt]
\multicolumn{4}{c}{HD 116458}\\[2pt]
\hline\\[-6pt]
46548.595& 6026& 639&13\\
46894.597& 4816& 546&12\\
46895.798& 3007&1175&14\\
47279.643& 6549& 699&11\\
47281.560& 6723& 580&12\\[2pt]
\hline\\[-6pt]
\multicolumn{4}{c}{HD 119419 (HR~5158)}\\[2pt]
\hline\\[-6pt]
46547.530&13748&3263& 9\\
46547.821&12784&4025& 8\\
46548.525&19481&2326&10\\
46548.823&20978&2977& 7\\
46549.836&18602&2216& 8\\
46894.552&18045&2434& 9\\
46894.862&20070&2027& 9\\
46895.517&17736&1665& 9\\
46895.622&16129&2052& 9\\
46895.755&15678&1947& 9\\
46895.886&17198&3049& 9\\
46896.669&19138&2628& 9\\
46897.718&18536&2004& 8\\
46897.828&18382&2235& 9\\
46897.906&19216&1648& 9\\
47279.728&18609&2015& 9\\
47280.794&16383&2081& 9\\
47281.669&21286&2139& 8\\
47281.781&18788&2761& 9\\
47281.877&22482&3005& 8\\[2pt]
\hline
\end{tabular}}
&\parbox[t]{5.6cm}{
\centering 
\begin{tabular}[t]
{@{}@{\extracolsep{-2.5pt}}lrrr@{\extracolsep{0pt}}@{}}
\hline
\hline\\[-4pt]
HJD&$\Hq$&$\sigma_{\rm q}$&$n$\\
$\null-2400000.$&(G)&(G)\\[2pt]
\hline
\hline\\[-6pt]
\multicolumn{4}{c}{HD 125248 (CS~Vir)}\\[2pt]
\hline\\[-6pt]
46219.586&10924&1270& 9\\
46547.627&10125& 481&10\\
46548.607& 9625& 903&10\\
46549.620& 7664& 815&11\\
46894.564& 7827& 674&10\\
46894.906& 7823& 640&10\\
46895.563& 7295& 662& 9\\
46895.902& 8572& 498&10\\
46896.600& 9757& 655& 8\\
46896.907& 9167& 252& 8\\
46897.686&10249& 641& 8\\
46897.896& 8857& 423& 7\\
47190.864& 8785& 734&18\\
47279.565&11588& 509& 7\\
47279.882&11362& 507& 8\\
47280.564&10300& 467& 8\\
47280.861&11527& 604& 8\\
47281.502&11492&1006& 7\\
47281.859&10110& 724& 8\\[2pt]
\hline\\[-6pt]
\multicolumn{4}{c}{HD 126515}\\[2pt]
\hline\\[-6pt]
46218.615&22006&1069& 7\\
46548.674&17158& 384&10\\
46894.787&21667& 855& 7\\
47190.877&21484&1482& 6\\
47279.694&22418&1480& 6\\
47281.715&22218&1824& 5\\[2pt]
\hline\\[-6pt]
\multicolumn{4}{c}{HD 128898 ($\alpha$~Cir)}\\[2pt]
\hline\\[-6pt]
46547.569& 6777& 795&12\\
46548.542& 6255& 915&12\\
46548.832& 5792&1210&12\\
46548.930& 4612& 753&12\\
46549.779& 6547& 581&12\\[2pt]
\hline\\[-6pt]
\multicolumn{4}{c}{HD 137509 (CPD $-$70 2069)}\\[2pt]
\hline\\[-6pt]
46547.549&31515&1506& 5\\
46547.905&28478&1288& 6\\
46548.551&26709&1788& 8\\
46548.857&24240&2226& 8\\
46549.586&24320&1830& 8\\
46549.848&25685&1385& 8\\
46897.878&28437&1877& 7\\
47279.860&29790& 794& 6\\
47281.835&22453&1813&10\\[2pt]
\hline\\[-6pt]
\multicolumn{4}{c}{HD 137909}\\[2pt]
\hline\\[-6pt]
46218.671& 6056& 892&12\\
46219.613& 7151& 724&11\\
46547.717& 5831& 973&12\\
46549.708& 5122& 871&13\\
46894.800& 5558& 688&14\\
46895.834& 6325& 574&14\\
46896.813& 6362& 476&14\\
46897.805& 6492& 446&14\\
47279.738& 6957& 788&13\\
47280.731& 7652& 745&13\\
47281.723& 7918& 542&13\\[2pt]
\hline
\end{tabular}}
&\parbox[t]{5.6cm}{
\centering 
\begin{tabular}[t]
{@{}@{\extracolsep{-2.5pt}}lrrr@{\extracolsep{0pt}}@{}}
\hline
\hline\\[-4pt]
HJD&$\Hq$&$\sigma_{\rm q}$&$n$\\
$\null-2400000.$&(G)&(G)\\[2pt]
\hline
\hline\\[-6pt]
\multicolumn{4}{c}{HD 147010 (BD $-$19 4359)}\\[2pt]
\hline\\[-6pt]
46219.691&16262&2223& 6\\
46547.783&11518&1593& 9\\
46548.795&15231&1066& 7\\
46549.767&11712& 901& 8\\
46894.826&12394& 766& 9\\
46895.649&10897&1378& 8\\
46895.864&10539&1386& 8\\
46896.841&11871& 997& 9\\
46897.848&14989&1445& 6\\
47189.873&11684&1641& 9\\
47190.846&14042&1677& 8\\
47279.598&11405& 784& 9\\
47279.818&11449&1010& 9\\
47280.593&12869&1306& 7\\
47280.817&11494& 908& 8\\
47281.633&14830&1730& 8\\
47281.803&15536&1341& 8\\[2pt]
\hline\\[-6pt]
\multicolumn{4}{c}{HD 153882 (BD $+$15 3095)}\\[2pt]
\hline\\[-6pt]
46547.798& 2879&1948&12\\
46548.872& 6150&1128&10\\
46549.788& 5105&1037&11\\
46894.760& 7881& 833&10\\
46894.920& 6835&1028&10\\
46895.723& 3074&2818& 8\\
46897.748& 6720&1455& 8\\
46897.918& 7251&1633& 9\\
47279.759& 4946&1300&10\\
47280.769& 3956&1575&10\\
47281.733& 4874&1566& 8\\
47281.888& 6606&1271& 9\\[2pt]
\hline\\[-6pt]
\multicolumn{4}{c}{HD 187474}\\[2pt]
\hline\\[-6pt]
46218.864& 6721& 613&12\\
46547.865& 7470& 359&13\\
46549.859& 6348& 385&13\\
46894.874& 7485& 582& 9\\
46896.871& 7046& 457& 9\\
47279.799& 8902& 468&10\\
47280.834& 9549& 451& 9\\[2pt]
\hline\\[-6pt]
\multicolumn{4}{c}{HD 201601}\\[2pt]
\hline\\[-6pt]
46218.905& 5614& 548&14\\
46219.830& 6406& 379&15\\
46547.827& 5981& 402&14\\
46549.903& 4467& 503&13\\
46895.921& 5417& 291&16\\
46896.888& 5596& 586&16\\
47280.870& 6409& 391&15\\
47281.900& 6419& 396&16\\[2pt]
\hline
\end{tabular}}\\
\end{tabular*}}
\end{table*}

In this Appendix, we present the revised values of the mean quadratic
magnetic field that we derived as described in Sect.~\ref{sec:moments}
for the stars studied by \citet{1995A&A...293..746M}. Namely, for each of
those stars that had been observed at least at three different epochs,
we computed for each diagnostic line the average $[\R{2}{I}]_{\rm av}$
of the values of $\R{2}{I}$ measured for this line by
\citet{1995A&A...293..746M} at the various epochs. We used those
averages to determine the parameters $a_1$ and $a_3$ of
Eq.~(\ref{eq:Hq}), then computed $[\R{2}{I}]_{\rm mag}$ by application
of Eq.~(\ref{eq:R2I_mag}), and finally derived $\Hq$ at each epoch by
application of Eq.~(\ref{eq:Hq_av}). 
 
The updated values of $\Hq$ that were obtained in that way are
presented in Table~\ref{tab:hquad_rev}, both for the stars with
resolved magnetically split lines that were included in the sample of
\citet{1995A&A...293..746M} and for the other stars of that study for
which improved values of the quadratic field could be
obtained ($n$ is the number of diagnostic lines from which the
corresponding value 
of $\Hq$ was determined). These values supersede those published by
\citet{1995A&A...293..746M}. 

Two of the stars for which \citet{1995A&A...293..746M} had
obtained observations at a sufficient number of epochs for application
of the above-described revised analysis procedure, HD~74521 and
HD~175362, were omitted from Table~\ref{tab:hquad_rev}. The quadratic
field of the former is below the detection limit, similar to HD~188041
in this paper (see Appendix~\ref{sec:hd188041}); the apparently
significant values derived by \citet{1995A&A...293..746M} most likely
were overestimates. The same limitation affects the determination of
the quadratic field of HD~153882 for the two observations obtained on
JD~2446896. For HD~175362, the assumption that the fit
parameters $a_1$ and $a_3$ have the same values at all epochs of
observation, which underlies the method applied for determination of
the revised values of $\Hq$, appears to break down, possibly because
the diagnostic lines that are used pertain to different elements and
ions, and/or because the distribution of those elements over the
stellar surface is very inhomogeneous \citep[see Table 3 and Figs.~26
to 31 of][]{1991A&AS...89..121M}.

\section{Pulsational crossover}
\label{sec:pulsxover}
In Sect.~\ref{sec:xdisc}, we introduced the pulsational crossover
mechanism as a possible interpretation of results of our crossover
measurements that cannot be explained by the classical rotational
crossover mechanism. The occurrence of pulsational crossover has been
related to the large variation of the radial velocity pulsation
amplitude with photospheric depth in roAp stars. The latter implies
that there are significant velocity gradients in the line-forming
region. In this Appendix, we show how the existence of such gradients
along the line of sight in the region of formation of spectral lines
that are used for determination of the magnetic field moments
generates a non-zero crossover in these lines, that is, non-zero values
of the second-order moments of their Stokes $V$ profiles about
their respective centres.

For the sake of simplicity, we assume that a single pulsation
mode is excited and that the rotational velocity of the star is
zero. These restrictions have no effect on the conclusions drawn
here. We also assume that observations are obtained with an
exposure time equal to one or several pulsation period(s) of the star,
so that the observed spectrum is averaged over this pulsation
period. This is appropriate since our purpose here is to demonstrate
the existence of a net, non-zero effect that does not average out over
a stellar pulsation period. 

Then, the second-order moment of the Stokes $V$ profile of a spectral
line about the central wavelength $\lambda_I$ of its observed Stokes
$I$ profile can be expressed as
\begin{eqnarray}
&&\R{2}{V}={\cal N}\,\intxy\int_0^{2\pi}d\psi\nonumber\\
&&\times\int V[x,y,\psi;\llo-\dlp;\Hvec(x,y)]\,(\llo)^2\,d\lambda\,,\nonumber\\
%&&\qquad\qquad \times(\llo)^2\,d\lambda\,,
\label{eq:R2Va}
\end{eqnarray}
where $(x,y)$ are the coordinates of a point on the stellar disk, in
units of the stellar radius; $\Hvec(x,y)$ is the local magnetic
vector at that point; $\dlp$ is the wavelength shift of the centre of
the emergent line in Stokes $I$ at that point caused by pulsation at
pulsation phase $\psi$, and ${\cal N}$ is a normalisation
constant. The $\lambda$ integration extends over the whole width of
the line \citep[for details, see][]{1988A&A...189..179M}. The shape of the local
emergent $V$ profile includes its distortion by the pulsation, which
varies across the stellar surface and with pulsation phase. The
velocity gradients that are present in the line-forming region, as a
result of the dependence of the pulsation amplitude on optical depth,
create departures from anti-symmetry in this profile \citep{1983SoPh...87..221L}. 

Performing the change of variable $\lambda\rightarrow\lambda+\dlp$,
Eq.~(\ref{eq:R2Va}) can be rewritten as
\begin{equation}
\R{2}{V}={\cal
  N}\left[I_0(\lambda_I)+2\,I_1(\lambda_I)+I_2(\lambda_I)\right]\,,
\label{eq:R2Vb}
\end{equation}
with
\begin{eqnarray}
I_0(\lambda_I)&=&\intxy\,\int_0^{2\pi}d\psi\nonumber\\
&\times&\int V[x,y,\psi;\llo;\Hvec(x,y)]\,(\llo)^2\,d\lambda\,,
\label{eq:I0}\\
I_1(\lambda_I)&=&\intxy\,\int_0^{2\pi}\dlp\,d\psi\nonumber\\
&\times&\int V[x,y,\psi;\llo;\Hvec(x,y)]\,(\llo)\,d\lambda\,,
\label{eq:I1}\\
I_2(\lambda_I)&=&\intxy\,\int_0^{2\pi}\Delta\lambda^2_{\rm
  p}(x,y;\psi) \,d\psi\nonumber\\
&\times&\int V[x,y,\psi;\llo;\Hvec(x,y)]\,d\lambda\,.
\label{eq:I2}
\end{eqnarray}

The non-antisymmetric emergent Stokes $V$ profiles can always be
expressed as the sum of a symmetric part and of an antisymmetric part as follows:
\begin{eqnarray}
V[x,y,\psi;\llo;\Hvec(x,y)]&=&V_{\rm  S}[x,y,\psi;\llo;\Hvec(x,y)]\nonumber\\
&+&V_{\rm A}[x,y,\psi;\llo;\Hvec(x,y)]\,,\nonumber\\
\end{eqnarray}
with
\begin{eqnarray}
V_{\rm S}[x,y,\psi;\llo;\Hvec(x,y)]&=&V_{\rm
  S}[x,y,\psi;\lambda_I-\lambda;\Hvec(x,y)]\,,\nonumber\\
\\
V_{\rm A}[x,y,\psi;\llo;\Hvec(x,y)]&=&-V_{\rm A}[x,y,\psi;\lambda_I-\lambda;\Hvec(x,y)]\,.\nonumber\\
\end{eqnarray}
Because of the integration over the wavelengths, only the symmetric
part of the Stokes $V$ profile can make a non-zero contribution to
$I_0(\lambda_I)$ and $I_2(\lambda_I)$. Conversely, this part cancels
out in the wavelength integration of $I_1(\lambda_I)$.

Pulsation phases separated by $\pi$ radians are characterised by
opposite velocity fields, so that 
\begin{equation}
\Delta\lambda_{\rm p}(x,y;\psi+\pi)=-\dlp\,.
\end{equation}
In the absence of velocity gradients, the Stokes $V$ profile is
anti-symmetric: the anti-symmetric part, $V_{\rm A}$, must be
independent of the pulsation phase. Therefore the term containing this
part cancels out in the integration on $\psi$ in $I_1(\lambda_I)$, so
that $I_1(\lambda_I)=0$. 

Thus only the symmetric part of the local Stokes $V$ profiles can make
non-zero contributions to the observable crossover via the terms
$I_0(\lambda_I)$ and $I_2(\lambda_I)$ of Eq.~(\ref{eq:R2Vb}). Accordingly
Eq.~(\ref{eq:R2Va}) reduces to
\begin{eqnarray}
\R{2}{V}&=&{\cal N}\,\intxy\,\int_0^{2\pi}d\psi\nonumber\\
&\times&\int V_{\rm S}[x,y,\psi;\llo;\Hvec(x,y)]\,(\llo)^2\,d\lambda\nonumber\\
&+&{\cal N}\,\intxy\,\int_0^{2\pi}\Delta\lambda^2_{\rm
  p}(x,y;\psi)\,d\psi\nonumber\\
&\times&\int V_{\rm S}[x,y,\psi;\llo;\Hvec(x,y)]\,d\lambda\,.
\end{eqnarray}
The symmetric part of the local $V$ profiles should not be expected to
average out by integration over the pulsation phases, as long as the
variation of the amplitude of pulsation across the photospheric layers
is not very small compared to the thermal Doppler width of the lines
\citep{1996SoPh..164..191L}. This condition is obviously fulfilled in
at least a fraction of the known roAp stars, since we do actually
observe a dependence in their pulsation amplitude on the line
formation depth. Thus the effect described in this Appendix can indeed
account, 
qualitatively, for the occurrence of crossover in pulsating Ap stars,
even if these stars have negligible rotation.  

\bibliographystyle{aa}
\bibliography{apres}
\end{document}